%% file: ms.tex
\documentclass[a4paper,11pt,twoside,fleqn]{book}
\pdfoutput=1
\usepackage{amsmath}
\usepackage{amstext}
\usepackage{amsfonts}
\usepackage{amssymb}
\usepackage{fixmath}
\usepackage[percent]{overpic}
\usepackage{amscd}
\usepackage{mathptmx}
\usepackage{bm}
\usepackage{hyperref}
\usepackage{multirow}

\usepackage{latexsym}
\usepackage[]{units}
\usepackage{graphicx}
\usepackage{color}
\usepackage{subfigure}
\usepackage{sidecap}
\usepackage[DIV14,BCOR=2.5cm,headinclude=true,footinclude=false]{typearea}
\usepackage{color}
\usepackage[english,german]{babel}

\usepackage{fancyhdr}
\include{command}

\begin{document}

\bibliographystyle{plain}   
\selectlanguage{english}
\input{titlepage}
\begin{small}
  \input{abstract}
\end{small}
\selectlanguage{english}
\pagenumbering{roman}
\tableofcontents
\thispagestyle{empty}
\cleardoublepage
\pagenumbering{arabic}
% \include{chapters/empty}
% \newpage\ \newpage
\setcounter{page}{1}
\include{chapters/introduction}
\include{chapters/methods}

\include{chapters/theory_section}
\include{chapters/bd_section}

\include{chapters/md_section}
\include{chapters/summary_outlook}
\include{chapters/biblio}

%\include{chapters/lysozyme}

 %\include{chapters/semi_analytical_polyakov}
%\include{chapters/ewald}
%\include{chapters/logarithmic_action_untilde}
%\include{chapters/summary}
  \appendix
  \input{chapters/appendixA}
  \input{chapters/appendixB}
  \input{chapters/appendixC}
% \input{appendixC} 
% \bibliography{biblio}
% \appendix
% \include{chapters/appendixA}

\cleardoublepage
\pagestyle{empty}
\input{acknowledgements}
%\selectlanguage{german}
%\cleardoublepage
%\input{selbstaendigkeitserklaerung}
%\selectlanguage{english}
\end{document}

%% file: command.tex
\newcommand{\fig}[1]{Fig.~#1}
\newcommand{\tab}[1]{Tab.~#1}
\newcommand{\eq}[1]{Eq.~(#1)}
\newcommand{\ch}[1]{Ch.~#1}
\renewcommand{\sec}[1]{Sec.~#1}
\newcommand{\app}[1]{App.~#1}
\renewcommand{\l}{\left}
\renewcommand{\r}{\right}
\renewcommand{\d}{\text{d}}
\newcommand{\Vext}{V_{\text{ext}}}
\newcommand{\Fext}{F_{\text{ext}}}
\newcommand{\Fexc}{\mathcal F_{\text{exc}}}
\newcommand{\muexc}{\mu_{\text{exc}}}
\newcommand{\bFexc}{\overline{\mathcal F}_{\text{exc}}}
\newcommand{\gammalv}{\gamma_{\ell\text{v}}}
\newcommand{\lambdaD}{\lambda_{\text{D}}}
\newcommand{\kB}{k_{\text{B}}}
\newcommand{\DT}{D_{\text{T}}}
\newcommand{\ST}{S_{\text{T}}}

\newcommand{\const}{\text{const}}
\newcommand{\NN}{\nonumber}

\makeatletter
\newcommand{\vast}{\bBigg@{4}}
\newcommand{\Vast}{\bBigg@{5}}
\makeatother

 \newcommand{\p}[2]{\frac{\partial #1}{\partial #2}} %erzeugt Darstellung einer partiellen Ableitung aus den nächsten zwei parametern
 \newcommand{\deriv}[2]{\frac{\d #1}{\d #2}} %erzeugt Darstellung einer partiellen Ableitung aus den nächsten zwei parametern
 \newcommand{\ptwo}[2]{\frac{\partial^2 #1}{\partial^2 #2}} %Darstellung von 
\newcommand{\rot}{\,\text{rot}\,}

\renewcommand{\v}[1]{\boldsymbol{#1}}

%% file: titlepage.tex
\begin{titlepage}
  \begin{center}
    
    \begin{figure}[ht]
      \begin{center}
        \includegraphics[width=\linewidth]{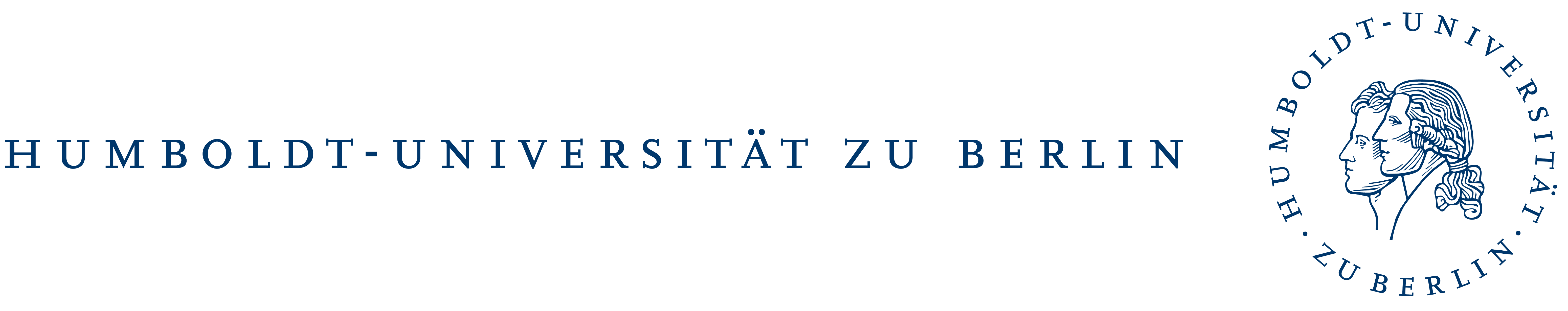}
      \end{center}
    \end{figure}

    %\vspace{2.0cm}
    \begin{Large}
      \begin{center}
% 	  \vspace{10mm}
% 	  \includegraphics[width=15cm]{HUB-Emblem.pdf} \\
	  \vspace{3mm}
		\noindent\hrulefill \\
		\Large{\rmfamily{
			\textsc{Mathematisch-Naturwissenschaftliche Fakult\"at I\\Institut f\"ur Physik\\}
			}}
		\vspace{-3mm}
		\noindent\hrulefill \\
      \end{center}
      \vspace{0.8cm}

      \begin{center}
        \Large{\textbf{
         Thermophoresis in Liquids and its Connection to Equilibrium Quantities
        }}\\
      \end{center}
     \end{Large}
    
    \begin{large}
      \vspace{0.7cm}
      \begin{center}
	\vspace{0.5cm} \textsc{Masterarbeit}\\ 
	\vspace{0.5cm} zur Erlangung des akademischen Grades \\ 
	M. Sc. im Fach Physik
      \end{center}
    \end{large}

    \vspace{1.0cm}
    
    \begin{large}
      \begin{center}
	eingereicht von \\
        Benjamin F. Maier \\
        \vspace{0.5cm}
         %geboren am\\
         %24. Dezember 1988 in Berlin
      \end{center}
    \end{large}
    
    \vspace{1.0cm}

    \begin{large}
      \begin{center}
        \begin{tabular}{rl}
	  Gutachter:\qquad
          1. & Prof. Dr. J. Dzubiella \\
          2. & Dr. S. Angioletti-Uberti \\
             & \\ 
          verteidigt am:     & 23. Juni 2014
        \end{tabular}
      \end{center}
    \end{large}
    
  \end{center}

\end{titlepage}

\newpage
\thispagestyle{empty}
\quad
\newpage
\setcounter{page}{1}

%% file: abstract.tex
\thispagestyle{empty}
\vspace{2.0cm}
\begin{center}
  {\large Abstract}
\end{center}
\noindent
%Thermophoresis 
  Thermophoresis is the process of particles moving along a temperature gradient in thermodynamic systems. Even though it has been studied for over 150 years, there is no complete theoretical description of thermophoresis in liquids and no complete analytic formula for the Soret coefficient, the quantifiying measure of the Soret equilibrium. However, recent studies connect its nature to equilibrium properties of the system, namely the excess enthalpy and the excess entropy, while there is still a debate over which of both describes the Soret coefficient more accurately and if it can even be represented using those quantities.
  
  In this work I present a theoretical derivation for both of the cases based on density analysis by means of Brownian motion and dynamical density functional theory, where I assume local equilibrium and the validity of the Einstein relation for the diffusion coefficient in temperature gradients. The interpretation of the Brownian stochastic differential equation (SDE) as an It\^o or Stratonovich SDE is shown to have an influence on the outcome of the density. I provide an indication that a Soret coefficient proportional to the excess enthalpy is connected to a system in a thermal gradient following its equation of state. Furthermore I derive that for the systems in thermal gradients the Boltzmann distribution law for external potentials does not hold but has to be replaced by a more general quantity.
  
  The theoretical predictions are consequently tested by means of BD simulations of systems in thermal gradients, where a one-dimensional (1D) system of Gaussian particles and a 1D system of hard rods is shown to follow their equation of state. A binary mixture of an ideal gas and a Gaussian particle produce simulations where the Soret coefficient of the Gaussian particle is proportional to its equilibrium solvation enthalpy. 1D and 2D ensembles of an ideal gas show the behavior of the generalized Boltzmann distribution. 
  
  I furthermore attempt to verify the theoretic derivations by means of MD simulations, investigating single noble gas solutes in SPC/E water. An MD setup is introduced which generates the desired temperature profiles. The Soret equilibrium SPC/E water density is shown to follow its equation of state. Soret coefficients and Soret equilibrium solute densities measured by this approach do however not entail a coherent agreement with any of the theoretical predictions. The measured Soret coefficient is shown to be consistently positive in the temperature region between $300\,\unit{K}$ and $400\,\unit{K}$, meaning that the solutes generally preferred the cold regions over the hot regions, while the enthalpy prediction of the Soret coefficient, which was shown to be valid in the BD simulations, predicted a sign change from positive to negative in that temperature regime. An answer to the question of the interpretation of the Brownian SDE as an It\^o or Stratonovich SDE was not found, either.

% \selectlanguage{german}
% \vspace{2.0cm}
% \begin{center}
%   {\large Zusammenfassung}
% \end{center}
% \noindent
% Ganz schoen abstrakt.
\newpage
\thispagestyle{empty}
\quad
\newpage
\setcounter{page}{1}

%% file: chapters/introduction.tex
\chapter{Introduction}

\section{Thermophoresis}
\enlargethispage{-1cm}
\textit{Thermophoresis} or \textit{thermodiffusion} describes the process of a system of particles travelling along a temperature gradient, often it is explicitly referred to as the process of the separation of chemicals in suspensions due to the presence of thermal gradients. Even though this process has been investigated since the very beginnings of thermodynamical studies, with descriptions of the effect by C. Ludwig in 1856 \cite{Ludwig:1856} and C. Soret in 1879 \cite{Soret:1879}, there is still a lively debate about its underlying principles. One of the key questions regarding thermophoresis is, given a certain chemical or gaseous system coupled to two heat baths of different temperatures, can one predict the motion of the particles both qualitatively and quantitatively. Formulated in a different way, is it possible to find an explicit expression for the so-called \textit{Soret coefficient} $\ST$, which is a measure of the thermodiffusive equilibrium given by the 
phenomenological equation
\begin{align}
 0 = -\nabla\rho-\rho\ST\nabla T,
\end{align}
where $\rho$ is the density and $\nabla T$ is the temperature gradient.

\begin{figure}[t!]
 \centering
    \includegraphics[width=\textwidth]{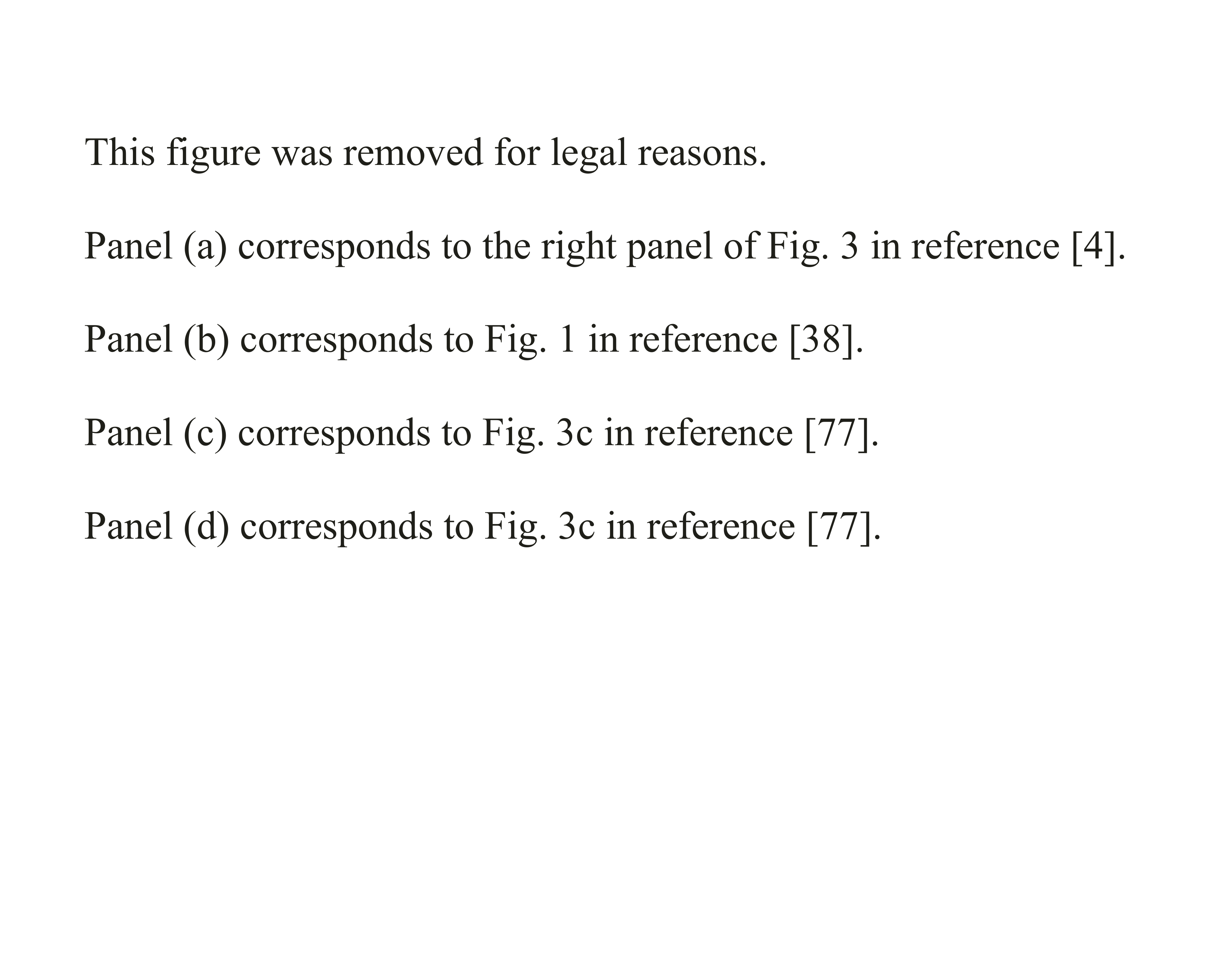}
 \caption{Temperature dependence of the Soret coefficient $\ST$ for various solutes in aqueous solutions. \textbf{(a)} Temperature dependence of $\ST$ for polystyrene (PS) beads and different sized DNA, taken from \cite{Braun:2006}. \textbf{(b)} Temperature dependence of $\ST$ for  various biological macromolecules. The solid curves are fits using the empirical formula \eq{\ref{eq:piazza_fit}}. Taken from \cite{Piazza:2008} with the measurements performed in \cite{Piazza:2006}. \textbf{(c)} Non-ionic part of the DNA Soret coefficient fitted with \eq{\ref{eq:piazza_fit}}. The DNA length $L\propto N$ dependence of the non-ionic Soret coefficient seems to be linear $\ST\propto L\propto N$ (inset). \textbf{(d)} A model from Braun \cite{Braun:2014} (solid lines) seems to explain the measurements of the total $\ST$ rather well, if one ignores the fact that the non-ionic part depicted in (c) is an empirical fit. (c) and (d) were taken from \cite{Braun:2014}. }
 \label{fig:ST_T_dependence}
\end{figure}

While a qualitative description for gaseous systems is relatively simple to achieve by considering a net momentum acting on a single particle from the thermal difference, driving the particle to the cold, this picture is not sufficient anymore for liquids due to the strong interactions and structural changes in fluid systems exposed to a thermal gradient.  Remarkably, studying colloidal suspensions, it has been found that not only does the Soret coefficient vary for different mean temperatures (see \fig{\ref{fig:ST_T_dependence}}), sizes of the colloids (see \fig{\ref{fig:ST_R_dependence}}), polarity of the colloids and salt concentration of the solvent (see \fig{\ref{fig:ST_R_dependence}c}), often it even switches its sign and the direction of the particles' motion is inverted \cite{Piazza:2008}. Hence, the behavior of liquid systems is explicitly of interest.

During the last two decades with the improvement of experimental techniques, thermophoresis of colloidal suspensions became an object of increasing interest, as one can specifically use the effect and the varying Soret coefficient as a tool to separate suspensions and to migrate solutes in a solvent, which can be a powerful enhancement to electrophoresis, the state-of-the-art tool for that purpose. Using infrared laser heating, thermodiffusion has been investigated for DNA strains, polystyrene spheres, polymers, proteins and biological macromolecules  \cite{Braun:2004,Braun:2006,Piazza:2003,Piazza:2008}.
\footnote{Note that there is a variety of studies involving the thermal behavior of the different kinds of colloids and the here cited papers only treat some of the mentioned -- a great overview of the studies is given by Piazza in \cite{Piazza:2008}.}

Even though thermophoresis is a non-equilibrium effect, the studies have shown that the driving forces can often be described using full equilibrium quantities by assuming \textit{local equilibrium}, where the assumption is that the length scale of the temperature gradient is significantly larger than all solvent-solute interactions. Explicitly the free energy of solvation and its thermodynamic relatives solvation enthalpy \cite{Wuerger:2013,Wuerger:2014} and solvation entropy \cite{Braun:2006,Dhont:2007} seem to play a major role. Nevertheless, there are positions questioning the local equilibrium view as per se unreasonable and doubt their applicability, especially for large solutes where hydronamics become more and more important \cite{Wuerger:2013,Piazza:2008}.

In this work I want to investigate the connection of those equilibrium quantities to the Soret coefficient and thus the underlying principles of thermophoresis in liquid systems. 

To this end we will proceed as follows. The remaining sections of this
chapter will provide an overview of the different approaches to
calculate and measure the Soret coefficient, the current state of the
research and a short introduction on hydrophobicity and interfacial
effects in microscopic systems, which is suspected to play a role in
thermophoresis of solutes in aqueous systems
\cite{Braun:2006,Dhont:2007}. \ch{\ref{ch:methods}} contains an overview
of the theoretical and computational methods and quantities needed to
pursue the investigations described in \ch{\ref{ch:results}}, which
contains the actual results of my studies. In \sec{\ref{sec:theory}} I
show the theoretical derivation of a transport equation in temperature
gradients by means of Brownian motion and the methods of dynamical
density functional theory (DDFT) and discuss the Application of the classical fluctuation-dissipation theorem (FDT) in temperature gradients briefly. We will see that the equilibrium property of the Soret coefficient is just an approximation in this context and that one can introduce two formulae which connect it to the enthalpy or the entropy, respectively. I proceed to test the findings from \sec{\ref{sec:theory}} in \sec{\ref{sec:bdsim}} by means of Brownian dynamics (BD) simulations. We will see that the simulation results indicate a strong connection between the Soret coefficient and the local enthalpy. Subsequently, in \sec{\ref{sec:mdsim_section}}, I will test my hypotheses by means of molecular dynamics (MD) simulations, where I try to predict the solute density in a thermophoretic simulation of noble gases in water. However, no quantitatively significant results will be found. \ch{\ref{ch:summary}} sums up the results of this work.

\section{Soret Coefficient}
\label{sec:soret_coefficient}
\subsection{Definition}
Imagine a thermodynamic system which is coupled to two heat baths at its boundaries, one ``hot'' bath at temperature $T_h$ and one ``cold'' bath at temperature $T_c$ at two boundaries. The particles may not escape into the heat baths, implying reflective boundary conditions. For simplicity, we imagine the temperature profile to be one-dimensional and in the $x$-direction. The process of thermophoresis is then explicitly described by the phenomenological equation for a particle (or mass) flux
\begin{align}
 \label{eq:soret_equation}
 \v j = -D\nabla\rho-\rho\DT\nabla T,
\end{align}
based on Fick's law $\v j= -D\nabla\rho$ with an additional drift term due to the temperature gradient $\nabla T$. Here, $D$ is the diffusion coefficient, $\DT$ is the thermal diffusion coefficient and $\rho$ is the particle (or mass) density.
%Often, especially for suspensions, $\rho$ is replaced by a concentration $c$ or a molar density $n$, however, we will stick with $\rho$ throughout this work.
The Soret coefficient emerges as soon as we look at stationary sytems $\v j=\const$, explicitly at the ``equilibrium'' case $\v j=0$, yielding
\begin{align}
 \label{eq:soret_equilibrium}
 \nabla\rho &= -\rho\frac{\DT}{D}\nabla T.
\end{align}
Here, using the term ``equilibrium'' refers to the vanishing particle flux  $\v j=0$, which does not necessarily mean that the system is at a true thermodynamic equilibrium. Since the two heat baths hold up a constant temperature profile there most likely still acts a heat flux $\v j_Q$ from the hot bath to the cold bath, casting the local thermodynamic equilibrium assumption in general invalid. We will therefore call this special equilibrium ``Soret equilibrium'' in the following \cite{Wuerger:2013}. The Soret coefficient is now the proportionality factor in \eq{\ref{eq:soret_equilibrium}}
\begin{align}
 \label{eq:soret_coefficient}
 \ST = \frac{\DT}{D}.
\end{align}
This coefficient manipulates the shape of the Soret equilibrium density $\rho$, which we can find from \eq{\ref{eq:soret_equilibrium}} as
\begin{align}
 \frac{\nabla\rho}{\rho} &= -\ST\nabla T\qquad\Rightarrow\qquad \nabla \ln\rho = -\nabla\int\limits^TS_{\tilde T}\d\tilde T\\
        \rho &= \rho_0\exp\l(-\int\limits^TS_{\tilde T}\d\tilde T\r).
\end{align}
In experiments the temperature difference between the thermostats $\Delta T$ is rather small compared to the Soret coefficient, $S_T\Delta T\ll1$, s.t. the Soret coefficient $S_{\overline{\text T}}$ of the mean temperature $\overline T=(T_c+T_h)/2$ is taken to be approximately constant \cite{Braun:2004}, which allows the approximations
\begin{align}
 \frac{\rho(\v r)}{\rho_0} &\simeq \exp\big(-S_{\overline{\text T}} (T(\v r)-\overline T)\big)\simeq 1-S_{\overline{\text T}} (T(\v r)-\overline T)
\end{align}
The density or concentration, respectively, is then measured, for example by means of beam deflection methods \cite{Piazza:2003,Piazza:2008} or by fluorescent activity \cite{Braun:2004} and can be fitted accordingly.

The method of choice in this work will be to find a Soret equilibrium density $\rho(\v r)$ and calculate the Soret coefficient exactly as
\begin{align}
 \ST = -\frac{\nabla\rho}{\rho \nabla T}.
\end{align}
Still, the question of the nature of the Soret coefficient remains open. 

\subsection{Theoretical Approaches}
Neglecting the influence of hydrodynamics, there exist two major approaches to quantify the Soret coefficient, both derived in the context of colloidal suspensions. The first connects the Soret coefficient of a colloidal particle to its enthalpy of solvation (or excess enthalpy), which has recently been derived by W\"urger \cite{Wuerger:2013,Wuerger:2014}. A second approach identifies the Soret coefficient as proportional to the colloid's entropy of solvation, a view which has already been proposed in the 1920s by Eastman \cite{Wuerger:2013} and is currently promoted by Braun and Dhont \cite{Dhont:2007,Braun:2006}. In the following I give a brief summary of both approaches.

Investigating the mechanisms behind the Seebeck-Peltier effect of electrothermophoresis, Onsager found his famous ``reciprocal relations of irreversible processes'' \cite{Onsager:1931}. Following his notion, a quantity called entropy production can be defined which is responsible for the relaxation of the system to equilibrium. For any thermodynamic observable $X$ it is proportional to the temporal derivative of $X$ times the thermodynamic force of $X$, which is the derivative of the entropy with respect to $X$. For continuous systems, one can introduce the dependency of both on space \cite{Katchalsky:1975}. Onsager's ansatz is then that the spatial flux of $X$ is linearly dependent on the force itself with a proportionality factor $L>0$, introducing his linear phenomenological equations. In our case, we are interested in the fluxes of heat and particles. The driving force for the dissipation of heat can be found to be $\nabla(T^{-1})$ \cite{Katchalsky:1975}, while that of the motion of particles can be found to be $\nabla(\mu/T)$ with $\mu=\mu(T,\rho)$ being the particles' chemical potential \cite{Katchalsky:1975,Wuerger:2013}. For dilute systems one can assume the chemical potential to be equal to its excess Gibbs free energy and an ideal gas contribution
\begin{align}
 \label{eq:mu_dilute}
 \mu(T,\rho) = \Delta G(T) + \kB T \ln\frac{\rho}{\rho_0}
\end{align}
where $\kB$ is Boltzmann's constant. W\"urger now assumes that the Soret equilibrium can be connected to a vanishing thermodynamic force  $\nabla(\mu/T)=0$ \cite{Wuerger:2013}. This yields
\begin{align}
 0 = \nabla\l(\frac{\Delta G}{T}\r) + \kB \nabla \ln\frac{\rho}{\rho_0} = +\kB\frac{\nabla\rho}{\rho} - \frac{\Delta H}{T^2}\nabla T
\end{align}
where we applied the thermodynamic identities
\begin{align}
 \Delta H &= \Delta G + T\Delta S\\
 \p{\Delta G}{T} &= -\Delta S
\end{align}
with $\Delta H$ being the excess enthalpy and $\Delta S$ being the excess entropy. Comparing the result to \eq{\ref{eq:soret_equilibrium}} we find
\begin{align}
 \label{eq:soret_wuerger}
 \ST = -\frac{\Delta H}{\kB T^2}.
\end{align}

The second route takes the total drift velocity in \eq{\ref{eq:soret_equation}}
\begin{align}
\v v = -D \frac{\nabla\rho}{\rho} - \DT\nabla T 
\end{align}
and suspects its causing force $\v F$ to be in balance with the force of friction given by Stokes' law $\v F^{\text{fric}}=-\gamma\v v=-\v F$, such that
\begin{align}
 \v F = \v v\gamma = -D\gamma\nabla\ln\rho-\DT\gamma\nabla T
\end{align}
with $\gamma$ being the friction coefficient which is given by the hydrodynamics of the problem. Using Einstein's relation for the diffusion coefficient $D = \kB T/\gamma$ we find
\begin{align}
\label{eq:force_from_drift}
 \v F = -\kB T\nabla\ln\rho - \frac{\DT\kB T}{D}\nabla T = -\kB T\nabla\ln\rho - \ST \kB T\nabla T.
\end{align}
\begin{figure}[t!]
 \centering
    \includegraphics[width=\textwidth]{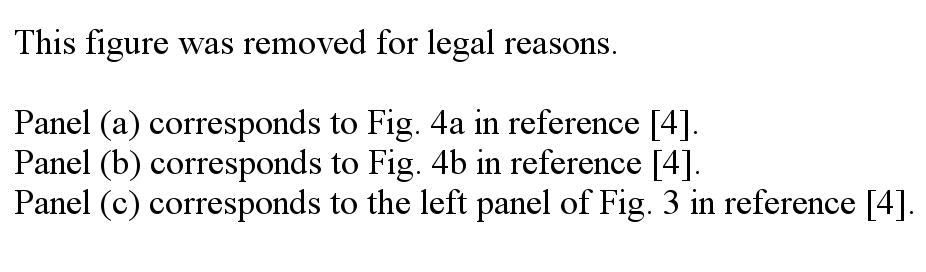}
 \caption{The dependence of the Soret coefficient $\ST$ on the size of the solutes at equal mean temperature. All figures were taken from \cite{Braun:2006}. \textbf{(a)} Polystyrene beads. The Soret coefficient scales with the surface area of the beads. \textbf{(b)} DNA of increasing length $L$ (increasing number of base pairs $N$). $\ST$ scales with the square root of $L$, $N$, respectively. \textbf{(c)} The dependence of $\ST$ on the Debye-H\"uckel length $\lambdaD$ (screening length of the electrostatic potential) for PS beads of diameter $d=1.1\mu\unit{m}$ at different mean temperatures. }
 \label{fig:ST_R_dependence}
\end{figure}
Braun and Dhont \cite{Dhont:2007} assume that the energy landscape which accounts for the force on the solute is given partially as the free energy of solvation, s.t.~$\v F_G = -\nabla\Delta G=\Delta S\nabla T$. Another contributing term however, is the force needed to overcome the osmotic ideal gas pressure $\Pi = \rho\kB T$ per particle
\begin{align}
 \v F_\Pi = -\frac{\nabla\Pi}{\rho} = -\frac \kB \rho\l(T\nabla\rho + \rho\nabla T \r) = -\kB\l(T\nabla\ln\rho + \nabla T\r)
\end{align}
Adding the forces and equating with \eq{\ref{eq:force_from_drift}} yields 
\begin{align}
 &\Delta S\nabla T - \kB T\nabla\ln\rho -\kB\nabla T = -\kB T\nabla\ln\rho - \ST\kB T\nabla T \\
 &\ST = \frac1T - \beta\Delta S
 \label{eq:soret_braun}
\end{align}
with $\beta=(\kB T)^{-1}$ being the inverse thermal energy. While Braun and Dhont adapted their notion by replacing the free energy of solvation by a more general ``reversible work'', there is still a conceptual difference to the Soret coefficient derived by W\"urger, as that reversible work is still defined as the work needed to remove a solute particle from an isothermal system of temperature $T$ and insert it into an isothermal system of temperature $T+\d T$, thus making it effectively a free energy of solvation. Following \cite{Wuerger:2013}, a similar argument has been proposed by Eastman. The Soret coefficient can be identified to be proportional to a quantity called ``heat of transfer'' $Q^*$ \cite{Katchalsky:1975} which in turn can be expressed in terms of an entropy of transfer $Q^*=S^*T$. Eastman identifies this entropy of transfer as the local entropy of solvation and thus reaches \eq{\ref{eq:soret_braun}}, dismissing the additional term $1/T$. A third route to find the entropy of solvation as the driving force of thermophoresis is to identify the thermodynamical force as $(\nabla\mu)/T$ and postulate that the Soret equilibrium corresponds to a mechanical equilibrium $\nabla\mu=0$ \cite{Wuerger:2013}. Connecting this condition with \eq{\ref{eq:mu_dilute}} and dismissing a term $\ln\rho\,\nabla T$, one finds again the entropy proportional to the Soret coefficient.

Braun always explicitly includes electrostatic considerations in his approaches. For instance for the measurements of PS beads in \fig{\ref{fig:ST_R_dependence}a} he neglects non-ionic contributions to the Soret coefficient and states that an effective charge of the beads scales with the spherical surface area, which is why the Soret coefficient scales with surface area as well \cite{Braun:2006}. Explaining the connection between $\ST$ and the DNA size dependence $\ST\propto\sqrt{L}\propto\sqrt{N}$ with $L$ being the length and $N$ the number of base pairs, respectively, he again neglects the non-ionic influence and states that the size dependence of the Soret coefficient can be estimated by considering an effective charge density inversely proportional to the surface area of a sphere from an hydrodynamic model where the radius is $R = L^{3/2}$, multiplied with the surface area $4\pi L^2$ of a sphere of radius $L$, yielding $\ST\propto\sqrt{L}$ \cite{Braun:2006}. However, W\"urger argued that the hydration free energy of DNA strains is an extensive quantity, s.t. it is given as $\Delta G = N \Delta g$ with $\Delta g$ being the hydration free energy per base pair. Thus, even with electrostatic considerations the Soret coefficient should be proportional to the length $L$ if the local equilibrium assumption holds \cite{Wuerger:2013}.

In a recent publication \cite{Braun:2014}, Braun and co-workers propose a modified model where charged solutes induce an electric field in the solvent and locally act as spherical capacitors, casting the whole process to a Seebeck-Peltier effect of electrothermophoresis. \footnote{In general, due to the Seebeck-Peltier effect, considering elecrostatic effects seems like a plausible approach to attack the problem of thermophoresis, for instance it has been found in MD simulations that thermal gradients induce an electric field in aqueous solutions by manipulating the orientation of water dipoles \cite{Armstrong:2013}}  Indeed, the predictions of Braun's model seem to accurately predict the Soret coefficient for DNA and RNA of different sizes, however the non-ionic part of the Soret coefficient is theoretically untreated and only fitted with an empirical formula by Piazza \cite{Piazza:2003}
\begin{align}
 \label{eq:piazza_fit}
 \ST = \ST^\infty \l[1-\exp\l(\frac{T^*-T}{T_0}\r)\r].
\end{align}
Remarkably though the fit coefficient $\ST^\infty$ scales linearly with $N$, indicating that the non-ionic part may be connected to the solvation free energy in some way.

Yet, W\"urger argues \cite{Wuerger:2013} that in general, the applicability of equilibrium quantities (in his case the enthalpy) in thermophoretic contexts is limited to small solutes, where the hydrodynamics of normal diffusion equals the hydrodynamics of thermal diffusion, thus cancelling each other. For larger solutes, the hydrodynamics of both cases are not equal, generating different velocity fields around the solute, thus dominating the equilibrium contribution. Still, the Soret equilibrium is reached when all contributing particle fluxes cancel each other, yet the different velocity fields lead to a constant energy dissipation and thus steadily increase the system's entropy, casting the whole situation a non-equilibrium one.

\section{The Solvation Free Energy of Hydrophobic Solutes}
\label{sec:solute_hydrophobicity}
Even though W\"urger proposed a strong argument for the non-applicability of equilibrium quantities for larger solutes, we want to investigate the hypothetical influence of those, as in general connecting the Soret coefficient to solvation quantities seems plausible since both share similar properties. Soret coefficient, solvation entropy and solvation enthalpy vary with temperature, solvent density, the solute size and its electrostatic properties and can switch signs with varying temperature. A significant part of the experimental research on thermophoresis is done for colloidal suspensions in aqueous solutions, hence it is worth
to recall the mechanisms quantifying the hydration free energy here, mainly following \cite{Chandler:2005}.
\begin{figure}[t!]
 \centering
    \includegraphics[width=8cm]{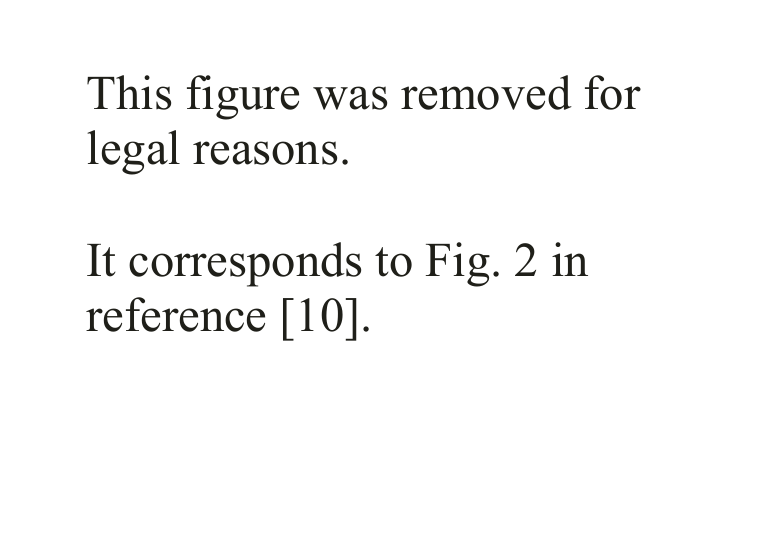}
 \caption{The change of hydration free energy $\Delta G$ per solute surface area $A=4\pi R^2$ for growing solute sizes (circles) as results from ``detailed microscopic calculations'' \cite{Chandler:2005}. For small solutes it scales linearly with the volume, showing entropic behavior, whereas it reaches a constant value equal to the water liquid-vapor surface tension $\gammalv$. The figure was taken from \cite{Chandler:2005}. }
 \label{fig:hydrophobic_crossover}
\end{figure}

Water is an interesting solvent, as the hydrogen bond network connecting single water molecules depicts a very strong interaction compared to other bonding forces in liquids.  Disturbing or breaking hydrogen bonds by the solvation of hydrophobic molecules and thus forcing the reordering of the hydrogen network comes with a price to pay in the hydration free energy $\Delta G =\Delta H - T\Delta S$. In general, the composition of the hydration free energy is governed by two effects at different solute sizes $R$ in relation to a crossover length $\ell$. Small solutes $R\lesssim\ell$ are not large enough to break hydrogen bonds. Hence the free energy is mainly composed of an entropic part which comes from the change of available space for the water and scales with the solute volume. Within the crossover region $R\simeq\ell$ few hydrogen bonds are broken but the network does not reorder to build an interface between water and solute, as the curvature of the solute still prevents that formation. In this case, the free energy is a mixture of enthalpic contributions due to the change of the total interaction energy by breaking some hydrogen bonds and entropic changes by the non-accessibility of the cavity volume. For large solutes $R\gtrsim\ell$, the curvature reduces and the water molecules can reorder, building a water interface around the solutes, thus shaping the whole situation an enthalpic one where the free energy is given as the interaction energy between solute and interface. This can be quantified using a macroscopic picture, where the interaction energy is given as the product of the interfacial surface area $A$ and the liquid-vapor surface tension of water $\gammalv$. Thus, for large spherical solutes of radius $R$, the hydration free energy at standard conditions is given as
\begin{align}
 \Delta G \simeq \Delta H \simeq 4\pi R^2\gammalv.
\end{align}
The behavior of $\Delta G/(4\pi R^2)$ for the crossover from small solutes to large solutes is depicted in \fig{\ref{fig:hydrophobic_crossover}} following the expectation of reaching $\gammalv$ for $R>\ell\simeq0.8\unit{nm}$. Remarkably, using this result in \eq{\ref{eq:soret_wuerger}} and comparing the outcome to the Soret coefficient in \fig{\ref{fig:ST_R_dependence}a}, we find the same scaling behavior but a different sign.

A sophisticated model for the calculation of solvation free energies by means of interfaces, especially for non-spherical models, is the class of variational implicit solvent models (VISM) \cite{Dzubiella:2006,Dzubiella:2006b,Dzubiella:2011}. Within this framework, the free energy of solvation is expressed as a functional of the interface, where the real interface is defined as the geometrical structure which minimizes the free energy. Given a roughly spherical solute of radius $R$ we find \cite{Dzubiella:2006}
\begin{align}
\label{eq:vism_free_energy}
 \Delta G(R) = PV + 4\pi R^2\gammalv\l(1-2\delta H\r) + \int\limits_R^\infty\d r\ 4\pi r^2\rho_{\text{solvent}} V_\text{LJ}(r) 
                  + \frac{Z^2e^2}{8\pi\epsilon_0 R}\l(\frac{1}{\epsilon_\ell}-\frac1{\epsilon_{\text v}}\r)
\end{align}
with the average mean curvature $\bar H = \int\d A\ \mathrm{H}/A$, where the integration is done over the interface's surface $A$ and the integrand is the mean curvature $\mathrm H$, a generally non-constant property of the interface's geometry which is explained in detail in \cite{Barrat:2003}. For a perfect sphere, the average mean curvature is $\bar H = 1/R$. Other quantities in this equation are the solute volume $V$, the Lennard-Jones interaction potential $V_{\text{LJ}}$, the liquid and vapor relative permittivities of solvent and gas phase, ${\epsilon_\ell}$ and $\epsilon_{\text v}$, and the Tolman length $\delta$ which is a parameter for the first order curvature correction of the interfacial tension \cite{Dzubiella:2006}. All the water properties $\gammalv$, $\rho$, $\delta$ and $\epsilon_\ell$ are temperature dependent and will be modeled in \sec{\ref{sec:thermodynamics}} and \app{\ref{app:tolman_length}}. Note that the molecule's net charge $Z$ can be temperature dependent for aqueous solutions too, e.g. for lysozyme \cite{Kuehner:1999}.

%We will use this model and the theory of hydrophic solvation to check basic assumptions by comparing their results with experimental or simulation values.

%\begin{\figure}

%% file: chapters/methods.tex
\chapter{Theoretical, Simulation and Free Energy Calculation Methods}
\label{ch:methods}
%\section{}

\section{Thermodynamic Quantities and Statistical Mechanics}
\label{sec:thermodynamics} 
Throughout this work we will use various thermodynamic quantities and relations as well as quantities from statistical mechanics, which are listed here.

\subparagraph{Statistical Operators, Expectation Values / Ensemble Averages} We will use the standard notation of angular brackets for an ensemble average $X =\l\langle\hat X\r\rangle$, the expectation value of a statistical operator $\hat X$. The notation stands for an ensemble average. Important statistical operators are the one-particle density operator
\begin{align}
\label{eq:oneparticle_dens_operator}
 \hat\rho_i(\v r) = \delta(\v r-\v r_i)
\end{align}
and the ensemble density operator
\begin{align}
\label{eq:ensemble_dens_operator}
 \hat\rho(\v r) = \sum_{i=1}^N \hat\rho_i(\v r) = \sum_{i=1}^N\delta(\v r-\v r_i).
\end{align}

\subparagraph{Interaction Potentials}
\begin{figure}[t!]
 \centering
    \includegraphics[width=12cm]{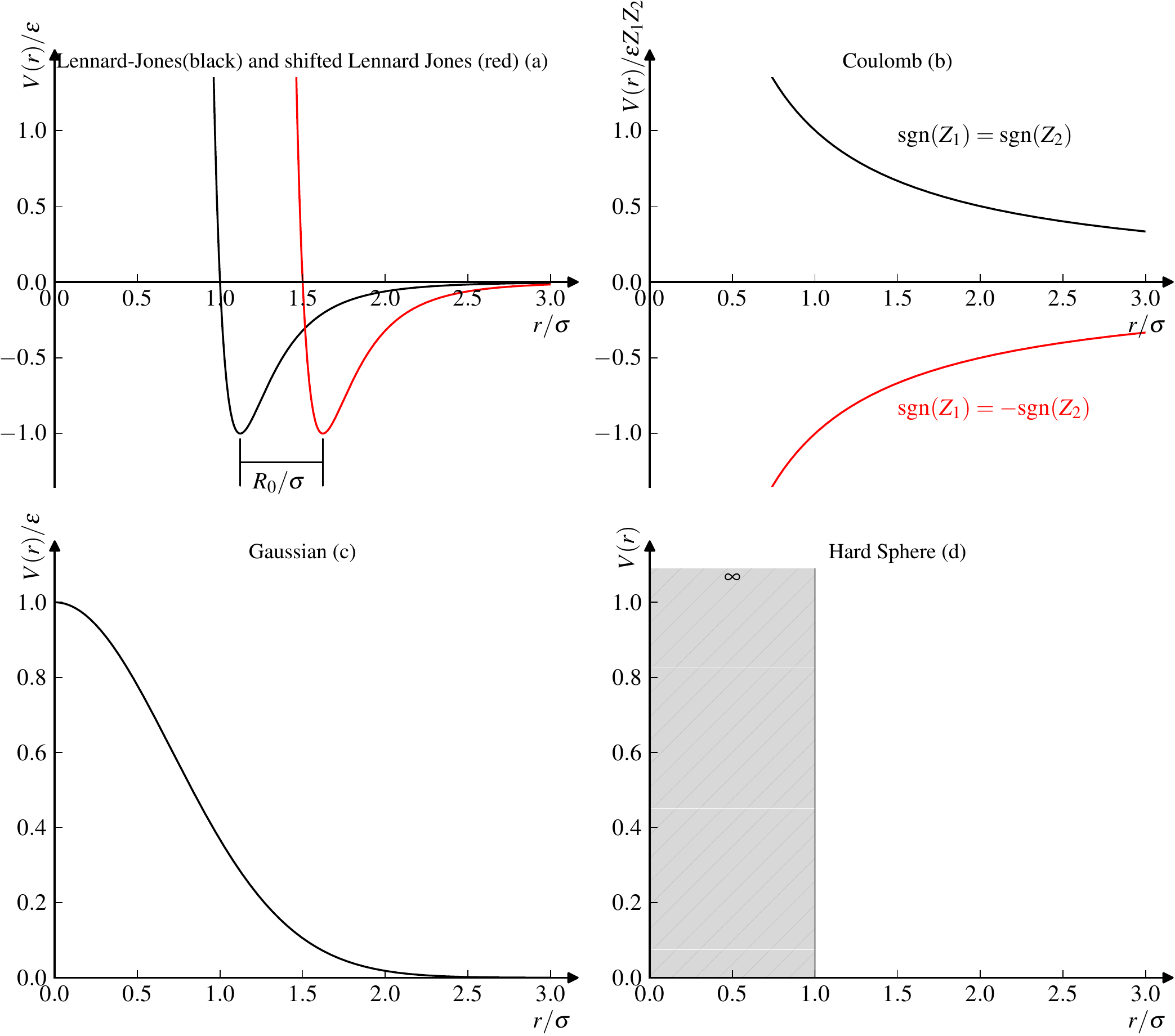}
 \caption{The pair interaction potentials used in this work. \textbf{(a)} The 12-6 Lennard-Jones potential and the shifted potential, used to create larger solutes without changing the solvation shell's dimension too much. \textbf{(b)} The electrostatic Coulomb potential. The sign of the potential is the product of the sign of the relative charges $Z_1$ and $Z_2$. \textbf{(c)} Gaussian interaction. \textbf{(d)} The hard sphere potential. }
 \label{fig:interaction_potentials}
\end{figure}
We will limit the discussion of interactions between particles or atoms in distance $r$ to radially symmetric pair interaction potentials $V(r)$. For molecular dynamics (MD) simulations we will use the normal 12-6 Lennard-Jones (LJ) potential and an LJ potential shifted in $r$-direction by a constant $R_0$, as well as the electrostatic Coulomb potential.  For Brownian dynamics (BD) simulations we will use Gaussian and hard sphere / hard rod interactions. The potentials are discussed in a more detailed way in \cite{Barrat:2003}. All potentials can be seen in \fig{\ref{fig:interaction_potentials}}.

\textit{LJ Potential}\ The Lennard-Jones potential is defined as
\begin{align}
 V_{LJ}(r) = 4\epsilon\l[\l(\frac\sigma r\r)^{12}-\l(\frac\sigma r\r)^6\r].
\end{align}
It is a semi-empirical potential which is composed of two parts. The negative (attractive) part models van-der-Waals dispersion, whereas the positive part accounts for repulsion due to electron orbital overlap. Typical properties of the potential are that it changes its sign at $\sigma$ and the minimum at $2^{1/6}\sigma$ has an energy value of $\epsilon$. Whenever it is necessary to calculate the pair potential of two different species $i$ and $j$, with parameters $\epsilon_i,\epsilon_j$ and $\sigma_i,\sigma_j$, we will use the Lorentz-Berthelot mixing rules which read
\begin{subequations}
\label{eq:mixing_rules}
\begin{align}
 \epsilon_{ij} &= \sqrt{\epsilon_i\epsilon_j}\\
 \sigma_{ij} &= \frac12\l(\sigma_i+\sigma_j\r).
\end{align}
\end{subequations}

\textit{Shifted LJ Potential}\ The shifted LJ potential includes an additional parameter $R_0$ shifting the LJ potential in $r$-direction. This is an artificial potential which allows to study curvature effects without increasing the decay length of the potential (for instance by increasing $\sigma$) \cite{Weiss:2013} and is defined as
\begin{align}
 \label{eq:shifted_LJ}
  V_{R_0}(r)= V_{LJ}(r-R_0) = 4\epsilon\l[\l(\frac\sigma {r-R_0}\r)^{12}-\l(\frac\sigma {r-R_0}\r)^6\r].
\end{align}

\textit{Coulomb Potential} The long range electrostatic Coulomb potential between two point charges $q_1 =Z_1e$ and $q_2=Z_2 e$ is given as
\begin{align}
 V_\text{elec}(r) = \epsilon Z_1Z_2\frac \sigma r
\end{align}
with $\epsilon\sigma = {e^2}/{4\pi\epsilon_0}$, the electron unit charge $e$ and the vacuum permittivity $\epsilon_0$. It is ubiquitous in natural systems.
% \textit{Yukawa potential} The long range electrostatic Coulomb potential between two point charges $q_1 =Z_1e$ and $q_2=Z_2 e$ is given as
% \begin{align}
%  V(r) = \epsilon Z_1Z_2\frac \sigma r
% \end{align}
% with
% \begin{align}
%  V(r) = \frac{1}{4\pi\epsilon}\frac{\sigma}{r}
% \end{align}

\textit{Gaussian Potential}\ Gaussian particles defined by the interaction potential
\begin{align}
 V_\text{Gauss}(r) = \epsilon \exp\l(-\frac{r^2}{\sigma^2}\r)
\end{align}
are of ``soft'' nature as they have no attraction but the repulsion is finite and the interaction force even vanishes at zero distance. The interaction is used as a model for the interaction of polymer cores \cite{Dzubiella:2003}.

\textit{Hard Sphere Interaction}\ A model system which is often used as a reference system but can also model e.g. spherical colloids is the hard sphere model. The interaction potential is defined as
\begin{align}
\label{eq:hard_sphere_interaction}
 V_{\text{HS}}(r) = \begin{cases}
         \infty, &r \leq \sigma\\
         0,      &r > \sigma\\
        \end{cases}.
\end{align}
This definition implies that hard spheres do generally not affect each other, unless they touch each other at distance $r=\sigma$ in which case they perform an elastic collision.

\subparagraph{Partition Function and Averages}
In strict equilibrium contexts we will work with the classical canonical partition function $Z$ defined for canonical ensembles of constant particle number $N$, constant temperature $T = (\kB\beta)^{-1}$ and constant volume $V$ as
\begin{align}
 Z &= \frac{1}{h^{3N}N!}\int\d^3\v r_1\dots\int\d^3\v r_N\int\d^3\v p_1\dots\int\d^3\v p_N \exp\l[-\beta\l(\sum_{i=1}^N\frac{\v p^2}{2m_i}+U(\{\v r_i\})\r)\r]\\
   &= \frac1{\Lambda^{3N}N!} \int\d^3\v r_1\dots\int\d^3\v r_N\exp\l[-\beta U(\{\v r_i\})\r]\\
   &= Q/(\Lambda^{3N}N!)
\end{align}
where the integration is done over the particle momenta $\v p_i$ and positions $\v r_i$ and we use Planck's constant $h$, the complete interaction energy of the system $U$, the inverse thermal energy $\beta=\l(\kB T\r)^{-1}$, the excess partition function $Q$, and the thermal wavelength
\begin{align}
 \label{eq:thermal_wavelength}
 \Lambda = \sqrt{\frac{h^2}{2\pi m \kB T}}.
\end{align}
Averages in an equilibrium context can be expressed as 
\begin{align}
 \l\langle \hat X \r\rangle =\frac 1{Zh^{3N}N!} \int\d^3\v r_1\dots\d^3\v r_N\int\d^3\v p_1\dots\d^3\v p_N\ \hat X \exp\l[-\beta\l(\sum_{i=1}^N\frac{\v p^2}{2m_i}+U(\{\v r_i\})\r)\r].
\end{align}

\subparagraph{State Variables}
Macroscopic observables describing the state of an equilibrated thermodynamic system of $N=\const$ particles are the pressure $P$, the volume $V$ and the temperature $T$. All of the equilibrium systems we treat are either NVT or NPT ensembles, thus follow the canonical partition function. For MD simulations we will connect the temperature to the total kinetic energy of the particles as
\begin{align}
 T = \l\langle\frac{1}{\kB N_{\text{df}}}\sum_{i=1}^N\frac{\v p_i^2}{m_i}\r\rangle,
\end{align}
for $N_\text{df}=3N-N_c$ degrees of freedom where $N_c$ is the total number of constraints and $N_c$ is the total number of rotational degrees of freedom, following \cite{Huenenberger:2005}

\subparagraph{Pair Correlation Function}
The pair correlation function
\begin{align}
 \label{eq:pair_corr_function}
 g(\v r',\v r) = \frac{\l\langle\hat \rho(\v r) \hat\rho(\v r')\r\rangle}{\l\langle\hat\rho(\v r)\r\rangle\l\langle\hat\rho(\v r')\r\rangle}
\end{align}
is a key quantity for the treatment of fluids. Looking at isotropic systems and at a single particle located at the origin, $g$ is a quantity describing the structure of particles surrounding said particle at distance $r=|\v r'-\v r|$. For the correlation between a particle and all other particles in an homogeneous equilibrated system of density $\rho_0=N/V$, $g$ is given as
\begin{align}
    g(r=\l|\v r\r|) = \frac1{V\rho_0^2}\,\l\langle\sum_{i\neq j}^{N,N}\delta\big(\v r -(\v r_i-\v r_j)\big)\r\rangle.
\end{align}
An often used limit to model $g$ is the high temperature or low density limit
\begin{align}
 \label{eq:high_density_limit}
 g(r) = \exp(-\beta V(r)).
\end{align}

\subparagraph{Virial Expansion}
The virial expansion for the equation of state up to second order is \cite{Barrat:2003}
\begin{align}
\label{eq:virial_eq_of_state}
 \frac{P}{\kB T} = \rho + B_2\rho^2,
\end{align}
where $B_2$ is the second order virial coefficient. The virial expansion is performed for low densities and uses the second virial coefficient
\begin{align}
\label{eq:low_dens_B2}
 B_2 = \frac12 \int\d^d\v r \l(1-e^{-\beta V(r)}\r) = \frac12 \int\d^d\v r h(r)
\end{align}
with $V(r)$ being the pair interaction potential and the integration done over the volume of the system (of dimension $d$). We also made use of the Meyer function $h(r) = 1-\exp(-\beta V(r))$.

\subparagraph{Hard Rod Equation of State}
The equation of state for a system of $N$ hard rods of length $\sigma$ in a one-dimensional box of length $L$ has been derived in \cite{Tonks:1936} as
\begin{align}
 \label{eq:eq_of_state_hard_rods}
 P = \frac{N\kB T}{L-(N-1)\sigma}.
\end{align}
A hard rod is the one-dimensional version of a hard sphere with interaction \eq{\ref{eq:hard_sphere_interaction}}.

\subparagraph{Thermodynamic Potentials and Related Quantities} 
The thermodynamic potentials of interest are the excess chemical potential $\muexc$, the Gibbs free energy of solvation per particle $\Delta G$ and the Helmholtz free energy of solvation per particle $\Delta F$. Note that the difference between the Gibbs solvation free energy $\Delta G$ in an NPT-ensemble and the Helmholtz solvation free energy $\Delta F$ in an NVT-ensemble is an additional term of $P\Delta V$ with $P$ being the system's constant pressure and $\Delta V$ being the change in volume due to the solvation. Since we will only treat systems where $\Delta V$ is negligibly small we will neglect the difference and use the terms $\Delta G$ and $\Delta F$ equivalently in the following. Furthermore, these quantities are equivalent to the excess chemical potential $\muexc$ for one solute particle. We can find the solvation enthalpy per particle $\Delta H$, the solvation entropy per particle $\Delta S$ and the specific heat of solvation $\Delta C_p$ by the thermodynamic identities
\begin{subequations}
\label{eq:thermodynamic_relations}
\begin{align}
\label{eq:entropy}
 \Delta S &= -\p{\muexc}{T}\\
\label{eq:enthalpy} 
 \Delta H &= \muexc+T\Delta S\\
\label{eq:specific_heat} 
 \Delta C_p &= \ptwo{\Delta H}{T}.
\end{align} 
\end{subequations}

\begin{figure}[tb]
 \centering
    \includegraphics[width=10cm]{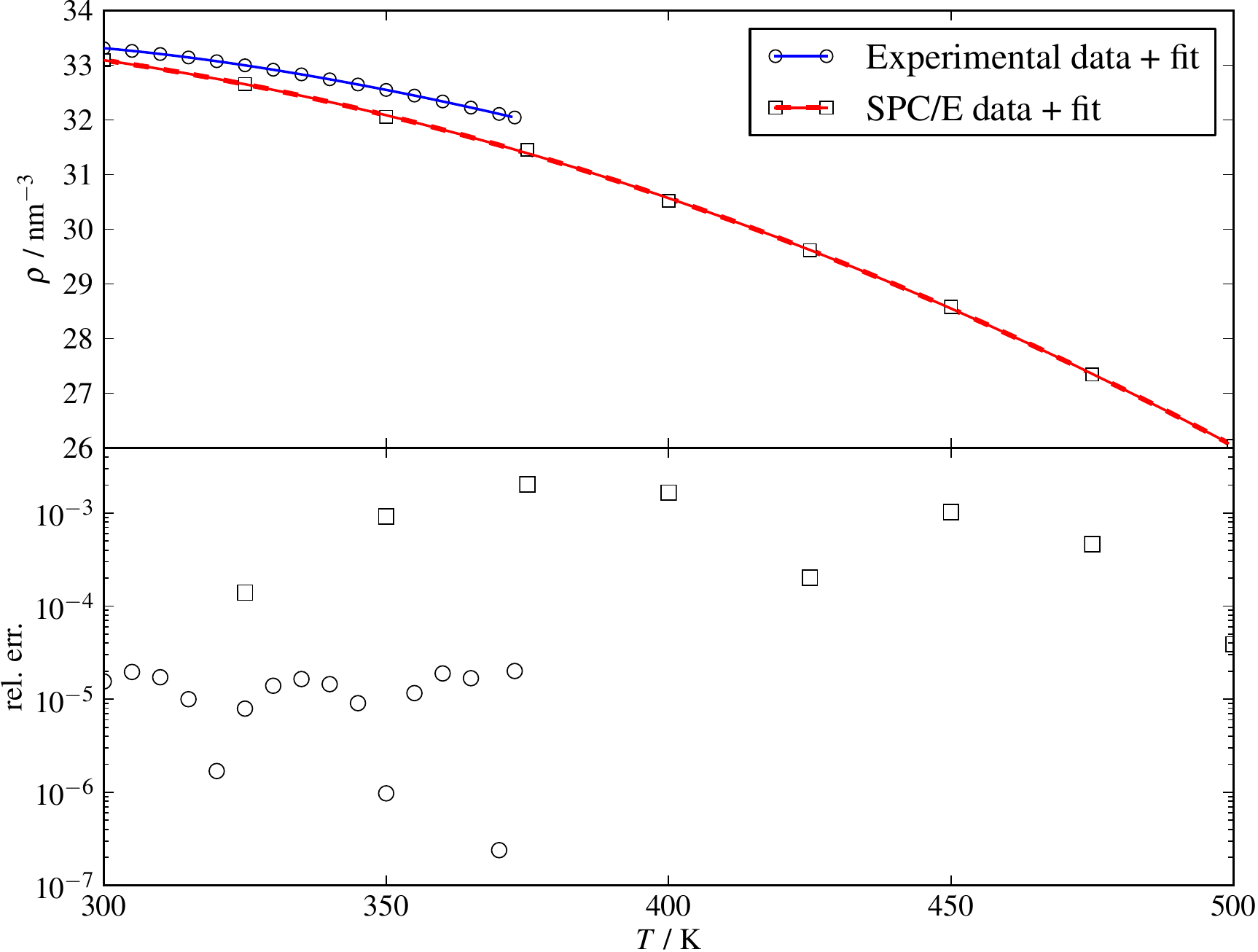}
 \caption{Number density dependence on temperature, experimental data from \cite{Wagner:2002}, simulation data from \cite{Ismail:2006} and polynomial fits of fourth and third order, respectively.}
 \label{fig:number_density}
\end{figure}

\subparagraph{Equation of State for Water}
For a constant pressure of $p=0.1\,\text{MPa}$, experimental data for the density of water has been tabled in \cite{Wagner:2002}. A polynomial fit of fourth order represents the experimental data satisfactorily, as can be seen in \fig{\ref{fig:number_density}},                                                                                                                                                                                                                                                                                                                                                                                                                                                                                                                                                                                                                                                                                                                                                                                                                                                                                                                                                                                                                                                                                                                                                                                                                                                                                                                                                                                                                                                                                                                                                                                                                                                                                                                                                                                                                                                           
\begin{align}
 \label{eq:number_density_fit}
 \rho_N(T) &= A_0 + A_1\,\tilde T + A_2\,\tilde T^2 + A_3\,\tilde T^3 + A_4\,\tilde T^4 \\
 A_0 &= -38.18575995 \,\text{nm}^{-3}\\
 A_1 &= 239.39492116\,\text{nm}^{-3}\\
 A_2 &= -296.02330806\,\text{nm}^{-3}\\
 A_3 &=162.46048481\,\text{nm}^{-3}\\
 A_4 &=-34.31668687 \,\text{nm}^{-3}                                                             
\end{align}
with $\tilde T = T/298.15\text{K}$.

We will furthermore need the number density of SPC/E water (explained in \sec{\ref{sec:mdsim_intro}}) for constant atmospheric pressure to compare it with the density obtained in our thermophoretic simulations. It has been measured in \cite{Ismail:2006} with the data displayed in \fig{\ref{fig:number_density}}. Here, a polynomial fit of third order seems sufficient, yielding \eq{\ref{eq:number_density_fit}} with parameters
\begin{subequations}
\label{eq:spce_density}
\begin{align}
 A_0 &= 27.97124672\,\text{nm}^{-3}\\
 A_1 &= 14.93831575 \,\text{nm}^{-3}\\
 A_2 &= -10.07488855 \,\text{nm}^{-3}\\
 A_3 &= 0.28637107\,\text{nm}^{-3}\\
 A_4 &= 0.
\end{align}
\end{subequations}

\begin{figure}[htb]
 \centering
    \includegraphics[width=10cm]{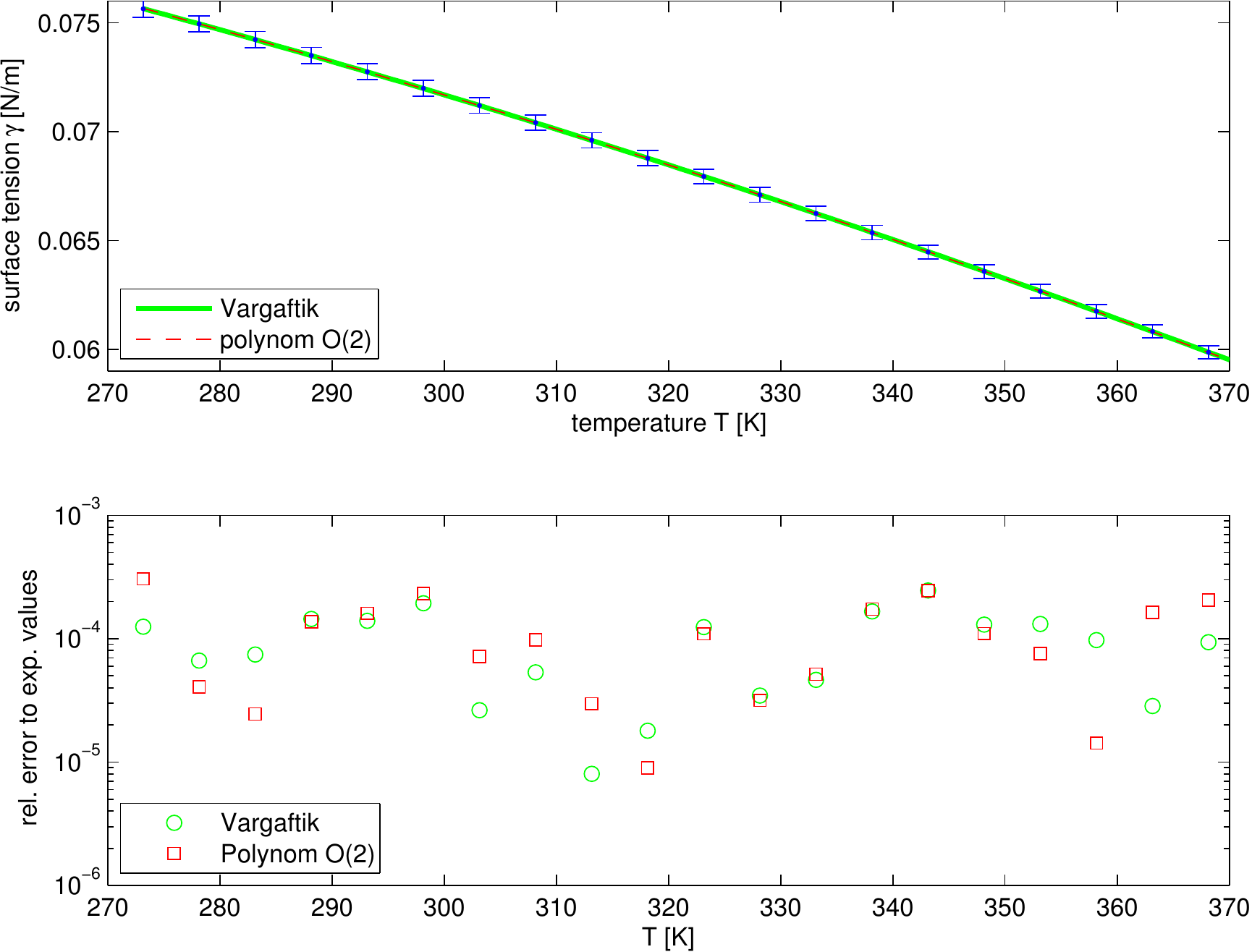}
 \caption{Surface tension of water, the two hypotheses \eq{\ref{eq:gamma}} and \eq{\ref{eq:gamma_poly}} versus the experimental data from \cite{Vargaftik:1983}.}
 \label{fig:srfc_tension}
\end{figure}

\subparagraph{Surface Tension}
The surface tension of water was parametrized in \cite{Vargaftik:1983} to be
\begin{equation}
 \label{eq:gamma}
 \gammalv(T) = B\,\l[1-\frac{T}{T_c}\r]^p\,\l[1+b\l(1-\frac{T}{T_c}\r)\r],
\end{equation}
with the parameters $T_C = 647.15\text K$, $B=235.8\times10^{-3}\text{N}/\text{m}, b=-0.625$ and $p=1.256$. However, for a simplified use, a quadratic fit seems to be sufficient to represent the experimental data accurately in the given temperature range (cf. \fig{\ref{fig:srfc_tension}}), with
\begin{align}
 \label{eq:gamma_poly}
 \gammalv(T) = a_2T^2+a_1T+a_0,
\end{align}
where
\begin{align}
a_2&=-2.6768917\times10^{-7}\text N\text m^{-1}\text K^{-2},\\                
a_1&=\ \ \,5.2815621\times10^{-6}\text N\text m^{-1}\text K^{-1},\\
a_0&=\ \ \,9.4194397\times10^{-2}\text N\text m^{-1}.
\end{align}
The surface tension for SPC/E water was found to follow \eq{\ref{eq:gamma}} with the parameters
\begin{subequations}
\label{eq:spce_gamma}
\begin{align} 
 B &= 205.32\times10^{-3}\text{N}/\text{m}\\
 b &= -0.6132\\
 p &= 11/9\\
 T_c &= 625.7\unit{K}.
\end{align}
\end{subequations}

\subparagraph{Static Permittivity}
According to \cite{Uematsu:1983}, the static water permittivity $\epsilon_\ell$ at constant density of $\rho_0 = 1000\text{kg}/\text m^3$ is found to be 
\begin{align}
 \epsilon_\ell(\tilde T) &= A_0 + A_1 \tilde T + A_2\tilde T^2 +\frac{A_3}{\tilde T} + \frac{A_4}{\tilde T^2}, \\
A_0 &= -1.754334\times10^2 \\
A_1 &= \ \ \,69.575  \\
A_2 &= -1.02099\times10^1 \\
A_3 &= \ \ \,2.3998771\times10^2 \\
A_4 &= -45.2059 
\end{align}
with $\tilde T=T/298.15\text K$. \cite{Uematsu:1983} gives a formula for varying densities, as well. However, taking into account the temperature dependence of the density for atmospheric pressure ($\approx 0.1\text{MPa}$), there exists a simpler formula with similar accuracy, given in \cite{Hamelin:1998}
\begin{align}
 \epsilon_\ell(T) &= A_0 + A_1\,(T-T_0) + A_2\,(T-T_0)^2 + A_3\,(T-T_0)^3 \\
            A_0 &= \ \ \,87.9144\\
            A_1 &= -0.404399\,\text K^{-1}\\
            A_2 &= \ \ \,9.58726\times10^{-4}\,\text K^{-2}\\
            A_3 &= -1.32802\times10^{-6}\,\text K^{-3}
\end{align}
with $T_0 = 273.15\text{K}$.

\subparagraph{Diffusion coefficient, Einstein Relation, Stokes' Law and SPC/E Viscosity}
The time dependent diffusion coefficient for a Brownian walker in equilibrated systems is given as \cite{Dzubiella_lecture:2013}
\begin{align}
  \label{eq:D_of_t}
  D(t) = \frac{\l\langle(\Delta\v r(t))^2\r\rangle}{2td},
\end{align}
with the system's dimensionality $d$ and the distance from its initial position $\Delta\v r(t)$. Einstein's relation can be found as the limit $\lim\limits_{t\rightarrow\infty} D(t)$ and is usually expressed as
\begin{align}
 \label{eq:einstein_relation}
  D = \frac{\kB T}{\gamma}
\end{align}
with $\gamma$ being the friction coefficient between the particle and the solvent, which is given by the hydrodynamics of the problem. We will make the usual assumption that the particle is a perfect sphere of radius $R$, s.t. the friction coefficient is given as
\begin{align}
 \label{eq:stokes_friction}
  \gamma = 6\pi\eta R
\end{align}
with $\eta$ being the viscosity of the solvent. Stokes' law gives the friction force of a spherical particle travelling with velocity $\v v$ as
\begin{align}
  \v F^{\text{fric}} = -\gamma \v v.
\end{align}
The viscosity of the solvent is generally temperature dependent and has been measured both experimentally and for SPC/E water \cite{Markesteijn:2012} to follow
\begin{align}
  \label{eq:viscosity}
  \eta(T) = (1\unit K)^{b}\times(T-T_0)^{-b}\,\unit{mPa\,s}.
\end{align}
with the parameters $T_0=212.4\unit K$ and $b=1.633$ for SPC/E water, $T_0=225.4\unit K$ and $b=1.637$ for the experimental values, respectively.

\subparagraph{Measuring the Soret Coefficient}
The Soret coefficient was described in detail in \sec{\ref{sec:soret_coefficient}}. We found it to be connected to the Soret equilibrium density as
\begin{align}
 \label{eq:soret_from_density}
  \ST = -\frac{\nabla \rho}{\rho\nabla T}.
\end{align}
Therefore, measuring the Soret equilibrium solute density resolved in the direction of $\nabla T$, we find $\ST$.

Another route to measure the Soret coefficient for a solute in a solvent at a roughly constant temperature $T$ is to restrain it at a point of the system where $T(x) = T$ using an external potential only acting on the solute, for instance an harmonic potential with force $\Fext(\Delta x)=-k\Delta x$. Then in Soret equilibrium and applying \eq{\ref{eq:force_from_drift}} to the equilibrium condition $\Fext = -F$ we find approximately
\begin{align}
  \label{eq:measure_soret_from_force}
  \ST &= -\frac{k}{\kB \nabla T} \l\langle\Delta x\r\rangle.
\end{align}
This method has been used for instance in \cite{Galliero:2008} for MD simulations.

\subparagraph{Heat Equation and Temperature Profile}
\label{sec:heat_equation}
The general setting for creating a system in a temperature gradient used in this work is to consider two heat baths at the boundaries in the $x$-direction of the system, one at the ``cold'' temperature $T_c$ and the other at the ``hot'' temperature $T_h$, while particles may not escape into the baths, implying reflective boundary conditions. To obtain the supposed temperature profile  of this setting we consider the heat equation
\begin{align}
 \partial_t T - \lambda \nabla^2 T = 0
\end{align}
with the thermal conductivity $\lambda$, which is an ordinary second order partial differential equation (PDE) that can be solved by common means. As mentioned, we are interested in the stationary solution of a fluid in a spatial volume of box length $L$ in the $x$-direction that has temperature $T_{c}$ at $x=-L/2$ and $T_{h}$ at $x=L/2$ and periodic boundary conditions (PBC) in $y$- and $z$-direction, hence it is sufficient to solve the one-dimensional boundary value problem
\begin{align}
 \ptwo{T}{x} = 0;\qquad T(-L/2) = T_{c};\qquad T(L/2) = T_{h}.
\end{align}
Trivially, the solution is a linear function, such that in the following, whenever it is necessary, we assume
\begin{align}
 T(x) = T_0(1+\epsilon x)
\end{align}
and for $\epsilon\rightarrow0$
\begin{align}
 \label{eq:invers_T_approx}
 [T(x)]^{-1} \simeq T_0^{-1}(1-\epsilon x).
\end{align}

\section{Free Energy Calculations}
In order to evaluate the free energy $\Delta G$ needed to insert one solute particle in a solvent bath of density $v$ and temperature $T$ quantitatively, we will make use of three methods in this work, all well known and often employed. The first is thermodynamic integration (TI), where the interaction between solute and solvent is slowly ``turned on'' via a coupling parameter $\lambda$, then the free energy is evaluated per integration over the change of the ensemble's mean interaction energy. The method of Bennett's acceptance ratio (BAR), closely related to TI, evaluates the free energy differences between steps of $\lambda$ by means of an estimation of the phase-space overlap between ensembles adjacent in $\lambda$-space. The third is a perturbation method called Widom's test particle insertion (TPI), where the solute is inserted as a ``ghost particle'' at random positions of a bulk solvent ensemble, followed by the evaluation of the free energy dependent on the mean of the interaction energy 
between ghost and solvent. All methods are well explained in the literature \cite{Allen:1987,Christ:2010,Bennett:1976,Widom:1963,Barrat:2003} but will be introduced briefly below for the sake of completeness.

Since the evaluation of $\Delta G$ is often done for discrete values of the temperature $T$ while $\Delta G$ is actually needed as a continuous function in $T$, we will furthermore introduce a fit function for $\Delta G(T)$ at the end of this section which will be used several times throughout this work.

\subsection{Thermodynamic Integration}
\label{sec:TI}
The main idea of thermodynamic integration is to consider the total ensemble interaction energy depending on an arbitrary parameter $\lambda$ denoting the state of the system. While the method is applicable in a variety of contexts, here we will think of $\lambda$ as a parameter determining the kind of interaction between a solute particle and an arbitray solvent particle. We denote the interaction potential at state $\lambda$ as $V_\lambda(r)$ with $V_{\lambda_A}(r)$ being the interaction potential in a known reference system at $\lambda_A$ and the full interaction potential (between solute and solvent as soon as it is fully inserted) as $V_{\lambda_B}(r)$. The full interaction energy at state $\lambda$ then reads
\begin{align}
 U(\lambda) = \sum_{j=1}^{N_v}V_\lambda(|\v r_u-\v r_j|) + \sum^{N_v-1}_{i=1}\sum^{N_v}_{j>i}V_{vv}(|\v r_i-\v r_j|).
\end{align}
At every state $\lambda$, the excess free energy of an NVT ensemble is given via the excess partition function $Z_\text{exc}\equiv Q$
\begin{align}
 \label{eq:DeltaF_via_Z}
 \beta F(\lambda) = -\ln Q(\lambda)=-\ln\l(\int\d^3\v r_u\int\d^3\v r_1 ...\int\d^3\v r_{N_v} \exp\l[-\beta U(\l\{\v r\r\};\lambda)\r]\r)
\end{align}
Deriving with respect to $\lambda$ and integration yields
\begin{align}
 \p{\beta F(\lambda)}{\lambda} &= \frac{\beta}{Q(\lambda)}\int\d^3\v r_u\int\d^3\v r_1 ...\int\d^3\v r_{N_v} \frac{\d U(\l\{\v r\r\};\lambda)}{\d\lambda}\exp\l[-\beta U(\l\{\v r\r\};\lambda)\r]\\
 \beta\Delta F& = \beta\int\limits_{\lambda_A}^{\lambda_B}\d\lambda \l\langle\frac{\d U}{\d \lambda}\r\rangle_\lambda.
\end{align}
The angled brackets $\l\langle\cdot\r\rangle_\lambda$ denote an ensemble mean for ensembles following the partition function $Q(\lambda)$. In equilibrium computer experiments, the integral abocve is approximated using trapezoidal integration over a finite set of $\lambda_i$ with $\lambda_A < \lambda_i<\lambda_B$. At every step $\lambda_i$, the mean $\l\langle\d U/\d\lambda\r\rangle$ can be evaluated using two methods. The first is a pure evaluation of the interaction energy's $\lambda$-derivative and subsequent evaluation of the mean over the ensemble's temporal evolution. The second route makes use of another statistical quantity, the pair correlation function \eq{\ref{eq:pair_corr_function}}. For the correlation between the solute and all solvent particles equilibrated system of solvent density $v_0$, $g$ is given as
\begin{align}
    g_\lambda(|\v r|) = \frac{2}{v_0}\,\l\langle\sum_{j=1}^{N_v}\delta\big(\v r -(\v r_u-\v r_j)\big)\r\rangle_\lambda.
\end{align}
Using the definition of the mean we find
\begin{align}
 \l\langle\frac{\d U}{\d\lambda}\r\rangle_\lambda &= \l\langle\sum_{j=1}^{N_v}\frac{\d V_\lambda(|\v r_u-\v r_j|)}{\d\lambda}\r\rangle_\lambda = \l\langle\sum_{j=1}^{N_v}\int\d^3\v r\frac{\d V_\lambda(r)}{\d\lambda}\delta(\v r -\big(\v r_u-\v r_j)\big)\r\rangle_\lambda\\
                                          &= \frac{v_0}{2}\int\d^3\v r\  g_\lambda(r)\frac{\d V_\lambda(r)}{\d\lambda}
\end{align}
This method has the advantage that with one key quantity $g_\lambda(r)$ one is not only able to extract information about the structure of the fluid but it enables one also to evaluate a variety of statistical means (where the derivation of those means are similar to the derivation above). Furthermore, as in certain limits one can model $g_\lambda(r)$ by analytical means, this form is of particular use for the analytical treatment of problems. Using this framework and noticing that the partition function and thus the mean depends on the temperature $T$, the free energy is given as a function of $T$ via
\begin{align}
  \label{eq:free_energy_from_gofr}
  \Delta F(T) & = \frac{v_0(T)}{2}\int\limits_{\lambda_A}^{\lambda_B}\d\lambda \int\d^3\v r\  g_{\lambda,T}(r)\frac{\d V_\lambda(r)}{\d\lambda}.
\end{align}
In the high temperature / low density limit \eq{\ref{eq:high_density_limit}} and with a $\lambda$-potential $V_\lambda(r)=\lambda V(r)$ with $0\leq\lambda\leq1$, $g_{\lambda,T}$ is given as
\begin{align}
  \label{eq:g_of_r_high_temp}
  g_{\lambda,T} &= \exp(-\lambda\beta V(r))
\end{align}
and the excess free energy becomes
\begin{subequations}
\label{eq:free_energy_from_gofr_high_temp}
\begin{align}
 \Delta F(T) &= \frac{v(T)}2 \int\limits_0^1\d\lambda\int\limits\d^d\v r\ V(r) g_{\lambda,T}(r) = \frac{v(T)}2\int\limits_0^1\d\lambda\int\limits\d^d \v r\ V(r) e^{-\lambda\beta V(r)} \\
          &= \frac{v(T)}{2}\int\limits\d^d \v r\l(-\frac1\beta\r)\l[e^{-\beta V(r)}-1\r] = \frac{v(T)}{2\beta}\int\limits\d^d \v r\ h(r)
\end{align}
\end{subequations}
with the Meyer function $h(r) = 1-\exp(-\beta V(r))$.

\subsection{Bennett's Acceptance Ratio}
\label{sec:BAR}
The derivation below was originally given in \cite{Bennett:1976}. The method uses \eq{\ref{eq:DeltaF_via_Z}} to get the free energy difference between two states from the ratio of the excess partition functions
\begin{align}
 \beta\big[F(\lambda_1)-F(\lambda_0)\big] = -\ln\frac{Q(\lambda_1)}{Q(\lambda_0)}.
\end{align}
Let us simplify the notation and denote the excess partition function at state $\lambda_i$ as 
\begin{align}
 Q_i = \int \d^3\v r_u \d^3\v r_1\dots\d^3\v r_{N_v}\ \exp\l(-\overline U_i\r)
\end{align}
with the dimensionless interaction energy $\overline U_i = \beta U(\{\v r\},\lambda_i)$. In systems of same temperature the ratio of $Q$ is equal to the ratio of $Z$. Consider now the Metropolis accepting function $M(x) = \min\{1,\exp(-x)\}$ known from Monte Carlo simulations with Boltzmann sampling. This function has the property $M(x)/M(-x)=\exp(-x)$ \cite{Bennett:1976}, thus giving the equation
\begin{align}
 M(\overline U_1-\overline U_0) \exp(-\overline U_0) = M(\overline U_0-\overline U_1) \exp(-\overline U_1).
\end{align}
Integrating both sides over the configurational space and subsequent multiplication of the left side with $Q_0/Q_0$ and the right side with $Q_1/Q_1$ yields
\begin{align}
 \frac{Q_0}{Q_0} \int\d^3\v r_u \d^3&\v r_1\dots\d^3\v r_{N_v}\ M(\overline U_1-\overline U_0)\exp\l(-\overline U_0\r)\\
  &= 
 \frac{Q_1}{Q_1} \int\d^3\v r_u \d^3\v r_1\dots\d^3\v r_{N_v}\ M(\overline U_0-\overline U_1)\exp\l(-\overline U_1\r)  \\
 \frac{Q_1}{Q_0} &= \frac{\l\langle M(\overline U_1-\overline U_0)\r\rangle_0}{\l\langle M(\overline U_0-\overline U_1)\r\rangle_1}.
 \label{eq:BAR_metropolis}
\end{align}
Here, the angle brackets' subscript $0$ denotes that the mean is evaluated in the system following the partition function $Z_0$ and vice versa. The equation may be interpreted such that both ways of ``inserting a particle'' are considered, first the actual insertion (evaluating the influence of potential 1 in an ensemble of state 0), second removing the particle (evaluating the influence of potential 0 in an ensemble of state 1). Bennett remarks that the estimation of the free energy by his method works best if the overlap of the configurational space of both ensembles is sufficiently large. Furthermore, in his final formulation of the method, the Metropolis function is replaced by the Fermi function $f(x) = (1+e^{-x})^{-1}$ and an arbitrary constant is introduced, showing that this minimizes the variance of the evaluated means.

In order to work with ensembles of sufficiently large phase space overlap, we make use of the methods of TI, split the insertion in a problem of $N_\lambda+1$ states in $\lambda$-space and find for the total solvation free energy
\begin{align}
 \label{eq:free_energy_from_BAR}
 \beta\Delta F = \beta\sum_{k=1}^{N_\lambda}{\big[F(\lambda_k)-F(\lambda_{k-1})\big]} = - \sum_{k=1}^{N_\lambda}\ln\frac{Q_k}{Q_{k-1}}.
\end{align}

\subsection{Widom's Test Particle Insertion}
\label{sec:TPI}
Widom's TPI \cite{Widom:1963,Corti:1998,Barrat:2003} is a limiting case of the BAR method \cite{Bennett:1976} as will be indicated below. He starts with a partition function of the system with fully inserted solute
\begin{align}
 Z_B &= \frac{1}{\Lambda_u^3\Lambda_s^{3N_v}(N_v+1)!} \int \d^3\v r_u \d^3\v r_1\dots\d^3\v r_{N_v}\times\\
     &\qquad \times \exp\l(-\beta\underbrace{\sum_{j=1}^{N_v}V_{\lambda_B}(|\v r_u-\v r_j|)}_{=\Psi} -\beta \underbrace{\sum^{N_v-1}_{i=1}\sum^{N_v}_{j>i}V_{vv}(|\v r_i-\v r_j|)}_{=\Phi_{N_v}}\r).
\end{align}
Deviding by the partition function of the system without solute yields
\begin{align}
 \frac{Z_B}{Z_A} &= \frac{V}{\Lambda_u^3(N_v+1)}\frac{\int\d^3\v r_1\dots\d^3\v r_{N_v}\ e^{-\beta\Psi}\ e^{-\beta\Phi_{N_v}}}
                                                     {\int\d^3\v r_1\dots\d^3\v r_{N_v}\ e^{-\beta\Phi_{N_v}}}
                  = \frac{1}{v \Lambda_u^3}\Big\langle\exp\l(-\beta\Psi\r)\Big\rangle
\end{align}
where $v = N_v/V\simeq(N_v+1)/V$ is again the density of the solvent. The mean in the equation above can be interpreted to be taken over all possible solute positions $\v r_u$ in ensembles following the partition function of the pure solvent. Noticing that the prefactor $1/v \Lambda_u^3$ corresponds to the ideal gas contribution, the excess solvation free energy is given via the excess partition functions (omitting the prefactor) to yield
\begin{align}
 \beta\Delta F &= -\ln\frac{Q_B}{Q_A} = -\ln\Big\langle\exp\l(-\beta\Psi\r)\Big\rangle
\end{align}

Note that in the mean of \eq{\ref{eq:BAR_metropolis}} we find for this special case $\overline U_1 -\overline U_0=\beta\Psi$. We see that by considering only cases $\beta\Psi>0$, the mean of \eq{\ref{eq:BAR_metropolis}} becomes the mean considered in this subsection, indicating that the TPI is indeed only a limit of the BAR method. Hence, one can suspect that the TPI method works only well for solutes which do not change the configurational space of the solvent too much, i.e. small solutes. Example given, comparisons between the usage of BAR and TPI showed a sufficient agreement for methane \cite{Paschek:2008}. A more elegant indication of the limit case argument is given in \cite{Bennett:1976}. As long as the results of BAR and TPI do not differ too much, the TPI method is the method of choice as it only requires simulation of bulk solvent ensembles, using which one can calculate the free energy of all kinds of solutes/interaction potentials in a short time, while for BAR one has to simulate ensembles for 
every step of $\lambda$.

\subsection{Fitting the Free Energy}
\label{sec:fitting}
Using the methods described above one can obtain the free energy only for discrete values of temperature $T$ as they are evaluated using simulations of equilibrated systems at $T$. However, the free energy is actually needed as a continuous function of $T$ in order to evaluate quantities such as the entropy and enthalpy, both depending on the temperature derivative of $\Delta G$. Hence it seems to be valuable to introduce a fit function not only capable of modeling the free energy but also its depending thermodynamic quantities. In this work we will deploy the fit function
\begin{align}
 \label{eq:free_energy_fit_function}
 \Delta G(T) = a + b T + c T^2 + d T\ln T,
\end{align}
which is widely used in the literature \cite{Sedlmeier:2011,Paschek:2008} because it yields reasonable fit functions for the solvation entropy $\Delta S$, solvation enthalpy $\Delta H$ and specific heat of solvation $\Delta C_p$, given via \eq{\ref{eq:thermodynamic_relations}a-c} as
\begin{subequations}
\label{eq:thermodynamic_fit_functions}
 \begin{align}
 \Delta S(T) &= - (d+b) - 2cT-d\ln T\\
 \Delta H(T) &= a -dT - cT^2\\
 \Delta C_p(T) &= -d - 2cT.
\end{align}
\end{subequations}

\section{Brownian Dynamics}
\subsection{Theoretical Basis}
An often appropriate picture to investigate the behavior of thermodynamic systems is a macroscopic one described by Brownian Dynamics (BD).  Within this framework an ensemble consists of $N$ explicit particles and an underlying implicit solvent permanently colliding with the explicit particles resulting in an intrinsic random force. This implicit solvent is embedded in the equations of motion as a noise term whose intensity can be locally dependent and is connected to the diffusion coefficient $D:=D(\v r)$. Note that in this context, $D(\v r)$ is \textit{not} necessarily the diffusion coefficient given by the Einstein relation, but has yet to be determined. The solvent furthermore acts as an inertia damper leading to the negligibility of all acceleration. Hence the only time dependent observables of the system are the particle's location with their temporal evolution given by the stochastic differential equation (SDE)
\begin{align}
 \label{eq:brownian}
 \d\v r_i = -\l[\frac{\nabla_i\Vext(\v r_i)}{\gamma}+\frac1\gamma\nabla_i\sum_{j=1}^NV(\v r_i-\v r_j)\r]\d t + \sqrt{2D(\v r_i)}\ \d \v B_{t,i}.
\end{align}
Here, $\gamma$ is the (in our context constant) friction coefficient or inverse mobility, $\Vext$ is an external potential acting on the particles, $V$ is a two-particle interaction potential and $\d\v B_t$ describes a Brownian process at time $t$, adapting the notation of \cite{Oksendal:2010}. This Brownian process can be interpreted to be connected to a Gaussian white noise $\v\eta_i(t)=\frac{\d\v B_{t,i}}{\d t}$ fulfilling
\begin{align}
 \l\langle\v\eta_i(t)\r\rangle &= 0,\\
 \l\langle\eta^\mu_i(t)\eta^\nu_j(t')\r\rangle (\d t)^2 &= \delta_{\mu\nu}\delta_{ij}\delta(t-t')\ \d t.
\end{align}

Now it is of crucial importance how we interpret \eq{\ref{eq:brownian}}, as there exist two major approaches to solve SDEs \cite{Oksendal:2010}. The first interpretation is that we read it as an SDE as described by It\^o
\begin{align}
 \label{eq:Ito_SDE}
 \d X_t = \sigma(X_t,t)\ \d t + b(X_t,t)\ \d B_t
\end{align}
with $X_t$ a random variable dependent on time $t$ and $\sigma$ and $b\geq0$ being non-anticipating functions (meaning that their value is not known for times $\tau>t$ when they explicitly depend on $X_t$; note that in general $b$ should be a tensor, however we limit it to a scalar function which is sufficient for our problem). In the It\^o case the integration of \eq{\ref{eq:Ito_SDE}} is done by taking the value of the integrand at each time step as that of the beginning of the step.
%and $\d B_t$ a Brownian process of It\^o character. 
The second interpretation is to read it as a Stratonovich SDE. In the Stratonovich case, the integrand for each time step of the integration \eq{\ref{eq:Ito_SDE}} of is the average of the evaluation at the beginning and at the end of the step \cite{Oksendal:2010,Kosztin:2000_chp2}. 

It can be shown \cite{Kosztin:2000_chp2} that the Stratonovich SDE may be transformed into an It\^o SDE using the transformation rules
\begin{align}
 \tilde \sigma &= \sigma + \frac12 b\nabla b,\\
 \tilde b &= b,
\end{align}
yielding the It\^o SDE
\begin{align}
 \d X_t = \l(\sigma(X_t,t)  + \frac12 b(X_t,t)\nabla b(X_t,t)\r) \d t + b(X_t,t)\ \d B_t
\end{align}

Applying this reasoning to \eq{\ref{eq:brownian}}, we find a more general equation of motion for the location of the particle $i$,
\begin{align}
\label{eq:adapted_brownian}
\d\v r_i &= \alpha \nabla_iD(\v r_i)\ \d t - \l[\frac{\nabla_i\Vext(\v r_i)}{\gamma} +\frac1\gamma \nabla_i\sum_{j=1}^NV(\v r_i-\v r_j)\r]\ \d t + \sqrt{2D(\v r_i)}\ \d \v B_{t,i}
\end{align}
with $\alpha=0,\frac12$ describing the It\^o, Stratonovich interpretation, respectively. In general, $\alpha$ can be any real number $0\leq\alpha\leq1$, with its value referring to the weighting of evaluating a mean of the two evaluation points of the integrand in a stochastic integral. The It\^o case is a weighting which ignores the integrand's evaluation at the end of a time step, whereas the Stratonovich case is a symmetric evaluation. In the following, we will consider \eq{\ref{eq:adapted_brownian}} including the arbitrary parameter $\alpha$ and intrepret it as an It\^o SDE, so we can apply his calculus which has simpler rules than Stratonovich's calculus \cite{Oksendal:2010}.

As one can see in \eq{\ref{eq:adapted_brownian}}, the choice of $\alpha$ leads to the inclusion (or exclusion) of the gradient of the local position dependent diffusion coefficient. This is related to the fact that a position dependent diffusion coefficient can have different microscopic origins \cite{Schnitzer:1993,Lancon:2001}. For real systems, $\alpha$ must have a certain value.

\subsection{Simulations}
\label{sec:meth_bd}
We can make use of the definitions from the last subsection and investigate thermodynamic systems by means of BD computer simulations, widely known and often applied \cite{Allen:1987,Dzubiella:2003,Marconi:1999}. Within the framework of BD simulations, the equations \eq{\ref{eq:adapted_brownian}} are integrated numerically for both the It\^o and Stratonovich interpretation, i.e.~the position change of the particle $i$ travelling in between timestep $n$ and timestep $n+1$ is approximated by the discrete time equations
\begin{align}
 \v r_i^{(n+1)} = \v r_i^{(n)} + \l[\frac 1\gamma\v F_\text{ext}\l(\v r^{(n)}_i\r)+ \frac1\gamma\sum_{j=1}^{N}\v F\l(\big|\v r^{(n)}_i-\v r^{(n)}_j\big|\r)
                   + \alpha\nabla D\r]\Delta t
                   + \sqrt{2D\l(\v r^{(n)}_i\r)\Delta t}\ \Delta\v B_i.
\end{align}
The forces are given as $\v F_\text{ext} = -\nabla \Vext$ and $\v F = -\nabla V$. The Brownian noise is given via spatial differences taken from a Gaussian distribution with zero mean and variance $\l\langle\Delta B^\mu_i \Delta B^\nu_j\r\rangle=\delta_{\mu\nu}\delta_{ij}$. As we are interested in systems with temperature gradients, we will work with reflective boundary conditions in the direction of the gradient in order to keep the net flux zero. The simulations can be performed in an arbitray number of dimensions but for the sake of feasibility we will mostly work with one-dimensional systems which are sufficient to test certain assumptions. 
%Furthermore for a first test of the principles established in \ch{\ref{ch:theory}} it suffices to investigate one-dimensional systems, with the exception of one investigation of a two-dimensional system where we will look at the influence of a non-conservative force field on the equilibrium density.
The choice of the integration parameter $\Delta t$ depends on the system's spatial resolution and has to be chosen such that the contributing forces do not change the current position too much in one time step, while ``too much'' has to be defined for the purpose of the simulation. The discussion for choosing $\Delta t$ is therefore done explicitly for the single simulations in \sec{\ref{sec:bdsim}}.

\section{Molecular Dynamics Simulations}
\label{sec:mdsim_intro}
In the last section we introduced computer simulations using an implicit underlying solvent that locally produced the desired temperature for an explicit solvent and the solute by means of noise, thus working like a local thermostat. An approach closer to experimental setups is to work with microscopic simulations, i.e. molecular dynamics (MD) simulations. In this framework, the equations of motion for the atomic locations and momenta contain only external and interaction forces, dismissing the noise. Usually, for a system of $N$ atoms in a spatial volume of size $L_x\times L_y\times L_z$, an Hamiltonian is defined as
\begin{align}
 \mathcal H = \sum_{i=1}^N \Vext^{(i)}(\v r_i) + \frac12\sum_{i\neq j}V^{(i,j)}(|\v r_i-\v r_j|)
\end{align}
where on every atom $i$ there can act a different external potential (denoted by the superscript $(i)$) and every pair potential for a pair $(i,j)$ can be defined seperately, for instance as (non-) rigid bonds building molecules from atoms or as an electric potential between different kinds of atoms. Usually the Hamiltonian can include triplet and quadruplet interactions, as well, accounting for multipole/orientation or torsion interactions. However, in this work we will only consider pair interactions. With the Hamiltonian defined above the equations of motions are given by the canonical equations. During the simulations those equations are integrated using the leapfrog algorithm where the particles' positions are updated over a discrete timestep $\Delta t$. The simulation accuracy is limited by the machine precision and underlies the propagation of the error induced by the temporal discretization, which can have fatal consequences if working in a microcanonic ensemble. As we will be looking at canonical ensembles of NVT and NPT, the implementation of thermostats and barostats is necessary, coupling the ensemble to external baths of temperature $T$ and/or pressure $P$, which inhibits this behavior. In this work this is done using the Berendsen thermostat and barostat \cite{Berendsen:1984}  as well as the Nos\'e-Hoover thermostat \cite{Nose:1984,Hoover:1985} and the Parrinello-Rahman barostat \cite{Parrinello:1981} in its extended version \cite{Nose:1983}. Since the Berendsen methods do not produce the canonical ensembles, they are only used for equilibration. We also employ the particle mesh Ewald summation algorithm \cite{Essmann:1995} where the electric field is splitted in a fast decaying short range part and a long range part which decays fast in Fourier space.

Usually simulations are performed with three-dimensional periodic boundary conditions. However, as we will be working with setups including two thermostats, we will often only apply boundary conditions in $x$- and $y$-directions, with special boundary conditions for $z$ explained later. For those boundary conditions, an adapted Ewald summation method for two dimensions is employed.
\begin{figure}[t!]
 \centering
    \includegraphics[width=5cm]{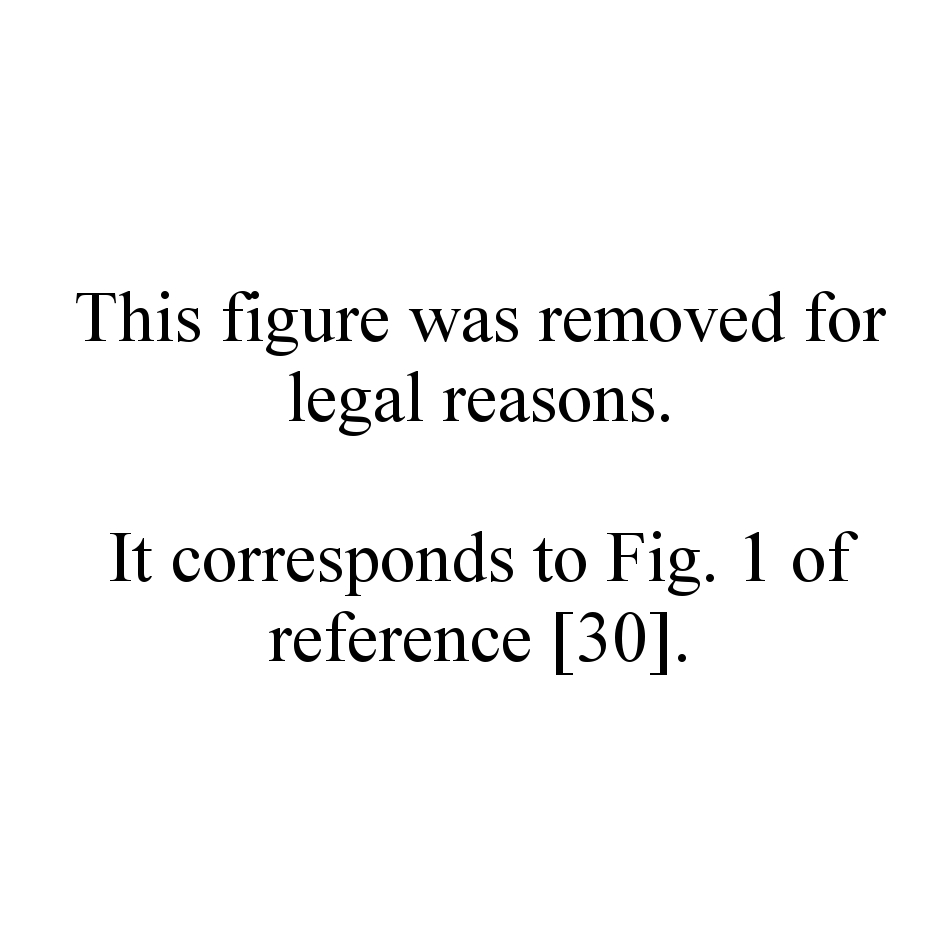}
 \caption{Schematic depiction of an SPC/E water molecule with the parameters given in the text. The dashed grey line represents the zero of the LJ potential originating from the oxygen atom. The picture was taken from \cite{nist}.}
 \label{fig:spce}
\end{figure}

The simulations are performed using the MD simulation package Gromacs \cite{Lindahl:2013} which provides all standard MD simulation algorithms. We used the GROMOS87 forcefield. For feasibility of the simulations it is necessary to introduce four cut-off radii which define a sphere in which forces are evaluated. The first defines the neighbor list (sphere of neighbors), the second the short range Coulomb interaction radius, the third defines at which a smoothening of the LJ-Potential (to approach zero) starts and the fourth defines the LJ- (or vdW-) cut-off. We used either leapfrog- (for thermophoretic simulations) or Langevin-integration (for free energy evaluations) of the equation of motions with a time integration constant of $\Delta t=0.002\unit{fs}$.

As a forcefield for water we used the extended simple point charge model (SPC/E), which is a standard model first proposed in \cite{Berendsen:1987}. An SPC/E water molecule consists of two hydrogen atoms each carrying a charge of $q$ and an oxygen atom carrying a charge of $-2q$ with $q=0.42380 e$. The oxygen atom is the origin of a Lennard-Jones potential with parameters $\sigma=0.315900072062\unit{nm}$ and $\epsilon/\kB = 78.19698466 \unit{K}$ and both hydrogen atoms are connected to the oxygen atom with a rigid bond of 0.1nm distance. A third constraint fixes the bond angle to $109.47^\circ$. A schematic picture of an SPC/E molecule is shown in \fig{\ref{fig:spce}}. Even though the SPC/E model is a crude approximation, it represents a variety of properties of real water correctly, although some only qualitatively \cite{Chaplin:2014}. There exist more advanced models, but for the purpose of cross checking the results of equilibrium simulations with thermophoretic simulations SPC/E is certainly sufficient, as we are not interested in comparisons with experiments and it provides faster simulations due to the constraints.

We will work with the LJ pair potential and the Coulomb potential. Furthermore we will use the shifted-LJ potential, mainly for the application in one of the thermostat setups for the thermophoretic simulations. Since the shifted-LJ is not part of Gromacs, we make use of the \textit{User} option to choose an interaction potential between species of molecules and provide tabulated values of the shifted force field to the program. The LJ parameters for the solutes of interest for this work are given in \tab{\ref{tab:LJ_parameters}}. Sometimes, molecules will be position restrained by an harmonic force of spring constant $k$, where for every direction there can be defined another spring constant.

\begin{table}
 \centering
 \input{bilder/md_ti/LJ_parameters}
 \caption{LJ parameters and masses for the solutes used in this work, taken from \cite{Dzubiella:2006,nist} and from the GROMOS87 force field.}
 \label{tab:LJ_parameters}
\end{table}
Temperature coupling is performed with a time constant of 0.2ps. Pressure coupling is performed with a time constant of 0.5ps with a compressibility of $4.5\times10^{-5}\unit{bar^{-1}}$ and a reference pressure of 1atm. For 2D PBC, we will use semi-isotropic pressure coupling where the coupling for the $z$-direction is turned off. If position restraints are used in an NPT ensemble, the reference coordinates of the restraints will be rescaled accordingly. A dispersion correction of the energy and pressure will be employed to account for the vdW cut-off.

All screenshots of simulations are rendered using VMD \cite{vmd}. The analysis of space resolved density and temperature was done using the MDAnalysis package for Python \cite{mda}.

\subsection{Free Energy Evaluations}
\label{sec:free_energy_MD_setups}
\subsubsection{BAR Method}

Simulations for the thermodynamic integration by means of the BAR method were performed for the solvation of a xenon-like LJ sphere in 466 SPC/E water molecules for a constant pressure of $p=1\unit{atm}$ and three-dimensional PBC using the Langevin integrator which adds thermal noise to the equation of motions much like in BD simulations (in fact, BD simulations are special cases of Langevin integration). As we are interested in the behavior of the free energy for varying temperature, we performed the simulations for 12 temperature values $290\unit K\leq T\leq 400\unit K$ with $\Delta T =10\unit K$. For every temperature $T$, the system was simulated for $0\leq\lambda\leq1$ (absence of solute $\rightarrow$ full insertion) with Gromacs' soft-core method for small $\lambda$. We simulated $21$ different values with a step size of $\Delta\lambda=0.05$. As cut-off radii we chose 1nm for the neighbor list, 1nm for the Coulomb interaction, 0.8nm for the beginning of the vdW-smoothening and 0.9nm for the vdW cut-off. For every $\lambda$, the initial system is minimized using the steepest descent method. Subsequently a 0.1ns NPT equilibration run is performed, followed by a 0.1ns NVT equilibration run and a 10ns production run where the interaction energy was calculated every 10ps. Using Gromacs' implementation of Bennett's method, for every transition $\lambda_i\rightarrow\lambda_{i+1}$ the difference in the free energy was calculated as $\Delta G_{i+1}$ and the total solvation free energy was calculated using \eq{\ref{eq:free_energy_from_BAR}}. After performing all $\lambda$ simulations for all values of $T$, the free energy can be fitted using \eq{\ref{eq:free_energy_fit_function}}.

\subsubsection{TPI method}

The TPI method for the evaluation of the excess free energy was used for more than one solute as it is more feasible than the BAR method. The results for xenon will be cross checked with the results from the BAR method to ensure that the solvation free energy for smaller solutes will be accurate. First, bulk SPC/E water simulations of 1070 molecules have been performed by first minimizing, subsequent 1ns NVT and 1ns NPT equilibriation runs and finally a 10ns NPT production run, saving the atoms' positions every 10ps.
Then for every solute, the simulations have been rerun, inserting the ghost particle 20\,000 times in each frame, each time evaluating the total interaction energy. For every randomly chosen point of insertion, the same neighbor list is used for 4 further insertions within a sphere of 0.05nm radius of the original insterion. By averaging over every frame, we can subsequently evaluate the solvation free energy and fit it using \eq{\ref{eq:free_energy_fit_function}}.

For both bulk simulations and TPIwe chose cut-off radii of 1.15nm for the neighbor list, 1.15nm for the Coulomb cut-off, 1.05nm for the vdW-smoothening and 1.15nm for the vdW cut-off.

The TPI method was performed for the solvation of xenon for 12 temperatures $290\unit{K}\leq T \leq400\unit K$ with $\Delta T = 10\unit K$. For the solvation of the other atoms, a temperature range of 11 temperatures $300\unit K\leq T\leq 500\unit K$ with $\Delta T = 20\unit K$ was investigated.

\subsection{Thermophoretic Simulations}

There already exist some algorithms to generate temperature gradients by means of simply adjusting particle momenta in thermostatted regions \cite{Mueller-Plathe:1999,Galliero:2008} or in slabs of the whole box to generate the desired temperature gradients (much like in a BD simulation) by heat exchange momentum change \cite{Ikeshoji:1994} or Nos\'e-Hoover thermostats \cite{Okumura:2006}. 

For the method investigated in this work we define thermostat regions in which position restrained molecules are thermostatted using the Nos\'e-Hoover algorithms. This is similar to experimental setups, where two regions are thermostatted to produce a steady temperature profile. The different approaches to actually perform measurements are described in \sec{\ref{sec:mdsim_section}} and some of the parameters given in the following will be introduced there.

The actual simulations were performed in a setup which will be introduced as the extended water thermostat setup, with $L_m=2.8\unit{nm}$, $L_{x,y}=2.6\unit{nm}$ and $\simeq671$ solvent molecules for the lighter noble gases and $L_m=2.5\unit{nm}$, $L_{x,y}=2.5\unit{nm}$ and $\simeq611$ solvent molecules for the xenon atom. The thermostat molecules were restrained with a spring constant of $k=5000\unit{kJ\,mol^{-1}\,nm^{-2}}$.

After steepest descent minimization, the systems equilibrated for 2ns in an NPT ensemble to obtain the right volume for the desired pressure. However the production run was in NVT. The reason for this is that we need a constant slab volume to obtain the density as a function of $z$ by means of a histogram. In hindsight, this problem may have been overcome by recording the simulation volume, too and consider its temporal evolution while evaluating the density. For the evaluation of observables, the first 10ns were omitted as additional NVT equilibration time.

The cut-off radii were chosen to be 1.15nm for the neighbor list, 1.15nm for the Coulomb cut-off, 1.05nm for the vdW-smoothening and 1.15nm for the vdW cut-off. The molecules' positions and velocities were recorded every 2ps.

After performing the simulations, the density was measured by dividing the simulation box in 200 slabs along the $z$-axis. Then for every recorded frame the center of mass of all $N_M$ molecules of molecule species $M$ was calculated and its position was binned into a histogram. Deviding by the total number of frames and $N_M$ as well as the constant slab volume yields the density in $z$-direction. The temperature was measured in a similar way, where again a molecule's $z$-value was identified as the $z$-value of its center of mass. After binning all the molecules for every frame, the temperature of molecule species $M$ in slab $S$ and spatial direction $\mu$ was calculated as a mean over all frames and all individual atoms
\begin{align}
 T_{S} = \l\langle\frac1{\kB N_{\text{df},M\in S}}\sum_{i\in (\text{atoms in }S)}\frac{(\v p_i)^2}{m_i}\r\rangle
\end{align}
following \cite{Okumura:2006,Huenenberger:2005}. Suppose there are $N_S$ SPC/E water molecules in slab $S$, then the number of degrees of freedom is $N_{\text{df},M\in S}=3\times(3N_S)-3N_S = 6N_S$, because we have $3N_S$ atoms in the slab which initially all have the 3 degrees of spatial freedom, however each water molecule has 3 constraints which have to be substracted.

% The second moment and uncertainty of the temperature in slab $S$ of direction $\mu$ is given as
% \begin{align}
%  (T^\mu_{S})^2 = \l\langle\l(\frac1{\kB N_{\text{df},M\in S}}\sum_{i\in (\text{atoms in }S)}\frac{(p^\mu_i)^2}{m_i}\r)^2\r\rangle
%   u_{T^\mu_S} &= \frac1{N_t}\sqrt{}
% \end{align}
% 
In previous works, results have been found showing that the energy flux in a system like ours is constant from ``hot'' to ``cold'' regions \cite{Ikeshoji:1994}, that a linear temperature profile is reached \cite{Galliero:2008,Armstrong:2013} and that the density is following the equation of state \cite{Okumura:2006,Armstrong:2013}, where this is connected to the condition of mechanical equilibrium \cite{Okumura:2006}.

%Temperature is induced via thermostats which couple the particle's momenta to an external heat bath of the desired temperature. 

%Another way to test our assumptions 

%% file: bilder/md_ti/LJ_parameters.tex
\begin{tabular}{l|cc|cc|c}
\hline\hline
        & \multicolumn{2}{c}{self interaction} & \multicolumn{2}{|c|}{interaction with SPC/E} & mass\\ 
        Solute & $\sigma [\unit{nm}]$ & $\epsilon [\unit{kJ/mol}]$ & $\sigma [\unit{nm}]$ & $\epsilon [\unit{kJ/mol}]$ & $m$ [u]\\
                \hline
Ar & 0.329000 & 0.817600 & 0.322450 & 0.729092 & 39.948000\\
Kr & 0.342000 & 0.951800 & 0.328950 & 0.786656 & 83.798000\\
Xe & 0.357000 & 1.071000 & 0.336450 & 0.834462 & 130.000000\\
SPC/E water & 0.315900 & 0.650166 & 0.315900 & 0.650166 & $15.9994 + 2\times1.008$\\
\hline
\end{tabular}

%% file: chapters/theory_section.tex
\chapter{Results}
\label{ch:results}

In this chapter, we will present the results of this work investigating the density of a system exposed to a temperature gradient using the methods described in the last chapter,  \ch{\ref{ch:methods}}. In \sec{\ref{sec:theory}}, we derive a theoretical description for the Soret equilibrium density by means of Brownian motion and dynamical density functional theory (DDFT), assuming that the Einstein relation \eq{\ref{eq:einstein_relation}} holds in temperature gradients. We will see that the result will depend on the choice of the Stratonovich/It\^o interpretation of the Brownian equation of motion, i.e. on the choice of the parameter $\alpha$. Furthermore we show that one can arrive at both the W\"urger \eq{\ref{eq:soret_wuerger}} and Braun \eq{\ref{eq:soret_braun}} result for the Soret coefficient by interpreting the dimensional scaling of the free energy functional differently. A hypothetical dependence of the density on a temperature dependent friction is discussed, as well. In order to test the results from this section, we will employ BD simulations in \sec{\ref{sec:bdsim}} for various systems and find that the Soret coefficient can be found to be proportional to the excess enthalpy, indicating systems in thermal gradients follow their equation of state. Taking a second route to test the theoretical derivations and to find an answer to the nature of the Soret coefficient in real systems is to employ MD simulations, which has been done in \sec{\ref{sec:mdsim_section}}, where we studied the solute density of noble gases in SPC/E water exposed to a temperature gradient. We will not find any significant results in this section.

\section{Dynamical Density Functional Theory of Local Diffusion}
\label{sec:theory}
\subsection{Temporal Evolution of Ensemble Density}
\label{sec:temporal_evolution}
Equations for the temporal dependence of the spatial distribution of interacting particles have been derived by Dean, Marconi and Tarazona in \cite{Dean:1996,Marconi:1999,Marconi:2000}. Starting with an ensemble of equations of motions for Brownian particles and using a density functional approach, their work resulted in the famous dynamical density functional theory (DDFT) for fluids, enabling the theoretical investigation of a variety of fluid systems \cite{Barrat:2003}. However, the derivations concentrated on systems of constant temperature, thus embedding a spatially constant white noise in the Brownian equations. As we are interested in equations of motions with white noise of varying intensity, it is necessary to derive the equations again, keeping in mind the diffusion coefficient's spatial dependence, $D:=D(\v r)$. However, we will stick closely to the derivation given by Dean, Marconi and Tarazona. We start with the stochastic differental equation (SDE) for a single particle $i$ of an ensemble of $N$ particles, known as the overdamped Langevin equation, given by \eq{\ref{eq:adapted_brownian}} as
\begin{align}
 \label{eq:theory_start}
\d\v r_i &=  \alpha \nabla_iD(\v r_i) -\l[ \frac{\nabla_i\Vext(\v r_i)}{\gamma} +\frac1\gamma \nabla_i\sum_{j=1}^NV(\v r_i-\v r_j)\r]\ \d t + \sqrt{2D(\v r_i)}\ \d \v B_{t,i},
\end{align}
which we remind has to be integrated following It\^o's rule, whereas the interpretation as It\^o's or Stratonovich's SDE is given by $\alpha$ being equal to 0 or 1/2, respectively.

It\^o's formula and his calculus now enable a straight-forward way to change the variables from position to an arbitrary function 
$f$ \cite{Oksendal:2010,Dean:1996}
\begin{align}
 \label{eq:change_of_variables}
 \d\l[f(\v r_i)\r] &=   \alpha D(\v r_i) -\l[ \frac{\nabla_i\Vext(\v r_i)}{\gamma} + \frac1\gamma\nabla_i\sum_{j=1}^NV(\v r_i-\v r_j)\r]\cdot \nabla_if(\v r_i)\ \d t +\NN\\
                   &\qquad + \sqrt{2D(\v r_i)}\nabla_i f(\v r_i)\ \d \v B_{t,i} +  D(\v r_i) \nabla_i^2 f(\v r_i)\ \d t.
\end{align}
In order to end with a description of the density for the whole system, we make use of the one-particle density operator \eq{\ref{eq:oneparticle_dens_operator}} and note that \cite{Dean:1996}
\begin{align}
 f(\v r_i(t)) &= \int\d^3\v r\ \delta(\v r-\v r_i(t))\ f(\v r) = \int\d^3\v r\ \hat\rho_i(\v r,t)\ f(\v r),\\
 \frac{\d f(\v r_i(t))}{\d t} &= \int\d^3\v r\p{\hat\rho_i(\v r,t)}{t} f(\v r).
\end{align}
Entering this into \eq{\ref{eq:change_of_variables}} (divided by $\d t$) and integration by parts with the conditions $\hat\rho_i=\nabla\hat\rho_i=0$ at the boundary yields the one-particle density operator evolution
\begin{align}
\p{\hat\rho_i(\v r,t)}{t} &= \nabla\l[\hat\rho_i(\v r,t)\l( - \alpha \nabla D(\v r) + \frac{\nabla\Vext(\v r)}{\gamma} + \frac{\nabla}{\gamma}\sum_{j=1}^NV(\v r-\v r_j)\r)\r] +\NN\\
                          &\qquad +\nabla^2\Big(D(\v r)\hat\rho_i(\v r,t)\Big) - \v\eta_i(t)\cdot\nabla\l(\hat\rho_i(\v r,t)\sqrt{2D(\v r)}\r).
                          \label{eq:one_particle_evolution}
\end{align}
Introducing the ensemble density operator \eq{\ref{eq:ensemble_dens_operator}} to the problem and summing \eq{\ref{eq:one_particle_evolution}} over all particles in the ensemble, we obtain
\begin{align}
\p{\hat\rho(\v r,t)}{t} &= \nabla \l[-\alpha \hat\rho(\v r,t)\nabla D(\v r) + \hat\rho(\v r,t)\frac{\nabla \Vext(\v r)}{\gamma} + \hat\rho(\v r,t)\int\d^3\v r'\ \hat\rho(\v r',t)\frac{\nabla V(\v r-\v r')}{\gamma}\r] +\NN\\
                          &\qquad +\nabla^2\Big(D(\v r)\hat\rho(\v r,t)\Big) - \nabla\cdot\sum_{i=1}^N\v\eta_i(t)\hat\rho_i(\v r,t)\sqrt{2D(\v r)}.
                          \label{eq:ensemble_density_1}
\end{align}
\enlargethispage{1cm}
The last term resembles a Gaussian noise field $\v\xi(\v x,t)$ with correlation
\begin{align}
 \l\langle\xi_\mu(\v x,t)\xi_\nu(\v y,t')\r\rangle(\d t)^2 &= \delta_{\mu\nu}\delta(t-t')\sum_{i=1}^N\nabla_x\nabla_y\l(\hat\rho_i(\v x,t)\hat\rho_i(\v y,t)\sqrt{2D(\v x)\cdot2D(\v y)}\r)\d t\NN\\
  &= 2\delta_{\mu\nu}\delta(t-t')\nabla_x\nabla_y\l(\delta(\v x-\v y)\hat\rho(\v x,t)\sqrt{D(\v x)D(\v y)}\r)\d t.\footnotemark
\end{align}
\footnotetext{As in \cite{Dean:1996} we made use of the delta function's property $\delta(\v x-\v r_i)\delta(\v y - \v r_i)=\delta(\v x-\v y)\delta(\v y - \v r_i)=\\\delta(\v x-\v y)\delta(\v x - \v r_i)$}
Similar to the argument given in \cite{Dean:1996}, one can introduce a second global noise field
\begin{align}
\label{eq:new_noise}
 \v\xi'(\v r,t) = \nabla\cdot\l(\v\eta(\v r,t)\sqrt{2\hat\rho(\v r,t)D(\v r)}\r)
\end{align}
with $\l\langle\eta^\mu_i(\v x,t)\eta^\nu_j(\v y,t)\r\rangle(\d t)^2=\delta_{\mu\nu}\delta_{ij}\delta(t-t')\delta(\v x-\v y)\d t$. Both $\v \xi$ and $\v \xi'$ have the same correlator, and therefore cast the sytem's behavior in the same manner. 

Using \eq{\ref{eq:new_noise}} to rewrite \eq{\ref{eq:ensemble_density_1}} and making use of $\rho(\v r,t)=\l\langle\hat\rho(\v r,t)\r\rangle$, we find the temporal evolution of the density to be
\begin{align}
\label{eq:density_evolution}
\p{\rho(\v r,t)}{t} &= \nabla \l[-\alpha \rho(\v r,t)\nabla D(\v r) + \rho(\v r,t)\frac{\nabla \Vext(\v r)}{\gamma} + \int\d^3\v r'\ \l\langle\hat\rho(\v r,t)\hat\rho(\v r',t)\r\rangle\frac{\nabla V(\v r-\v r')}{\gamma}\r] +\NN\\
                          &\qquad +\nabla^2\Big(D(\v r)\rho(\v r,t)\Big) .
\end{align}
The noise term in \eq{\ref{eq:ensemble_density_1}} vanished due to the independence of the diffusion coefficient field, the newly introduced noise field and the density operator. This equation is similar to the result in \cite{Marconi:1999} but includes the spatial dependence of the diffusion coefficient field as well as the choice of the It\^o or Stratonovich interpretation, respectively.
\subsection{Soret Equilibrium Density of an Ideal Gas}
\label{sec:density_id_gas}
In order to achieve a better understanding of the connection between diffusion and the Soret equilibrium density, it is useful to study a simple system, e.g. neglecting particle interaction. For the ideal gas, the interaction potential $V$ vanishes, reducing \eq{\ref{eq:density_evolution}} to
%essentially to an equation for a one particle density
%or a conditional transition probability
\begin{align}
 \label{eq:density_evolution_id_gas}
 \p{\rho(\v r,t)}{t} &= \nabla \Big[D(\v r)\nabla\rho(\v r,t) + (1-\alpha) \rho(\v r,t)\nabla D(\v r) + \rho(\v r,t)\frac{\nabla \Vext(\v r)}{\gamma} \Big].
\end{align}
In the following we will omit the spatial and temporal dependencies for the sake of readability.
Keeping in mind that the continuity equation reads
\begin{align}
  \dot \rho +  \nabla\v j = 0
\end{align}
the equation above gives the probability (or particle) flux
\begin{align}
  \label{eq:flux}
  \v j = -(1-\alpha) \rho \nabla D - D\nabla \rho - \rho\frac{\nabla \Vext}{\gamma}
\end{align}
Now, a stationary solution means a constant flux, however, in an equilibrated system (which does not necessarily have to mean thermal equilibrium) the total particle flux is zero. As we are interested in the system's behavior after equilibration, we use the equilibrium condition $\v j=0$ and find
\begin{align}
0 &= (1-\alpha) \rho \nabla D + D\nabla \rho + \rho\frac{\nabla \Vext}{\gamma}\\
    \frac{\nabla\rho}{\rho} &= -(1-\alpha)\frac{\nabla D}{D} - \frac{\nabla\Vext}{\gamma D}\\
   \nabla\ln\rho &= -(1-\alpha)\nabla\ln D -\nabla\int_{\v r_0}^{\v r} \d\tilde{\v r}\ \frac{\nabla\Vext}{\gamma D}.
\end{align}
The last term is a line integral from a reference point $\v r_0$ to the point $\v r$ over the external force field scaled by the spatially dependent diffusion coefficient. The Soret equilibrium solution reads
\begin{align}
 \label{eq:density_id_gas}
 \rho(\v r) = \frac{\mathcal N}{[D(\v r)]^{1-\alpha}}\,\exp\l(-\int_{\v r_0}^{\v r} \d\tilde{\v r}\ \frac{\nabla_{\tilde{r}}\Vext(\tilde{\v r})}{\gamma D(\tilde{\v r})}\r)
\end{align}
with $\mathcal N$ being a normalization constant. \footnote{In the following, whenever $\mathcal N$ is mentioned associated with a density $\rho(\v r)=\mathcal N f(\v r)$, it has to be read as a constant such that $\int\d^3\v r\ \rho(\v r)=N$ (with $N=1$ in association with a probability $p(\v r)$).} Indeed, this result has been derived in \cite{Oksendal:2010} in a different manner.

\subsubsection{Connection between Temperature, Diffusion Coefficient and Density}
For a vanishing external potential, the equilibrium density \eq{\ref{eq:density_id_gas}} reduces to
\begin{align}
 \label{eq:intrinsic_density}
 \rho(\v r) = \frac{\mathcal N}{[D(\v r)]^{1-\alpha}},
\end{align}
with different consequences for the choice of $\alpha$. As has been argued in \cite{Lancon:2001,Schnitzer:1993} a choice of $\alpha=1$ represents a thermal equilibrium, while the diffusion coefficient may still be spatially dependent via the mobility. This would yield the expected equilibrium density of equal probability of presence at any point in the system $\rho =\const$. However, we are more interested in non iso-thermal systems and so have to look at $\alpha=0,\frac12$. In these cases, a thermal gradient would employ a force on the particles, yielding an intrinsic temperature dependence of the density, through the temperature dependence of the diffusion coefficient. Unfortunately, there is no other obvious choice for the temperature dependence of the diffusion coefficient than the famous Einstein relation
\begin{align}
 \label{eq:Einstein_relation}
 D = \frac{\kB T}{\gamma},
\end{align}
which is derived under the condition of a spatially independent diffusion coefficient \cite{Kosztin:2000_chp4}. However, this condition arises from the general assumption that every equilibrium solution of \eq{\ref{eq:flux}} has to resemble a Boltzmann distribution
\begin{align}
 \label{eq:Boltzmann_distribution}
 \rho(\v r) = \mathcal N \exp\l(-\beta\Vext\r).
\end{align}
In order to bring this assumption in agreement with \eq{\ref{eq:density_id_gas}}, one has to find an expression for $D$, s.t.
\begin{align}
  \beta(\v r)\Vext(\v r) &= (1-\alpha)\ln D(\v r) +\int_{\v r_0}^{\v r} \d\tilde{\v r}\ \frac{\nabla_{\tilde{r}}\Vext(\tilde{\v r})}{\gamma D(\tilde{\v r})}\\
  \nabla D &= \frac{1}{1-\alpha} D \nabla(\beta\Vext)-\frac{1}{1-\alpha}\frac{\nabla\Vext}{\gamma}.  
\end{align}
This is an ordinary inhomogeneous partial differential equation of the form
\begin{align}
  \label{eq:D_PDE}
  \nabla D &= D \nabla g + \v h
\end{align}
where $g =\frac{\beta \Vext}{1-\alpha}$ and $\v h=- \frac{\nabla \Vext}{\gamma(1-\alpha)}$. Its solution is 
\begin{align}
  D(\v x) = \l[\int\limits^{\v{x}} \v h(\v x)\, e^{-g(\tilde{\v x})}\, \d \tilde{\v x} + K \r]\,e^{g(\v x)},
\end{align}
where $K$ is an arbitrary constant which will be set to zero in the following. Putting $g(\v x)$ in the equation yields
\begin{align}
  D(\v x) = \l[-\frac1{\gamma(1-\alpha)} \int\limits^{\v x} [\nabla \Vext(\tilde{\v x})]e^{-\frac{\Vext(\tilde{\v x})}{\kB T(\tilde{\v x})(1-\alpha)}}\, \d \tilde{\v x} \r]e^{-\frac{\Vext(\v x)}{\kB T(\v x)(1-\alpha)}}
\end{align}
which is the exact result. In order to investigate this result a bit further, we look at a one-dimensional system and make use of the linear nature of the temperature function as it was derived in \sec{\ref{sec:heat_equation}}. For a small temperature gradient $\epsilon\rightarrow0$ and with a prime denoting the derivative with respect to $x$ or $\tilde x$, respectively, we find
%\subsubsection{1D Diffusion Coefficient Boltzmann Distribution}
%Using this result Entering everything into the exact definition of the diffusion coefficient yields
\begin{align}
D(x) &= -\frac{1}{\gamma(1-\alpha)}\,e^{\frac{\beta\Vext}{1-\alpha}}\l[\int\limits^x\d\tilde x\ \Vext'\,e^{-\frac{\beta\Vext}{1-\alpha}}\r]\\
     &= -\frac{1}{\gamma(1-\alpha)}\,e^{\frac{\beta_0\Vext}{1-\alpha}}\l(1-\frac{\epsilon\Vext x\beta_0}{1-\alpha}\r)\times\NN\\
     &\qquad \times\l[\int\limits^x\d\tilde x\ \Vext'\,e^{-\frac{\beta_0\Vext}{1-\alpha}}+\int\limits^x\d\tilde x\ \frac{\epsilon\Vext'\Vext\tilde x\beta_0}{1-\alpha}\,e^{-\frac{\beta_0\Vext}{1-\alpha}}\r] +\mathcal O\l(\epsilon^2\r)
     \label{eq:D_inbetween}
\end{align}
Let us solve the integrals in the squared brackets first. The first term gives
\begin{align}
  \int\limits^{x} \Vext'\,e^{-\frac{\beta_0\Vext}{1-\alpha}}\d\tilde x &= -\frac{1-\alpha}{\beta_0}\,e^{-\frac{\beta_0\Vext}{1-\alpha}}
\end{align}
The second one can be rewritten and integrated by parts multiple times to obtain
\begin{align}
 \int\limits^{x} &\frac{\Vext'\Vext\epsilon\tilde x\beta_0}{1-\alpha}\,e^{-\frac{\beta_0\Vext}{1-\alpha}}\d\tilde x =
    -\Vext\epsilon x \,e^{-\frac{\beta_0\Vext}{1-\alpha}} + \epsilon\int\limits^{x} (\Vext \tilde x)'\,e^{-\frac{\beta_0\Vext}{1-\alpha}}\d\tilde x\\
    &= -\Vext\epsilon x \,e^{-\frac{\beta_0\Vext}{1-\alpha}} + \epsilon\int\limits^{x} \Vext'\tilde x\,e^{-\frac{\beta_0\Vext}{1-\alpha}}\d\tilde x +\epsilon\int\limits^{x} \Vext\,e^{-\frac{\beta_0\Vext}{1-\alpha}}\d\tilde x\\
    &= -\Vext\epsilon x \,e^{-\frac{\beta_0\Vext}{1-\alpha}} - \epsilon\frac{1-\alpha}{\beta_0}x\,e^{-\frac{\beta_0\Vext}{1-\alpha}} + \epsilon\frac{1-\alpha}{\beta_0}\int\limits^{x} e^{-\frac{\beta_0\Vext}{1-\alpha}}\d\tilde x+ \epsilon\int\limits^{x} \Vext\,e^{-\frac{\beta_0\Vext}{1-\alpha}}\d\tilde x
\end{align}
Putting everything in \eq{\ref{eq:D_inbetween}} and ignoring higher orders of $\epsilon$ gives the diffusion coefficient in linear order of $\epsilon$
\begin{align}
  D(x) &= -\frac{1}{\gamma(1-\alpha)}
          \Bigg[-\frac{1-\alpha}{\beta_0}-\epsilon x\,\frac{1-\alpha}{\beta_0}+\epsilon\,e^{-\frac{\beta_0\Vext}{1-\alpha}}\int\limits^x\d\tilde x        
              \l(\frac{1-\alpha}{\beta_0}+\Vext\r)\,e^{-\frac{\beta_0\Vext}{1-\alpha}} -\NN\\
       &\qquad\qquad\qquad\,       - \epsilon x \Vext e^{-\frac{\beta_0\Vext}{1-\alpha}} + \epsilon x \Vext e^{-\frac{\beta_0\Vext}{1-\alpha}}
          \Bigg] +\mathcal O\l(\epsilon^2\r)\\
       &= \frac{\kB T_0}{\gamma}\l[1+\epsilon\l(x-e^{\frac{\beta_0\Vext}{1-\alpha}}\int\limits^{x} \l(\frac{\beta_0\Vext}{1-\alpha}+1\r)\,e^{-\frac{\beta_0\Vext}{1-\alpha}}\d\tilde x\r)\r]+\mathcal O\l(\epsilon^2\r)
\end{align}
Note that the result enables us to investigate how this relation for the diffusion coefficient behaves in two important limits. First, it reduces to the known Einstein relation $D=\kB T_0/\gamma$ in the limit of $\epsilon\rightarrow0$. However, it reduces to the same relation with a vanishing external potential, thus impeding an intrinsic temperature dependence of the density of an ideal gas in a thermal gradient, setting it constant (this reflects the expectation from \eq{\ref{eq:Boltzmann_distribution}}). Since it is known that this should only be the case for $\alpha=1$ \cite{Lancon:2001,Schnitzer:1993}, we can conclude that \eq{\ref{eq:Boltzmann_distribution}} is not a good assumption for the Soret equilbrium density. 

It has been shown that the ideal gas Soret equilibrium density in a thermal gradient should take the form \cite{Schnitzer:1993,Dhont:2007}
\begin{align}
 \rho(\v r)=\frac{\mathcal N}{[T(\v r)]^{1-\alpha}}.
\end{align}
Using this we can assume a new Boltzmann-like distribution in case of an external potential, which includes the ideal gas term $T^{-(1-\alpha)}$ and the exponential factor from the Boltzmann distribution,
\begin{align}
 \rho(\v r)=\frac{\mathcal N}{[T(\v r)]^{1-\alpha}}\exp\big(-\beta(\v r)\Vext(\v r)\big).
\end{align}
By comparing with \eq{\ref{eq:density_id_gas}} and differentiating we find yet another defining equation for the diffusion coefficient,
\begin{align}
 -(1-\alpha)\ln D -\int\d\tilde{\v x}\frac{\nabla\Vext}{\gamma D} &= -(1-\alpha)\ln T - \frac{\Vext}{\kB T}\\
 \nabla D &= D\nabla\l(\ln T + \frac{1}{1-\alpha}\,\frac{\Vext}{\kB T}\r)-\frac{1}{1-\alpha}\frac{\nabla\Vext}{\gamma}.
\end{align}
We recognize a PDE of type \eq{\ref{eq:D_PDE}} with $g = \ln T + \frac{1}{1-\alpha}\,\frac{\Vext}{\kB T}$. For a one-dimensional ideal gas, the exact solution is
\begin{align}
  D( x) = \l[-\frac1\gamma \int\limits^{ x} \frac{\Vext'(\tilde{ x})}{T(\tilde{ x})}\, e^{-\frac{\Vext(\tilde{ x})}{\kB T(\tilde{ x})(1-\alpha)}}\, \d \tilde{ x} \r]\,T(x)\,e^{\frac{\Vext(x)}{\kB T(x)(1-\alpha)}}.
\end{align}
Again, we want to investigate this equation's behavior for small temperature gradients $\epsilon\rightarrow0$
\begin{align}
 D(x) &= \l[-\frac1{\gamma(1-\alpha)} \int\limits^{x} \frac{\Vext'}{T_0}\,\big(1-\epsilon \tilde x\big)\,e^{-\frac{{\Vext}}{\kB T_0}\frac{1-\epsilon \tilde x}{1-\alpha}}\,\d \tilde x\r]\,T_0\big(1+\epsilon x\big)\,e^{\frac{{\Vext}}{\kB T_0}\frac{1-\epsilon x}{1-\alpha}} +\mathcal O\l(\epsilon^2\r)\\
      &= \l[-\frac1{\gamma(1-\alpha)} \int\limits^{x} \frac{\Vext'}{T_0}\,\big(1-\epsilon \tilde x\big)\,e^{-\frac{{\Vext}}{\kB T_0(1-\alpha)}}\l(1+\frac{{\Vext}\epsilon \tilde x}{\kB T_0(1-\alpha)}\r)\,\d \tilde x\r]\times\NN\\
      &\qquad\qquad\qquad\times  T_0\big(1+\epsilon x\big)\,e^{\frac{{\Vext}}{\kB T_0(1-\alpha)}}\,\l(1-\frac{{\Vext}\epsilon x}{\kB T_0(1-\alpha)}\r) +\mathcal O\l(\epsilon^2\r)\\      
      &= \l[-\frac1{\gamma(1-\alpha)} \int\limits^{x} \l(\frac{\Vext'}{T_0}\,e^{-\frac{{\Vext}}{\kB T_0(1-\alpha)}}-\frac{{\Vext'}\epsilon \tilde x}{T_0}\,e^{-\frac{{\Vext}}{\kB T_0(1-\alpha)}}+\frac{{\Vext}'{\Vext}\epsilon \tilde x}{\kB T_0^2(1-\alpha)}\,e^{-\frac{{\Vext}}{\kB T_0(1-\alpha)}}
         \r)\d\tilde x\r]\times\NN\\
      &\qquad\qquad\qquad\times T_0 e^{\frac{{\Vext}}{\kB T_0(1-\alpha)}}\,\l(1+\l(1-\frac{{\Vext}}{\kB T_0(1-\alpha)}\r)\epsilon x\r) + \mathcal O\l(\epsilon^2\r).\label{eq:D_inbetween}
\end{align}
Let us solve the integrals in the squared brackets first. The first term gives
\begin{align}
  \int\limits^{x} \frac{{\Vext}'}{T_0}\,e^{-\frac{{\Vext}}{\kB T_0(1-\alpha)}}\d\tilde x &= -\kB(1-\alpha) e^{-\frac{{\Vext}}{\kB T_0(1-\alpha)}}.
\end{align}
The other two can be rewritten and yield after multiple integration by parts
\begin{align}
  \kB \epsilon\int\limits^{x}&\l(1-\alpha-\frac{{\Vext}}{\kB T_0}\r)\,\tilde x\,\l(-\frac{{\Vext}'}{\kB T_0(1-\alpha)}\r)e^{-\frac{{\Vext}}{\kB T_0(1-\alpha)}}\d\tilde x = \\
        &=-\frac{{\Vext}\epsilon x}{T_0}e^{-\frac{{\Vext}}{\kB T_0(1-\alpha)}} + \epsilon \kB \int\limits^x\frac{{\Vext}}{\kB T_0}e^{-\frac{{\Vext}}{\kB T_0(1-\alpha)}}\d\tilde x.
\end{align}
Putting everything in \eq{\ref{eq:D_inbetween}} and ignoring higher orders of $\epsilon$ gives the diffusion coefficient in linear order
\begin{align}
  D(x) &= \frac{\kB T_0}{\gamma}+\epsilon \Bigg[\frac{{\Vext}x}{\gamma(1-\alpha)}-\frac{\kB T_0{\Vext}x}{\kB T_0\gamma(1-\alpha)}+\frac{\kB T_0}{\gamma}x-\NN\\
       &\qquad\qquad\qquad\frac{\kB T_0}{\gamma(1-\alpha)}e^{\frac{{\Vext}}{\kB T_0(1-\alpha)}}\int\limits^x\frac{{\Vext}}{\kB T_0}e^{-\frac{{\Vext}}{\kB T_0(1-\alpha)}}\d\tilde x\Bigg] + \mathcal O\l(\epsilon^2\r) \\
       &=  \frac{\kB T_0}{\gamma}\l[1+\epsilon \l(x-e^{\frac{{\Vext}}{\kB T_0(1-\alpha)}}\int\limits^x\frac{{\Vext}}{\kB T_0(1-\alpha)}e^{-\frac{{\Vext}}{\kB T_0(1-\alpha)}}\d\tilde x\r)\r]   + \mathcal O\l(\epsilon^2\r).
\end{align}
Note that the result shows neat behavior in two important limits. First, it reduces to the known Einstein relation $D=\kB T/\gamma$ in the limit of $\epsilon\rightarrow0$. Second, with a vanishing external potential, it yields $D(x) = \kB T(x)/\gamma$, in contrast to the result of the plain Boltzmann distribution.

The virtual non-applicability of the Einstein relation due to the diffusion coefficient's spatial dependence, emerging from the fluctuation-dissipation-theorem (FDT) as \cite{Kosztin:2000_chp4},
\begin{align}
 \nabla D = \nabla\Vext\l(D\beta-\gamma^{-1}\r)
\end{align}
is almost generally ignored in the literature. Remarkably, there have been indications that it can be used nevertheless  \cite{Galliero:2008,Astumian:2007}, implying that the classical FDT is incomplete and rather a special case of a general FDT. Looking at \eq{\ref{eq:density_id_gas}}, we can interpret it as a transition probability for a particle at location $\v r_0$ to diffuse to location $\v r$. In thermal equilibrium, this probability is given in terms of the energy difference at both positions, $\exp\big(-\beta(\Vext(\v r)-\Vext(\v r_0)\big)$. In thermally non-equilibrated systems and assuming the Einstein relation to be valid, one obtains the exponential factor of \eq{\ref{eq:density_id_gas}},
\begin{align}
 \exp\l(-\int_{\v r_0}^{\v r} \d\tilde{\v r}\ \frac{\nabla_{\tilde{r}}\Vext(\tilde{\v r})}{\kB T(\tilde{\v r})}\r),
\end{align}
an integral over the external force field scaled by the temperature which reduces to the simple energy difference in case of spatially independent temperature. It can be shown that this may be generalized from a scaled force field and the location to arbitrary thermodynamic force fields $X$ and conjugated quantities $\alpha_X$ \cite{Astumian:2007}. However, just recently there has been published another view by the group of Kroy \cite{Kroy:2010,Kroy:2014}. Studying the motion of a Brownian particle carrying its own temperature field and treating the process as non-Markovian with time-dependent friction, they found an oscillation-dependent temperature field in the FDT in reciprocal time space, meaning that the Brownian motion is induced by different temperatures for ``fast'' and ``slow'' fluctuations, even at the same position. There are indications that this behavior can be generalized to static temperature fields \cite{Kroy:2014} which suggests that the simple assumption $D(x)=\kB T(x)/\gamma$ is not exactly valid.  

We will nevertheless stick with the assumption that the Einstein relation holds even in systems with temperature gradients in the following, hoping the error made thereby is not of critical value. Furthermore, to be consistent with the notation in the literature, we fix the physical scale by setting $\kB = \gamma = 1$ and thus work with
\begin{align}
 \label{eq:D_is_T}
 D(\v r) = T(\v r)
\end{align}
in the following (giving diffusion coefficient and temperature the dimension of energy). This yields an ideal gas equilibrium density of
\begin{align}
 \label{eq:density_id_gas_of_T}
 \rho(\v r) = \frac{\mathcal N}{[T(\v r)]^{1-\alpha}}\,\exp\l(-\int_{\v r_0}^{\v r} \d\tilde{\v r}\ \frac{\nabla_{\tilde{r}}\Vext(\tilde{\v r})}{T(\tilde{\v r})}\r),
\end{align}
which reduces to the classical Boltzmann distribution in thermally equilibrated systems. Note that in abscence of an external potential, this resembles exactly the equation of state of an ideal gas for $\alpha=0$, $\rho T=\Pi=\const$. In case of $\alpha=1/2$ we seem to obtain a rather odd dependence $\rho\sqrt T = \tilde \Pi =\const$. Indeed the exponential factor has been found under certain approximations in \cite{Wojnar:2002} for the gravitational field, whereas the author discusses the influence of the external potential on the heat equation and thus the temperature profile more detailed.

Remarkably, when applying the Einstein relation to a situation where the friction coefficient $\gamma$ depends on space, too, the derivation given in \sec{\ref{sec:temporal_evolution}} stays the same with the only change $\gamma\rightarrow\gamma(\v r)$. Usually for this case, the parameter $\alpha$ is set to $\alpha=1$ \cite{Schnitzer:1993,Lancon:2001,Chau:2010}, hence cancelling every dependence of $\rho$ on transport quantities. However, the friction can be suspected to be indirectly dependent on space through the spatially dependent temperature, for instance given by the temperature dependent viscosity of the solvent as in \eq{\ref{eq:stokes_friction}}. In this case, a choice of $\alpha=0,\frac12$ would be justified, with \eq{\ref{eq:density_id_gas}} yielding a Soret equilibrium density of 
\begin{align}
 \label{eq:density_id_gas_of_T_with_friction}
 \rho(\v r) = \mathcal N\l[\frac{\gamma(T(\v r))}{\kB T(\v r)}\r]^{1-\alpha}\,\exp\l(-\int_{\v r_0}^{\v r} \d\tilde{\v r}\ \frac{\nabla_{\tilde{r}}\Vext(\tilde{\v r})}{T(\tilde{\v r})}\r),
\end{align}

Note that while in a one-dimensional system the integral in this equation and \eq{\ref{eq:density_id_gas_of_T}} can be solved (at least numerically), it becomes path-dependent in higher dimensions, implying that there does not exist an equilibrium solution \cite{Schnitzer:1993}. However, there exists a framework to overcome this obstacle, as will be shown in \sec{\ref{sec:2D_HO}}. 

%\subsubsection{Density for Einstein-Relation Diffusion}

\subsection{Adjusted Ideal Gas Free Energy Functional}
The often cited Helmholtz free energy functional for the ideal gas,
\begin{align}
 \label{eq:energy_functional}
 \mathcal F_{\text{id,ext}}[\rho] = \kB T \int\d^3\v r\ \rho(\v r)\l\{\ln\l(\rho(\v r)\Lambda^3\r)-1\r\} + \int\d^3\v r\ \rho(\v r)\Vext(\v r),
\end{align}
alongside with its functional derivative, giving \cite{Marconi:1999,Marconi:2000,Dzubiella:2003}
\begin{align}
 \p{\rho(\v r,t)}{t} &= \nabla\l(\rho(\v r,t)D(\v r)\nabla\frac{\delta\beta\mathcal F[\rho]}{\delta \rho}\r),\\
 \v j (\v r,t) &= -\rho(\v r,t)D(\v r)\nabla\frac{\delta\beta\mathcal F[\rho]}{\delta \rho},
 \label{eq:functional_derivative}
\end{align}
breaks down as soon as we try to use it in the context of DDFT with spatially dependent diffusion to yield \eq{\ref{eq:density_evolution_id_gas},\ref{eq:flux}}. Hence, in this subsection I want to suggest an adjusted free energy functional which generates the right density equation of motion in the presence of a thermogradient and reduces to \eq{\ref{eq:energy_functional}-\ref{eq:functional_derivative}} in thermal equilibrium.

We notice that the gradient of the functional derivative yields
\begin{align}
 \label{eq:old_functional_derivative}
 \nabla\frac{\delta\beta\mathcal F_{\text{id,ext}}[\rho]}{\delta \rho}=- \frac{3}{2}\frac{\nabla T}{T} + \kB \frac{\nabla\rho}{\rho} + \nabla(\beta\Vext),
\end{align}
since the thermal wavelength depends on $T$ via $\Lambda^3\propto T^{-3/2}$ (c.f. \eq{\ref{eq:thermal_wavelength}}). In this equation, only the second term fits our expectation of \eq{\ref{eq:density_evolution_id_gas}}. The first term has the right structure but lacks the correct prefactor.
%In order to get rid of the first term, we notice the necessity of a dimensionless functional $\overline{\mathcal F}$ and thus scale it with $\beta$ before taking the derivative (implying that subsequently we multiply with $T$ to keep the structure of the result consistent).
Furthermore, the scaling of the functional with $\beta$ gives rise to an unwanted term of $\Vext\nabla\beta$ that can only be avoided if the scaling of the external part of the functional is done by integrating the force field scaled by $\beta$, yielding the external part of the functional 
\begin{align}
 \overline{\mathcal F}_\text{ext}[\rho] &= \int\d^3\v r\ \rho(\v r)\int\limits^{\v r}\d\tilde{\v r} \frac{\nabla\Vext(\tilde{\v r})}{T(\tilde{\v r})}.
\end{align}
To acknowledge this different scaling technique we replace the dimensionless functional $\beta\mathcal F$ by the general dimensionless functional $\overline{\mathcal F}$ in the following.

The first term in \eq{\ref{eq:old_functional_derivative}} is rather odd since it gives a completely different prefactor as the expectation $1-\alpha$ which is needed to represent \eq{\ref{eq:density_evolution_id_gas}} correctly, induced by the thermal wavelength. Therefore, following \cite{Dzubiella:2013}, I will shortly derive the temperature dependence of the ideal gas density from the classical partition function by means of weighting with a Boltzmann factor and show that this is not the right approach to obtain an insightful result. We start with a single particle in a three-dimensional box of volume $V=L^3$. The partition function is given as
\begin{align}
 Z = \frac{1}{h^3}\int\d^3\v r\int\d^3\v p \exp\l(-\frac{\v p^2}{2m\kB T(x)}\r).
\end{align}
Note that we assume local equilibrium at every box slice $x$ s.t. we can use the ansatz of a Boltzmann weighting which is only applicable for equilibrated systems. For the momentum part this a simple Gaussian integral. Note that the spatial integrals in $y$ and $z$ have the integrand 1, yielding $L^2$. Assuming a linear temperature profile $T=T_0(1+\epsilon x)$ we obtain
\begin{align}
    Z &= \frac{L^2}{h^3} \int\d x \l(\sqrt{2m\pi\kB T(x)}\r)^3 = \frac{L^2}{h^3} \l(\sqrt{2m\pi\kB T_0}\r)^3\int\d x \l(1+\epsilon x\r)^{3/2}\\
      &= \frac{L^2}{h^3}\frac{\l(\sqrt{2m\pi\kB T_0}\r)^3}{T_0}\frac{2}{5\epsilon}\l(T_\text{max}^{5/2}-T_\text{min}^{5/2}\r).
\end{align}
The density is then
\begin{align}
 \rho(\v r) &= \l\langle\delta(\v r -\v r')\r\rangle = \frac{1}{Zh^3}\int\d^3\v r'\int\d^3\v p\ \delta(\v r -\v r')\exp\l(-\frac{\v p^2}{2m\kB T(x')}\r)\\
            &= \frac{1}{Zh^3}\int\d^3\v r'\ \delta(\v r -\v r')\l(\sqrt{2m\pi\kB T(x)}\r)^3\\
            &= \frac{5\epsilon T_0}{2L^2}\frac{[T(x)]^{3/2}}{T_\text{max}^{5/2}-T_\text{min}^{5/2}}.
\end{align}
This would imply that an ideal gas in a temperature gradient has the tendency to accumulate at the hot boundary and to avoid the cold, which is completely counterintuitive and not in agreement with experiments. Most probably this indicates the failure of the local equilibrium assumption for the partition function. Therefore it does not seem to make sense to use the ideal gas equilibrium result in the functional either. I suggest to replace the original argument of the logarithmic term in \eq{\ref{eq:energy_functional}} as
\begin{align}
 \rho\Lambda^3 \rightarrow \frac{\rho T^{1-\alpha}}{[\Lambda(T_0)]^{-3}T_0^{1-\alpha}},
\end{align}
which has different consequences for the choice of $\alpha$.
\begin{itemize}
 \item With $\alpha=1$ or $T=T_0=\const$ the new normalization reduces to the original result $\rho\Lambda^3_0$, as it is necessary for thermal equilibrium.
 \item With $\alpha=0$, we obtain an argument of 
    \begin{align}
     \frac{\rho T}{[\Lambda(T_0)]^{-3} T_0} = \frac{\rho\kB T}{[\Lambda(T_0)]^{-3}\kB T_0} = \frac{\Pi}{\Pi_0},
    \end{align}
    where $\Pi$ is the current ideal gas osmotic pressure and $\Pi_0$ is a reference pressure for one ideal gas particle in thermal equilibrium at temperature $T_0$.
 \item With $\alpha=\frac12$, the argument becomes
    \begin{align}
     \frac{\rho \sqrt T}{[\Lambda(T_0)]^{-3} \sqrt{T_0}},
    \end{align}    
    referring to an hypothetical equation of state $\rho\sqrt{T}=\tilde \Pi=\const$.
\end{itemize}
In the following, we will omit the normalization and simply write $\ln\l(\rho T^{1-\alpha}\r)$.

Overall and using \eq{\ref{eq:D_is_T}} we obtain the adjusted functional and flux
\begin{align}
 \label{eq:energy_functional_mod}
 \overline{\mathcal F}_\text{id,ext}[\rho] &= \int\d^3\v r\ \rho(\v r)\l\{\ln\l(\rho(\v r)T^{1-\alpha}(\v r)\r)-1\r\} + \int\d^3\v r\ \rho(\v r)\int\limits^{\v r}\d\tilde{\v r} \frac{\nabla\Vext(\tilde{\v r})}{T(\tilde{\v r})},\\
 \v j (\v r,t) &= -\rho(\v r,t)T(\v r)\nabla\frac{\delta\overline{\mathcal F}[\rho]}{\delta \rho}.
 \label{eq:functional_derivative_mod}
\end{align}
\subsection{Soret Equilibrium Density of Interacting Particles}
\label{sec:interacting_particles}
\subsubsection{Homogeneous Systems}
A rather difficult task is to identify the behavior of the system as soon as one introduces the interaction potential $V(r)$. The time dependent two point density-density correlation function
\begin{align}
 \rho^{(2)}(\v r,\v r',t) = \l\langle\hat\rho(\v r,t)\hat\rho(\v r',t)\r\rangle
\end{align}
is an unknown quantity and cannot be derived using \eq{\ref{eq:density_evolution}}, as it is non-closed in $\hat\rho$. Solving for it would result in an equation dependent on the three point correlation function and so on, resulting in an infinite hierarchy \cite{Marconi:1999}. Nevertheless, it has been shown that the time independent equilibrium correlation function 
\begin{align}
 c^{(2)}(\v r,\v r') = -\beta\frac{\delta^2 \Fexc}{\delta \rho_0(\v r)\delta \rho_0(\v r')}
\end{align}
(at to $\rho(\v r,t)$ equivalent equilibrium density $\rho_0(\v r)$) is a reasonable approximation for the actual correlation function \cite{Marconi:1999,Dzubiella:2003}, where $\Fexc$ is the generally unknown excess free energy functional at equilibrium. We, too, will follow this road and assume that the equilbrium correlation function is a good approximation in Soret equilibrium. As shown in \cite{Marconi:1999}, the corresponding integral given in \eq{\ref{eq:density_evolution}} can be approximated as
\begin{align}
 \v j_\text{exc}(\v r,t) = -\int\d\v r' \l\langle\hat\rho(\v r,t)\hat\rho(\v r',t)\r\rangle \nabla V (\v r- \v r') = -\rho(\v r,t) T(\v r) \nabla \frac{\delta\bFexc}{\delta\rho(\v r,t)},
\end{align}
with $\bFexc$ being the dimensionless version of $\Fexc$ (scaled with $\beta$). This poses a problem, as it is not known how exactly to scale the functional. The natural choice were a simple multiplication with $\beta$ as it has been performed for the ideal gas free energy functional in the last section. However from the external potential contribution we know that a curve integral over a temperature scaled force field were another valid option. In formulae, we get 
\begin{align}
 \bFexc^{(H)}[\rho(\v r)] &= \beta(\v r)\Fexc[\rho(\v r)],\\
 \bFexc^{(S)}[\rho(\v r)] &= \int\limits^{\v r}\d\tilde{\v r}\beta(\tilde{\v r}) \nabla \Fexc[\rho(\tilde{\v r})],
\end{align}
(with the superscripts being explained below) and consequentially the flux contributions
\begin{align}
 \v j_\text{exc}^{(H)}(\v r,t) &= -\rho(\v r,t)T(\v r) \nabla \frac{\delta(\beta \Fexc)}{\delta\rho(\v r,t)} \\
                               &= -\rho(\v r,t)T(\v r) \l(\frac{\delta\Fexc}{\delta\rho(\v r,t)}\nabla\beta+\beta\nabla\frac{\delta\Fexc}{\delta\rho(\v r,t)}\r)\\
 \v j_\text{exc}^{(S)}(\v r,t) &= -\rho(\v r,t)T(\v r) \nabla \frac{\delta}{\delta\rho(\v r,t)}\int\limits^{\v r}\d\tilde{\v r}\beta(\tilde{\v r}) \nabla \Fexc[\rho(\tilde{\v r})] \\
 &= -\rho(\v r,t)T(\v r) \nabla \int\limits^{\v r}\d\tilde{\v r}\beta(\tilde{\v r}) \nabla \frac{\delta\Fexc}{\delta\rho(\v r,t)} \\
                               &= -\rho(\v r,t)T(\v r) \l(\beta\nabla\frac{\delta\Fexc}{\delta\rho(\v r,t)}\r).                       
\end{align}
Identifying the functional derivative as the excess chemical potential $\muexc=\delta\Fexc/\delta\rho$ at density $\rho(\v r,t)$ and temperature $T(\v r)$ and making use of the thermodynamic relations for excess entropy $\Delta S$ and excess enthalpy $\Delta H$ \eq{\ref{eq:thermodynamic_relations}} we find
\begin{align}
  \v j_\text{exc}^{(H)} &= -\rho T\l(-\muexc\frac{\nabla T}{T^2}+\frac1T\p{\muexc}{T}\nabla T\r)\\
                                &= \rho \beta\Delta H\nabla T,\\
  \v j_\text{exc}^{(S)} &= -\rho T\l(\frac1T\p{\muexc}{T}\nabla T\r)\\
                                &= \rho \Delta S\nabla T.
\end{align}
As one can see, the superscripts denote on which thermodynamic quantity the flux depends. Note that the crucial assumption here is \textit{local equilibrium}, i.e. that the system locally acts like a reference system with the same density and temperature at equilibrium. The total functionals then read
\begin{align}
   \overline{\mathcal F}^{(H)}[\rho] &= \int\d^3\v r\ \rho(\v r)\l\{\ln\l(\rho(\v r)T^{1-\alpha}(\v r)\r)-1\r\} + \int\d^3\v r\ \rho(\v r)\int\limits^{\v r}\d\tilde{\v r} \frac{\nabla\Vext(\tilde{\v r})}{T(\tilde{\v r})} +\NN\\
                                   &\qquad +\frac{1}{T}\Fexc[\rho(\v r)],\\
   \overline{\mathcal F}^{(S)}[\rho] &= \int\d^3\v r\ \rho(\v r)\l\{\ln\l(\rho(\v r)T^{1-\alpha}(\v r)\r)-1\r\} + \int\d^3\v r\ \rho(\v r)\int\limits^{\v r}\d\tilde{\v r} \frac{\nabla\Vext(\tilde{\v r})}{T(\tilde{\v r})} +\NN\\
                                   &\qquad +\int\limits^{\v r}\d\tilde{\v r} \frac{1}{T}\nabla \Fexc[\rho(\tilde{\v r})].
\end{align}
Note that the derivation in this section is purely based on standard DDFT. There exist frameworks which may treat the problem in a more natural way, by means of a power functional \cite{Schmidt:2013} or a generalized non-equilbrium functional \cite{Wittkowski:2012}.

\subsubsection{Binary Mixture of Solvent and Dilute Solutes}
Suppose we are interested in the flux and spatial density of a solute $u$ in an explicit solvent $v$. Both underlie a temperature gradient employed by the implicit solvent and the packing fraction is small enough to assume that the system consists of exactly one solute particle embedded in $N_v\gg1$ solvent particles. The different groups of particles interact via the interaction potentials $V_{uu}(r)$, $V_{uv}(r)$, $V_{vv}(r)$, where we can set $V_{uu} = 0$ since $N_u=1$. Then the density evolution of solute and solvent, with the densities denoted by $u(\v r,t)$ and $v(\v r,t)$, is given via the coupled PDEs
\begin{align}
  \p{v(\v r,t)}{t} &= \nabla \Big[(1-\alpha) v(\v r,t)\nabla T(\v r) + T(\v r)\nabla v(\v r,t) + v(\v r,t) \nabla \Vext(\v r)  + \NN\\
                        &\qquad\qquad     + \int\d^3\v r'\ \l\langle\hat v(\v r,t)\hat u(\v r',t)\r\rangle\nabla V_{uv}(\v r-\v r') + \NN\\
                        &\qquad\qquad     + \int\d^3\v r'\ \l\langle\hat v(\v r,t)\hat v(\v r',t)\r\rangle\nabla V_{vv}(\v r-\v r')  \Big] \label{eq:ev_of_v}\\                          
  \p{u(\v r,t)}{t} &= \nabla \Big[(1-\alpha) u(\v r,t)\nabla T(\v r) + T(\v r)\nabla u(\v r,t) + u(\v r,t) \nabla \Vext(\v r) + \NN\\
                        &\qquad\qquad     + \int\d^3\v r'\ \l\langle\hat u(\v r,t)\hat v(\v r',t)\r\rangle\nabla V_{uv}(\v r-\v r')  \Big]
\end{align}
Now, at every point of the system, since $N_v\gg N_u$, we assume that the evolution of $v$ is governed by the external potential and the temperature gradient and ignore the binary interaction term in \eq{\ref{eq:ev_of_v}}. This allows us to first evaluate the density profile of the solvent up to equilibrium $\v j_v = 0$, then approximate the correlator of solute and solvent via the equilibrium free energy functional derivative at solvent density $v$. The solute flux is approximately
\begin{align}
  \label{eq:solute_flux}
  \v j^{(\#)}_u(\v r, t) &= -u(\v r,t)T(\v r)\nabla \frac{\delta \overline{\mathcal F}_\text{id,ext}}{\delta u(\v r, t)} -u(\v r,t)T(\v r)\nabla \frac{\delta \overline{\mathcal F}^{(\#)}_\text{exc}}{\delta v(\v r)},
\end{align}
where now at every point $\v r$, the derivative $\delta \overline{\mathcal F}^{(\#)}_\text{exc}/\delta v$ is the free energy needed to insert one solute particle in an equilibrated system at reference density $v_0 = v(\v r)$ and temperature $T_0 = T(\v r)$.

Still, the choice of the scaling of $\mathcal F$ poses a problem to predict the thermal diffusion coefficient $\DT$. A possibility to investigate its nature is to measure the steady state equilibrium density $u(\v r)$. Since at Soret equilibrium the flux vanishes ($\v j_u = 0$) and the derivative of the ideal/external part of the functional is known, the only remaining unknown part in
\begin{align}
  u^{(\#)}(\v r) &= \frac{\mathcal N}{T^{1-\alpha}}\,\exp\l(-\int\limits^{\v r}\d\tilde{\v r} \frac{\nabla\Vext(\tilde{\v r})}{T(\tilde{\v r})} - \frac{\delta \overline{\mathcal F}^{(\#)}_\text{exc}}{\delta v}\Bigg|_{v = v(\v r)}\r)
\end{align}
is $\delta \overline{\mathcal F}^{(\#)}_\text{exc}/\delta v$. We can proceed as follows.

We can calculate the solvation free energy $\Delta G$ for a single solute particle (then equal to the excess chemical potential) in a solvent of density $v_0=v(\v r)$ at temperature $T_0 = T(\v r)$ in thermally equilibrated systems, yielding a curve $\Delta G(T)$. From this we can predict a curve for the equilibrium density for each case of $\bFexc^{(\#)}$, in abscence of an external potential given as
\begin{subequations}
\label{eq:solute_density_theory}
\begin{align}
   u^{(H)}(\v r) &= \frac{\mathcal N}{T^{1-\alpha}}\,\exp\Big(-\beta(\v r)\Delta G[T(\v r)]\Big)\\
   u^{(S)}(\v r) &= \frac{\mathcal N}{T^{1-\alpha}}\,\exp\l(-\int\limits^{\v r}\d\tilde{\v r} \frac{1}{T(\tilde{\v r})}\nabla_{\tilde{\v r}} \Delta G(T(\tilde{\v r}))\r)\NN\\
                 &= \frac{\mathcal N}{T^{1-\alpha}}\,\exp\l(-\int\limits^{T(\v r)}\d\tilde T \frac{1}{\tilde T} \p{\Delta G(\tilde T)}{\tilde T}\r),
\end{align}
\end{subequations}
which we can compare to the densities obtained from experiments. 

For a second route we can use \eq{\ref{eq:free_energy_fit_function}} to establish fit functions for the solute densities \eq{\ref{eq:solute_density_theory}}. First we find
\begin{align}
   \beta(\v r)\Delta G(T) &= \frac{a}{T} + b + c T + d \ln T\\
   \int\limits^{T}\d\tilde T \frac{1}{\tilde T} \p{\Delta G(\tilde T)}{\tilde T} &= (b+d)\ln T+2cT+\frac12 d\ln^2T.
\end{align}
Putting this in the solute density functions yields
\begin{subequations}
\label{eq:density_fit_functions}
    \begin{align}
    \label{}
    u^{(H)}(\v r) &= \frac{\mathcal N}{[T(\v r)]^{1-\alpha+d}}\,\exp\Big(-\frac{a}{T(\v r)} -cT(\v r) \Big)\\
    u^{(S)}(\v r) &= \frac{\mathcal N}{[T(\v r)]^{1-\alpha+b+d}}\,\exp\l(-2cT(\v r)-\frac12d\ln^2[T(\v r)]\r).
    \end{align}
\end{subequations}    
We see that for the enthalpic case the fit parameter $b$ disappears due to the normalization. The same happens for the fit parameter $a$ in the entropic case. Hence, both fits only provide enough information for the specific heat of solvation, as well as the enthalpy for the first case and the entropy for the second. The pure solvation free energy is not possible to obtain from density fits.

\subparagraph{Temperature Dependent Friction} If we hypothetically included a temperature dependent friction in the derivation and let it not cancel out by an additional $\alpha=1$ term in \eq{\ref{eq:theory_start}}, we would arrive at equilbrium densities that are proportional to said friction, c.f. \eq{\ref{eq:density_id_gas_of_T_with_friction}}, i.e.
\begin{subequations}
\label{eq:solute_density_theory_with_friction}
\begin{align}
   u^{(H)}(\v r) &= \mathcal N\l[\frac{\gamma(T(\v r))}{\kB T(\v r)}\r]^{1-\alpha}\,\exp\Big(-\beta(\v r)\Delta G[T(\v r)]\Big)\\
   u^{(S)}(\v r) &= \mathcal N\l[\frac{\gamma(T(\v r))}{\kB T(\v r)}\r]^{1-\alpha}\,\exp\l(-\int\limits^{T(\v r)}\d\tilde T \frac{1}{\tilde T} \p{\Delta G(\tilde T)}{\tilde T}\r).
\end{align}
\end{subequations}
Admittedly, this has the inelegance of a density involving a transport quantity, which is a rather non-physical picture. Nevertheless, we will investigate this form, as it gives an additional hypothetical term for the Soret coefficient.

\subsubsection{Soret Coefficient}
Applying \eq{\ref{eq:soret_from_density}} to the found densities \eq{\ref{eq:solute_density_theory}}, we find the Soret coefficients
\begin{subequations}
\label{eq:soret_definitions}
\begin{align}
  \label{eq:soret_enthalpy}
   \ST^{(H)} &= \frac {1-\alpha}{T} - \frac{\beta\Delta H}{T}\\
  \label{eq:soret_entropy}
   \ST^{(S)} &= \frac {1-\alpha}{T} - \beta\Delta S
\end{align}
\end{subequations}
For $\alpha=0$, those are the hypotheses found by W\"urger \eq{\ref{eq:soret_wuerger}} and Braun \eq{\ref{eq:soret_braun}}, where the W\"urger hypothesis now includes the intrinsic ideal gas term.

Looking at the artificial densities \eq{\ref{eq:solute_density_theory_with_friction}}, we find another contributing hypothetical term from the friction
\begin{align}
 \label{eq:soret_fric}
 \ST^{(\text{fric})} = -\frac {1-\alpha}\gamma \p{\gamma}{T}.
\end{align}

\subsubsection{Estimation of the Local Force on a Solute in an Ideal Gas Solvent}
We can try to predict the behavior of the solute density by investigating a model system consisting of a solute and an ideal gas solvent in a one-dimensional box. The system is set up with the solute being located at $x=0$ and a temperature profile of $T(x) = 1+\epsilon x$. We assume that the interaction potential $V(r)$ and the interaction force $\nabla V(r)$ between solute and solvent are short ranging and vanish in sufficient distance to the boundaries. The interaction potential acts like an external potential on the solvent, which is why we assume the solvent density $v(x)$ to follow \eq{\ref{eq:density_id_gas_of_T}}. We furthermore limit ourselves to $\alpha=0$. The force on the solute $F$ is then
\begin{align}
    F &= -\int\limits_{-\infty}^{+\infty}\d x v(x)\nabla V(x) = -\mathcal N \int\d x \frac{\nabla V(x)}{T(x)}\exp\l(-\int\limits^x_0\d \tilde x\frac{\nabla V(\tilde x)}{T(\tilde x)}\r)\\
      &= \mathcal N \int\d x \nabla\exp\l(-\int\limits^x_0\d \tilde x\frac{\nabla V(\tilde x)}{T(\tilde x)}\r)
       = \mathcal N \exp\l(-\int\limits^x_0\d \tilde x\frac{\nabla V(\tilde x)}{T(\tilde x)}\r)\Bigg|_{-\infty}^{+\infty}.
\end{align}
For small gradients $\epsilon\rightarrow0$ we can expand this solution in orders of $\epsilon$, using the approximation of \eq{\ref{eq:invers_T_approx}}, which yields
\begin{align}
    F &= \mathcal N \exp\l(-\frac{V(x)}{T_0}+\frac{\epsilon}{T_0}\int\limits^x_0\d \tilde x\ \tilde x\nabla V(\tilde x)+\mathcal O(\epsilon^2)\r)\Bigg|_{-\infty}^{+\infty}\\
      &= \mathcal N e^{-V(x)/T_0}\l(1+\frac{\epsilon}{T_0}\int\limits^x_0\d \tilde x\ \tilde x\nabla V(\tilde x)+\mathcal O(\epsilon^2)\r)
         \Bigg|_{-\infty}^{+\infty}\\
      &= \mathcal N e^{-V(x)/T_0}\l(1+\frac{\epsilon}{T_0}\l(V(x)x\int\limits^x_0\d \tilde x\ V(\tilde x)\r)+\mathcal O(\epsilon^2)\r)
         \Bigg|_{-\infty}^{+\infty}\\
      &= - \mathcal N\frac{\epsilon}{T_0}\int V(x) \d x+\mathcal O(\epsilon^2).
\end{align}
In first order of $\epsilon$, the local force on the solute is proportional to the total interaction energy of a ghost solute in a thermally equilibrated solvent (meaning that the solvent is equally distributed around the solute). This indicates that the local force and thus the local flux is proportional to the enthalpy rather than the entropy as the first is a measure of interaction energy. However, the result is not the actual enthalpy which would be
\begin{align}
  H &= \int V(x) e^{-\beta_0V(x)}\d x
\end{align}

\subsubsection{Connection between \eq{\ref{eq:solute_density_theory}} and the Equation of State}
As we have seen in \sec{\ref{sec:density_id_gas}}, in abscence of an external potential and for $\alpha=0$, the supposed density just yields the equation of state $\rho T =\const$. One can wonder if that is still the case for interacting particles, i.e. \eq{\ref{eq:solute_density_theory}} for a homogenous system (where the solute density $u$ becomes $\rho$ and the excess free energy becomes $\Delta G(v,T)\rightarrow\Delta G(\rho,T)$). Here, we will indicate that this is indeed the case. Suppose we have a homogeneous thermodynamic system of low density and/or high temperature, such that we can express the equation of state as \eq{\ref{eq:virial_eq_of_state}} and have
\begin{align}
\label{eq:rho_from_virial}
 \frac1\rho = \frac{\kB T}{P}\l(1+B_2\rho\r).
\end{align}
If we now take the inverse of \eq{\ref{eq:solute_density_theory}a} and expand the exponential factor subsequently (because $\beta\rightarrow0$), we get
\begin{align}
\label{eq:rho_from_thermo}
 \frac1\rho = \frac{\kB T}{\mathcal N}\exp(\beta\Delta G) = \frac{\kB T}{\mathcal N}(1+\beta\Delta G).
\end{align}
Since $\mathcal N$ is a constant, we identify it as the constant pressure $P$. Now we recall that in a high temperature limit, we found the excess free energy \eq{\ref{eq:free_energy_from_gofr_high_temp}}, 
\begin{align}
 \beta\Delta G = \frac{\rho(T)}{2}\int\d^3\v r\l(1- e^{-\beta V(r)}\r)
\end{align}
Comparing this with the definition of the virial coefficient \eq{\ref{eq:low_dens_B2}}, 
\begin{align}
\label{eq:low_dens_B2}
 B_2 = \frac12 \int\d^3\v r \l(1-e^{-\beta V(r)}\r),
\end{align}
indeed we find
\begin{align}
 \rho B_2 = \beta\Delta G,
\end{align}
indicating that both expressions \eq{\ref{eq:rho_from_virial}} and \eq{\ref{eq:rho_from_thermo}} are equivalent and the thermopho\-retic system follows the equation of state for $\gamma=\const$ and $\alpha=0$. Note that this is only the case for $\rho=\rho^{(H)}$, i.e. the Soret coefficient being proportional to the excess enthalpy.

\subsection{Summary}
In this section we found various analytical descriptions for a system's and a solute's Soret equilibrium density when exposed to a thermal gradient in the framework of Brownian motion and dynamical density functional theory, where we differentiated between the application of It\^o- and Stratonovich-calculus with a model parameter $\alpha=0,\frac12$. Both hypotheses for the Soret coefficient of a solute in a solvent described in \sec{\ref{sec:soret_coefficient}} are connected to a special case of scaling the solvation free energy with the temperature field and yield different Soret equilibrium densities. The ideal gas contribution to the Soret coefficient is found to be dependent on $\alpha$. An additional friction factor for the Soret coefficient was found in case of a temperature dependent mobility of a Brownian particle. Estimating the local force acting on a solute in an ideal gas solvent, we found a connection to the total interaction energy between solute and undisturbed system, indicating that the Soret coefficient may be proportional to the solvation free enthalpy. Assuming that the Soret coefficient is proportional to the solvation free enthalpy, we found indications that the derived Soret equilibrium density is following the equation of state.

%% file: chapters/bd_section.tex
\section{Validation of Analytical Soret Equilibrium Density Distributions along Thermal Gradients by Implicit-Solvent BD Simulations}
\label{sec:bdsim}
In this section, we want to test our findings from \sec{\ref{sec:theory}} by means of Brownian dynamics (BD) simulations, as described in \sec{\ref{sec:meth_bd}}. In the last section we found analytical Soret equilibrium density descriptions depending on the choice of the It\^o/Stratonovich parameter $\alpha$, the binary choice of considering temperature induced friction and the different functional scaling measures (yielding dependence on the solvation enthalpy $\Delta H$ or the the solvation entropy $\Delta S$, respectively). Because the first two choices have to be inserted in the equations of motions explicitly, this section only aims at answering the open question of the functional scaling. While we consider all the systems for both $\alpha=0$ and $\alpha=1/2$, the friction will be set constant. We will furthermore assume the validity of Einstein's relation. Additionally, we test the validity of the assumed ideal gas Soret equilibrium density as derived in \sec{\ref{sec:density_id_gas}}.

Due to the application of \eq{\ref{eq:D_is_T}}, we set $\kB =1$ and the friction $\gamma$=1 to get the discrete time equations
\begin{align}
 \label{eq:adapted_discrete_brownian}
 \v r_i^{(n+1)} = \v r_i^{(n)} + \l[\v F_\text{ext}\l(\v r^{(n)}_i\r)+ \sum_{j=1}^{N}\v F\l(\big|\v r^{(n)}_i-\v r^{(n)}_j\big|\r)
                   + \alpha\nabla T\r]\Delta t
                   + \sqrt{2T\l(\v r^{(n)}_i\r)\Delta t}\ \Delta\v B_i.
\end{align}
We will work with a linear $x$-dependent temperature profile as given in \sec{\ref{sec:heat_equation}}, while in this section the temperature will be given as
\begin{align}
 T(x) = T_0\,\l(C_{x_0} + x\frac{\nabla T}{T_0}\r).
\end{align}
In order to keep the net flux in the direction of the gradient zero, we will work with reflective boundary conditions in $x$-direction. Furthermore for a first test of the principles established in \sec{\ref{sec:theory}} it suffices to investigate one-dimensional systems, with the exception of one investigation of a two-dimensional system where we will look at the influence of a non-conservative force field on the equilibrium density.
%Furthermore for a first test of the principles established in \ch{\ref{ch:theory}} it suffices to investigate one-dimensional systems, with the exception of one investigation of a two-dimensional system where we will look at the influence of a non-conservative force field on the equilibrium density.

As indicated, for every system we are interested in the Soret equilibrium densities to compare them with our expectations, as it is done in actual experiments, too. Therefore we will simulate $N$ particles in regions $x\in[-L/2,L/2]$, $(x,y)\in[-L/2,L/2]\times[-L/2,L/2]$ and record a histogram of the particles' positions with a resolution of $n_b$ bins ($n_b^2$ for the two-dimensional case). The choice of the integration constant $\Delta t$ will be discussed for every simulation individually.

At first we will only simulate ideal gas particles (without interaction) in one and two dimensions to test the basic findings of the last sections, finding agreement. Afterwards we will look at a system where an ideal gas is the solvent for a Gaussian particle, trying to find an answer to the question whether the local excess entropy drives the system's relaxation or the local excess enthalpy or if the local equilibrium assumption is in general not sufficient to describe the situation. We will find that the relaxation is roughly driven by the local excess enthalpy and so the system seems to follow the equation of state. Subsequently we will investigate two homogeneous systems, one only containing Gaussian particles and the other only consisting of hard rods. Both seem to follow their equation of state, as well.
\begin{figure}[t!]
 \centering
    \includegraphics[width=\textwidth]{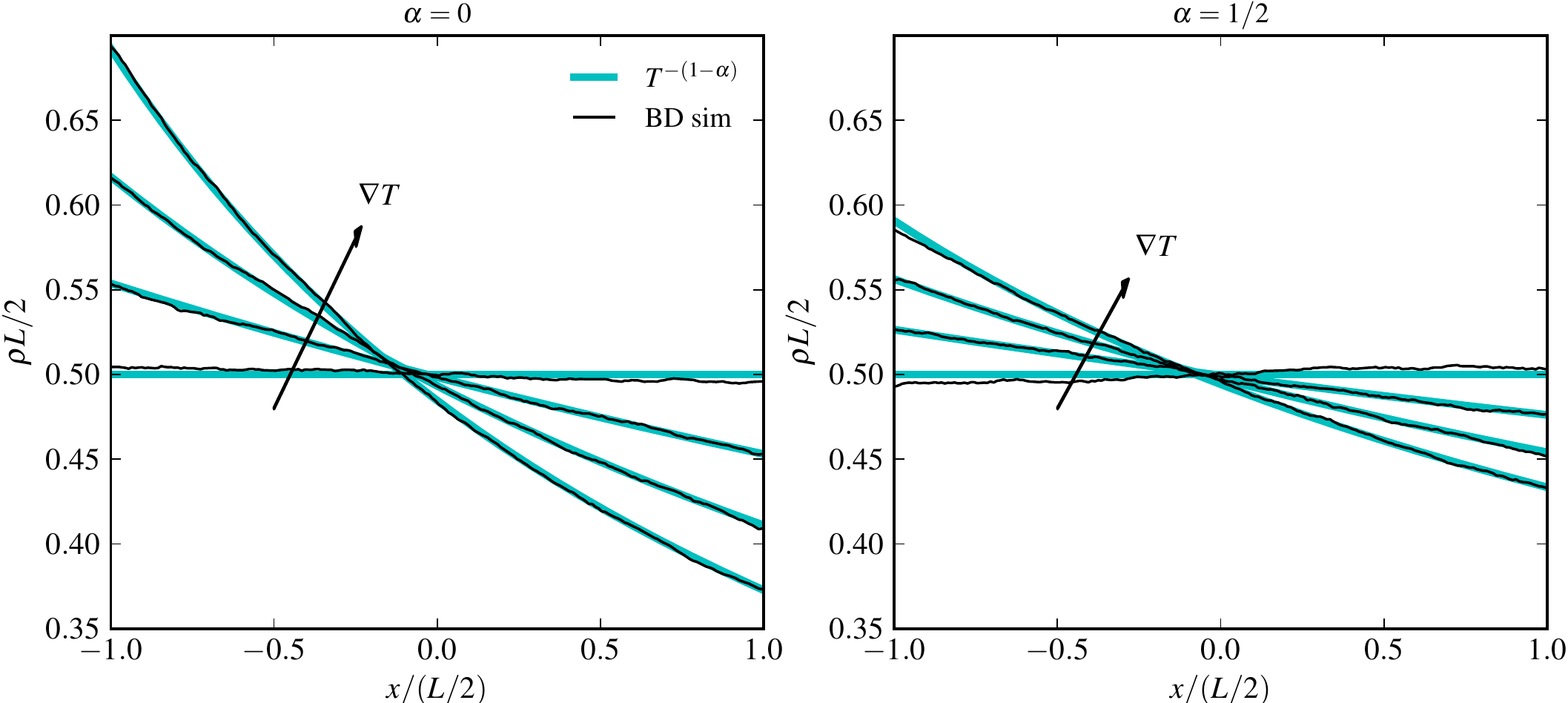}
 \caption{Comparison between BD simulation results and the expected equilibrium density of a one-dimensional ideal gas in a box with reflective boundary conditions in a temperature gradient. The simulation details are given in \sec{\ref{sec:id_gas_free}}.}
 \label{fig:id_gas_free}
\end{figure}
\subsection{Ideal Gas}

\subsubsection{Free Ideal Gas}
\label{sec:id_gas_free}
The first simulation was done for $N = 100$ free ideal gas particles in a one-dimensional box of dimension $L=2.0$ with a binning $n_b=200$, giving a spatial resolution of $\Delta x=0.01$ for the density histogram. We deployed a temperature gradient with parameters $T_0=1.0$, $C_{x_0}=1.0$ and $\nabla TL/(2T_0) = 0.0,0.1,0.2,0.3$. The maximum force in this system is given via the diffusion coefficient at the right boundary $T(L/2)/T_0 = 1.3$. We define that on average a particle should be able to travel a distance of $\Delta x_t = 2\Delta x$ (2 bins) in one timestep. Using the definition of the diffusion coefficient \eq{\ref{eq:D_of_t}} we find the timestep $\Delta t = \Delta x_t^2/(2T_\text{max}) = 4\times0.01^2/(2\times1.3)\simeq0.0001$ which was used here. The particles' initial positions were equally distributed over the box size, then the simulation was performed over $n_t=10^7$ timesteps while recording of the histogram has been done for all timesteps. From the last chapter we expect an equilibrium density of 
\begin{align}
\rho(x) = \frac{\mathcal N}{T^{1-\alpha}}.
\end{align}
The results of the simulations are shown in \fig{\ref{fig:id_gas_free}} and show excellent agreement with the expectation.
\newpage
\subsubsection{Ideal Gas in External Potential}
\begin{figure}[t!]
 \centering
    \includegraphics[width=\textwidth]{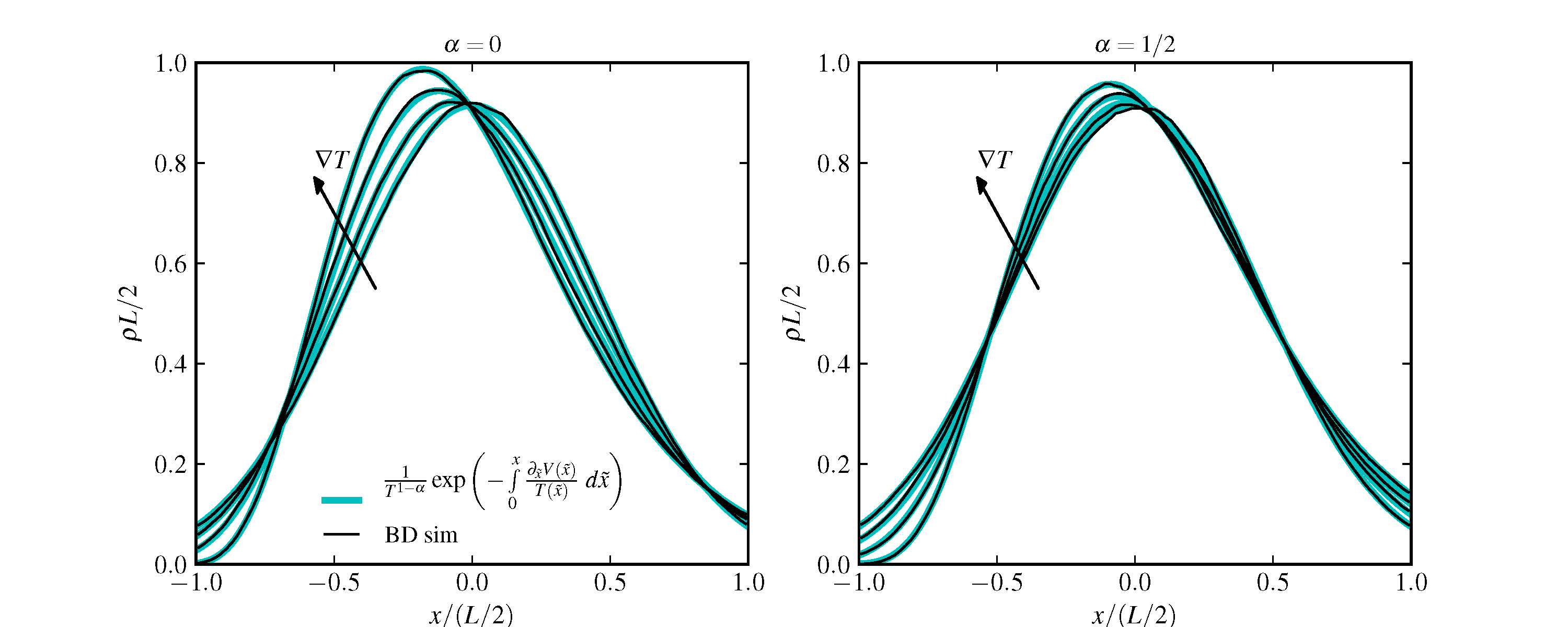}
 \caption{Comparison between BD simulation results and the expected equilibrium density of a one-dimensional ideal gas in a box with reflective boundary conditions in a temperature gradient and harmonic external potential. The simulation details are given in \sec{\ref{sec:1D_gas_ext}}.}
 \label{fig:1D_HO}
\end{figure}
\subparagraph{One-Dimensional System}
\label{sec:1D_gas_ext}
The next step is to check how the system behaves in presence of an external potential. To this end we used a similar setup as in \sec{\ref{sec:id_gas_free}} with larger gradients $\nabla TL/(2T_0) = 0,0.3,0.6,0.9$ and an external potential
\begin{align}
 \Vext(x) = \frac k2 x^2,
\end{align}
with $k = 5.0$. Note that for $\nabla T=0$ the simulation resembles a standard Ornstein-Uhlenbeck process \cite{Uhlenbeck:1930,Oksendal:2010}. The histogram was binned using $n_b=100$. For the diffusive force the timestep of $\Delta t = 0.0001$ is still appropriate, as well as for the external force, considering \newline $\Delta t = (L/100)^2/F_\text{ext,max} = 4\times0.01^2/5\simeq0.0001$. The simulations were performed with $N=100$ particles with initial conditions and maximum timesteps equal to those in \sec{\ref{sec:id_gas_free}}. From the last chapter we expect an equilibrium density of 
\begin{align}
\rho(x) &= \frac{\mathcal N}{[T(x)]^{1-\alpha}}\,\exp\l(-\int\limits_{x_0}^{x} \d\tilde{x}\ \frac{\partial_{\tilde{x}}\Vext(\tilde{x})}{T(\tilde{x})}\r)\\
        &= \frac{\mathcal N}{[T(x)]^{1-\alpha}}\,\exp\l(-\int\limits_{0}^{x} \d\tilde{x}\ \frac{k\tilde x}{T_0\l(1+\tilde{x}\frac{\nabla T}{T_0}\r)}\r)\\
        &= \frac{\mathcal N}{[T(x)]^{1-\alpha-\frac{kT_0}{\nabla T^2}}}\,\exp\l(-\frac{kx}{\nabla T}\r)
\end{align}
The results of the simulations are shown in \fig{\ref{fig:1D_HO}} and agree well with the expectation again. Note that the lower limit of the integration is set to $x_0=0$ arbitrarily, as it cancels out due to normalization. However, this reasoning is not applicable in higher-dimensional systems, the matter of interest in the next subsection.

\subparagraph{Two-Dimensional System}
\label{sec:2D_HO}
As remarked in \cite{Schnitzer:1993}, for dimensions larger than one the equilibrium density given by \eq{\ref{eq:density_id_gas_of_T}} may not be obtained trivially via integration because the integration in the exponential factor is path dependent, indicating that a theoretical description of the equilibrium density may not be achieved. Here, we want to show that this is not the case and that one can model the equilibrium density similar to a path integral where only the classical paths of minimum integrated force field are considered. We will work in a two-dimensional system.

The first thing to notice is that the equilibrium density induced by \eq{\ref{eq:brownian}} in abscence of interaction is the same as produced by a system in thermal equilibrium exposed to a scaled force field, described by the equation
\begin{align}
 \d\v r_i = \v F_\text{scaled}\d t + \sqrt{2}\ \d \v B_{t,i},
\end{align}
with
\begin{align}
 \v F_\text{scaled}(\v r) = \v F_\text{ext}+\v F_T = -\frac{\nabla\Vext(\v r)}{T(\v r)}- (1-\alpha)\frac{\nabla T}{T(\v r)}.
\end{align}
Considering a linear temperature gradient $T \equiv T(x)$  we calculate the rotation of the contributing fields to check whether they are conservative or not, as for conservative force fields we have $\rot \v F =0$.
\begin{align}
\rot \v F_\text{ext} &= \p{F^{(y)}_\text{ext}}{ x}\-\p{F^{(x)}_\text{ext}}{ y}
                   = \p{ }{y}\frac1{T(x)}\p{\Vext}{x}-\p{ }{x}\frac1{T(x)}\p{\Vext}{y} 
                   = \frac1{[T(x)]^2}\p{\Vext}{y}\p{T}{x}.
\end{align}
This expression only vanishes for $\Vext\equiv\Vext(x)$, an unvalid assumption for the general case. The second part (where per construction $F^{(y)}_T=0$) gives
\begin{align}
\rot \v F_T &= \p{F^{(x)}_T}{y} = 0.
\end{align}
Hence the temperature induced part is conservative which will be used later.

As has been indicated in the last chapter, the time dependent density can be interpreted as the transition probability for one particle to diffuse from a reference point $\v r_0$ to the point $\v r$ in time $t$. There exists a framework which treats these kind of problems by means of a path integral \cite{Chaichian:2001}. In the path integral framework, the probability is given by means of an integral over all paths $\v c(\tau)$ starting at $\v r_0$ and ending at $\v r$ after time $t$. The integrand is a Boltzmann factor $\exp(-S[\v c(\tau)])$ where $S$ usually is the action, a time integral of the system's Lagrangian.
%action external potential over the route $\gamma$, represented by the Feynman-Kac formula
However, an obvious Lagrangian does not exist for Brownian systems (especially not being exposed to a non-conservative force field). This problem may be overcome by introducing a dissipated energy term to the Lagrangian \cite{Lin:2013} or an additional force dependent term \cite{Chaichian:2001}, while the latter is used in the Feynman-Kac formula for the transition probabilities in Brownian systems by means of Wiener path integrals.
% \begin{align}
%  W_B(\v r,t|\v r_0,0) = \int_{\mathcal C} \mathcal D\v c
%                         \exp \l[   -\int_r      \r]
% \end{align}

However, since we are interested in the equilibrium density at $t\rightarrow\infty$ and due to the structural equivalence of \eq{\ref{eq:density_id_gas_of_T}} to the Boltzmann factor for a conservative force in thermal equilibrium
\begin{align}
 \exp(-\beta_0\Vext)=\exp\l(\int\d\v x \beta_0\v F_\text{ext}(\v x)\r),
\end{align}
we assume that the quantity of importance for the solution of the problem is indeed the integrated force field rather than the action. A further approximation is that we limit our integration to the classical paths for a particle following strictly the stream lines of the given force field, as those will be the paths of minimal energy with the maximal weight from the Boltzmann factor. For every $\v r$, let us denote the path following the force field's stream lines as $\v c_m(\tau)$ with $\v c_m(0)=\v r_0$ and $\v c_m(t\rightarrow\infty)=\v r$. Then our postulation for the equilibrium density in the style of an approximated path integral is
\begin{align}
 \label{eq:postulated_path_integral}
 \rho(\v r) = \mathcal N \exp\l[\int\limits_{\v c_m}\d\v x \cdot\v F_\text{scaled}(\v x) \r].
\end{align}
Indeed, comparing the postulation to the path integral given in \cite{Chaichian:2001}, \eq{1.2.93}, we find a structurally similar equation with an additional time dependence and an additional prefactor of $1/2$ for the integration of the force field, while there are indications that it reaches \eq{\ref{eq:postulated_path_integral}} in the limit of $t\rightarrow\infty$, as one can see for the Ornstein-Uhlenbeck process in \cite{Chaichian:2001}, example 1.10, or \eq{1.2.134}. Note that in the derivation of \cite{Chaichian:2001}, \eq{1.2.93}, the assumption of $\v F$ being conservative has been made, such that it can be written as a prefactor of the integral, while in fact the path integration has to be done over the exponential of the integrated force field as well (c.f. \cite{Chaichian:2001}, problem 1.2.11).

The choice of the reference point of the integration $\v r_0$ is naturally arbitrary or depends on the initial conditions, respectively. However, without presence of a Brownian force, from every initial point $\v r_0$ a particle will end its path in a fix point $\v r_f$ where the force field vanishes $\v F_\text{scaled}(\v r_f) = 0$. Therefore, in presence of a Brownian force, we assume that from every initial point $\v r_0$, the particle will follow the stream lines to the fix point $\v r_0\rightarrow \v r_f$, then continue its path along the stream lines to the final destination $\v r_f\rightarrow\v r$. Since for every initial condition $\v r_0$ the first part of the path $\v r_0 \rightarrow\v r_f$ is the same, we can omit that first part and set $\v r_0 = \v r_f$. Of course this argumentation is limited to force fields with exactly one fix point, the discussion of other force fields will not be held in this work.

\begin{figure}[t!]
 \centering
    \includegraphics[width=\textwidth]{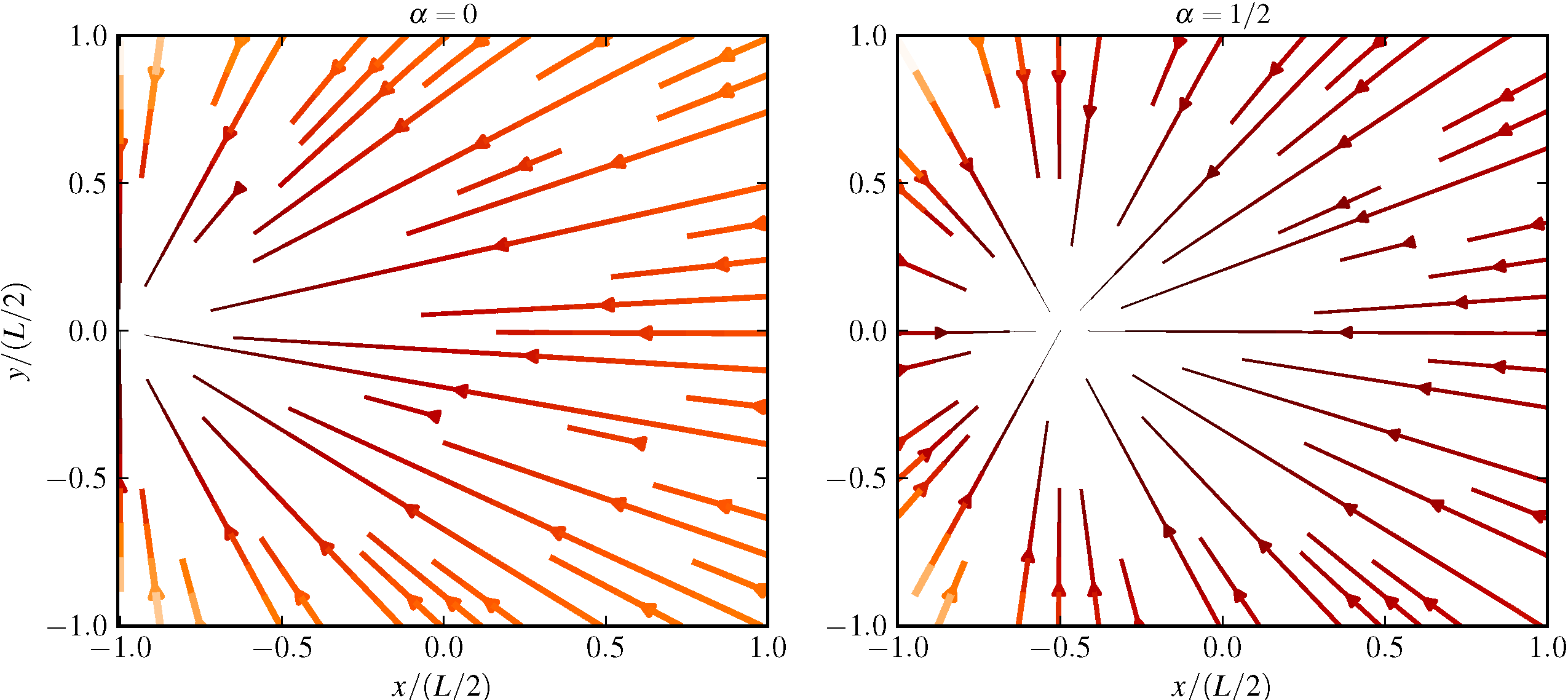}
 \caption{Scaled force field $\v F_\text{scaled}$ with an harmonic external potential $\Vext(x,y) = (x^2+y^2)/4$ and a temperature of $T(x)=1+x/10$. The stream lines of the field are straight lines leading to the fix point \eq{\ref{eq:fixpoint}}. Thickness and brightness of the lines are proportional to the absolute value of the force. }
 \label{fig:2D_HO_stream}
\end{figure}

We want to test the postulation of \eq{\ref{eq:postulated_path_integral}} in the two-dimensional case given an harmonic external potential of
\begin{align}
 \Vext(\v r) &= \frac k2 \v r^2 = \frac k2 \l(x^2 + y^2\r)
\end{align}
and a temperature gradient of
\begin{align}
 T(x) &= T_0\l(1+\epsilon x\r)
\end{align}
with $\epsilon = \nabla T/T_0$, s.t. the force field is given as
\begin{align}
 \label{eq:Fscaled}
 \v F_\text{scaled}(\v r) = - \frac{1}{T_0(1+\epsilon x)} \l( \begin{array}{c}
                                                                kx+(1-\alpha)T_0\epsilon\\
                                                                ky
                                                              \end{array}\r).
\end{align}
An illustration of the force field for a specific parameter set is given in \fig{\ref{fig:2D_HO_stream}}. The first task is to find the fix point of the forcefield, $\v F_\text{scaled}(\v r_f) = 0$
\begin{align}
\label{eq:fixpoint}
 \v r_f = \l( \begin{array}{c}
                -T_0\epsilon(1-\alpha)/k\\
                0
                \end{array}
          \r).
\end{align}
Second, we parametrize the trajectories of the system starting at $\v r_f=(x_f,y_f)$ and ending at $\v r=(x,y)$ with a parameter $\tau$. From the structure of $\v F_\text{scaled}$ we see that those are of linear form, s.t. in polar coordinates we have
\begin{align} 
 \tilde x(\tau) &= x_f + \tau \cos\phi(\v r, \v r_f),\qquad \d\tilde x = \cos\phi(\v r, \v r_f)\ \d\tau,\\
 \tilde y(\tau) &= y_f + \tau \sin\phi(\v r, \v r_f),\qquad\,\, \d \tilde y = \sin\phi(\v r, \v r_f)\ \d\tau
\end{align}
with the orientation
\begin{align}
 \phi(\v r, \v r_f) = \arctan2(y-y_f,x-x_f)
\end{align}
and the upper bound
\begin{align}
\tau_\text{max} = \sqrt{\l(x-x_f\r)^2+\l(y-y_f\r)^2}. 
\end{align}
In order to solve the integral in the exponential factor of \eq{\ref{eq:postulated_path_integral}}, we make use of the fact that the temperature part $\v F_T$ of the scaled force is conservative, s.t. that part of the integration is path independent and yields the by now well-known prefactor of $1/T^{1-\alpha}$.
The remainder of the integral over the trajectory is
\begin{align}
 \int\limits_{\v c_m(\tau)}\d\tilde{ \v r} \cdot \v F_\text{ext}(\tilde{\v r})
   &= -\frac k{T_0} \int\limits_{(x_f,0)}^{(x,y)}
        \frac{\tilde x\d\tilde x+\tilde y\d\tilde y}{1+\epsilon \tilde x}\\
   &= -\frac k{T_0} \int\limits_{0}^{\tau_\text{max}}
        \frac{x_f \cos\phi+ \tau \cos^2\phi + \tau \sin^2\phi}{1+\epsilon (x_f + \tau \cos\phi)}\d\tau    \\
   &= -\frac k{T_0} \int\limits_{0}^{\tau_\text{max}}
        \frac{x_f \cos\phi+\tau}{1+\epsilon (x_f + \tau \cos\phi)}\d\tau     \\
   &= -\frac{k}{T_0}\frac{(\epsilon x_f\sin^2\phi-1) \ln(\epsilon x_f+\epsilon \tau \cos\phi+1)+\epsilon \tau \cos\phi}{\epsilon^2 \cos^2\phi} \Bigg|_0^{\tau_\text{max}}.
    \label{eq:integral}
\end{align}
\begin{figure}[tph]
 \centering
 \begin{overpic}[width=\textwidth]{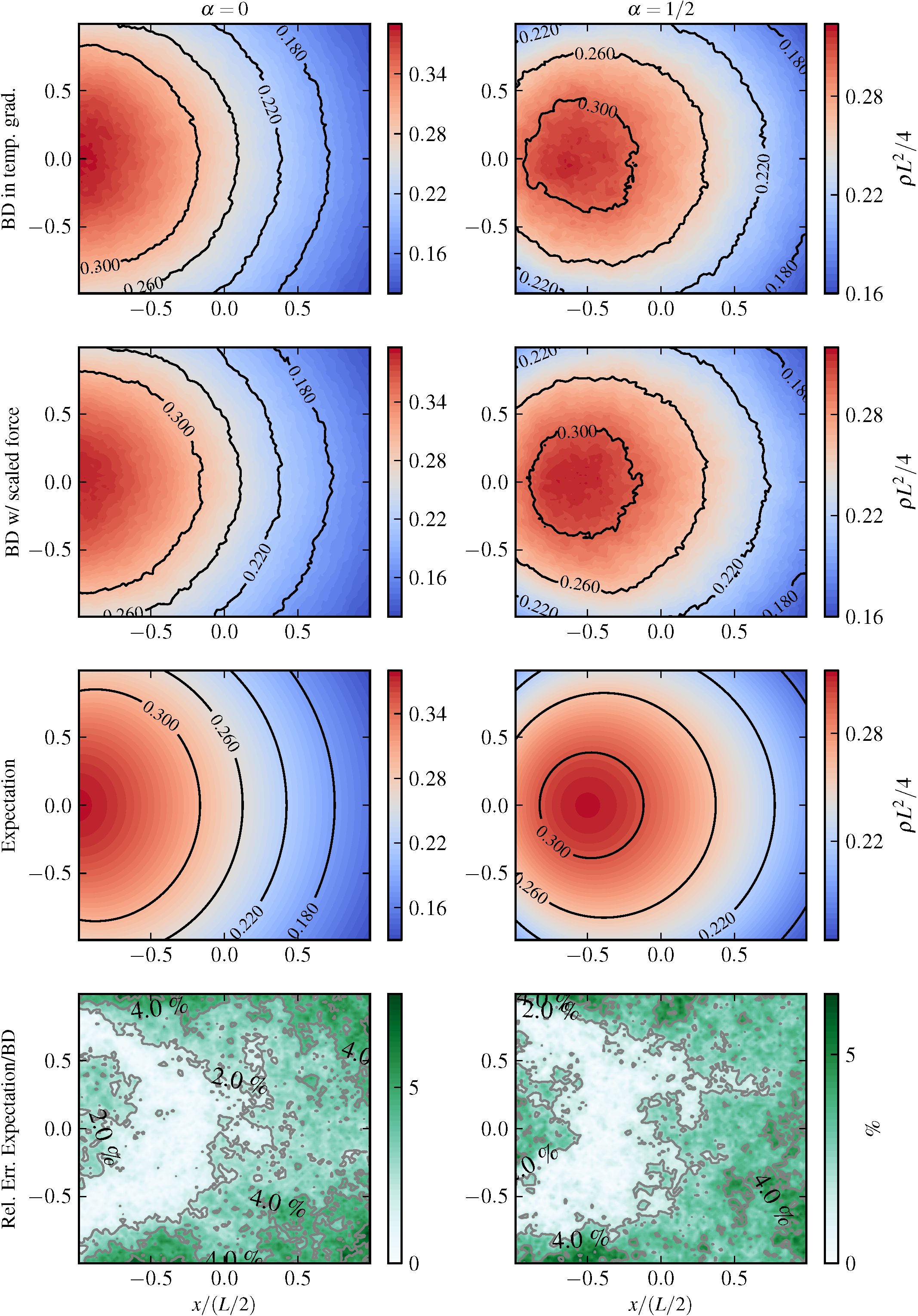}
 \put (2.1,34.5) {\tiny \rotatebox{90}{\eq{\ref{eq:postulated_path_integral}}~/~\eq{\ref{eq:expectation_2D}}}}
\end{overpic}
 \caption{Equilibrium densities of an ideal gas in a two-dimensional box. \textbf{Top}: BD simulation with harmonic external potential and spatially dependent noise. \textbf{Center Top}: BD simulation with a temperature scaled harmonic external force of \eq{\ref{eq:Fscaled}} and spatially independent thermal noise. \textbf{Center Bottom}: Expectation of \eq{\ref{eq:expectation_2D}}. \textbf{Bottom}: Relative error of \eq{\ref{eq:expectation_2D}} compared to the BD simulation in the temperature gradient (top). Setup details are given in \sec{\ref{sec:2D_HO}}.}
 \label{fig:2D_HO_densities}
\end{figure}

\begin{figure}[t!]
 \centering
 \begin{overpic}[width=\textwidth]{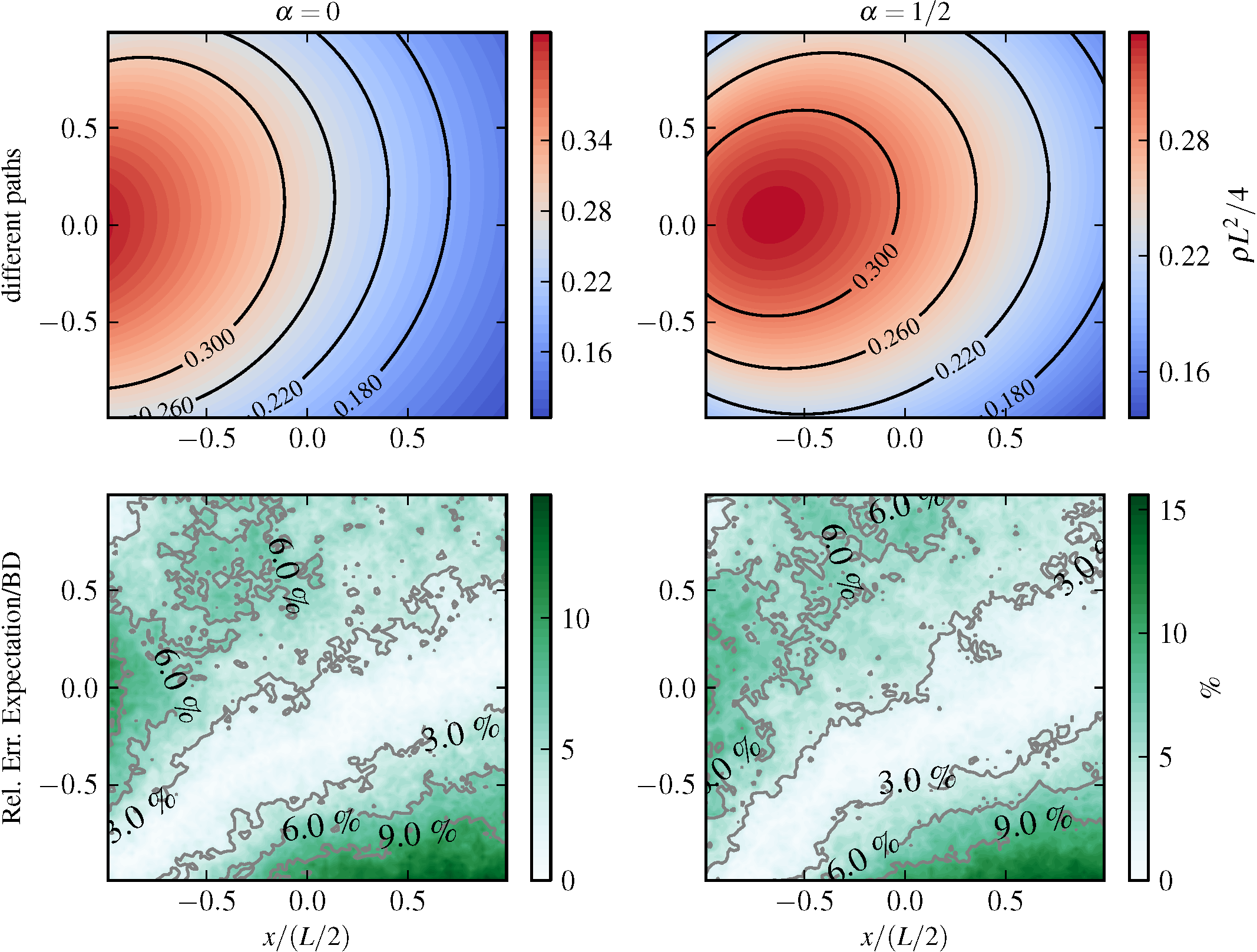}
 \put (2.1,54) {\tiny \rotatebox{90}{\eq{\ref{eq:postulated_path_integral}}}}
\end{overpic}
 \caption{The expectation \eq{\ref{eq:postulated_path_integral}} with other paths (straight lines starting at $x_0=1$, $y=-1$) and the relative error to the BD simulation.}
 \label{fig:2D_HO_other_paths}
\end{figure}

Entering everything into \eq{\ref{eq:postulated_path_integral}} yields the expected equilibrium density
\begin{align}
&\rho(\v r) = \frac{\mathcal N}{\l[T(x)\r]^{1-\alpha}}\exp\Vast[-\frac{k\sqrt{\l(x+\frac{T_0\epsilon}{k}(1-\alpha)\r)^2+y^2}}{T_0\epsilon\cos\phi} - \l(\frac k{T_0\epsilon^2 \cos^2\phi}-(1-\alpha)\tan^2\phi\r) \times\NN\\
 &\qquad \qquad\qquad \qquad\ \times \ln\l[1-\frac{T_0\epsilon^2}{k}(1-\alpha)+\epsilon\cos\phi\sqrt{\l(x+\frac{\epsilon}{k}(1-\alpha)\r)^2+y^2} \r]\Vast].
 \label{eq:expectation_2D}
\end{align}
Note that the result seems to be indefinite for $\cos\phi=0$ which is not true, since setting $\cos\phi=0$ before the integration yields a definite result.

Now to test \eq{\ref{eq:expectation_2D}} we take $k=1/2$, $T_0=1$, $\epsilon=1/10$, a squared box of length $L=2$ and perform a BD simulation. Boundary conditions are reflective in $x$-direction and periodic in $y$-direction. We simulated a single particle over $10^9$ time steps of $\Delta t=0.0001$ (which is appropriate, because the maximum forces are lower than in the last problems) and recorded from the beginning to obtain the density as a histogram of $100\times100$ bins. The first simulation was done with the force being solely $\v F_\text{ext}$ and a Brownian noise with variance $\sqrt{2T(x)}$, while the second was performed with the scaled force $\v F_\text{scaled}$ and a Brownian noise with variance $\sqrt{2}$. The results are shown in \fig{\ref{fig:2D_HO_densities}} and compared to the expectation. First, one notices the equivalence of both simulations. While they reflect different situations, both yield the same equilibrium density, as expected. Second, one sees that \eq{\ref{eq:expectation_2D}} 
does not reflect the experimental density completely but is an acceptable approximation, as the difference between the simulation and the expectation is maximally $\simeq6\%$.

Another check can be done by performing the integration over other paths. We do so by using \eq{\ref{eq:expectation_2D}} with another starting point than $(x_f,y_f)$, namely $x_0=1$ and $y_0=-1$. This yields the results given in \fig{\ref{fig:2D_HO_other_paths}}. As can be seen, the difference between the simulation and the approximated density is significantly worse with a maximal relative error of $\simeq15\%$.

\subsection{Thermophoresis of Gaussian Solute in Ideal Gas}
\label{sec:gaussian_solvation}
In this section we want to check whether the assumptions made in \sec{\ref{sec:interacting_particles}} are valid, namely if the two point density-density correlator can really be modeled by the derivative of the equilibrium free energy functional and if that is the case, whether the thermal diffusion coefficient is proportional to the solvation enthalpy or solvation entropy. To this end we both simulate and analytically treat an ensemble consisting of an ideal gas solvent (meaning that the solvent-solvent interaction is $V_{vv}=0$) and a single Gaussian solute, referring to the interaction between solute and solvent as
\begin{align}
\label{eq:gaussian_potential}
V_{uv}(r)\equiv V(r) = \varepsilon\exp\l(-\frac{r^2}{\sigma^2}\r).
\end{align}
We will proceed as follows. First we need to evaluate the solvation free energy for the Gaussian particle solvated in the ideal gas solvent as a smooth function of the temperature. To this end, the knowledge of the solvent density's temperature dependence $v(T)$ is necessary and will be assumed to resemble $v(T)\propto 1/T^{1-\alpha}$ as was shown in the previous sections. Knowing this we make use of the methods described in \sec{\ref{sec:TI}}, i.e. thermodynamic integration by means of the pair correlation function $g_{\lambda,T}(r) \equiv g(r)$ and the $\lambda$-dependent interaction energy \begin{align}
\label{eq:lambda_interaction_gaussian}
V_\lambda(r) =\lambda V(r)
\end{align}
with $\lambda$ running from 0 to 1.  We obtain $g(r)$ by equilibrium simulations at the right temperature and solvent density as well as modeling it theoretically. Subsequently, enthalpy and entropy are calculated and the expectations for the solute density $u(T)$ in a temperature gradient are obtained with \eq{\ref{eq:solute_density_theory}}. Second, we perform simulations in temperature gradients and compare the results with the expectation.

\subsubsection{Thermophoretic Simulation Setup}
\label{sec:thermo_sim_setup}
In order to calculate the free energy at the right conditions, we first have to set the parameters which will be used later in the thermophoretic simulations. We will work with an ideal gas of $N_v=20$ particles and one solute particle. The reference ideal gas osmotic pressure is set to $\rho T_0 = 1$, such that for a one-dimensional system of $N_v+N_u=21$ particles and the reference temperature $T_0$ we find the box length $L=21$. We will work with a temperature profile
\begin{align}
 T(x) = 2.5+x\frac{\nabla T}{T_0}
\end{align}
and a gradients of $\nabla TL/T_0=0.0,1.05,2.10,3.15,4.20$, s.t. we have a minimum temperature of $T(-L/2)=0.4T_0$ and a maximum temperature of $T(L/2)=4.6T_0$ for $\nabla TL/T_0=4.20$. Regarding the interaction potential we set $\varepsilon = 3.0$ and $\sigma=0.5$ s.t. the packing fraction is $\Phi=\sigma/L=0.024$. 

\subsubsection{Free Energy Calculation}
\subparagraph{Numerical Evaluation}
\begin{figure}[t!]
 \centering
    \includegraphics[width=\textwidth]{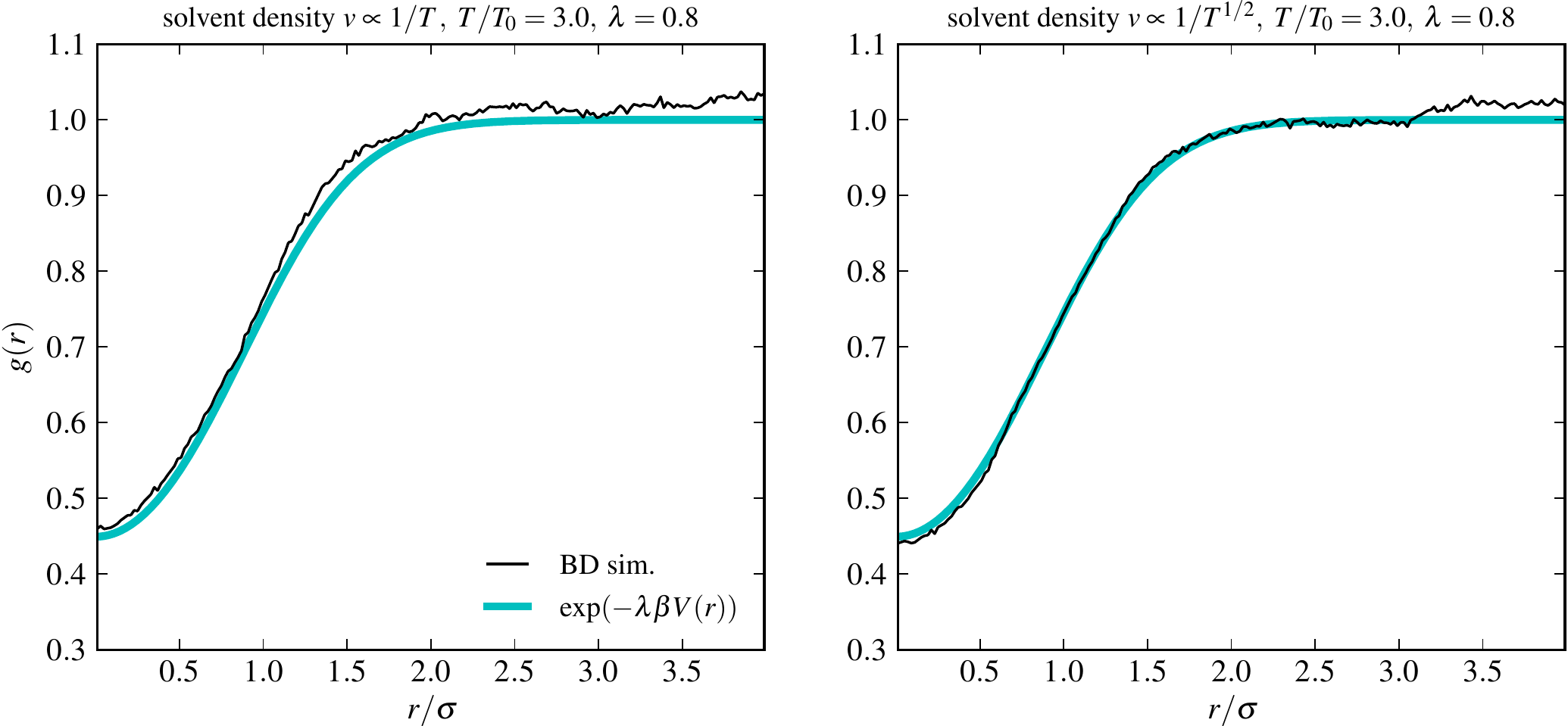}
 \caption{A typical $g(r)$ for pairs of solvent and solute particles, here taken from the equilibrium simulations performed at $T/T_0=3.0$, $\lambda=0.8$, at the reference solvent densities $v(T)$ for $\nabla TL/T_0 = 4.20$. It is compared to the high temperature limit given in \eq{\ref{eq:g_high_temp}}.}
 \label{fig:g_of_r_gaussian}
\end{figure}
\begin{table}
 \centering
 \input{tables/norm_const}
 \caption{Values for the normalization constant used to model the solvent density dependent on $T$ for the equilibrium simulations.}
 \label{tab:norm_const}
\end{table}
For the numerical evaluation of the free energy we first have to determine the solvent density in dependence of the gradient and the local temperature. We assume the density to be following $1/T^{1-\alpha}$, however the prefactor depends on the gradient due to the normalization and is given as
\begin{align}
\mathcal{N}_{\nabla T,\alpha} = N_v\l(\int\limits_{-L/2}^{+L/2}\!\d x\ \frac{1}{[T(x)]^{1-\alpha}}\r)^{-1}.
\end{align}
\begin{figure}[t!]
 \centering
    \includegraphics[width=\textwidth]{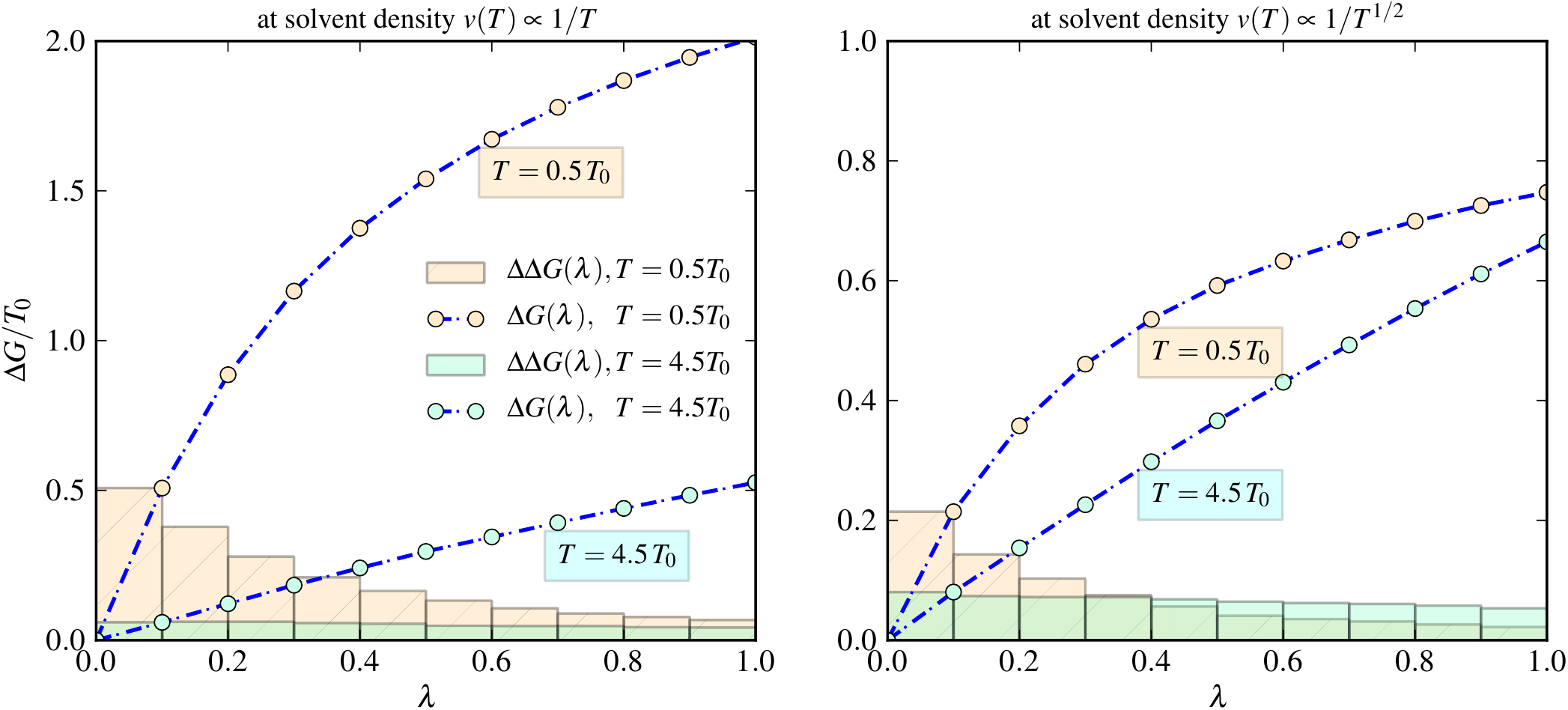}
 \caption{Four typical curves for the thermodynamic integrations taken from the equilibrium simulations performed at $T/T_0=0.5$ (cyan) and $T/T_0=4.5$ (orange), at the reference solvent densities $v(T)$ for $\nabla TL/T_0 = 2.10$.
 %The free energy between steps of lambda $\Delta\Delta G$ is displayed as a bar plot, while the integrated free energy is plotted as curve.
 }
 \label{fig:lambda_integration}
\end{figure}
The values are reported in \tab{\ref{tab:norm_const}}. Knowing this we performed equilibrium simulations for $T/T_0\in[0.5,4.5]$ with a step size of $\Delta T/T_0 = 0.5$ and the same particle numbers as for the thermophoretic simulations later. For every temperature and temperature gradient the one-dimensional box length was adapted to follow $L(T)=N_v/v(T)$ and 11 simulations have been performed, every simulation for a discrete value of $\lambda_i$ between $\lambda_0=0$ and $\lambda_{10}=1$ and a step size of $\Delta \lambda =0.1$ using the solute-solvent interaction potential given by \eq{\ref{eq:gaussian_potential}-\ref{eq:lambda_interaction_gaussian}}. Considering the same conditions for a position change $\Delta x$ per time step as in the sections before and the interaction force $F(r) = 2r\varepsilon\sigma^{-2}\exp(-(r/\sigma)^2)$ we find a maximum force of $F_\text{max} \simeq 5$, for which a temporal step size of $\Delta t = 0.0001$ would still be appropriate, however the maximum temperature is $T_\text{max} = 4.6$, giving a $\Delta t=4\times0.001^2/(2\times4.6)\simeq 5\times10^{-5}$, which is used in the simulations. Since we perform thermodynamic integration by means of \eq{\ref{eq:free_energy_from_gofr}}, a histogram of $g(r)$ was recorded following the technique described in \cite{Allen:1987} using 200 bins, a histogram cutoff radius of $r_{c,g}=4\sigma$ and a force cutoff radius of $8\sigma$. Every simulation run for a total of $10^7$ time steps with recording of the histogram starting after $10^3$ time steps.

In \fig{\ref{fig:g_of_r_gaussian}} one can see a typical example of $g(r)$ obtained from simulations for one specific parameter set, compared to the analytical model of a high temperature limit, explained in the next subsection. Note that the often used limit $\lim\limits_{r\rightarrow\infty} g(r)=1$ is only valid in the thermodynamic limit $N\rightarrow\infty$ and thus does not have to be reached here. For every step $\lambda_i\rightarrow\lambda_{i+1}$ we calculated the difference in the free energy as
\begin{align}
    \Delta\Delta G(T,\lambda_{i}) = \frac{v(T)\Delta\lambda}2 \l(\int\limits_0^{r_{c,g}}\!\d r g_{\lambda_i,T}(r) V(r)+\int\limits_0^{r_{c,g}}\!\d r g_{\lambda_{i-1},T}(r) V(r)\r)
\end{align}
where the integration over $r$ was done numerically using the trapezoidal method. The free energy for $\lambda_0\rightarrow\lambda_i$ is then given as
\begin{align}
 \Delta G(T,\lambda_{i}) = \sum_{j=1}^i \Delta\Delta G(T,\lambda_j).
\end{align}
Both quantities are displayed as an example in \fig{\ref{fig:lambda_integration}} for $\nabla TL/T_0=2.10$ and two temperature values. 

\begin{figure}[t!]
 \centering
    \includegraphics[width=\textwidth]{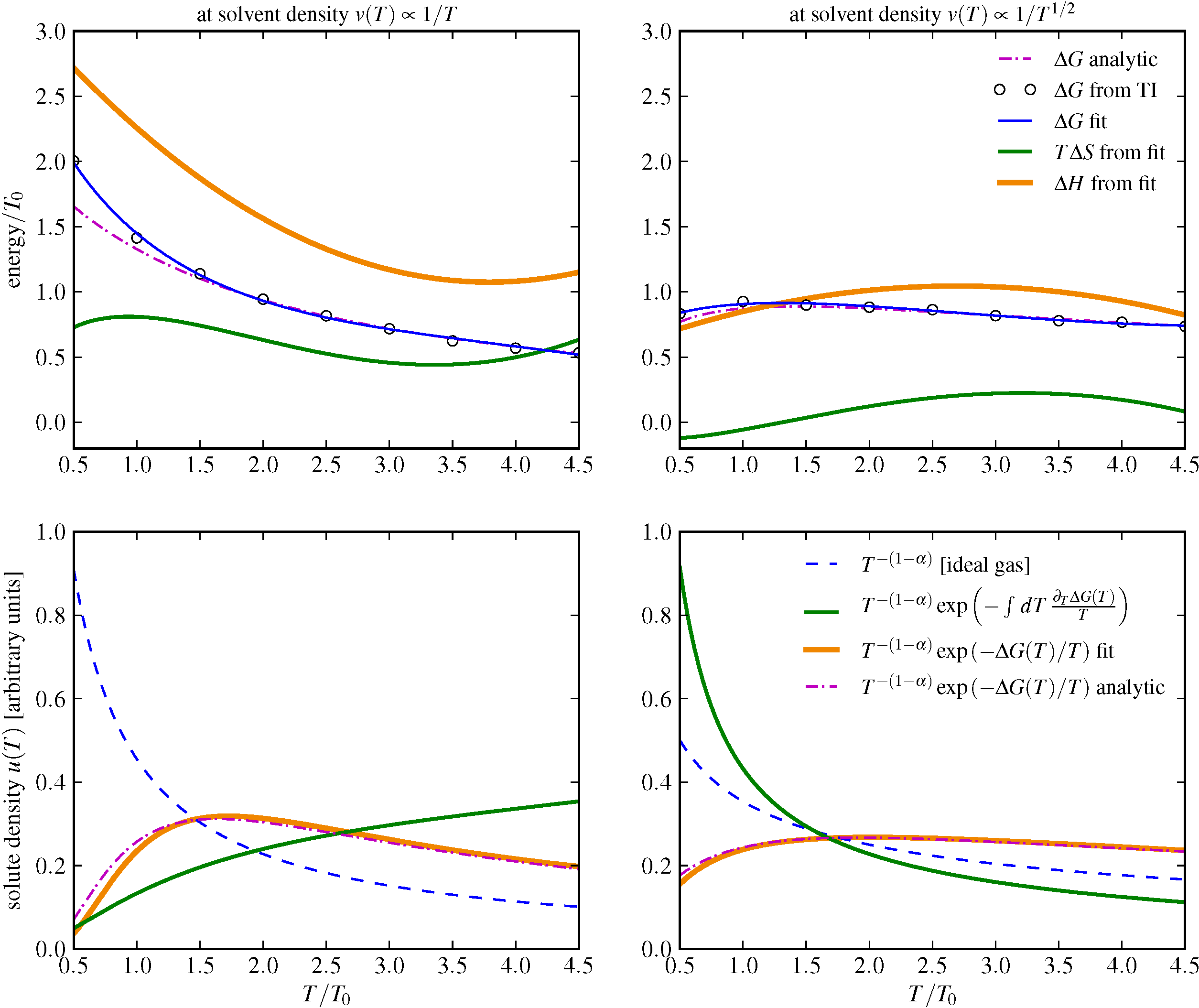}
 \caption{Free energy and supposed solute densities as a function of the temperature simulated at the solvent density $v(T)$ for  $\nabla TL/T_0 = 2.10$. \textbf{Top:} Numerical data, fits and analytic evaluation as well as the thermodynamic quantities solvation enthalpy and solvation entropy. \textbf{Bottom:} The supposed solute densities for later simulations in thermogradients. Note the obvious difference between the two hypotheses $u^{(H)}$ (orange) and $u^{(S)}$ (green). The ideal gas density is plotted for comparison.}
 \label{fig:deltaG_polymer_in_ideal_gas}
\end{figure}

Taking $\Delta G(T) = \Delta G(T,\lambda_{10})$, we find curves as displayed in \fig{\ref{fig:deltaG_polymer_in_ideal_gas}} for every value of $\nabla T$ and perform least square fits according to \eq{\ref{eq:free_energy_fit_function}}, from which we can conclude enthalpy and entropy from \eq{\ref{eq:thermodynamic_fit_functions}} as well as the supposed densities in thermogradients from \eq{\ref{eq:density_fit_functions}}. The fit parameters are given in \tab{\ref{tab:fit_params_bd}}. As one can see in \fig{\ref{fig:deltaG_polymer_in_ideal_gas}} the results show a very different behavior in the solute density for the two cases $u^{(H)}(T)$, $u^{(S)}(T)$, s.t. if the simulations performed in thermogradients reveal a close connection to either of the two it will seem reasonable to dismiss the other.

\begin{table}[t!]
 \centering
 \input{tables/fit_params_bd}
 \caption{Fit parameters found by means of a least square fit from the numerical thermodynamic integration data to fit $G(T)$ according to \eq{\ref{eq:free_energy_fit_function}}. The temperature gradient dependence refers to the chosen solvent density $v(T)$ which has a different prefactor for the different gradients.
 %Note that the temperature gradient is solely for the purpose of simulating at the right solvent density $v(T)$.
 One fit (for $\nabla TL/T_0 = 2.10$) along with its consequences for thermodynamic quantities as well as the supposed solute densities is shown in \fig{\ref{fig:deltaG_polymer_in_ideal_gas}}.}
 \label{tab:fit_params_bd}
\end{table}

\subparagraph{Analytical Treatment}
In order to check the numerical evaluations from the last sections we will consider an analytical derivation in the following. We use the high temperature limit \eq{\ref{eq:g_of_r_high_temp}}
\begin{align}
\label{eq:g_high_temp}
 g(r) = \exp\Big[-\lambda\beta V(r)\Big] = \exp\l[-\lambda\beta\varepsilon\exp\l(-\frac{r^2}{\sigma^2}\r)\r],
\end{align}
where a comparison between this formula and a BD simulation for one specific parameter set is drawn in \fig{\ref{fig:g_of_r_gaussian}}, indicating the validity of \eq{\ref{eq:g_high_temp}}. As a matter of fact, for the case of an ideal gas solvent \eq{\ref{eq:g_high_temp}} is exact as the solute potential $V$ acts like an external potential, casting $g(r)$ in the form of an ideal gas thermal equilibrium density. We use this to analytically evaluate the free energy of solvation. Recalling \eq{\ref{eq:free_energy_from_gofr_high_temp}} for a one-dimensional system, we have the free energy
\begin{align}
 \Delta G(T) &= \frac{v(T)}{\beta}\int\limits_0^\infty\d r\l[1- e^{-\beta V(r)}\r].
\end{align}
Expanding the exponential function yields
\begin{align}
 \Delta G(T) &= \frac{v(T)}{\beta} \int\limits_0^\infty\d r\ \l[1-\l(1+\sum_{n=1}^\infty (-1)^{n+1}\frac{\l(\beta V(r)\r)^n}{n!}\r)\r]\\
             &= \frac{v(T)}{\beta} \int\limits_0^\infty\d r\ \sum_{n=1}^\infty (-1)^{n+1}\frac{\l(\beta \varepsilon\exp\l(-\frac{r^2}{\sigma^2}\r)\r)^n}{n!}\\
             &= \frac{v(T)}{\beta} \sum_{n=1}^\infty (-1)^{n+1}\frac{\l(\beta \varepsilon\r)^n}{n!}\int\limits_0^\infty\d r\ \exp\l(-n\frac{r^2}{\sigma^2}\r)\\
             &= \frac{v(T)}{\beta}\,\frac{\sigma\sqrt{\pi}}{2}\sum_{n=1}^\infty (-1)^{n+1}\frac{\l(\beta \varepsilon\r)^n}{n!\sqrt{n}}.
\label{eq:Gaussian_DeltaG_analytic}             
\end{align}
The enthalpic version of the solute density is then given as
\begin{align}
 u^{(H)}(x) = \frac{\mathcal N}{[T(x)]^{1-\alpha}}\,\exp\l(-\frac{\mathcal N_{\nabla T,\alpha}}{[T(x)]^{1-\alpha}}\,\frac{\sigma\sqrt{\pi}}{2}\sum_{n=1}^\infty (-1)^{n+1}\frac{1}{n!\sqrt{n}}\l(\frac{\varepsilon}{T(x)}\r)^n\r).
\end{align}
Both are shown alongside the curves from the numerical evaluation in \fig{\ref{fig:deltaG_polymer_in_ideal_gas}} and show reasonable agreement, especially for the higher temperatures as one would suspect from the application of a high temperature limit. As we will see in the next section the enthalpic version of the density suffices to be evaluated.

\subsubsection{Gaussian Solute in Temperature Gradient}
Simulations have been run using the setup described in \sec{\ref{sec:thermo_sim_setup}} where again the densities were obtained as a histogram by binning the particle locations in $200$ bins. The temporal time step was again $\Delta t= 5\times10^{-5}$ and the simulation run for $10^{10}$ time steps, starting the recording of the histogram after $10^3$ time steps. To prevent a static flux between the two walls we worked with reflective boundary conditions in $x$-direction.

\begin{figure}[tph]
 \centering
    \includegraphics[width=\textwidth]{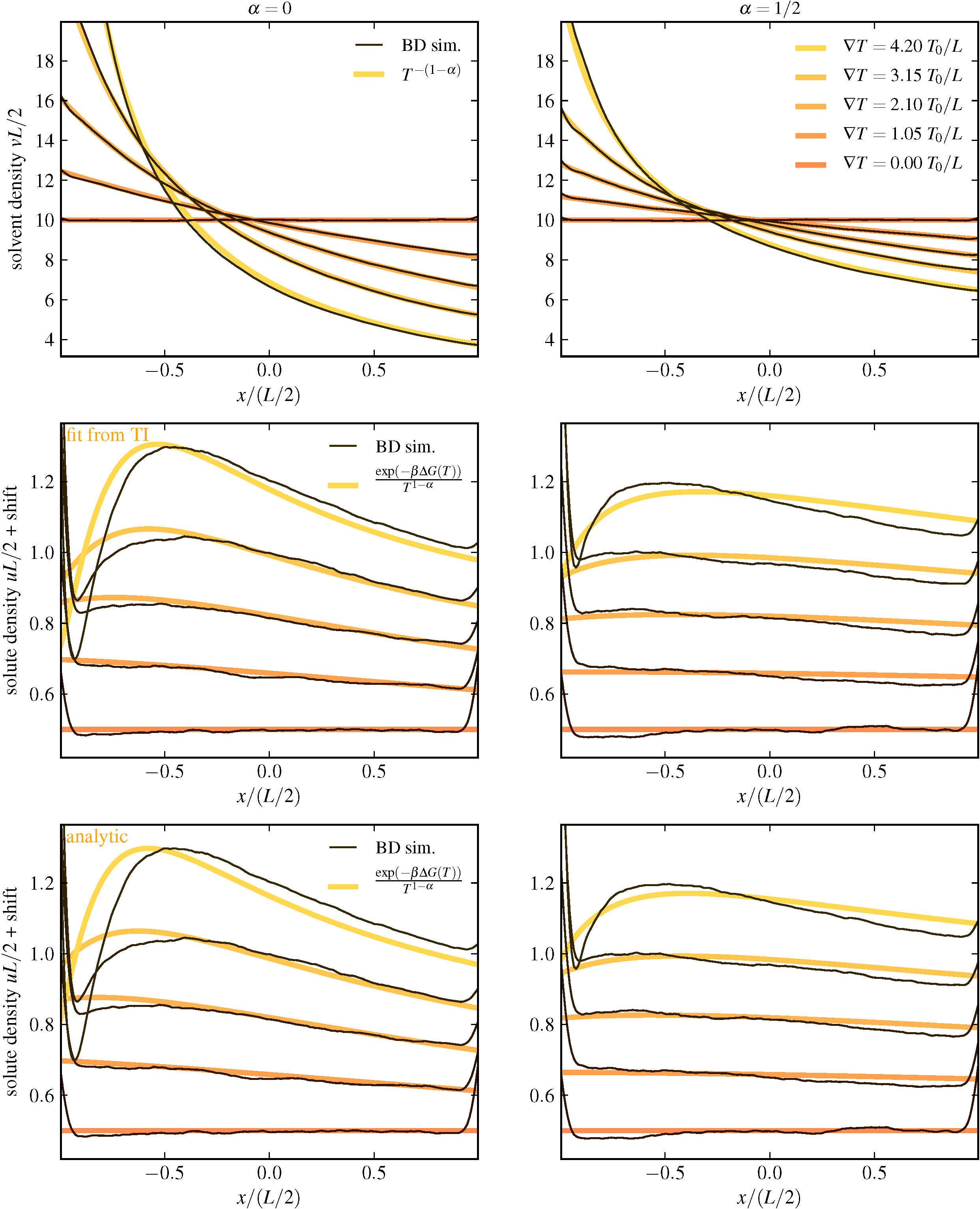}
 \caption{Thermophoretic simulation of the Gaussian solute in the ideal gas solvent in a one-dimensional box of size $L=21$ compared to expectations. \textbf{Top:} The ideal gas solvent is following the expectation $v(t) = \mathcal N_{\nabla T,\alpha}/T^{1-\alpha}$. \textbf{Center:} The solute density compared to the expectation of the enthalpic density assumption (obtained from the numerical TI). The curves are shifted for better visibility. \textbf{Bottom:} The solute density compared to the expectation of the enthalpic density assumption (obtained by analytic means). The solute density curves are shifted for better visibility with an increasing positive shift in $y$-direction for increasing $\nabla T$.}
 \label{fig:polymer_in_ideal_gas_expectation}
\end{figure}

\begin{figure}[tph]
 \centering
    \includegraphics[width=\textwidth]{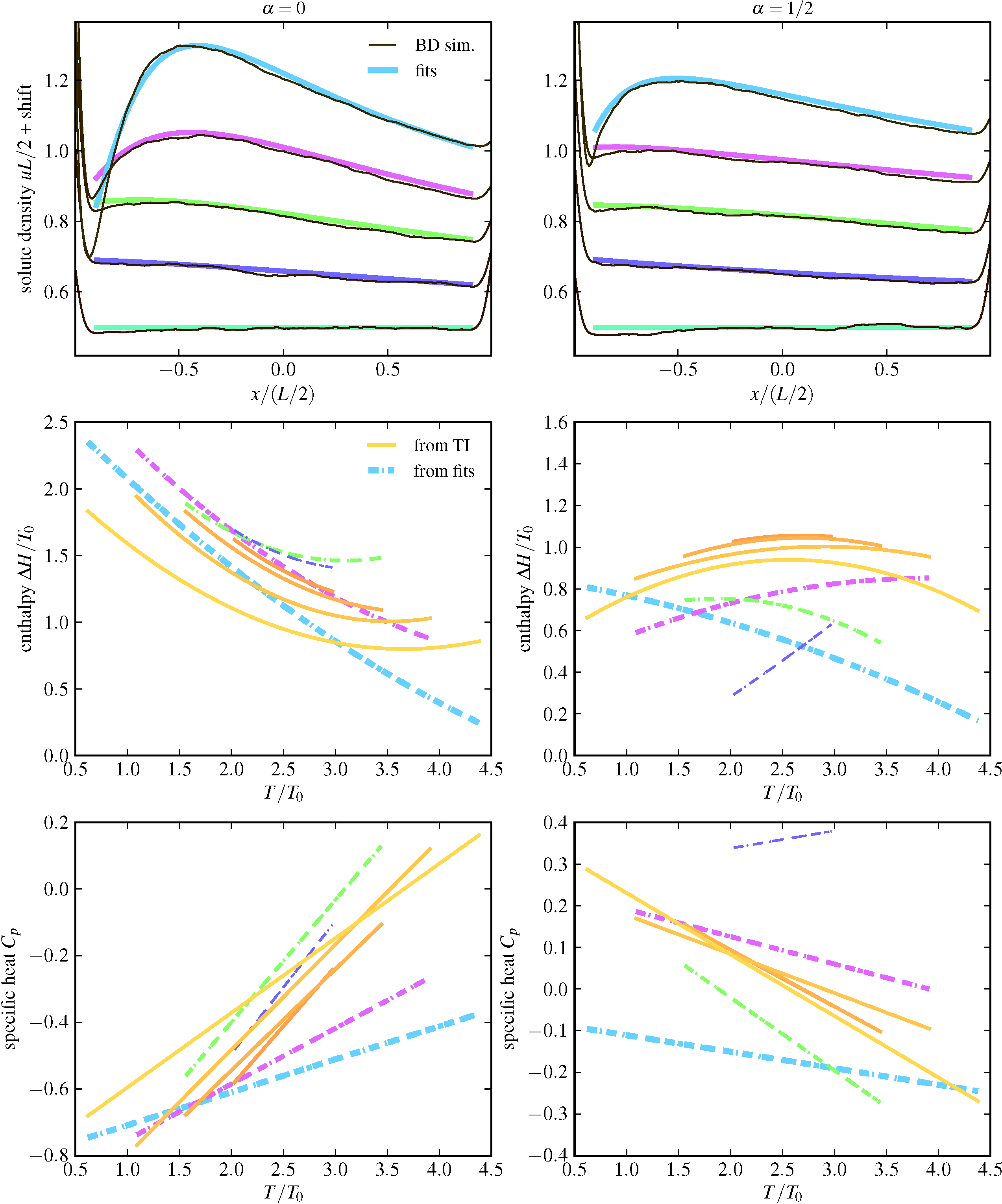}
 \caption{Thermophoretic simulation of the Gaussian solute in the ideal gas solvent compared to fits according to \eq{\ref{eq:density_fit_functions}}. \textbf{Top:} Solute density from simulations and fits. Again, the density curves are shifted for better visibility with an increasing positive shift in $y$-direction for increasing $\nabla T$. \textbf{Center:} Solvation enthalpy obtained from the fits compared with the enthalpy obtained from the equilibrium simulations (numerical TI). \textbf{Bottom:} Specific heat of solvation obtained from the fits compared with the specific heat obtained from the equilibrium simulations (numerical TI). }
 \label{fig:polymer_in_ideal_gas_fit}
\end{figure}

\begin{figure}[tph]
 \centering
    \includegraphics[width=\textwidth]{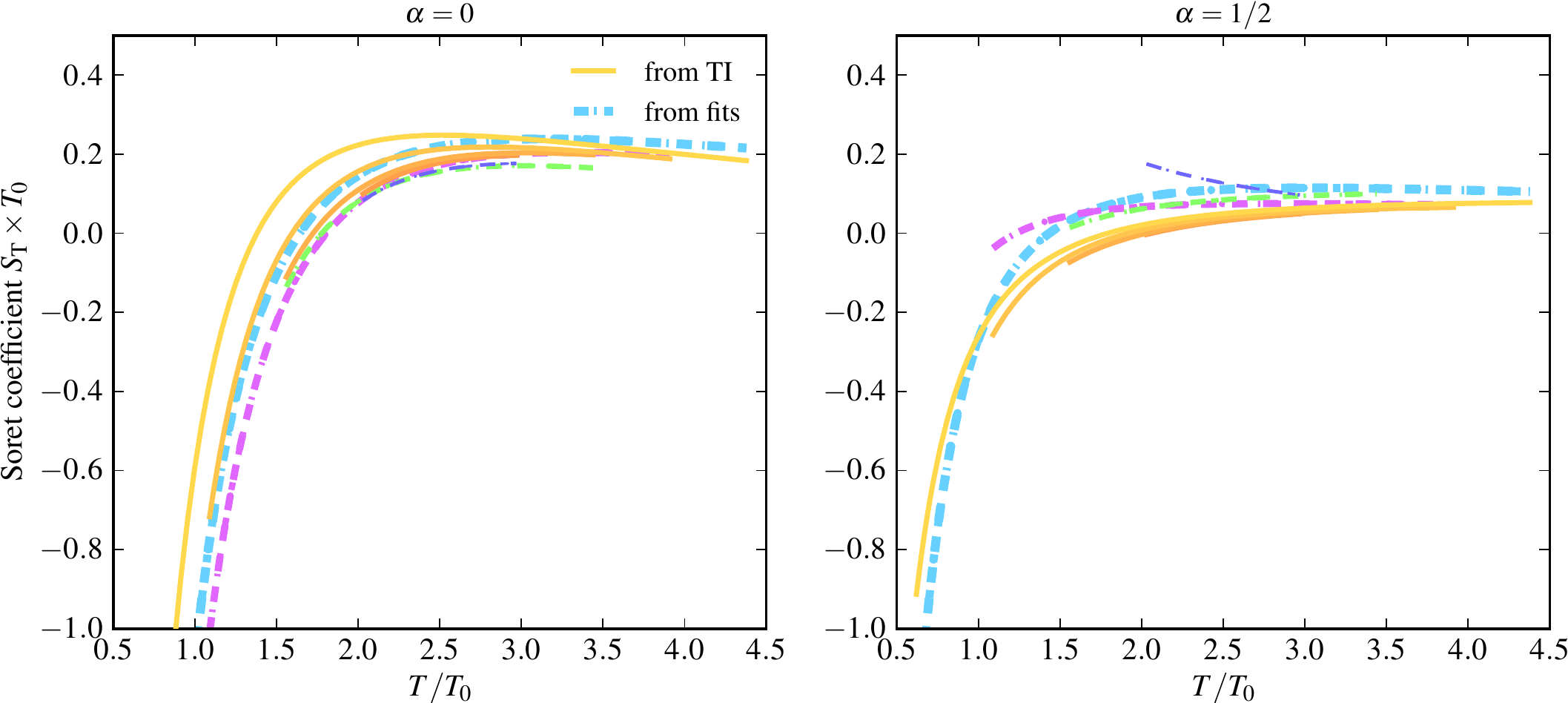}
 \caption{The Soret coefficient according to \eq{\ref{eq:soret_enthalpy}}, from equilibrium simulations (solid lines) and from the fits from the thermophoretic simulations (dashed lines). }
 \label{fig:polymer_in_ideal_gas_soret}
\end{figure}

The results are shown in \fig{\ref{fig:polymer_in_ideal_gas_expectation}}. As one can see, the ideal gas is following the expectaction almost exactly indicating the validity of the assumption that the solute does not disturb the solvent density too much. Furthermore we see the reasonable agreement of the solute density from the simulation and our expectation of $u^{(H)}(T(x))$, \eq{\ref{eq:solute_density_theory}a}
%for $\alpha=0$
both from the fit of the numerical TI as well as the analytical evaluation. Naturally it is not reasonable to assume exact agreement as we model a thermal non-equilibrium situation using quantities of thermal equilibrium. However the results indicate a link between the thermal diffusion coefficient $\DT$ and the local solvation enthalpy. Consequently this means that the system is roughly following the equation of state.

Using \eq{\ref{eq:density_fit_functions}a} we can fit the density from the simulation to find the enthalpy and specific heat of solvation which is shown in \fig{\ref{fig:polymer_in_ideal_gas_fit}}. Again one notices the tendency of agreement between thermophoretic simulations and the local equilibrium assumption, giving curves in the same range and the same appearance. However the results for the Stratonovich case $\alpha=1/2$ seem to be of a larger discrepancy with one fit even having a different signed slope compared to the other curves.

Another observation is that the solute density has a steep rise at the boundaries. This probably happens because once the solute is near  e.g. the left boundary, there will be more solvent particles on its right, thus deploying a net force towards the boundary, indicating that we assume a wrong solvent density at the boundary and thus the non-applicibality of the equilibrium assumption there. The boundary effects may alter the density appearance over the whole box and hence lead to further differences between the simulation and the local equilibrium assumptions.

Even though the results already show reasonable agreement with the hypotheses of \eq{\ref{eq:solute_density_theory}}, we want to compare the Soret coefficient that we can obtain from the Soret equilibrium density with the enthalpic hypothesis \eq{\ref{eq:soret_enthalpy}}. Both are depicted in \fig{\ref{fig:polymer_in_ideal_gas_soret}} for every temperature gradient. We see that the equilibrium and thermophoretic case differ, but follow the same trend.

%Nevertheless one has to notice a larger disagreement between the expectation and the simulation for $\alpha=1/2$ with the , while the reason for this discrepancy is unknown. 

\subsection{Other Homogeneous Systems}
\subsubsection{Ensemble of Gaussian Particles}

\subparagraph{Analytical Considerations}
As we have seen in the last subsection, the system of one solute particle in an ideal gas solvent seems to follow the equation of state, where it was rather simple to obtain the solute density since we knew the solvent density and the excess free energy was solely depended on that.  Another system interesting to study is one of pure Gaussian particles using the interaction energy of \eq{\ref{eq:gaussian_potential}}, where we limit ourselves to one-dimensional systems again. Here, the derivation of an equilibrium density $\rho$ is more difficult since the excess free energy per particle depends on the density itself, giving a transcendent equation. We will therefore model the density by means of the virial equation of state up to second order from \eq{\ref{eq:virial_eq_of_state}} and solve for $\rho$, giving
\begin{align}
\label{eq:density_from_B2}
 \rho = -\frac{1}{2B_2} + \sqrt{\frac1{4B_2^2}+\frac{P}{\kB T B_2}}
\end{align}
(omitting the negative solution), while we have to find the right virial coefficient $B_2$. Usually the virial expansion is performed for low densities which enables us to use \eq{\ref{eq:low_dens_B2}} and the derivation of \eq{\ref{eq:Gaussian_DeltaG_analytic}} to find 
\begin{align}
 B^{(\text{low})}_2 &= \frac 12 \int\limits_{-\infty}^{+\infty}\d r \l(1-e^{-\beta V(r)}\r) = \frac{\sigma\sqrt{\pi}}{2}\sum_{n=1}^\infty (-1)^{n+1}\frac{\l(\beta \varepsilon\r)^n}{n!\sqrt{n}}. 
\end{align}
Another route is to make use of the mean-field approach which is often an appropriate approximation for Gaussian particles \cite{Dzubiella:2001}. Within this approach the pair correlation function is taken to be $g(r) = 1$. Since the Gaussian interaction potential does not contain singularities with infinite repulsion, a significant overlap of particles can occur, flattening out the total interaction energy landscape and thus justifying the approximation. Following \cite{Dzubiella:2001}, we find
\begin{align}
 B^{(\text{mean})}_2 &= \frac12 \int\limits_{-\infty}^{+\infty}\d r V(r) = \frac{\varepsilon}2\sqrt{\pi}\sigma.
\end{align}
Furthermore, since the ideal gas follows $\rho\sqrt{T}=\const.$ for $\alpha=1/2$, we assume a virial expansion of
\begin{align}
 \frac{P_{1/2}}{\kB\sqrt{T}}= \rho + B_2 \rho^2
\end{align}
which admittedly is kind of an odd assumption but is investigated here nevertheless. For the density this yields \eq{\ref{eq:density_from_B2}} with $P\rightarrow P_{1/2}$ and $T\rightarrow\sqrt{T}$. Summed up, we have two approximations for the densities
\begin{subequations}
\label{eq:densities_gaussian_system}
\begin{align}
 \rho_\alpha^{\text{(low)}} &= -\frac{1}{2B^{\text{(low)}}_2} + \sqrt{\frac1{4(B^{\text{(low)}}_2)^2}+\frac{P_\alpha}{\kB T^{1-\alpha} B^{\text{(low)}}_2}}\\
 \rho_\alpha^{\text{(mean)}} &= -\frac{1}{2B^{\text{(mean)}}_2} + \sqrt{\frac1{4(B^{\text{(mean)}}_2)^2}+\frac{P_\alpha}{\kB T^{1-\alpha} B^{\text{(mean)}}_2}}.
\end{align} 
\end{subequations}
The pressure $P_\alpha$ has to be fixed by external conditions.

\begin{figure}[t!]
 \centering
    \includegraphics[width=\textwidth]{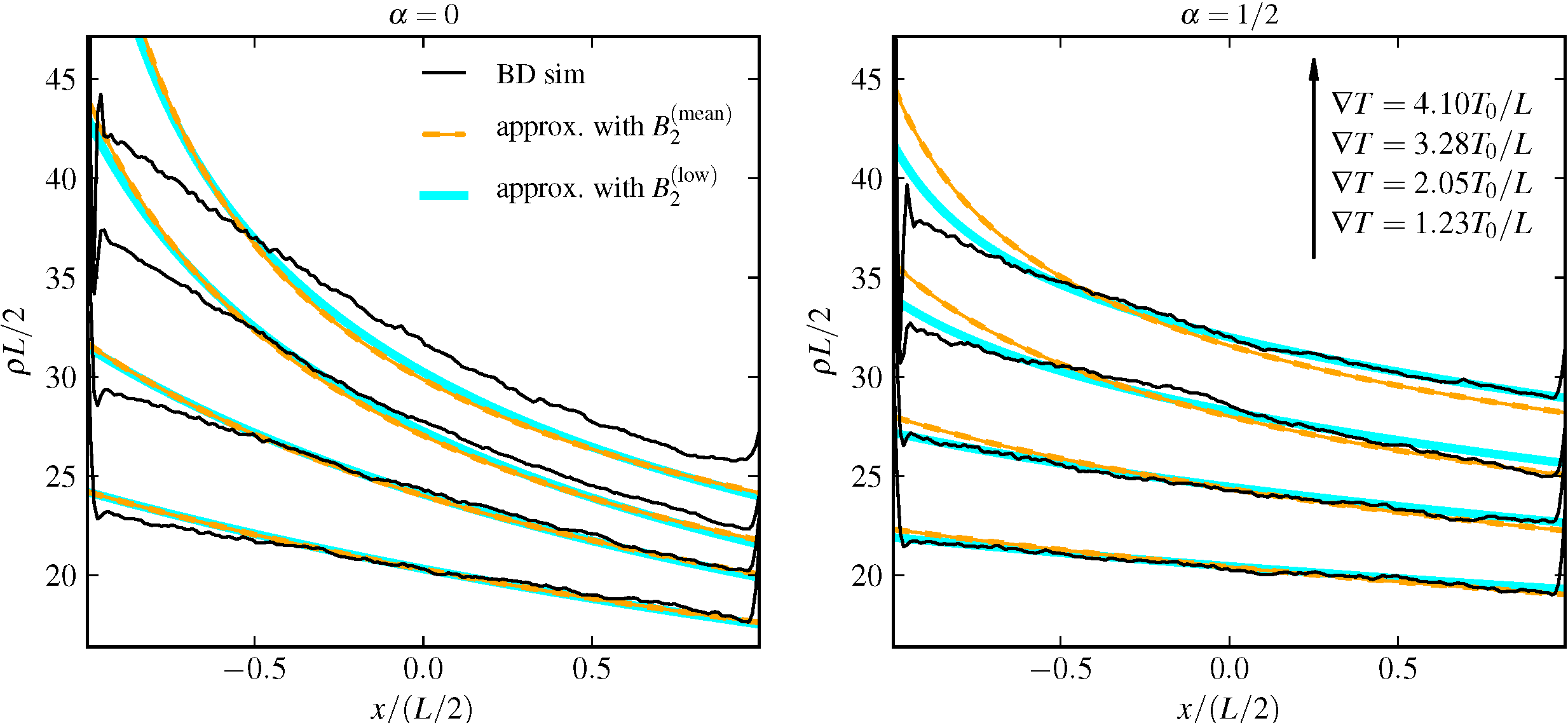}
 \caption{Equilibrium densities for a system of Gaussian particles in temperature gradients, simulations compared to the assumptions \eq{\ref{eq:densities_gaussian_system}}. The curves are shifted in positive $y$-direction with increasing temperature gradient.}
 \label{fig:polymer_solvent}
\end{figure}

\subparagraph{Simulations}
We performed one-dimensional BD simulations to test the approximations \eq{\ref{eq:densities_gaussian_system}} with a setup of $N=41$ particles of Gaussian interaction with $\sigma=0.5$ and $\epsilon=3.0$. The box length was set to $L=41$ in order to have a thermal equilibrium density of $\rho_0 =1.0$ and the boundary conditions were reflective. We choose a temperature profile of
\begin{align}
 T(x) = 2.5+x\frac{\nabla T}{T_0}
\end{align}
with $T_0=1.0$ and temperature gradients of $\nabla TL/T_0=1.23,2.05,3.28,4.10$ and a time integration constant $\Delta t=5\times10^{-10}$ which is still appropriate. Starting with an equally distributed system, the equations of motion were integrated over $3\times10^8$ steps with recording of the density histogram starting after $10^3$ steps and binning in 200 bins.
The results are shown in \fig{\ref{fig:polymer_solvent}} and compared to the approximations of \eq{\ref{eq:densities_gaussian_system}}, where the pressure $P_\alpha$ was fixed using $\rho_0=1.0$ and $T(0) = 2.5$.

We see reasonable agreement with \eq{\ref{eq:densities_gaussian_system}}, especially for lower densities, the regime in which the virial expansion is justified. Again, the system seems to follow roughly the equation of state.

\begin{figure}[t!]
 \centering
    \includegraphics[width=\textwidth]{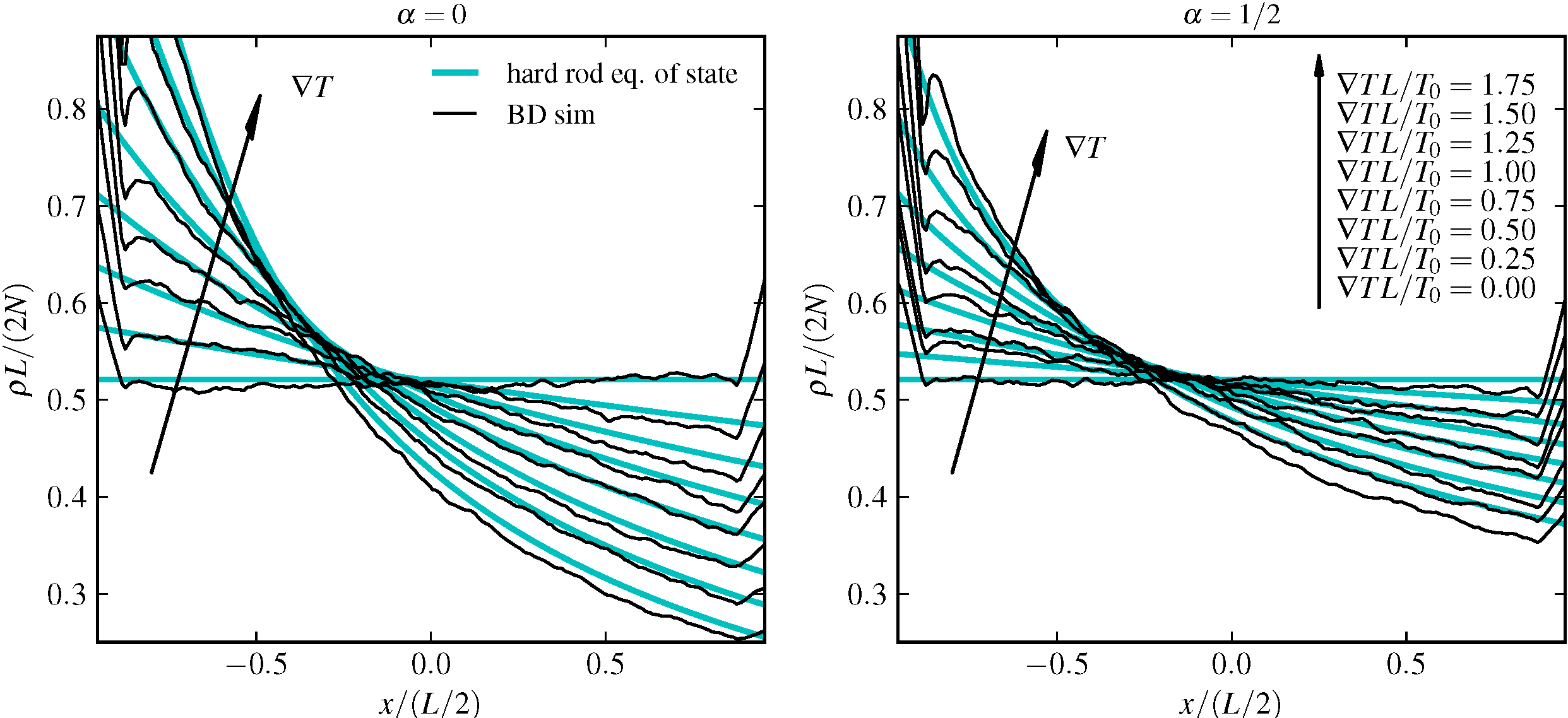}
 \caption{Comparison between Equilibrium densities of a system with hard rods in a linear temperature profile from simulations and from the equation of state \ref{eq:density_hard_rods} for increasing temperature gradients.}
 \label{fig:hard_rods}
\end{figure}

\subsubsection{Ensemble of Hard Rods}
\subparagraph{Analytic Considerations}
A system extensively studied is the one of hard rods in a one-dimensional box of length $L$ because it is one of the rare systems whose thermodynamic properties can be found exactly. Therefore it is a neat reference system to check the assumption of a system following its equation of state while being in a temperature gradient. The hard rod system uses the interaction potential \eq{\ref{eq:hard_sphere_interaction}} with $r$ being the distance between the rods' geometrical centers. We can use the equation of state for $N$ hard rods of length $\sigma$ \eq{\ref{eq:eq_of_state_hard_rods}} to get the equilbrium density $\rho = N/L$ as 
\begin{align}
 P&= \frac{N\kB T}{L - (N - 1)\sigma} = \frac{\rho\kB T}{1-\rho \sigma +\sigma/L }\\
 \rho &= \frac{P(1+\sigma/L)}{T+P\sigma}.
 \label{eq:density_hard_rods}
\end{align}
Again, for $\alpha=1/2$ we will use the odd equation of state with $T\rightarrow\sqrt{T}$ yielding the more general relation
\begin{align}
 \rho_\alpha &= \frac{P_\alpha(1+\sigma/L)}{T^{1-\alpha}+P_\alpha\sigma}.
 \label{eq:density_hard_rods}
\end{align}

\subparagraph{Simulations}
We want to check whether we can apply \eq{\ref{eq:density_hard_rods}} in a system of a linear temperature profile by means of BD simulations. Here, the numerical integration of motion can not directly be performed using \eq{\ref{eq:adapted_discrete_brownian}} because the interaction potential causes infinite forces. Therefor we proceed as follows. The forces can be modeled by taking every collision between particles as elastic. After updating the particles' positions due to the noise, we check for overlap of particles. If overlap occured, we determine the point of collision in between the discrete timesteps and update the position according to the laws of energy and momentum conversation.

We chose a one-dimensional box of length $L=5.0$, containing $N=5$ particles of length $\sigma=0.2$ and used a temperature profile of
\begin{align}
 T(x) = 1.0 + x\frac{\nabla T}{T_0}
\end{align}
with $\nabla T L/T_0 =0,0.25,0.50,0.75,1.00,1.25,1.50,1.75$ and $T_0=1.0$ s.t. the maximum temperature was $T_\text{max} \simeq 1.9$. Using the definition of the diffusion coefficient and an approximate variation of $\Delta x=0.02$ we find a temporal step size of $\Delta t= 4\times0.01^2/(2\times1.9)\simeq0.0001$. Starting with an equally distributed system, the equations of motion were integrated over $6\times10^7$ and recording of the equilibrium density started after $2\times10^5$ time steps by binning the location of the particles' geometrical center in 200 bins. The results of the simulations are shown in \fig{\ref{fig:hard_rods}} and compared to \eq{\ref{eq:density_hard_rods}}, where we fixed the pressure to $P_\alpha=1.2$ using $\rho_0=N/L=1.0$ and $T(0) = 1.0$.

Again, the system seems to roughly follow the equation of state.

\subsection{Summary}
In this section we investigated the behavior of systems exposed to a linear temperature gradient by means of BD simulations, assuming that the Einstein relation \eq{\ref{eq:einstein_relation}} holds in temperature gradients. We found that the ideal gas follows \eq{\ref{eq:density_id_gas_of_T}} for one- and two-dimensional systems, where we indicated that in the two-dimensional case the density can be approximated by means of a path integral over a non-conservative force-field considering only classical paths via \eq{\ref{eq:postulated_path_integral}}. A binary system of a Gaussian solute in an ideal gas solvent showed that the local solvation enthalpy drives the relaxation of the system, indicating that it follows the equation of state. Two further studies of homogeneous ensembles consisting of a) Gaussian particles and b) hard rods showed agreement with predictions from the equation of state, too.

%% file: tables/norm_const.tex
\begin{tabular}{ccc}
\hline\hline
$\nabla TL/T_0$ & $\mathcal{N}_{\nabla T,\alpha=1}$ & $\mathcal N_{\nabla T,\alpha=1/2}$\\
\hline
0.00 & 2.381 & 1.506\\
1.05 & 2.321 & 1.490\\
2.10 & 2.183 & 1.454\\
3.15 & 1.937 & 1.391\\
4.20 & 1.477 & 1.271\\
\hline
\end{tabular}

%% file: tables/fit_params_bd.tex
\begin{tabular}{c|cccc}
\hline\hline
                & \multicolumn{4}{c}{$\alpha=0$}\\
                      $\nabla TL/T_0$ & $a$ & $b$ & $c$ & $d$\\
    \hline
    1.05 & +3.532060 & -1.823576 & -0.178895 & +1.305741\\
2.10 & +3.255565 & -1.656150 & -0.151676 & +1.150322\\
3.15 & +2.967123 & -1.516854 & -0.157620 & +1.112006\\
4.20 & +2.296931 & -1.155239 & -0.111882 & +0.818776\\
\hline
                & \multicolumn{4}{c}{$\alpha=1/2$}\\
                      $\nabla TL/T_0$ & $a$ & $b$ & $c$ & $d$\\
    \hline
    1.05 & +0.571734 & +0.283355 & +0.067326 & -0.361376\\
2.10 & +0.551252 & +0.287375 & +0.068122 & -0.366826\\
3.15 & +0.609651 & +0.218671 & +0.046887 & -0.271598\\
4.20 & +0.455396 & +0.286463 & +0.073849 & -0.378253\\
\hline
\end{tabular}

%% file: chapters/md_section.tex
\section{Validation of Analytical Soret Equilibrium Density Distributions along Thermal Gradients by Explicit-Solvent MD Simulations}
\label{sec:mdsim_section}
The second approach to test the hypotheses established in \sec{\ref{sec:theory}} is to apply MD simulations, introduced in \sec{\ref{sec:mdsim_intro}}. As mentioned we will not apply the existing non-equilibrium algorithms but stick with a classical MD simulation in which two thermostatted regions at defined boundaries provide the system with heat, or take it out of the system, respectively. The solvent is chosen to be SPC/E water. As we are mainly interested in a proof of principle, the investigations will be limited to single noble gas atoms treated as Lennard-Jones spheres, leaving electrostatic effects aside for simplicity. We will proceed as follows.

In \sec{\ref{sec:thermophoretic_setup}} we will introduce different setups to induce thermal gradients and comment on their usefulness to generate meaningful results, finding that special thermostats made of water molecules in two-dimensional boundary conditions are the best choice. \sec{\ref{sec:mdsim_free_energy}} deals with the calculation of the solvation free energy of the solutes given in \tab{\ref{tab:LJ_parameters}}. Finally, \sec{\ref{sec:mdsim_results}} includes the results of thermophoretic simulations and draws a comparison to the free energy calculations from  \sec{\ref{sec:mdsim_free_energy}}.

\subsection{Setups for Thermophoretic Simulations}
\label{sec:thermophoretic_setup}
The main idea is to divide a cuboid simulation volume of dimensions $L_x\times L_y\times L_z$ into three regions along the $z$-axis, one region called $h$ (thermostatted with the ``hot'' temperature $T_h$), one region called $c$ (thermostatted with the ``cold'' temperature $T_c<T_h$) and the region of interest $m$, in which water molecules and solutes are located. We investigated three different setups to generate the temperature gradient. All of them have in common that the thermostats are made of position restrained molecules which are thermostatted, rather than thermostatting regions. The first has two advantages over the latter. First, one can control the total particle flux by defining the interaction potentials between the thermostats and the solvent/solute such that one achieves the Soret equilibrium condition $\v j=0$. Second, the algorithm does not have to evaluate the particles' position everytime it applies the thermotat algorithm; on the contrary, it is always applied to the same particles, which saves simulation time.

The setups are explained in the following.

\subsubsection{hcp Thermostats}
\begin{figure}[t!]
 \centering
    \includegraphics[width=\textwidth]{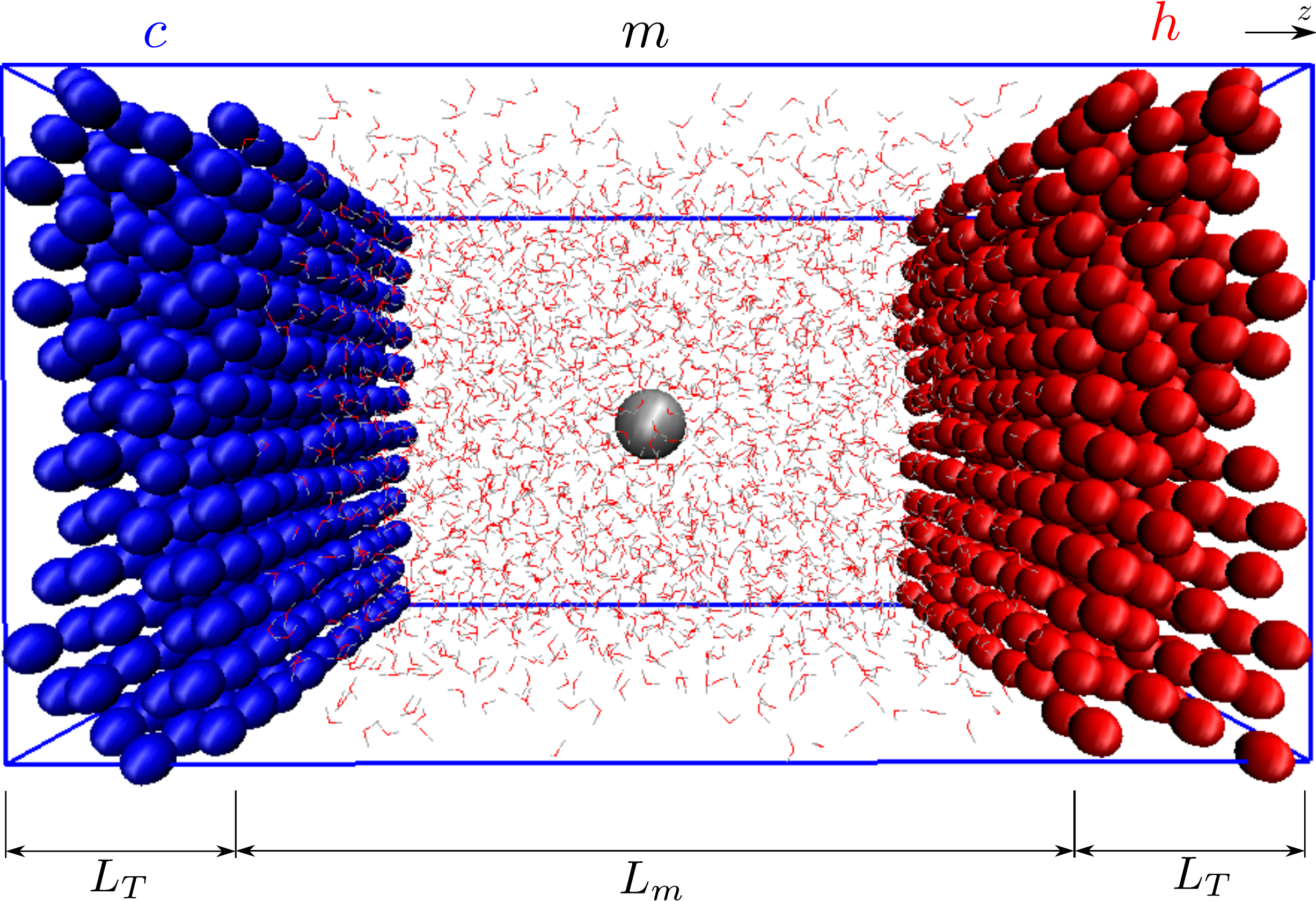}
 \caption{The hcp thermostat setup in which the thermostat molecules are noble gas atoms arranged in an hcp lattice. The noble gas atoms are thermostatted and provide the solvent molecules (lines, SPC/E water) with heat (coming from the red ``hot'' region), take out heat, respectively (in the blue ``cold'') region.}
 \label{fig:setup1}
\end{figure}
The first setup to be investigated is one where the thermostats are made of LJ spheres of radius $R$ ordered in an hcp-lattice. First a simulation box is defined with a cuboid region $m$ of volume $V = L_xL_yL_m$. Subsequently, a number $n_z$ of thermostat layers is definded. The spatial difference between two layers can be obtained by trigonometrical means to be $\Delta L = \frac23R\sqrt6$. Then the thermostat region has a length of $L_T = (n_z-1)\Delta L + \Delta z=(n_z-1)\frac23R\sqrt6$ with $\Delta z$ a global offset which is set to prevent thermostat molecules to leave the box too far in $z$-direction. The regions $h$ and $c$ are added to $L_m$ s.t. the new box length in $z$ is $L_z = L_m+2L_T$ and the new volume is $V=L_xL_yL_z$. After filling the regions $h$ and $c$ with noble gas-like atoms, the dimensions in $x$ and $y$ are adapted, s.t. we have a true periodic lattice when applying two-dimensional periodic boundary conditions (PBC). The third boundary condition (in $z$) is set to be reflective in order to prevent any mass flux. After preparing the walls, the solute is positioned at the center of the box and the solvent is added using GROMAC's \textit{genbox} program. Note that one should use the option \textit{-rvdw} which defines a radius around which solvent molecules will not be placed near the solute or the wall molecules. This prevents water molecules from being placed in the voids of the walls and makes minimization faster. One can see an illustration of this setup in \fig{\ref{fig:setup1}}.

\begin{figure}[t!]
 \centering
    \includegraphics[width=12cm]{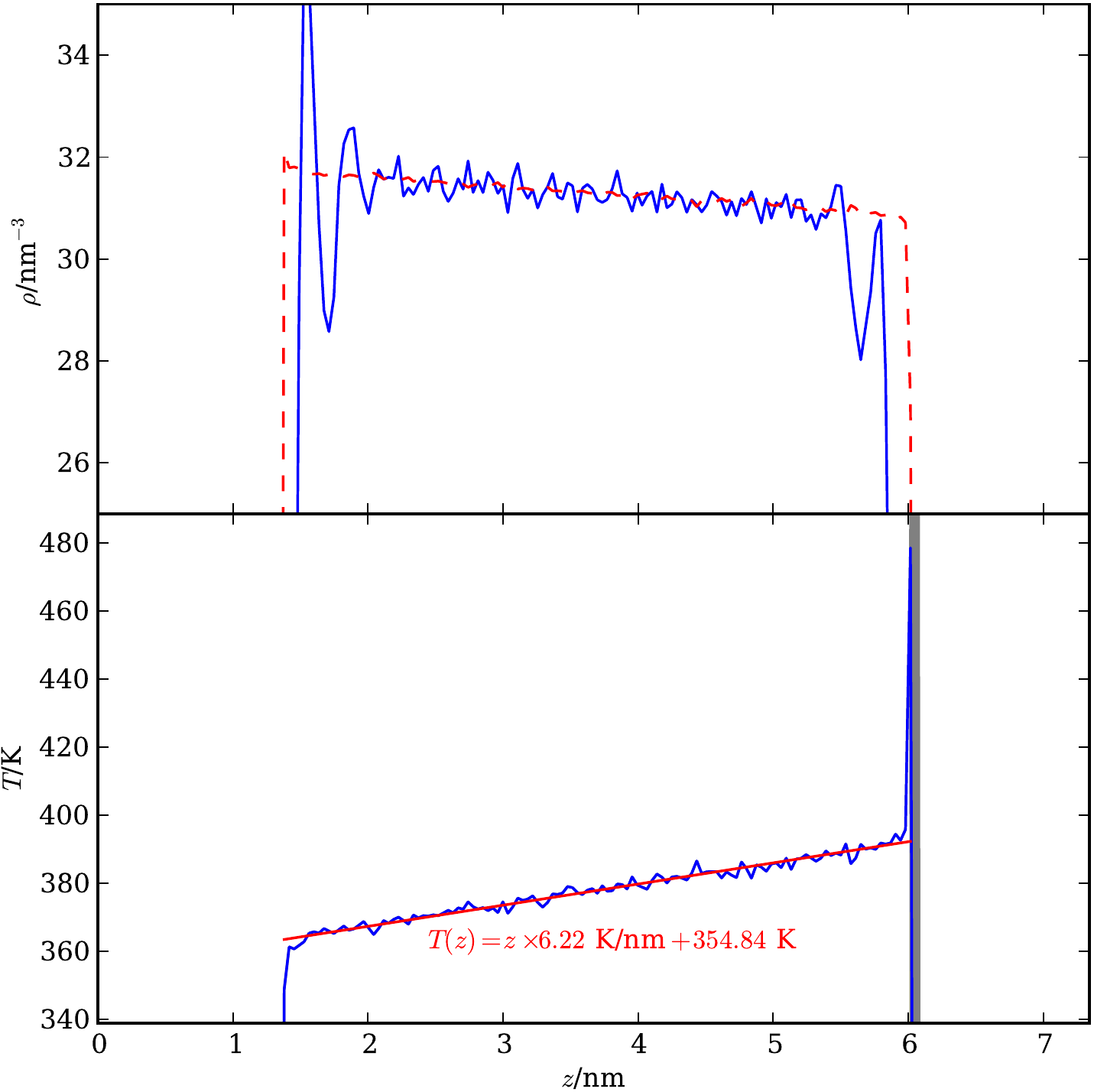}
 \caption{Results of an example bulk simulation using the hcp thermostats. \textbf{Top:} The density obtained from the simulation (solid blue line) fits remarkably well with the expectation from the equation of state (dashed red line), \eq{\ref{eq:number_density_fit}}. \textbf{Bottom:} The produced temperature profile (blue line) is linear but does not reach the boundary temperatures. The solid red line is a linear fit.}
 \label{fig:setup1_bulk}
\end{figure}

For a first impression of this thermostat, we performed bulk simulations using an absurdly large temperature difference with $T_c=100\unit{K}$, $T_h=700\unit{K}$ and restrained the wall particles with an harmonic potential of spring constant $k=8000\unit{kJ\,mol^{-1}\,nm^{-2}}$. After minimization and equilibration in an NPT ensemble for $0.1\unit{ns}$, the production run was performed in NVT for $20\unit{ns}$ with saving position and velocity values of the solvent atoms every $10\unit{ps}$. Density and temperature profiles in $z$-direction were calculated using the methods described in \sec{\ref{sec:mdsim_intro}}. We find the results shown in \fig{\ref{fig:setup1_bulk}}. While the temperature profile is linear as was derived in \sec{\ref{sec:heat_equation}}, it does not show the desired behavior to reach $T_c$ and $T_h$ at the boundaries. In fact, for the large temperature difference applied, the resulting gradient is remarkably small. This result has been reached by other simulations using the same thermostats and even lower spring constants without changing. It shows that the hcp setup gives little control over the desired temperature profile. The density profile shows remarkable behavior, as it follows almost exactly the SPC/E equation of state, given by \eq{\ref{eq:number_density_fit}} as $\rho(T(z))$ with the SPC/E parameters.

Solute simulations in which we tried to measure the Soret coefficient by means of \eq{\ref{eq:measure_soret_from_force}} failed, because the small temperature gradients did not impose a force strong enough to produce significant results in feasible simulation time.

\subsubsection{Water Thermostats with 3D PBC}

The second setup to be investigated is one where the regions $h$ and $c$ are made of thermostatted position restrained water molecules. We suspect the problem of the thermostats in the last section to be that the heat transfer between the noble gas hcp walls and the water solvent is rather low. To overcome this problem, one suspects that a water thermostat will couple better to the solvent and thus provides better control over the temperature gradient. Furthermore we want to check if it is possible to perform simulations with 3D PBC where the $h$-slab is centered at $z=L_z/2$, s.t. we have two regions $m$, one with a supposed temperature gradient $\nabla T$ and one with a temperature gradient $-\nabla T$. A schematic depiction of the setup is shown in \eq{\ref{fig:setup2}}.
\begin{figure}[t!]
 \centering
    \includegraphics[width=\textwidth]{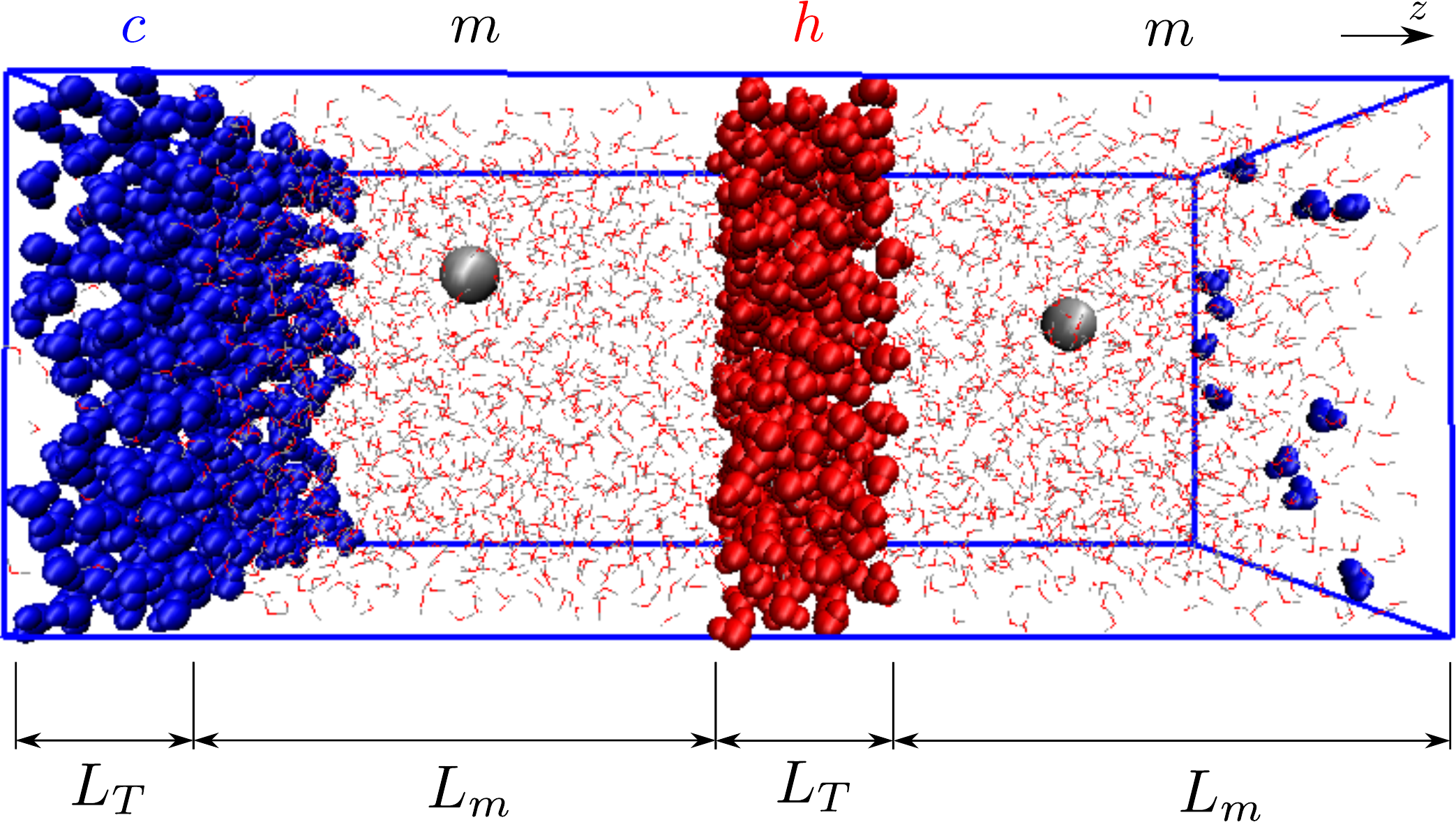}
 \caption{The water thermostat setup with 3D PBC. The thermostats are made of position restrained SPC/E water molecules. }
 \label{fig:setup2}
\end{figure}
\begin{figure}[t!]
 \centering
    \includegraphics[width=12cm]{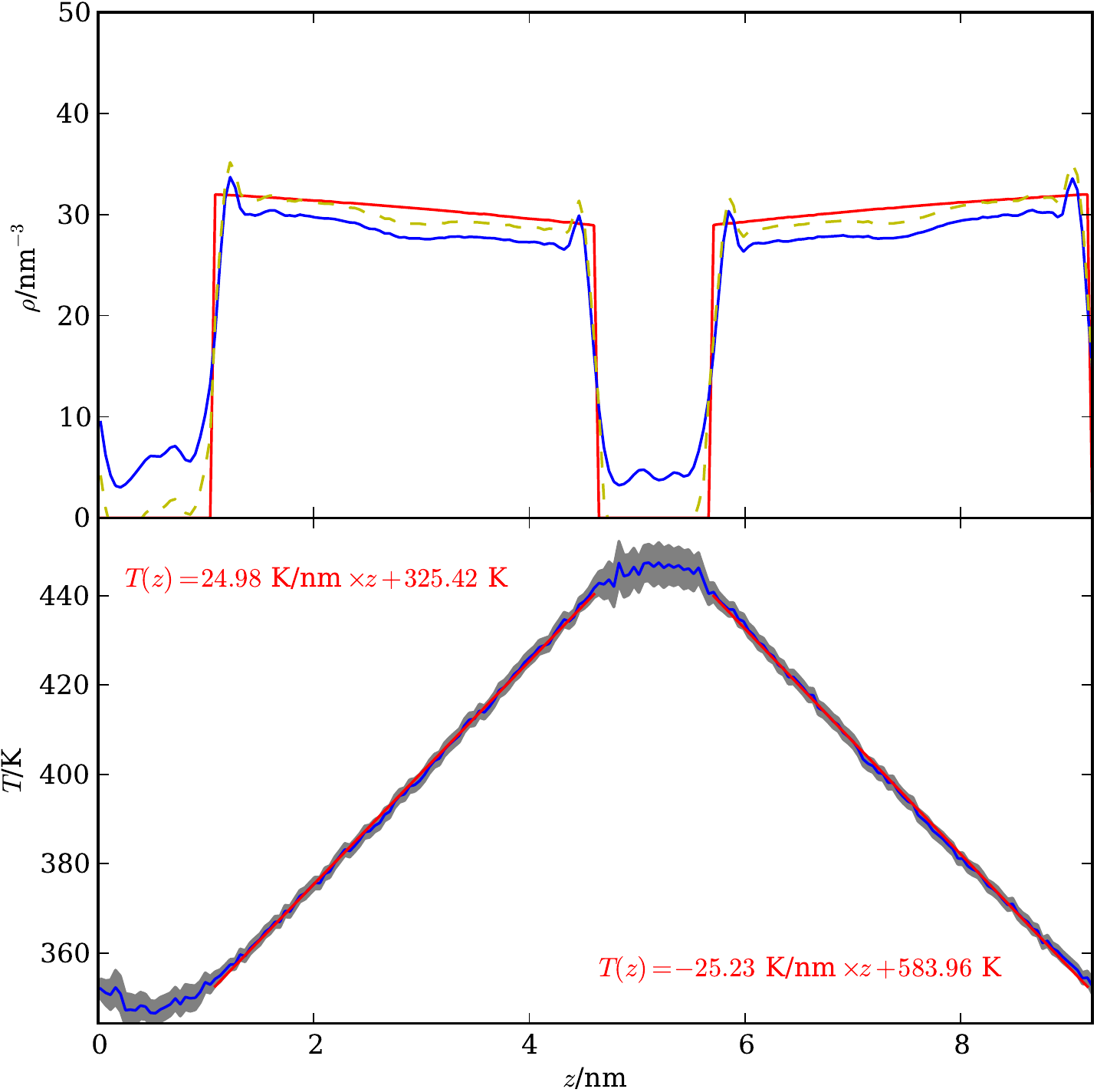}
 \caption{Results of an example bulk simulation using the water thermostats with 3D PBC with $T_c=350\unit K$ and $T_c=450\unit K$. \textbf{Top:} The density obtained from the simulation (solid blue line) does not fit with the expectation from the equation of state \eq{\ref{eq:number_density_fit}} (red line). Even after applying a correction where the integrated density from the thermostat regions is added to solvent region (dashed yellow line), eq. of state and simulation result differ. \textbf{Bottom:} The produced temperature profile (blue line) is linear and reaches the desired boundary temperatures. The solid red line is are linear fits.}
 \label{fig:setup2_bulk}
\end{figure}

We tested our assumptions by performing simulations with $T_c=350\unit K$ and $T_h=450\unit K$ and restrained the thermostat molecules with an harmonic potential of a spring constant of $k=5000\unit{kJ\,mol^{-1}\,nm^{-2}}$. After minimization and NPT equilibration of 1ns, an NVT production run was performed for 40ns where position and velocity of the molecules were recorded every 10ps. After the run, density and temperature were averaged over the frames and are displayed in \fig{\ref{fig:setup2_bulk}}. As one can see, the temperature profiles follow exactly the expectation from \sec{\ref{sec:heat_equation}}, with the solvent reaching the desired temperatures at the boundaries to the regions $h$ and $c$. However, the simulation density (blue line in \fig{\ref{fig:setup2_bulk}}) differs significantly from the equation of state because water molecules from the regions $m$ diffuse into the thermostat regions $h$ and $c$. Not only does that change the desired density, but one can  suspect that this effect produces a non-vanishing net flow of solvent molecules in $z$-direction. The yellow dashed line in \fig{\ref{fig:setup2_bulk}} shows a correction, where the integrated solvent density from the thermostat region is substracted and added to the $m$ regions. Even with the correction, the simulation density does not exactly follow the desired expectation of the equation of state, indicating that a net flow in $z$-direction is possible to change the outcome.

We tried to measure the Soret coefficient of a xenon solute by means of \eq{\ref{eq:measure_soret_from_force}}, however the results indicated that a non-vanishing mass flow biased the outcome.

\subsubsection{Extended Water Thermostats with 2D PBC}
We see that in order to obtain the right solvent density, it is indeed useful to employ two-dimensional PBC. Furthermore the solvent molecules must be prevented to diffuse too far into the thermostats, yet they have to be allowed to interact with the thermostat on a short length scale in order to carry heat sufficiently back to the region $m$, give heat to the region $c$, respectively.  We therefore propose a more complicated setup to produce the desired conditions.

We divide the simulation box in five parts in $z$-directions. The region which keeps its purpose as the region carrying solvent and solute is $m$. The region of one of the thermostats $x=h,c$ with length $L_T$ in $z$-direction is devided in two regions $xA$ and $xB$ with length $L_{T,A}=(1-\phi) L_T$ and $L_{T,B}=\phi L_T$; $xB$ shares a boundary with $m$ whereas $xA$ is at the box boundary. The regions are filled with water-like molecules called $SxA$ and $SxB$. They carry the SPC/E properties to interact with each other, but interact differently with the solvent molecules $SOL$ and the solute dummy particles $XEM$. In order to prevent $SOL$ molecules to diffuse too far into the thermostats, the molecules $SxB$ act with a shifted-LJ potential, thus building a repulsive interface in the thermostat region. The $SxA$ molecules interact normally with $SOL$, s.t. the heat can be transmitted properly. We want to avoid solute molecules $XEM$ to diffuse too far into the thermostat regions, too, such that $SxA$ acts with a shifted-LJ on $XEM$, building an interface which $XEM$ can not overcome. The only difference between the molecules $ShA$ and $ScA$ is that they are thermostatted with the different temperatures (which is the only difference between $ShB$ and $ScB$, too). The interaction parameters $\epsilon$ and $\sigma$ are taken from \tab{\ref{tab:LJ_parameters}}, only the division parameter $\phi$, the thermostat length $L_T$ and the shift parameters $R_0$ are fixed as
\begin{subequations}
\begin{align}
 \phi &= 0.35\\
 L_T &= 1.00\unit{nm}\\
 R_0(SxA\leftrightarrow SOL) &= 0.15\unit{nm}\\
 R_0(SxB\leftrightarrow SOL) &= 0.00\unit{nm}\\
 R_0(SxA\leftrightarrow XEM) &= 0.30\unit{nm}\\
 R_0(SxB\leftrightarrow XEM) &= 0.00\unit{nm}.
\end{align}
\end{subequations}
In \sec{\ref{sec:mdsim_results}} we will see that this thermostat setup gives the control over the system's conditions that the other setups were lacking. 

For measuring the Soret coefficient we will not restrain the solute and follow the route given by \eq{\ref{eq:soret_from_density}}, hoping that this will yield clearer results than using the force measurement.

\begin{figure}[t!]
 \centering
    \includegraphics[width=\textwidth]{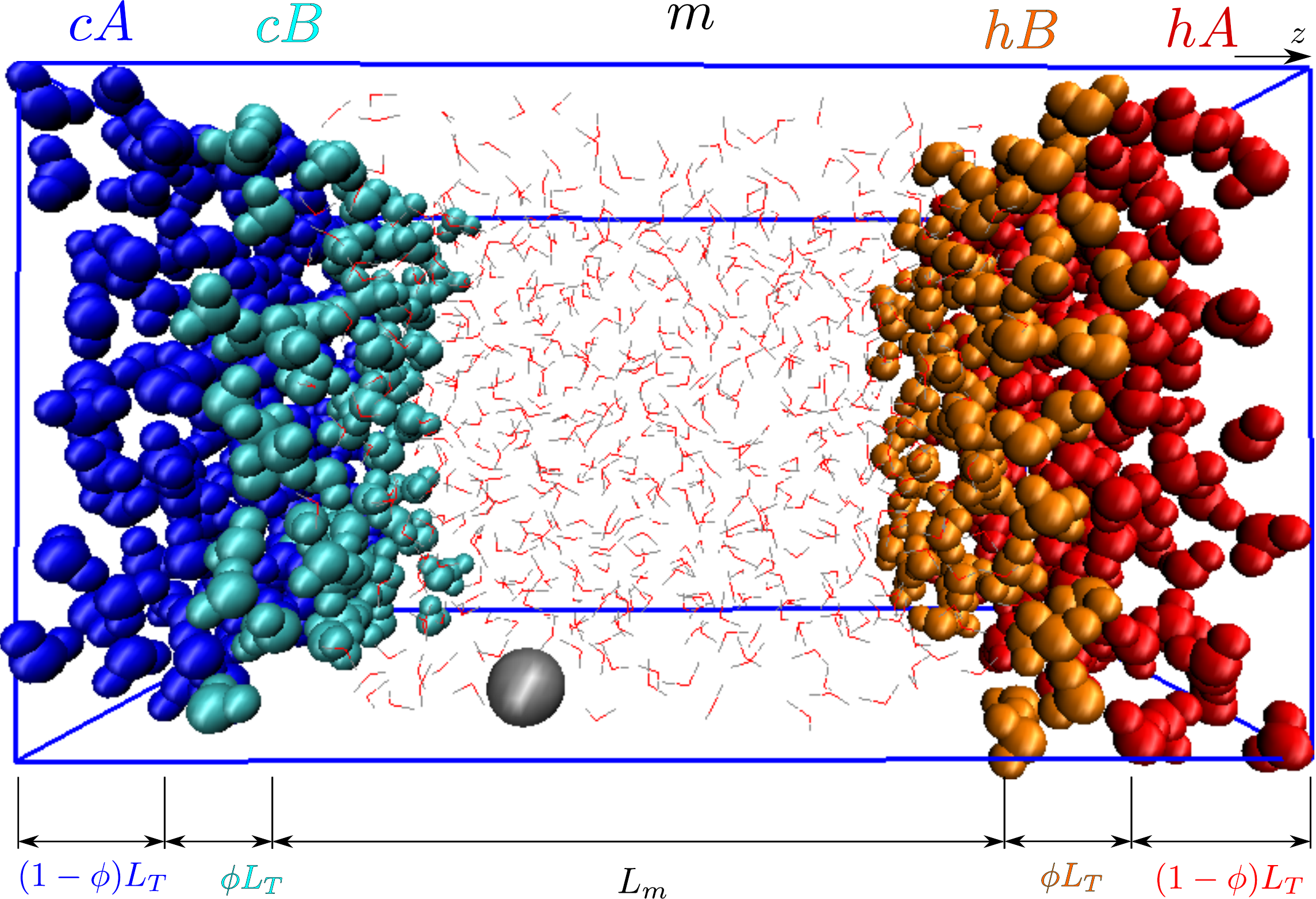}
 \caption{The extended water thermostat setup with three-dimensional PBC. The $ShA$ molecules are red, the $ShB$ molecules orange, the $ScA$ molecules blue and the $ScB$ molecules cyan. The solute $XEM$ is the silver sphere and the water molecules $SOL$ are depicted as lines.}
 \label{fig:setup2}
\end{figure}

\subsection{Free Energy Calculations}
\label{sec:mdsim_free_energy}
Before the thermophoretic simulations can be performed, we need the hydration free energy of the solute in SPC/E water as a function of $T$, s.t. we are able to interpret the thermophoretic results correctly. To this end we used TI over a large region of temperature values with the results given in the following.

\subsubsection{Solvation of a Xenon Atom in SPC/E Water with the BAR Method}
\begin{figure}[p]
 \centering
    \includegraphics[width=\textwidth]{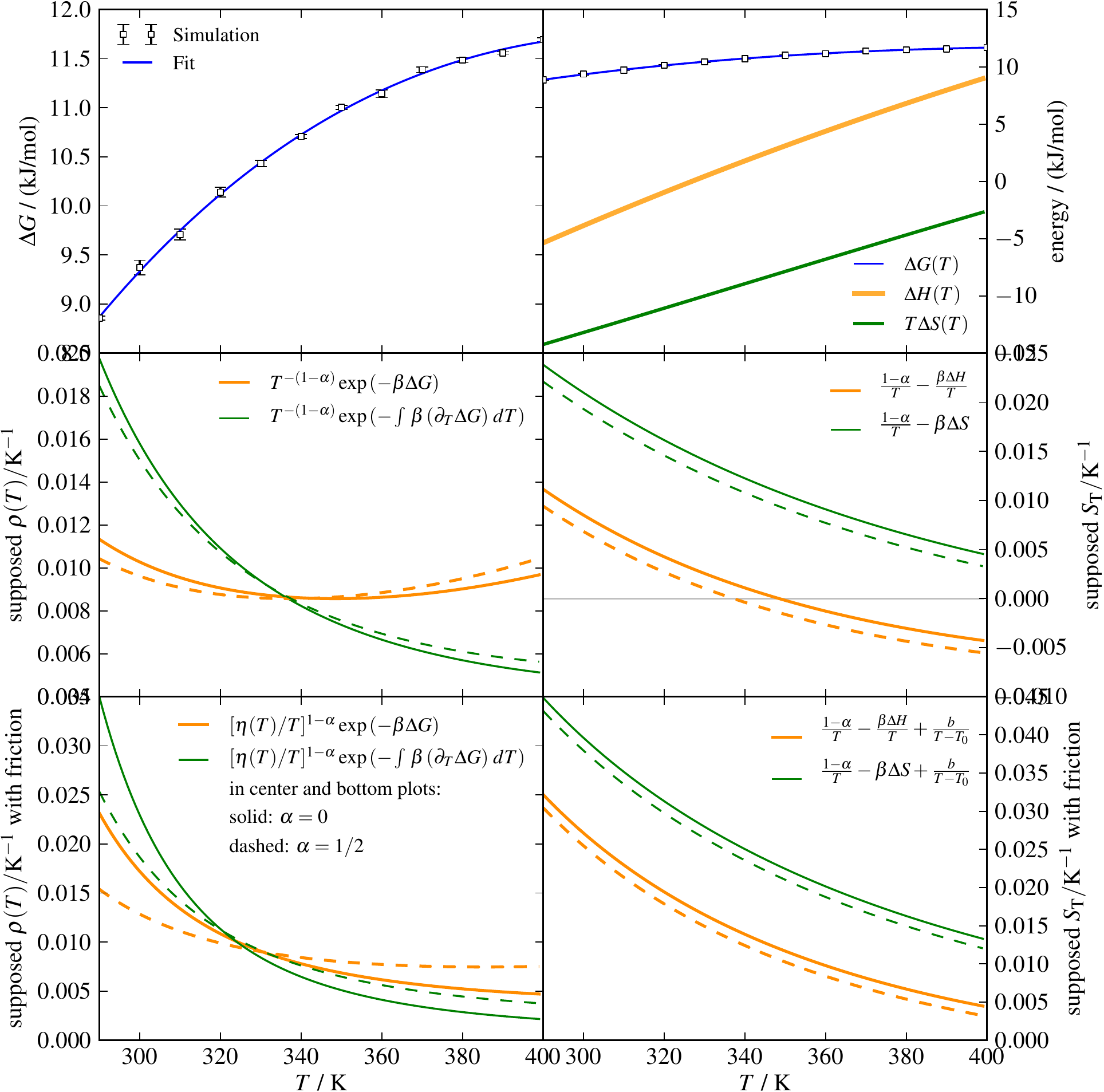}
 \caption{Results of the free energy $\Delta G$ for the hydration of xenon in SPC/E water by means of the BAR method. \textbf{Top left: } Simulation results and fit with \eq{\ref{eq:free_energy_fit_function}}. \textbf{Top right: } Solvation free energy, solvation enthalpy and solvation entropy (\eq{\ref{eq:thermodynamic_fit_functions}}). \textbf{Center left:} The supposed densities in thermophoretic simulations, following \eq{\ref{eq:density_fit_functions}} and \textbf{center right:} the corresponding supposed Soret coefficients \eq{\ref{eq:soret_definitions}}. \textbf{Bottom left: } Supposed densities in thermophoretic simulations with additional friction dependence \eq{\ref{eq:solute_density_theory_with_friction}}, with the friction proportional to the SPC/E viscosity \eq{\ref{eq:viscosity}} and \textbf{bottom right:} the corresponding supposed Soret coefficients (\eq{\ref{eq:soret_definitions}} with \eq{\ref{eq:soret_fric}}).}
 \label{fig:BAR_xe}
\end{figure}

The simulations were done as described in \sec{\ref{sec:free_energy_MD_setups}} and the results are shown in \fig{\ref{fig:BAR_xe}} together with a fit using the fit function \eq{\ref{eq:free_energy_fit_function}}, whose parameters are given in \tab{\ref{tab:fit_params_md}}. Additionally, \fig{\ref{fig:BAR_xe}} shows the supposed behavior for the densities \eq{\ref{eq:density_fit_functions}} for a system exposed to a temperature gradient. We see that the two proposed hypotheses for the Soret coefficient \eq{\ref{eq:soret_definitions}} result in crucially different densities, as was already the case in \sec{\ref{sec:gaussian_solvation}}. Furthermore, we show the density in case that the friction is temperature dependent, too. Throughout this section, we assume that the friction is \eq{\ref{eq:stokes_friction}} $\gamma(T) = 6\pi\eta(T)R$ with a temperature dependent viscosity and a constant radius $R$, such that it suffices to include the SPC/E viscosity \eq{\ref{eq:viscosity}} in our considerations. Wee see that especially for temperatures $T<400\unit K$ it is difficult to distinguish the friction and enthalpy induced density from the entropy induced density without friction. As in  \fig{\ref{fig:BAR_xe}} the hypothesized Soret coefficient is plotted, too, we furthermore see that in general the Soret coefficient derived in section \sec{\ref{sec:theory}} seems to decrease with increasing temperature. Comparing this behavior with the behavior of biological macromolecules, PS beads or the non-ionic part of DNA's Soret coefficient, shown in \fig{\ref{fig:ST_T_dependence}}, we notice a completely different behavior. We see however that using the hypothesized definitions of the Soret coefficient we obtain a sign change with increasing temperature which is a crucial property of the Soret coefficient of real molecules. The sign change happens at the temperature where the density is extremal due to their connection \eq{\ref{eq:soret_from_density}}.

By using Widom's TPI for different noble gas like atoms, we want to study the influence of the atom size on the hypothesized Soret coefficient and see whether we find a sign change for increasing particle size.

\subsubsection{Solvation of Argon, Krypton and Xenon in SPC/E Water with the TPI Method}
\begin{figure}[t!]
 \centering
    \includegraphics[width=10cm]{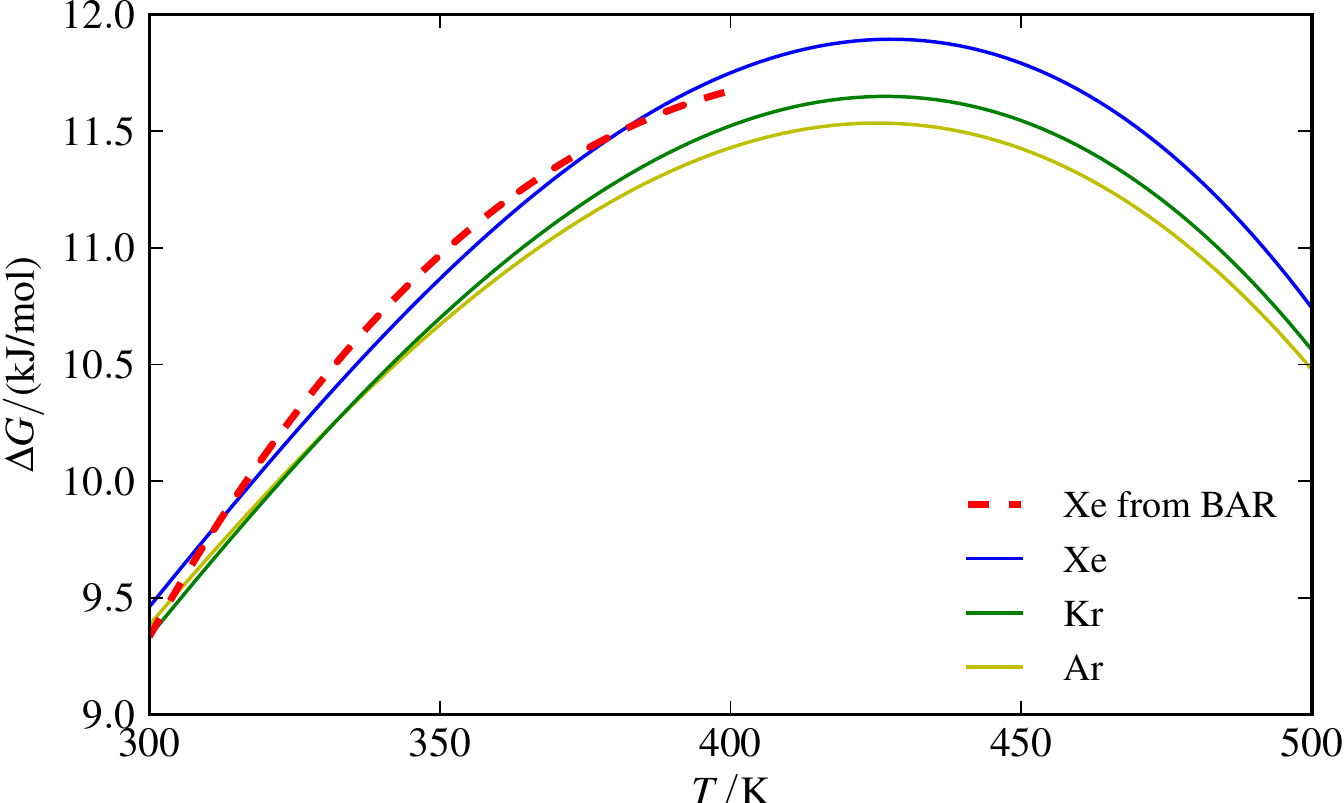}
 \caption{Results for the evaluated solvation free energies of this section, plotted via \eq{\ref{eq:free_energy_fit_function}} using the parameters given in \tab{\ref{tab:fit_params_md}}}
 \label{fig:delta_G_results_md}
\end{figure}

\begin{figure}[p]
 \centering
    \includegraphics[width=\textwidth]{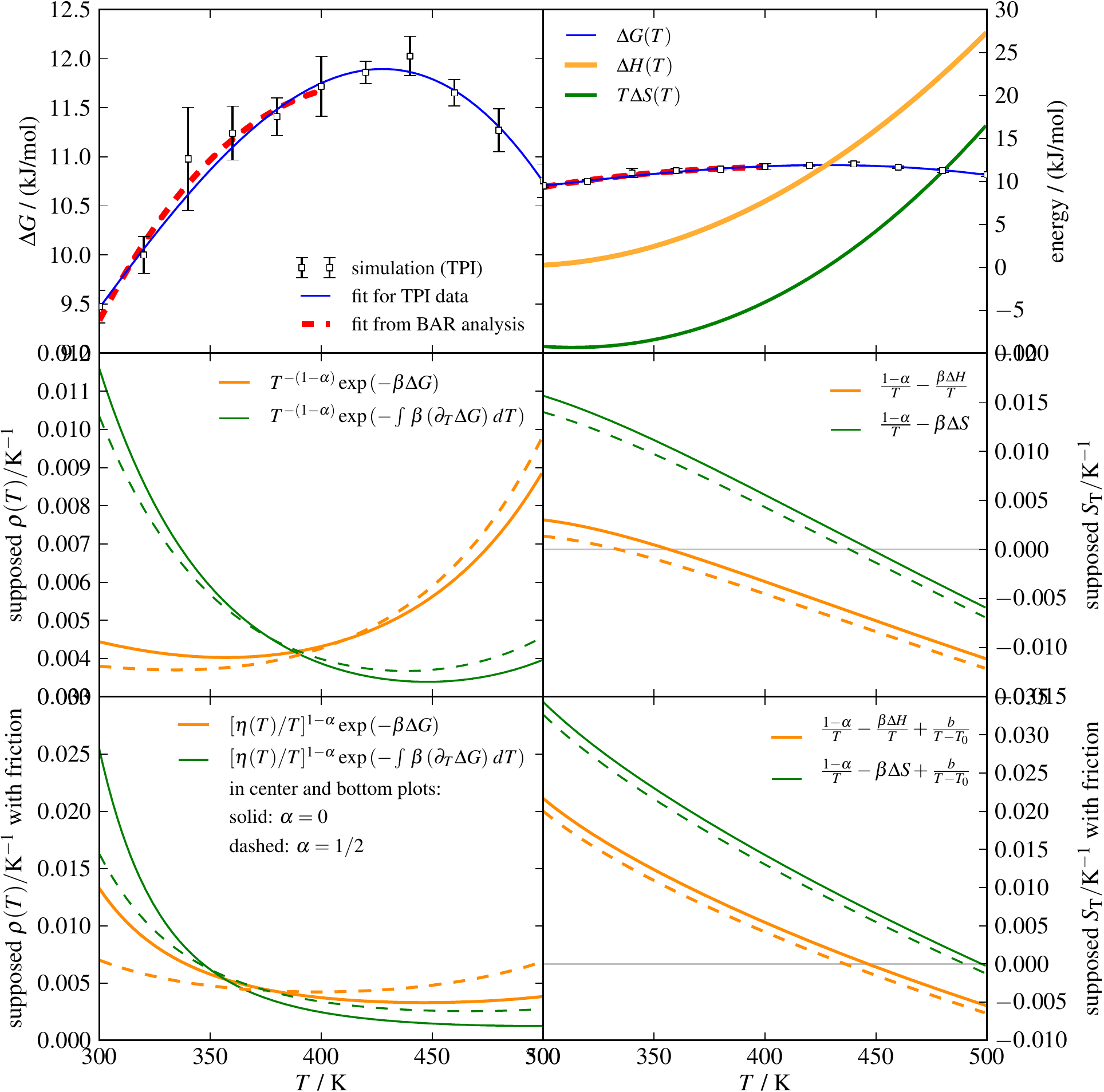}
 \caption{Results of the free energy $\Delta G$ for the hydration of xenon in SPC/E water by means of the TPI method. \textbf{Top left: } Simulation results and fit with \eq{\ref{eq:free_energy_fit_function}} and the comparison to the results from the BAR solvation, \fig{\ref{fig:BAR_xe}} \textbf{Top right: } Solvation free energy, solvation enthalpy and solvation entropy (\eq{\ref{eq:thermodynamic_fit_functions}}). \textbf{Center left:} The supposed densities in thermophoretic simulations, following \eq{\ref{eq:density_fit_functions}} and \textbf{center right:} the corresponding supposed Soret coefficients \eq{\ref{eq:soret_definitions}}. \textbf{Bottom left: } Supposed densities in thermophoretic simulations with additional friction dependence \eq{\ref{eq:solute_density_theory_with_friction}}, with the friction proportional to the SPC/E viscosity \eq{\ref{eq:viscosity}} and \textbf{bottom right:} the corresponding supposed Soret coefficients (\eq{\ref{eq:soret_definitions}} with \eq{\ref{eq:soret_fric}}).}
 \label{fig:widom_xe}
\end{figure}

Widom's TPI has been used to obtain the solvation free energy for three noble gas atoms, as described in \sec{\ref{sec:free_energy_MD_setups}} with the LJ parameters given in \tab{\ref{tab:LJ_parameters}}. The results for xenon are shown in \fig{\ref{fig:widom_xe}}, whereas the results for argon and krypton are presented in \app{\ref{app:noble_gas_free_energy}}, with the values of all fits according to \eq{\ref{eq:free_energy_fit_function}} given in \tab{\ref{tab:fit_params_md}} and shown in \fig{\ref{fig:delta_G_results_md}}.

We generally see the behavior already discussed for the BAR solvation over a larger temperature range. However, the TPI fit of the xenon $\Delta G$ yields a different curvature in the hydration enthalpy $\Delta H$, meaning that the TPI fit parobala has a positive prefactor with a minimum at a temperature $T<350\unit K$ whereas the the BAR fit parabola yields a negative prefactor and a maximum for a temperature $T>400\unit K$. This shows that all results obtained solely by fits should be treated with care and not extrapolated to regions where the temperature was not simulated. However, studying trends such as the general Soret coefficient temperature dependenc and scaling with solute size is still possible as those properties do not crucially dependend on the curvature of the $\ST$ curves.

\begin{figure}[t!]
 \centering
    \includegraphics[width=\textwidth]{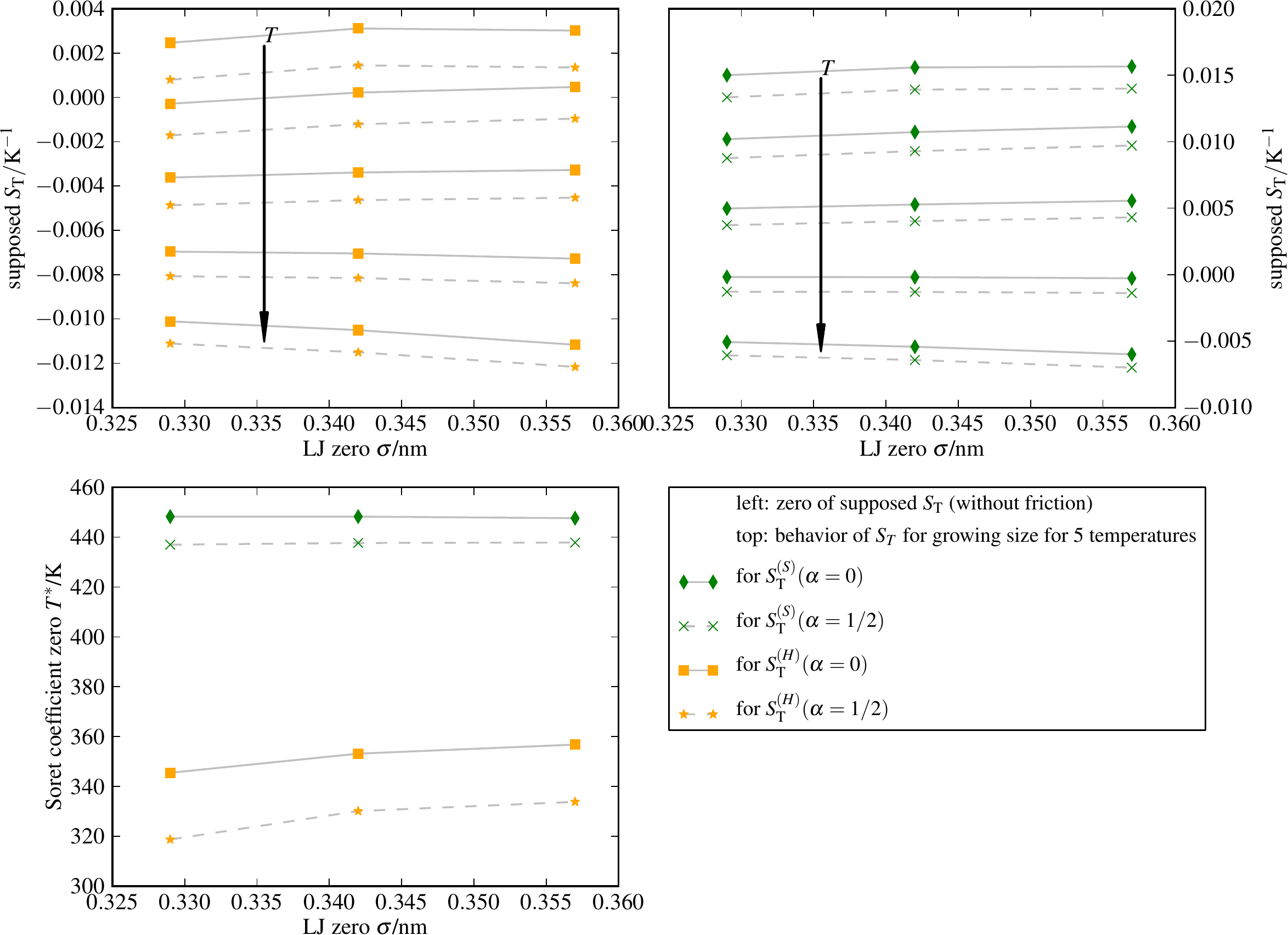}
 \caption{\textbf{Top: } Hypothesized (\eq{\ref{eq:soret_definitions}}) Soret coefficient's dependence on the particle size for 5 equidistant temperatures $300\unit K\leq T \leq  500\unit K$. We used the fits from the TPI method to obtain the values for $\ST$. \textbf{Bottom: }The zeros $T^*$ (crossover temperature as in \eq{\ref{eq:piazza_fit}}) of the supposed Soret coefficient \eq{\ref{eq:soret_definitions}} plotted against the LJ parameter $\sigma$ as a further measure of the atom size dependence of the Soret coefficient.}
 \label{fig:soret_zeros}
\end{figure}

Given the three different particle sizes we notice that the zero $T^*$ (crossover temperature as in \eq{\ref{eq:piazza_fit}}) of the Soret coefficient hypotheses \eq{\ref{eq:soret_definitions}} seems to shift to a lower temperature with decreasing size. This behavior is investigated in the lower part of \fig{\ref{fig:soret_zeros}}, where a Newton search for a zero was performed on the six functions given by the three noble gas fits and the two hypotheses for the Soret coefficient, for both cases of $\alpha$. The result shows that in our size range, the zero from the entropy hypothesis seems not to be shifted, whereas the zero of the enthalpy hypothesis increases with increasing particle size. A more general analysis of the hypothetical Soret coefficient's solute size behavior is shown in the upper part of \fig{\ref{fig:soret_zeros}}. We notice that the Soret coefficient seems to be generally size dependent. In the temperature region where the Soret coefficient is positive, it increases with particle size and decreases with particle size in the temperature region where it is negative. The Soret coefficient of the entropy hypotheses seems to show a weaker size dependence.

We should emphasize again that these results are only hypothetical due to them being based on the equilibrium assumptions from \sec{\ref{sec:theory}}.

\begin{table}[t!]
 \centering
 \input{bilder/md_ti/fit_params_md}
 \caption{Fit parameters for the temperature dependence of the hydration free energy \eq{\ref{eq:free_energy_fit_function}} found by means of a least square fit from the TPI data according to \eq{\ref{eq:free_energy_fit_function}}. The fits are depicted in \fig{\ref{fig:delta_G_results_md}} and in \app{\ref{app:noble_gas_free_energy}}.}
 \label{tab:fit_params_md}
\end{table}

\subsubsection{Solvation Free Energy for Shifted-LJ Spheres}
Even though the following results have not been used to check the thermophoretic simulations (meaning that we did not investigate the thermophoresis of shifted-LJ solutes), we want to include them here to show how the hydration free energy of such solutes can be obtained by means of TI.

\begin{figure}[t!]
    \centering
    \subfigure[\textbf{Left:} The pair correlation functions between the xenon like shifted-LJ solute and the SPC/E oxygen obtained from the simulations at increasing values for $R_0$ with $0\leq R_0\leq0.204\unit{nm}$. \textbf{Right: } The mean force acting on the solute, as obtained from integration by means of $g(r)$, \eq{\ref{eq:shifted_LJ_exp_force}}.]{
     \includegraphics[width=\textwidth]{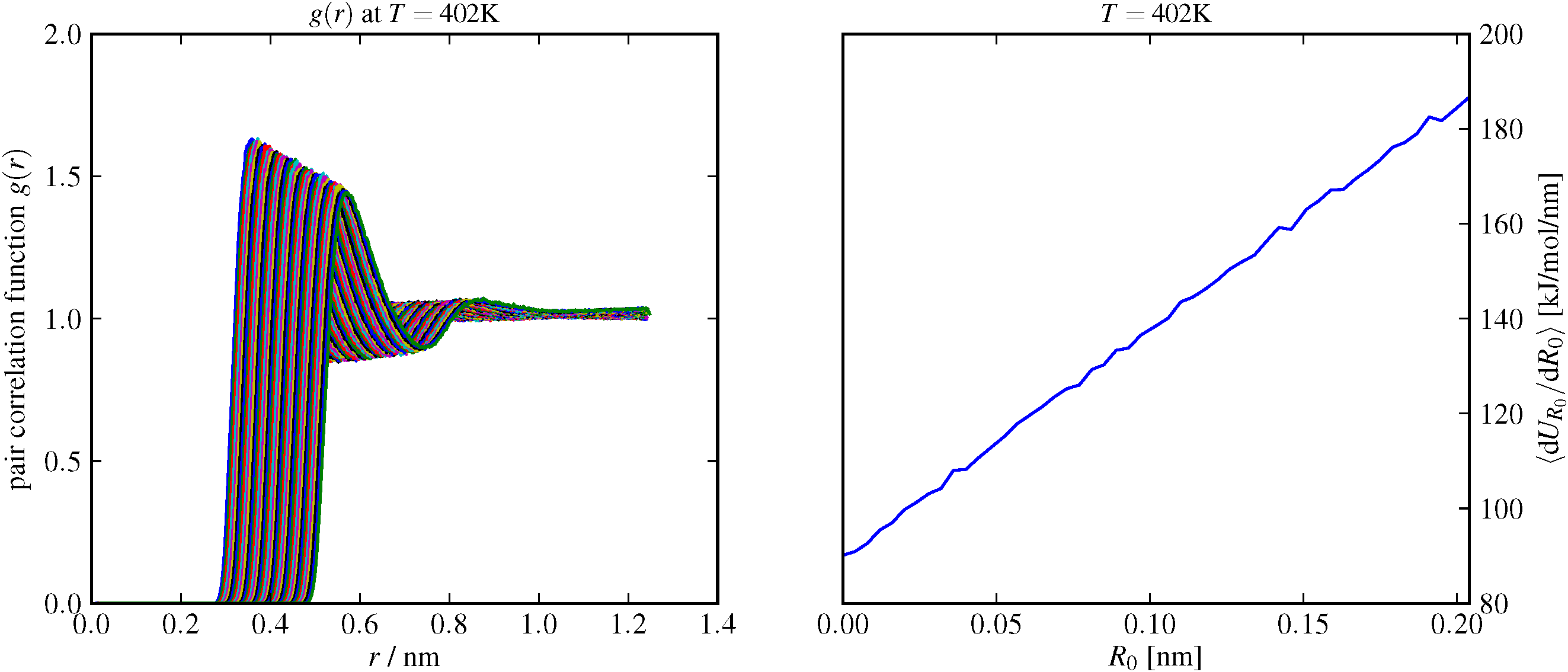}
     \label{fig:shifted_LJ_rdf_force}
    }%
    \\
    \subfigure[The solvation free energy $\Delta G$ for $R_0$. Compared are the numerical evaluation described in this section (green) and the macroscopic model \eq{\ref{eq:vism_free_energy}} (red) ]{
    \includegraphics[width=\textwidth]{{{bilder/delta_G_RisR0.5}}}
    \label{fig:shifted_LJ_delta_G}
    }%
    \label{fig:shifted_LJ_TI}
    \caption{Results of the thermodynamic integration for the solvation of a shifted-LJ solute in SPC/E water.}
\end{figure}

Thermodynamic integration for the solvation of a shifted-LJ solute (pair potential \eq{\ref{eq:shifted_LJ}}) of radius $R_0^{\text{max}}$ can be performed easily  by identifying the shift parameter $R_0$ as $\lambda$. The integration path for the whole solvation is chosen to first perform standard TI by slowly turning on the LJ potential with $\lambda:0\rightarrow1$, yielding the excess free energy $\Delta F_{LJ}$. Afterwards the integration is performed for $\lambda\equiv R_0$ from $0$ to $R_0^{\text{max}}$. The total interaction energy for a solute at step $R_0$  and its derivative read
\begin{align}
    U_{R_0} &=  U_0 + \sum_{i=1}^{N_{\text{solvent}}} {V_{R_0}(|\v r_{\text{solute}}- \v r_i|)} 
             =  U_0 + \sum_{i=1}^{N_{\text{solvent}}} {V_{LJ}(|\v r_{\text{solute}}- \v r_i|-R_0)}\\
    \deriv{ U_{R_0}}{R_0} &= \sum_{i=1}^{N_{\text{solvent}}} f_{LJ} (|\v r_{\text{solute}}- \v r_i|-R_0).
\end{align}
with $f_{LJ}(r)=-\d V_{LJ}(r)/\d r$ being the force induced by the standard LJ potential. Then using \eq{\ref{eq:free_energy_from_gofr}} we find
\begin{align}
    \label{eq:shifted_LJ_exp_force}
     \l\langle\deriv{U_{R_0}}{R_0}\r\rangle &=2\pi\rho\int\limits_0^\infty\text d r\ r^2\ f_{LJ}(r-R_0)\ g_{R_0}(r)\\
    \Delta F_{R_0^{\text{max}}} &= \Delta F_{LJ} + \int\limits_0^{R_0^{\text{max}}}\text d R_0\l\langle\deriv{U_{R_0}}{R_0}\r\rangle\\
                                &= \Delta F_{LJ} + 2\pi\rho\int\limits_0^{R_0^{\text{max}}}\text d R_0\int\limits_0^\infty\text d r\ r^2\ f_{LJ}(r-R_0)\ g_{R_0}(r)   . 
\end{align}
This derivation has been used to perform shifted-LJ TI for the solvation of a xenon like atom (with interaction parameters from \tab{\ref{tab:LJ_parameters}}) at $T=402\unit K$. The solvation free energy for $R_0=0$ was taken from the BAR integration in the last section as $\Delta G_{LJ}=11.7\unit{kJ/mol}$. Simulations were performed for potentials of $0\leq R_0\leq0.204\unit{nm}$ with a step size of $\Delta R_0=0.04\unit{nm}$. The cut-off radii were $1.0\unit{nm}+R_0$ for the neighbour list, $1.0\unit{nm}+R_0$ for the short range coulomb sphere, $0.9\unit{nm}+R_0$ for the vdW attraction and $0.8\unit{nm}+R_0$ for the vdW switch. For every $R_0$, after minimization, NVT equilibration of 0.1ns, NPT equilibration of 0.1ns and an NPT production run of 10ns, the pair correlation function $g_{R_0}(r)$ between solute and solvent has been obtained by Gromacs, where particle positions were recorded every 10ps. The integrator for the simulations was chosen to be of Langevin nature. The results of this evaluation can be seen in the left figure of \fig{\ref{fig:shifted_LJ_rdf_force}} and follow the behavior observed in \cite{Weiss:2013}, i.e. that $g(r)$ is shifted by $R_0$ in $r$-direction and its maxima get flattened with increasing $R_0$. Using \eq{\ref{eq:shifted_LJ_exp_force}} by numerical integration using the Simpson rule we find the mean force acting on the solute, displayed in \fig{\ref{fig:shifted_LJ_rdf_force} (right)}, which in turn can be integrated numerically to yield the hydration free energy, displayed in \fig{\ref{fig:shifted_LJ_delta_G}}. Note that $\epsilon$ and $\sigma$ needed for the integration are the mixing parameters of xenon and SPC/E water given in \tab{\ref{tab:LJ_parameters}}.

We find a quadratic behavior of $\Delta G(R_0)$, suggesting that the free energy can be modeled by the macroscopic interface model described in \sec{\ref{sec:solute_hydrophobicity}}. To test the simulations, we use \eq{\ref{eq:vism_free_energy}} with the SPC/E liquid-vapor surface tension \eq{\ref{eq:spce_gamma}} $\gammalv=45.6\unit{mN/m}$, SPC/E number density \eq{\ref{eq:spce_density}} $\rho=30.497\unit{nm^{-3}}$, the mean mean curvature for a perfect sphere $\bar H = 1/R$ and the Tolman length $\delta = 0.0537\unit{nm}$ which was taken from the xenon model described in \app{\ref{app:tolman_length}}. The term coming from volume change is neglected, as well as the electrostatic term (since $q=0$ for xenon). The choice of $R$ poses a problem as that will significantly scale the outcome. Following \cite{Weiss:2013} we will define $R$ as the distance from the xenon atom where the pair correlation function is 1/2, $g(R)=1/2$. Both the surface tension part and the dispersion part of $\Delta G$ as well as their sum are displayed in \fig{\ref{fig:shifted_LJ_delta_G}}. They seem to display the same behavior but differ quantitatively. The reason may lie in assuming a wrong curvature correction $\bar H\delta$ or in a poor choice of $R$.

\newpage
\subsection{Thermophoretic Simulations}
\label{sec:mdsim_results}
We performed 7 significant thermophoretic simulations using the extended water thermostat setup, all as described in \sec{\ref{sec:free_energy_MD_setups}} with different simulation times and temperature regimes for xenon, krypton and argon.
%The purpose of the weak interacting ``id'' solute is to create an environment where the hydration part of the Soret coefficient is probably small, s.t. one can hopefully exclude one of the $\alpha$ choices.
All of the results from the simulations are given as figures in the appendix, with only the xenon simulation discussed in detail here. As our main goal is to find an answer to the question whether we can isolate one of the eight hypotheses for the Soret coefficient to be consistently in agreement with simulations, we summarize the results in \tab{\ref{tab:thermophoresis_results}}, noting  which of the hypothesis yields the most satisfying fit of the data there. By fit, we mean that for every solute, the eight hypothetical densities were calculated using the corresponding parameters from \tab{\ref{tab:fit_params_md}} and the only fit parameter was chosen to be the density's normalization constant.
%Given the xenon simulation as an example which is displayed in  \fig{\ref{fig:dens_xe_300_400}}.
\begin{table}[t!]
 \centering
    \begin{tabular}{c|ccrlllll}
    \hline\hline
%           & Temp. & \multicolumn{2}{|c|}{interaction with SPC/E} & mass\\ 
%                            & $T_h$ [K] &  \\run time [ns] & $\ST$ hypotheses with lowest cum.Err.\\
    \multirow{2}{*}{Solute} & $T_c$ [K] &  run    & \multicolumn{6}{c}{$\ST$ hypotheses fits with lowest cumulative Error}   \\
                            & $T_h$ [K] &  time [ns]  & \multicolumn{6}{c}{[cum.Err.$(T_{\text{solute}}(z))\times10^{2}$;cum.Err.$(T_{\text{solvent}}(z))\times10^{2}$]} \\
    \hline
    \multirow{4}{*}{Ar} &300& \multirow{2}{*}{260}  & 1) &$\Delta H$&$\gamma$&$\alpha=\frac12$& $[2.19;\, 2.08]$ & \fig{\ref{fig:dens_ar_300_350}}\\
                        &350&                      & 2) &$\Delta H$& &$\alpha=0$& $[2.13;\, 2.16]$&\\
     \cline{2-9}
                        &350& \multirow{2}{*}{200}  & 1) &$\Delta H$&$\gamma$&$\alpha=\frac12$& $[2.15;\, 2.21]$ & \fig{\ref{fig:dens_ar_350_400}}\\
                        &400&                      & 2) &$\Delta S$& &$\alpha=0$& $[2.60;\, 2.57]$& \\
    \hline                        
    \hline                        
    \multirow{10}{*}{Kr} &300& \multirow{2}{*}{350}  & 1)& $\Delta H$& &$\alpha=0$& $[1.51;\, 1.51]$ & \fig{\ref{fig:dens_kr_300_350}}\\
                        &350&                      & 2)  &$\Delta H$& &$\alpha=\frac12$& $[1.67;\, 1.67]$ & \fig{\ref{fig:soret_kr_300_350}}\\
     \cline{2-9}
                        &350& \multirow{2}{*}{434}  & 1) &$\Delta H$&$\gamma$&$\alpha=\frac12$& $[1.70;\, 1.72]$ & \fig{\ref{fig:dens_kr_350_400}}\\
                        &400&                      & 2) &$\Delta S$& &$\alpha=0$& $[2.50;\, 2.87]$&\\
     \cline{2-9}
                        &380& \multirow{3}{*}{307}  & 1) &$\Delta S$& &$\alpha=0$&$[1.74;\, 1.79]$&  \\
                        &420&                      & 2) &$\Delta H$&$\gamma$&$\alpha=0$& $[1.74;\, 1.78]$&\fig{\ref{fig:dens_kr_380_420}}\\
                        &   &                      & 3) &$\Delta S$&$\gamma$&$\alpha=\frac12$& $[1.87;\, 1.78]$&\\
     \cline{2-9}
                        &400& \multirow{3}{*}{200}  & 1) &$\Delta S$& &$\alpha=\frac12$& $[1.88;\, 1.83]$ & \\
                        &450&                      & 2) &$\Delta H$ &$\gamma$&$\alpha=0$& $[2.10;\, 1.99]$&\fig{\ref{fig:dens_kr_400_450}}\\
                        &   &                      & 3) &$\Delta S$ & &  $\alpha=0$& $[2.12;\, 2.00]$&\\
    \hline                        
    \hline                        
    \multirow{2}{*}{Xe} &300& \multirow{2}{*}{260}  & 1) &$\Delta H$&$\gamma$ &$\alpha=\frac12$& $[2.83;\, 2.77]$ & \fig{\ref{fig:dens_xe_300_400}}\\
                        &400&                      & 2) &$\Delta H$& &$\alpha=0$& $[4.01;\, 4.25]$&\\
    \hline
    \end{tabular}
    \caption{Simulation parameters for the thermophoretic simulations of noble gas solutes. }
    \label{tab:thermophoresis_results}
\end{table}
The eight hypotheses are given as all combinations of a) 2 choices for $\alpha$, b) 2 choices for the dependence on solvation properties (entropy or enthalpy) and c) the binary choice of considering friction or not, thus all functions \eq{\ref{eq:density_fit_functions}} and \eq{\ref{eq:solute_density_theory_with_friction}}, all depicted e.g. in the center left and lower left plots of \fig{\ref{fig:widom_xe}} for xenon. The fits depend on the temperature profiles $T(z)$ which are obtained via a linear fit of the solvent temperature profile and the solute temperature profile, respectively. We will test all eight hypotheses with both temperature profiles as a self consistency check. As a measure for the goodness of the fit $f(T(z))$ we chose the pythagoreic sum of all deviations from the simulation data $\rho^{\text{dat}}(z_i)$ over the slabs in the fit region and call the result the cumulative error
\begin{align}
 \text{cum.Err.} = \sqrt{\sum_i \l(f(T(z))-\rho^{\text{dat}}(z_i)\r)^2}.
\end{align}
\begin{figure}[t!]
 \centering
    \includegraphics[width=10cm]{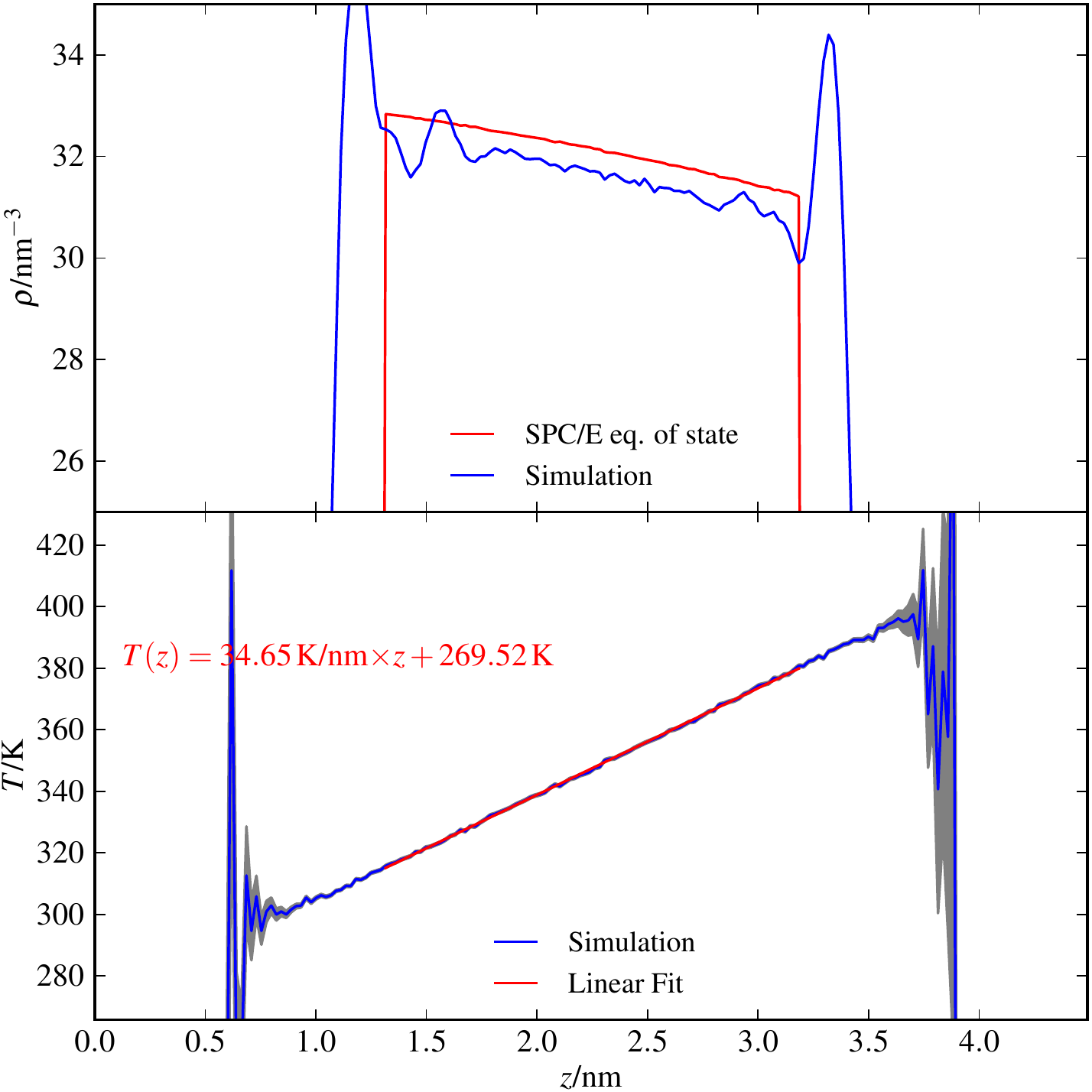}
    \includegraphics[width=10cm]{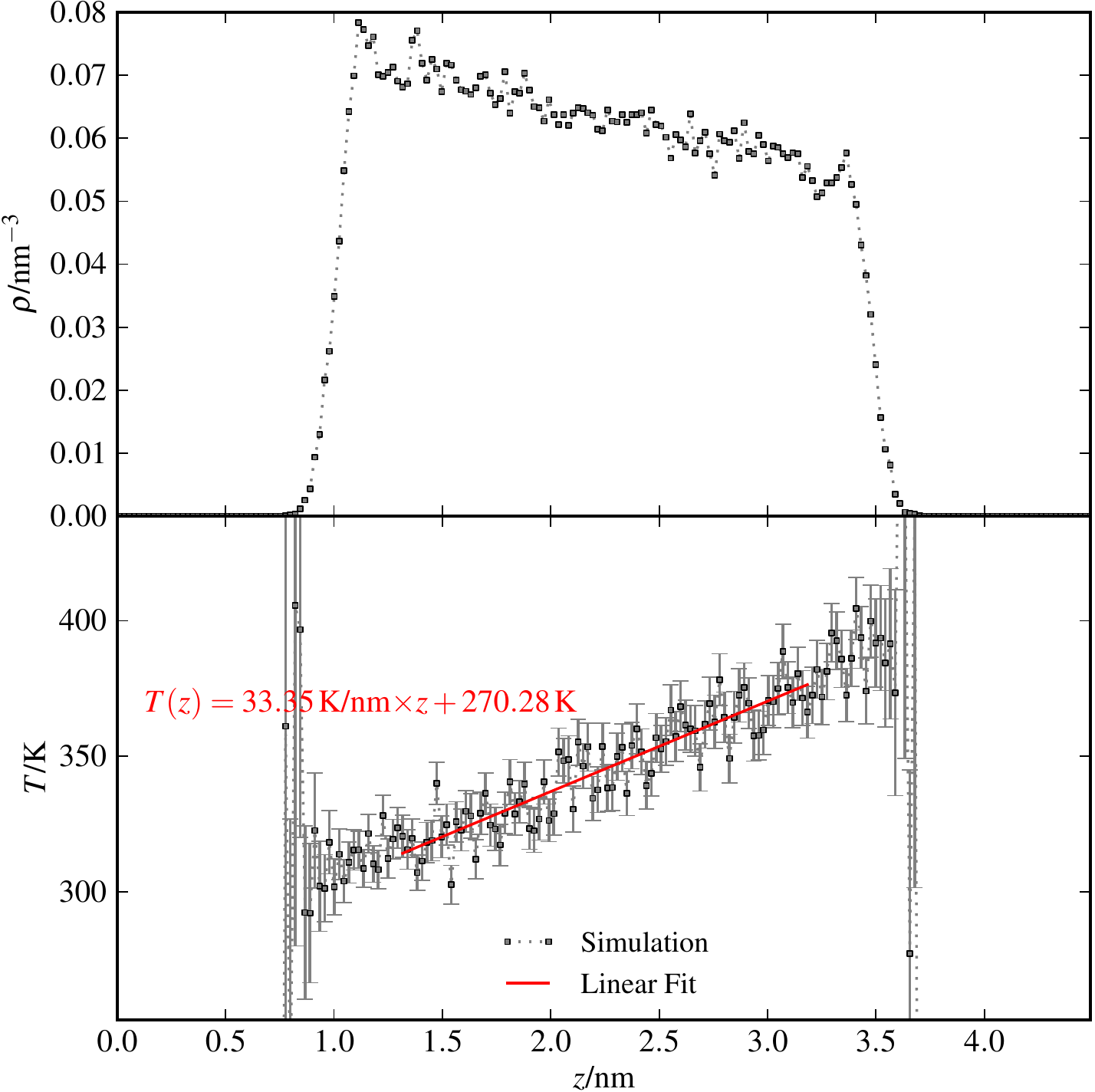}
 \caption{Solvent (\textbf{top}) and solute (\textbf{bottom}) densities and temperatures from a simulation of xenon in a thermostatted system of $T_c=300\unit K$, $T_h=400\unit K$ }
 \label{fig:temp_xe_300_400}
\end{figure}

\begin{figure}[t!]
 \centering
    \includegraphics[width=\textwidth]{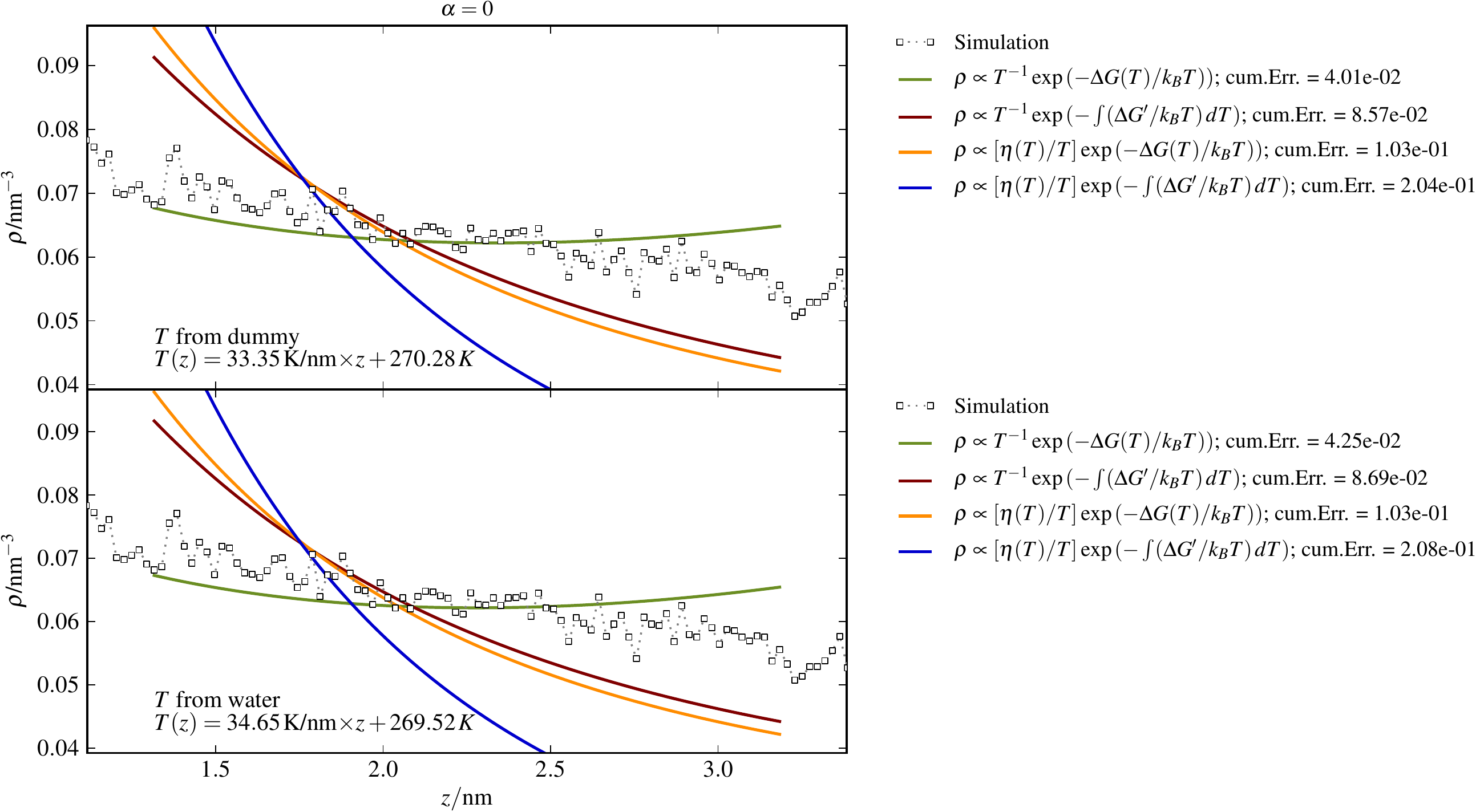}
    \includegraphics[width=\textwidth]{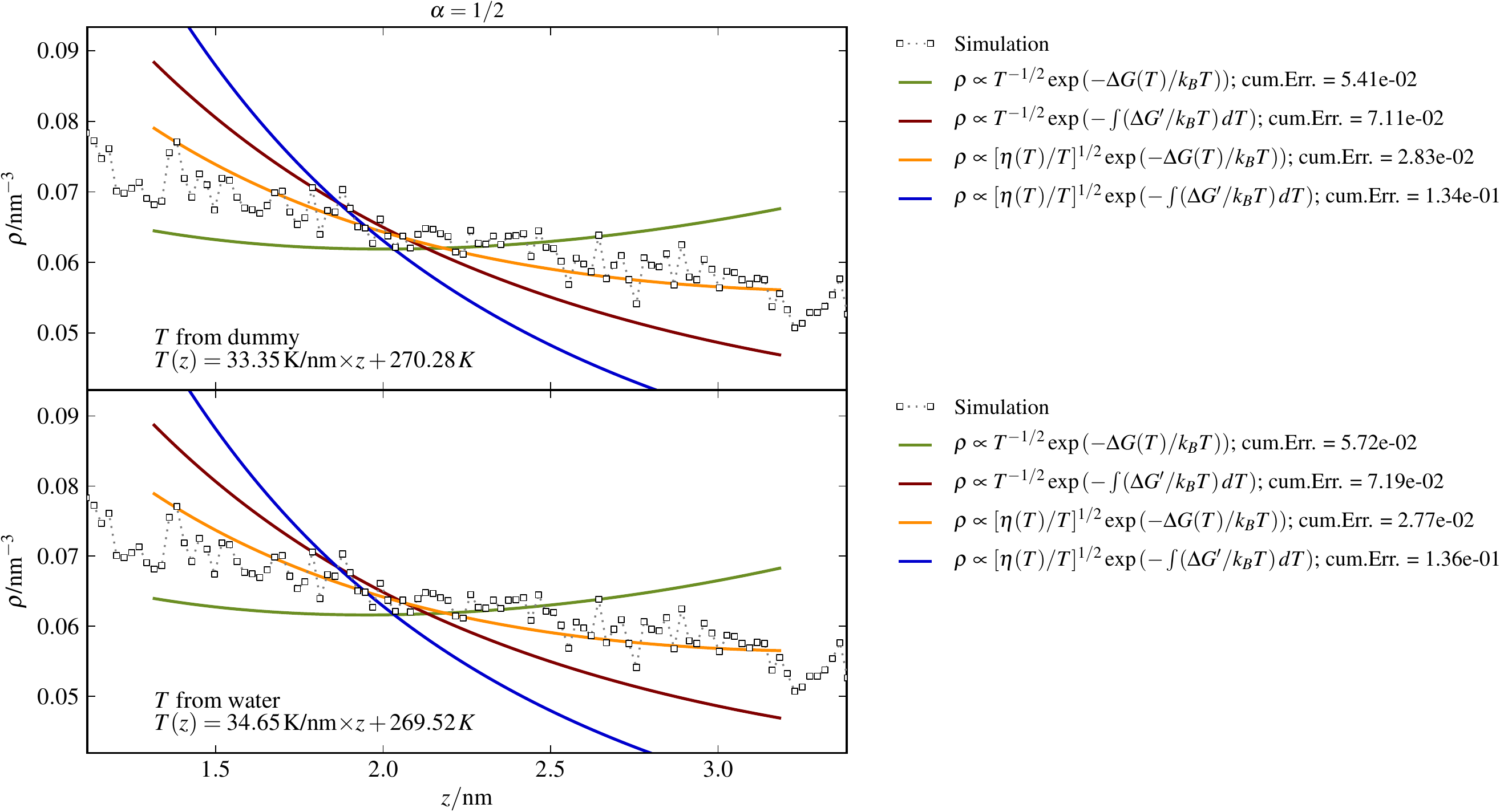}
 \caption{Density from a thermophoretic simulation of xenon running for $\simeq260$ns and temperature values of $T_c=300\unit K$, $T_h=400\unit K$ and all eight hypotheses. The only density to seem to represent the data accurately in some way is the enthalpy hypothesis with temperature dpendent friction and $\alpha=1/2$.}
 \label{fig:dens_xe_300_400}
\end{figure}

Since we already know that the SPC/E water is following its equation of state and we connected this behavior to the enthalpy hypothesis of the Soret coefficient in \sec{\ref{sec:interacting_particles}}, we actually expect the solute density to follow $u^{(H)}(T(z))$ with $\alpha=0$ (\eq{\ref{eq:density_fit_functions}}, solid orange curve in center left plot of \fig{\ref{fig:widom_xe}}). Thus, we performed thermophoretic simulations in the temperature region $T\in[300\unit K,400\unit K]$ to confirm a possible sign change of $\ST$ which would be reflected in the density as a minimum. Simulation parameters are listed in \tab{\ref{tab:thermophoresis_results}}.

The first thing to notice from the simulation results (one shown in \fig{\ref{fig:temp_xe_300_400}} and \fig{\ref{fig:dens_xe_300_400}}, the rest shown in \app{\ref{app:mdsim_app}}) is that we obtain better control over temperature and solvent density than with the other thermostats. Yet, in some of the simulations the solvent density is not completely correct. We can suspect that the equilibration in NPT was not run long enough or that in general, performing NPT production runs would be a better choice over NVT. The difference between the SPC/E equation of state and simulation is however not too large ($\simeq3\%$) compared to the densities obtained with the 3D PBC water thermostats (\fig{\ref{fig:setup2_bulk}}).

\begin{figure}
 \centering
    \includegraphics[width=\textwidth]{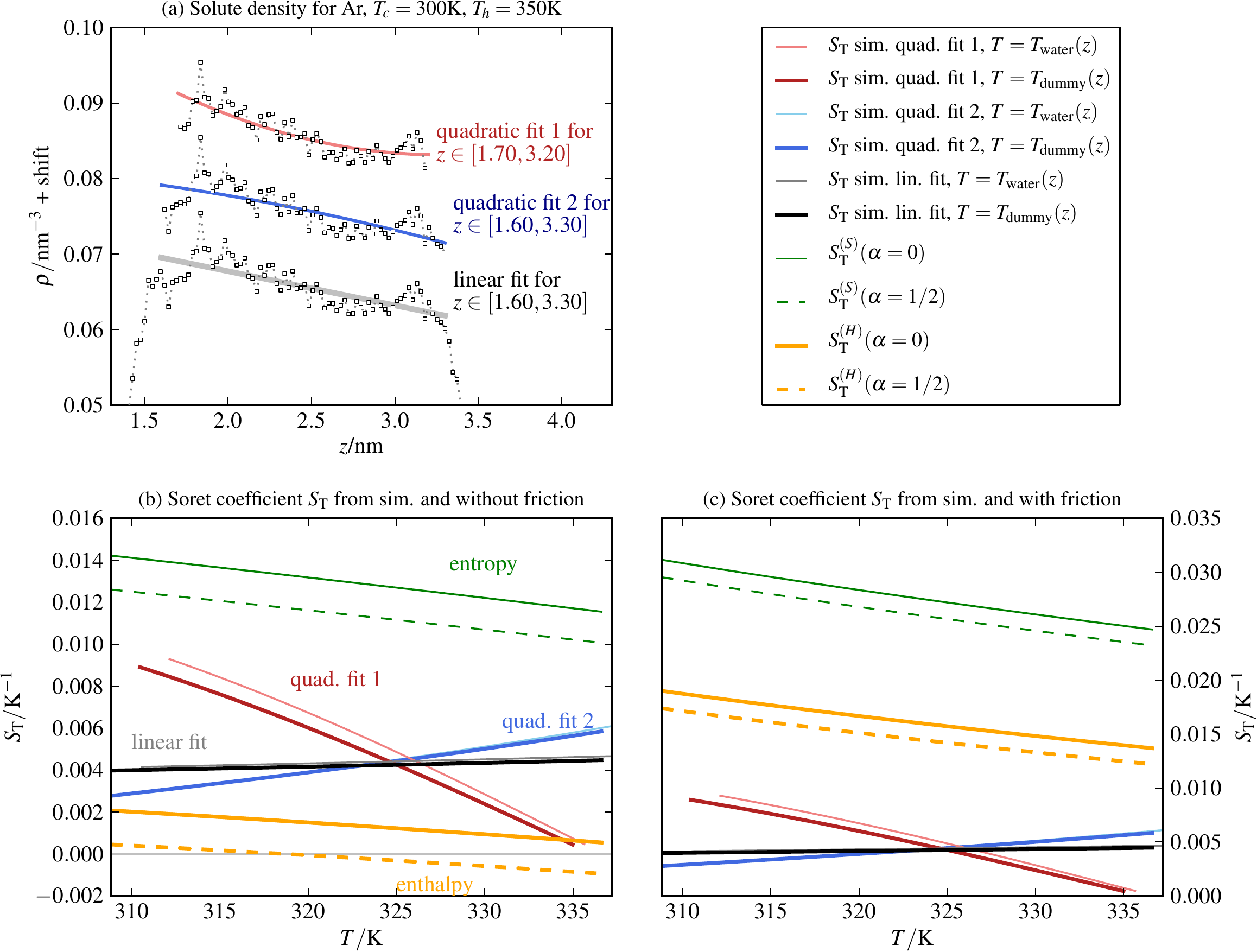}
 \caption{Soret coefficient analysis for a thermophoretic simulation of krypton running for $200$ns and temperature values of $T_c=300\unit K$, $T_h=350\unit K$. \textbf{(a)} Solute density and possible fits. \textbf{(b)} and \textbf{(c)} Soret coefficient by means of \eq{\ref{eq:soret_from_density}} for both $T=T_\text{water}(z)$ and $T=T_\text{krypton}(z)$ for all fits compared to all eight theoretical hypotheses of $\ST$ \textbf{(b)} with \eq{\ref{eq:soret_definitions}} and \textbf{(c)} \eq{\ref{eq:soret_definitions}} + \eq{\ref{eq:soret_fric}}.}
 \label{fig:soret_ar_300_350}
\end{figure}

\begin{figure}
 \centering
    \includegraphics[width=\textwidth]{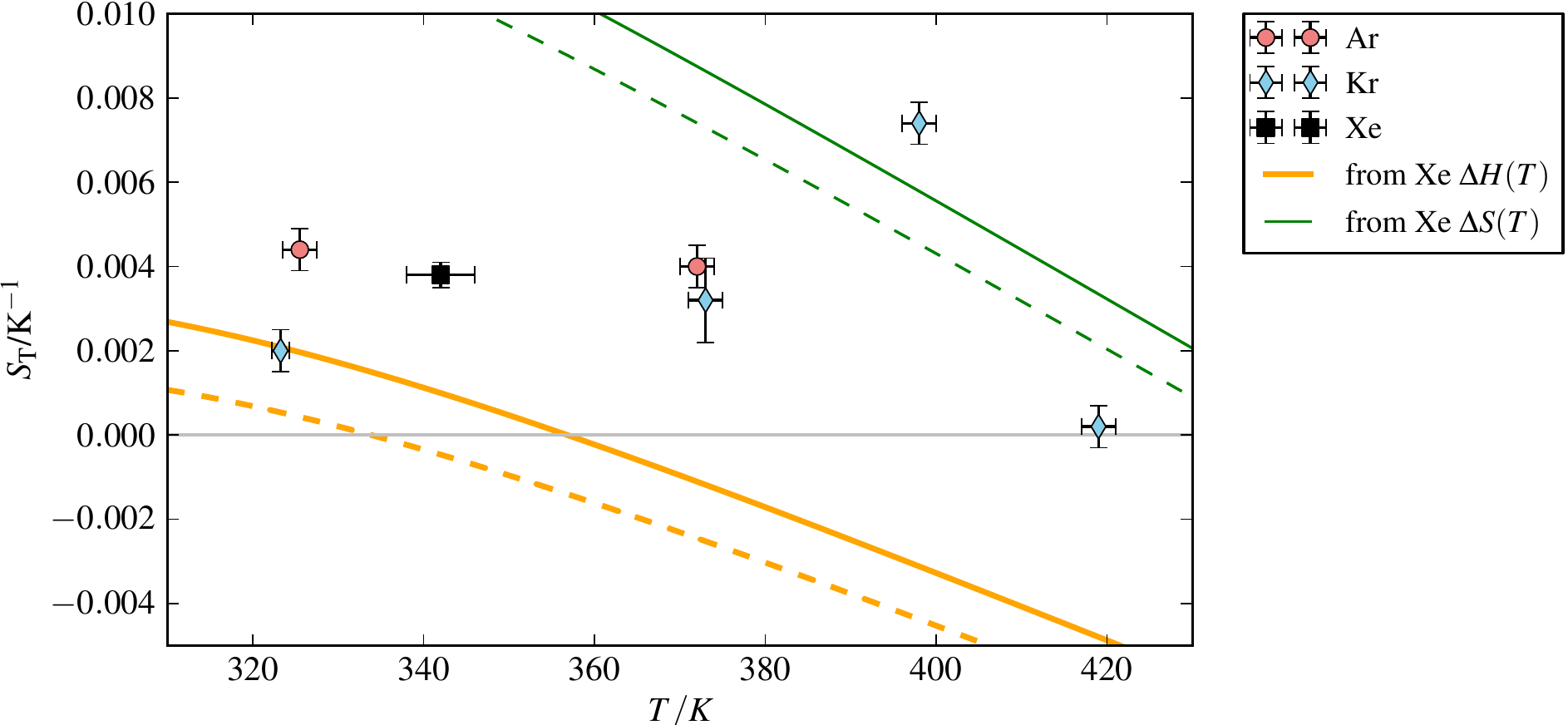}
 \caption{Soret coefficient from all simulations extracted as the regions where the Soret coefficients from all density fits coincided, compared to \eq{\ref{eq:soret_definitions}} for xenon.}
 \label{fig:soret_from_md}
\end{figure}

In general we notice that the obtained solute density data is subject to significant noise, even though the simulations have been run for a rather long time. The results of the density fits, shown in \tab{\ref{tab:thermophoresis_results}}, show no coherent behavior of the density following consistently one of the hypotheses. One can also obtain the density as polynomial fits of the simulation data, shown in \fig{\ref{fig:soret_ar_300_350}}, top left. As one can see, the Soret coefficient calculated from those fits by means of \eq{\ref{eq:soret_from_density}} is not very reliable. Depending on the choice of the fit region the data generates quadratic fits of different curvature. Since the Soret coefficient is proportional to the temperature derivative, it depends on the curvature on the fit. As soon as the curvature switches its sign, the Soret coefficient switches its behavior with increasing temperature, too. This can be seen in the lower part of \fig{\ref{fig:soret_ar_300_350}}. Hence, we decided to extract the Soret coefficient of the simulations as that region in which $\ST$ from all polynomial density fits coincides. The fits of the other simulations can be seen in \app{\ref{app:mdsim_app}}. The extracted Soret coefficients from this definition are shown in \fig{\ref{fig:soret_from_md}}

%We see that the data is rather poor.

%In \fig{\ref{fig:soret_ar_300_350}} one can see an example for a measured Soret coefficient by means of \eq{\ref{eq:soret_from_density}} compared to the eight hypotheses, here for krypton. In this case the fits suggest the validity of the enthalpy hypothesis, as was stated in \tab{\ref{tab:thermophoresis_results}}. This result is to be taken with care, as other simulations did

%coincide with this prediction. 

It is thus only possible to judge over trends rather than to obtain an answer for the actual Soret coefficient, which is that in general the solutes seem to prefer the cold region over the hot region of the simulation volume, as all Soret coefficients extracted from the region of density fit agreement shown in \fig{\ref{fig:soret_from_md}} are positive. This is contrary to the equilibrium predictions from \eq{\ref{eq:soret_definitions}a} which demands a sign switch in the simulated temperature region. However, one may have to reconsider how to extract a Soret coefficient from the data, as the Soret coefficient should definitely vary significantly over the simulated temperature regions, s.t. one should be able to obtain more than one value from the data of one simulation. 

A general problem of the thermophoretic simulations is that the force induced from the temperature gradient is rather small. Hence, to see an effect in the density or in force measurements, it is necessary to increase the temperature gradient $\nabla T$. However, doing so increases the possibility of an increased significance of higher orders of $\nabla T$ which were not considered in this 
work. Keeping the balance is a difficult task which apparently has not been mastered here.

% \begin{figure}[t!]
%  \centering
%     \includegraphics[width=\textwidth]{{bilder/kr_300_350/soret_krypton}.pdf}
%  \caption{Soret coefficient analysis for a thermophoretic simulation of krypton running for $200$ns and temperature values of $T_c=300\unit K$, $T_h=350\unit K$. \textbf{(a)} Solute density and quadratic fit. We picked this as an example because the water density in this simulation was exactly following the SPC/E equation of state (see \fig{\ref{fig:temp_ar_350_400}}). \textbf{(b)} and \textbf{(c)} Soret coefficient by means of \eq{\ref{eq:soret_from_density}} for both $T=T_\text{water}(z)$ and $T=T_\text{krypton}(z)$ compared to all eight theoretical hypotheses of $\ST$ \textbf{(b)} with \eq{\ref{eq:soret_definitions}} and \textbf{(c)} \eq{\ref{eq:soret_definitions}} + \eq{\ref{eq:soret_fric}}.}
%  \label{fig:soret_kr_300_350}
% \end{figure}

\subsection{Summary}

In this section we evaluated the hydration free energy of argon, krypton and xenon over a large temperature range by means of Widom's TPI and compared one of the results to the hydration free energy obtained by means of the BAR method. We saw reasonable agreement between the two, however the curvature of the curves obtained by both methods varied. From the temperature dependence of the obtained $\Delta G$ curves we evaluated the hypothetical Soret coefficients \eq{\ref{eq:soret_definitions}} and \eq{\ref{eq:soret_definitions}}+\eq{\ref{eq:soret_fric}}, as well as the corresponding predictions for the solute densities in thermophoretic simulations. We furthermore showed that the sign change temperature (the zero) of $\ST$ increased with increasing temperature for the hypothesis \eq{\ref{eq:soret_definitions}a} and stayed constant for \eq{\ref{eq:soret_definitions}b}. Additionally we investigated the supposed scaling behavior for $\ST$ with increasing size  were we found that $\ST$ increases with particle radius $R$ in regions where $\ST$ is positive and decreases with $R$ in regions of negative $\ST$.

We further investigated three setups to produce temperature gradients in water MD simulations, where only the one called ``extended water thermostats with 2D PBC'' showed to entail the desired control over the system's conditions meaning that it creates the right temperature gradient (linear with the correct boundary values) and approximately the right solvent density (following the SPC/E equation of state). We performed thermophoretic simulations for unrestrained argon, krypton and xenon in a temperature range where we expected the resulting solute density to show a minimum (corresponding to a sign change of the enthalpy dependent Soret coefficient \eq{\ref{eq:soret_definitions}a}). However, no significant results have been found. Furthermore we extracted one value of an Soret coefficient for every MD simulation, with every found value being positive in the region $300\unit K\leq T\leq400\unit K$. These findings however depend on a crude definition of a Soret coefficient extraction and should be taken with care.

As a byproduct of this section, we introduced a way to evaluate the solvation free energy for a solute of shifted-LJ potential by means of thermodynamic integration using the solute-solvent pair correlation function $g(r)$.

%% file: bilder/md_ti/fit_params_md.tex
\begin{tabular}{c|cccc}
\hline\hline
 Solute & $a [\unit{kJ/mol}]$ & $b [\unit{kJ/(mol\,K)}]$ & $c [\unit{kJ/(mol\, K^2)}]$ & $d [\unit{kJ/(mol\,K\,ln(1\,K))}]$\\
                \hline
Ar & +35.44281675 & -1.42174094 & -0.00047718 & +0.25913155\\
Kr & +40.12291238 & -1.60261548 & -0.00052463 & +0.29057474\\
Xe & +51.28532570 & -1.97299662 & -0.00061159 & +0.35363533\\
Xe from BAR & -66.32384337 & +1.71761659 & +0.00019693 & -0.26727971\\
\hline
\end{tabular}

%% file: chapters/summary_outlook.tex
\chapter{Summary and Outlook}
\label{ch:summary}
In this work we tried to connect the quantifying measure of thermophoresis, the Soret coefficient, to equilibrium properties of the treated systems by means of density analysis of ensembles exposed to thermal gradients. In previous experimental works, the Soret coefficient was measured for dilute solutions of biological macromolecules and colloids in water and shows increasing behavior for increasing solute size and increasing temperature, often switching its sign in the temperature range of liquid water. This behavior was found to hold even for the non-ionic part of the Soret coefficient.

Using the methods of Brownian motion and dynamical density functional theory, we found hypothetic theoretical formulae for the density and thus the Soret coefficient of a thermophoretic system connecting it for one case to the equilibrium excess enthalpy per particle and for the other to the equilibrium excess entropy per particle. In case of dilute solutions these quantities are the solvation enthalpy and the solvation entropy. The need of a differentiation between the two cases arises from two possibilities to scale the free energy functional with the temperature field. In case of a homogeneous system, we showed that identifying the Soret coefficient with the enthalpy means that the system would be following the equation of state. We furthermore showed that when one exposes an ideal gas in a temperature gradient to an external potential the resulting density is not following the Boltzmann distribution but rather a quantity where the external potential is replaced by the spatial integration of the external force field scaled by the temperature field. All the theoretically derived Soret equilibrium densities were found to depend on the parameter $\alpha=0$ or $\alpha=1/2$, respectively, which arise from the It\^o, Stratonovich interpretation of the underlying Brownian stochastic differential equation and are suspected to be connected to different microscopic processes for a local diffusion coefficient.

A crucial assumption for those derivations was that the Einstein relation for the connection of the diffusion coefficient and the temperature holds, even in temperature gradients.

We proceeded to test our assumptions by means of BD simulations where the Einstein relation and a constant friction were used explicitly and found agreement with the theoretical derivations for all investigated systems. In the homogeneous case which included the discussion of a 1D and 2D ideal gas, a 1D system of Gaussian particles and a 1D system of hard rods, we found the systems to be approximately following their equation of state, implying that the Soret coefficient is connected to the excess enthalpy per particle. In case of the binary mixture of an ideal gas solvent and a Gaussian solute, we found the Soret coefficient connected to the solvation enthalpy.

Subsequently we aimed to test the theoretical derivations by means of MD simulations, treating systems of single noble gas solutes in water. We measured the hydration free energy of the solutes in thermal equilibrium and studied how the theoretical predictions \eq{\ref{eq:soret_definitions}} and \eq{\ref{eq:soret_definitions}} + \eq{\ref{eq:soret_fric}} for the Soret coefficient would behave with increasing temperature, finding that they would have a sign change in the same region as in experimental measurements when assuming the solute-solvent friction not to affect the Soret coefficient. However, the predicted overall temperature dependence was that the Soret coefficient would decrease with increasing temperature, contrary to experimental results. We furthermore found indications that the Soret coefficient would increase with growing solute size in temperature regions where it is positive and decrease with growing solute size in temperature regions where it is negative. Note that these findings were only predictions from the measurement of the equilibrium solvation free energy by means of \eq{\ref{eq:soret_definitions}}.

In order to produce MD simulations entailing thermal gradients, we investigated different methods. The setup ``extendend water thermostats with two-dimensional boundary conditions'' was found to give the best control over the system's properties in terms of the production of the thermal gradient and the right solvent density. The water density from the thermophoretic simulations was in agreement with the SPC/E equation of state. Using this setup we measured the solute density for the noble gas solutes depending on the induced temperature, hoping to find agreement from the predictions of the equilibrium simulations. Even though we found reasonable agreement between one of the hypotheses and the Soret equilibrium density for every simulation, no coherent picture of one of the hypotheses generally predicting the system's behavior could be drawn. We furthermore saw that several polynomial fits of the solute density and subsequent evaluation of $\ST$ by means of \eq{\ref{eq:soret_from_density}} from those fits did not yield reliable results, meaning that the calculated Soret coefficient had slopes with different signs. However, for every simulation we found that all simulations had regions in which the fits coincided. We consequentially defined this region of coincidence as the measured Soret coefficient at the temperature of this region. Doing so, we found Soret coefficients which where generally positive in the region $300\unit K\leq T\leq400\unit K$, which means that all of the solutes generally would have preferred the cold over the hot regions. However, the analytic prediction \eq{\ref{eq:soret_definitions}a}, which was found as the Soret coefficient in the BD simulations would have predicted a sign change. 

On the basis of this work, future studies can further investigate systems with thermal gradients. In order to confirm the validity of the BD simulations for more realistic systems, one should simulate 3D systems and possibly more realistic binary mixtures, e.g. SPC/E water and noble gases as explicit particles in an implicit solvent underlying a linear temperature profile, which would give the possibility to check the results from the MD simulations and possible differences between the two approaches to simulate systems in thermal gradients. Furthermore a detailed discussion of the fluctuation-dissipation theorem (FDT) is necessary, in order to better understand the nature of the diffusion coefficient in temperature gradients and to check the validity of the usage of Einstein's relation. To this end one could aim to measure the diffusion coefficient in thermophoretic simulations and to compare the solute-solvent pair correlation function $g(r)$ in thermophoretic simulations with those obtained in equilibrium simulations. One should furthermore study the works by Kroy \cite{Kroy:2010,Kroy:2014}, which seem to provide a coherent analysis of the FDT in temperature gradients.

Regarding MD simulations, more investigations are necessary on how to obtain more reliable results for the solute density in thermophoretic MD simulations, meaning that it is necessary to find the right balance between a sufficiently small temperature gradient and measuring a smooth Soret equilibrium density. As we found the SPC/E water to be in agreement with its equation of state it would be interesting to measure the excess chemical potential of one water molecule in dependence of the temperature in order to check if \eq{\ref{eq:solute_density_theory}} correctly gives the equation of state. Furthermore, a reliable method to extract the Soret coefficient from the simulation data is needed.

For the theoretical derivations, future studies should test the applicability of the power functional approach \cite{Schmidt:2013} and the generalized non-equilibrium density functional \cite{Wittkowski:2012} in order to obtain better theoretical models of the Soret equilibrium density and the Soret coefficient. A further open problem is that the value of the It\^o/Smoluchowski parameter $\alpha$ is still uncertain, as the MD simulations performed in this work did not give an answer for any of the theoretical hypotheses. Theoretic considerations of microscopic processes may help to obtain an answer to that question.

Furthermore, one should explicitly include the influence of electrostatics and -dynamics in theoretical considerations, as those seem to play a major role \cite{Braun:2014}. But, as the reason behind the difference of experimental measurements and the theoretical/simulation results of this work remains unknown, further research on the non-electrostatic effects behind thermophoresis is needed, too.

%% file: chapters/biblio.tex
%%%%%%%%%%%%%%%%%%%%%%%%%%%%%%%%%%%%%%%%%%%%%%%%%%%%%%%%%%%%%%%%%%%%%%%%%%%%%
%%%%%%%%%%%%%%%%%%%%   ARTICLES   %%%%%%%%%%%%%%%%%%%%%%%%%%%%%%%%%%%%%%%%%%%
%%%%%%%%%%%%%%%%%%%%%%%%%%%%%%%%%%%%%%%%%%%%%%%%%%%%%%%%%%%%%%%%%%%%%%%%%%%%%

\bibliography{thermophoresis}
\bibliographystyle{h-physrev}

%\begin{comment}

%% file: chapters/appendixA.tex
\chapter{TPI Free Energies for Ar and Kr}
\label{app:noble_gas_free_energy}
In the following the free energy evaluations of argon and krypton, are presented. Their results have been presented in \fig{\ref{fig:delta_G_results_md}} and \tab{\ref{tab:fit_params_md}}, but the graphical depiction of the consequences for density and Soret coefficient has been left out in \sec{\ref{sec:mdsim_free_energy}}, 

\begin{figure}[b!]
 \centering
    \includegraphics[width=\textwidth]{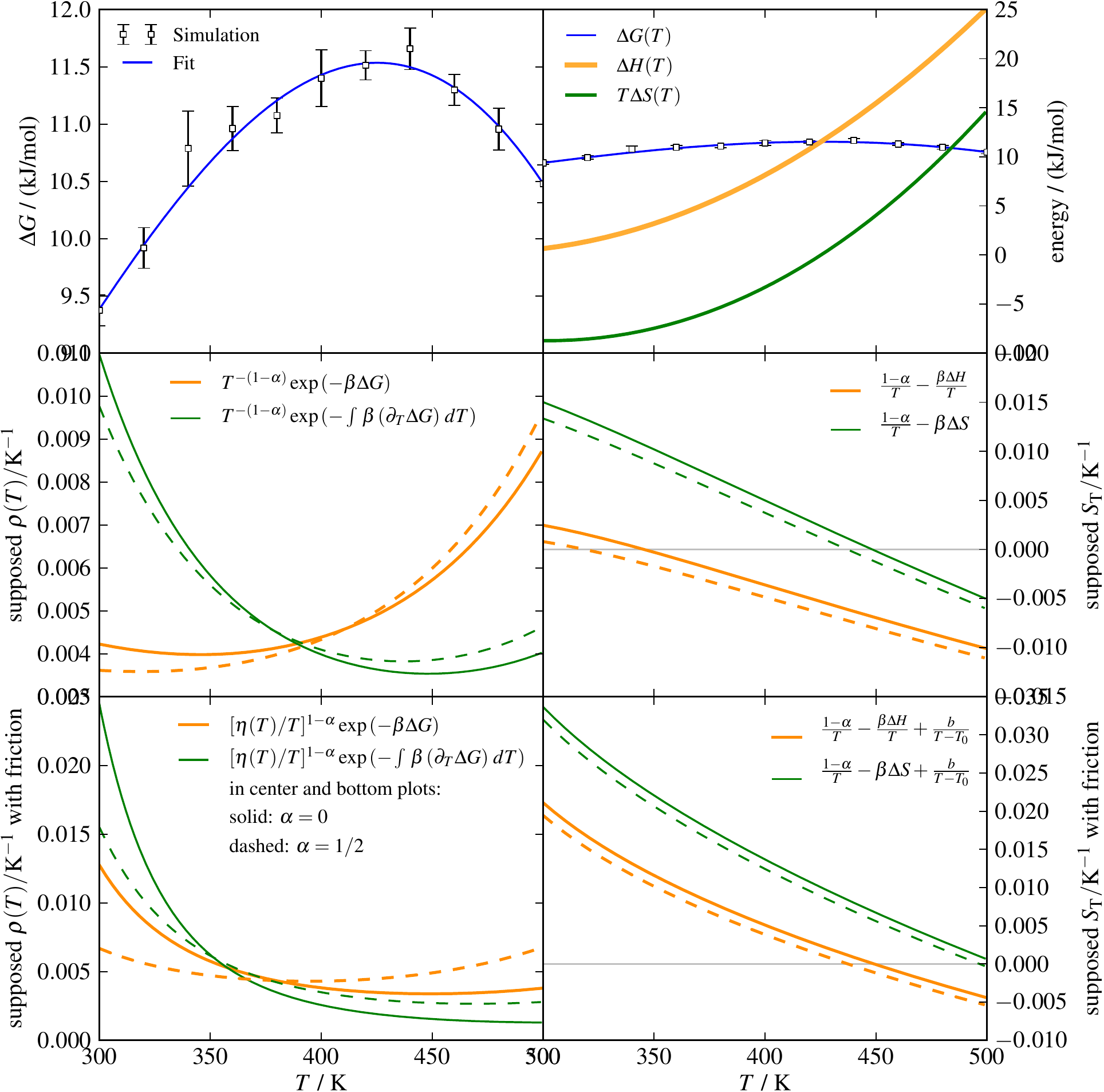}
 \caption{Results of the free energy $\Delta G$ for the hydration of \textit{\textbf{argon}} in SPC/E water by means of the TPI method. \textbf{Top left: } Simulation results and fit with \eq{\ref{eq:free_energy_fit_function}}. \textbf{Top right: } Solvation free energy, solvation enthalpy and solvation entropy (\eq{\ref{eq:thermodynamic_fit_functions}}). \textbf{Center left:} The supposed densities in thermophoretic simulations, following \eq{\ref{eq:density_fit_functions}} and \textbf{center right:} the corresponding supposed Soret coefficients \eq{\ref{eq:soret_definitions}}. \textbf{Bottom left: } Supposed densities in thermophoretic simulations with additional friction dependence \eq{\ref{eq:solute_density_theory_with_friction}}, with the friction proportional to the SPC/E viscosity \eq{\ref{eq:viscosity}} and \textbf{bottom right:} the corresponding supposed Soret coefficients (\eq{\ref{eq:soret_definitions}} with \eq{\ref{eq:soret_fric}}).}
 \label{fig:widom_ar}
\end{figure}

\begin{figure}[t!]
 \centering
    \includegraphics[width=\textwidth]{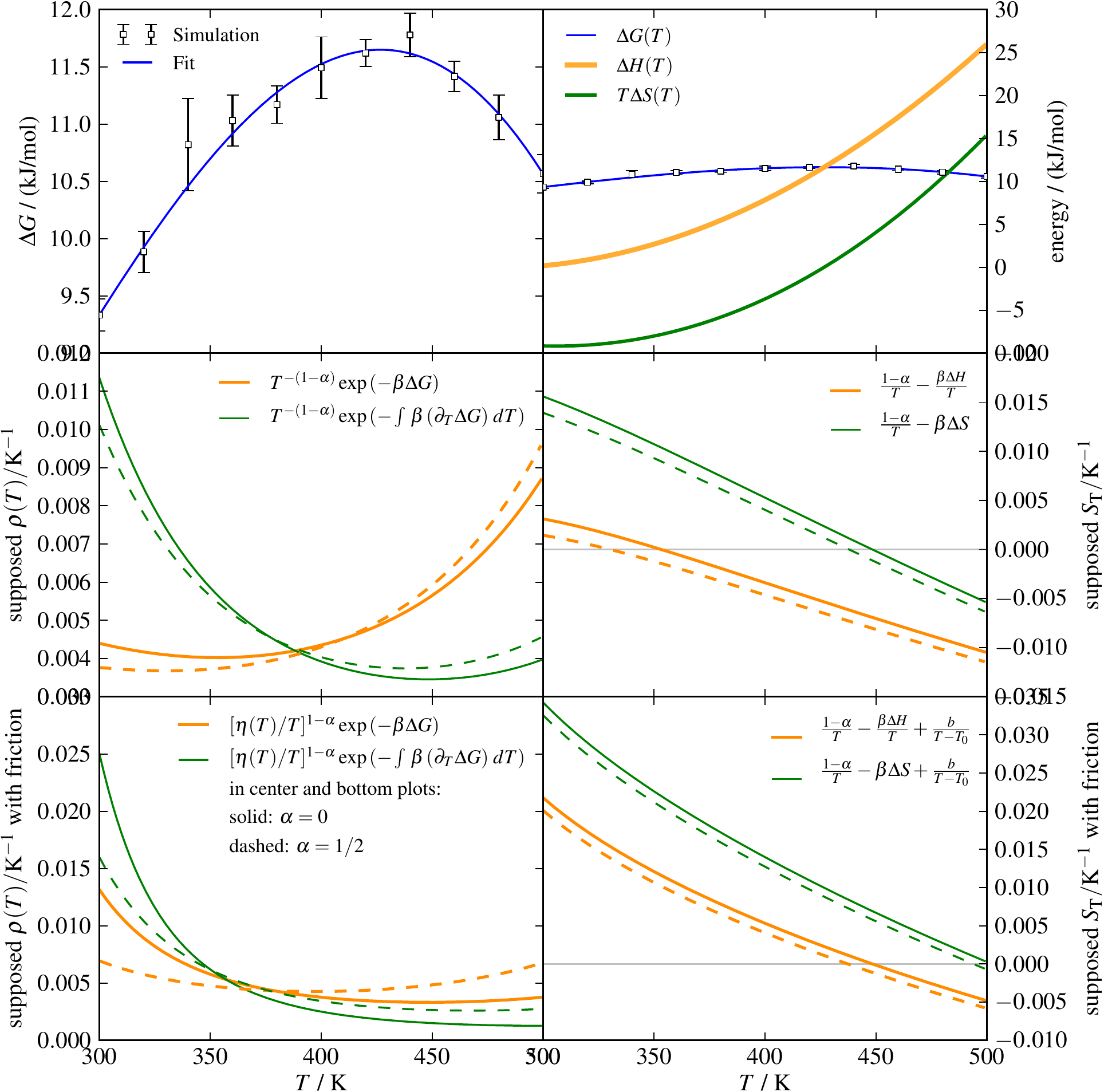}
 \caption{Results of the free energy $\Delta G$ for the hydration of \textit{\textbf{krypton}} in SPC/E water by means of the TPI method. \textbf{Top left: } Simulation results and fit with \eq{\ref{eq:free_energy_fit_function}}. \textbf{Top right: } Solvation free energy, solvation enthalpy and solvation entropy (\eq{\ref{eq:thermodynamic_fit_functions}}). \textbf{Center left:} The supposed densities in thermophoretic simulations, following \eq{\ref{eq:density_fit_functions}} and \textbf{center right:} the corresponding supposed Soret coefficients \eq{\ref{eq:soret_definitions}}. \textbf{Bottom left: } Supposed densities in thermophoretic simulations with additional friction dependence \eq{\ref{eq:solute_density_theory_with_friction}}, with the friction proportional to the SPC/E viscosity \eq{\ref{eq:viscosity}} and \textbf{bottom right:} the corresponding supposed Soret coefficients (\eq{\ref{eq:soret_definitions}} with \eq{\ref{eq:soret_fric}}).}
 \label{fig:widom_kr}
\end{figure}

%% file: chapters/appendixB.tex
\chapter{Remaining Results of Thermophoretic MD Simulations }
\label{app:mdsim_app}
In the following the remaining depictions of the thermophoretic simulations of noble gases are presented.

\section{Temperature and Densities}
\begin{figure}[t!]
 \centering
    \includegraphics[width=10cm]{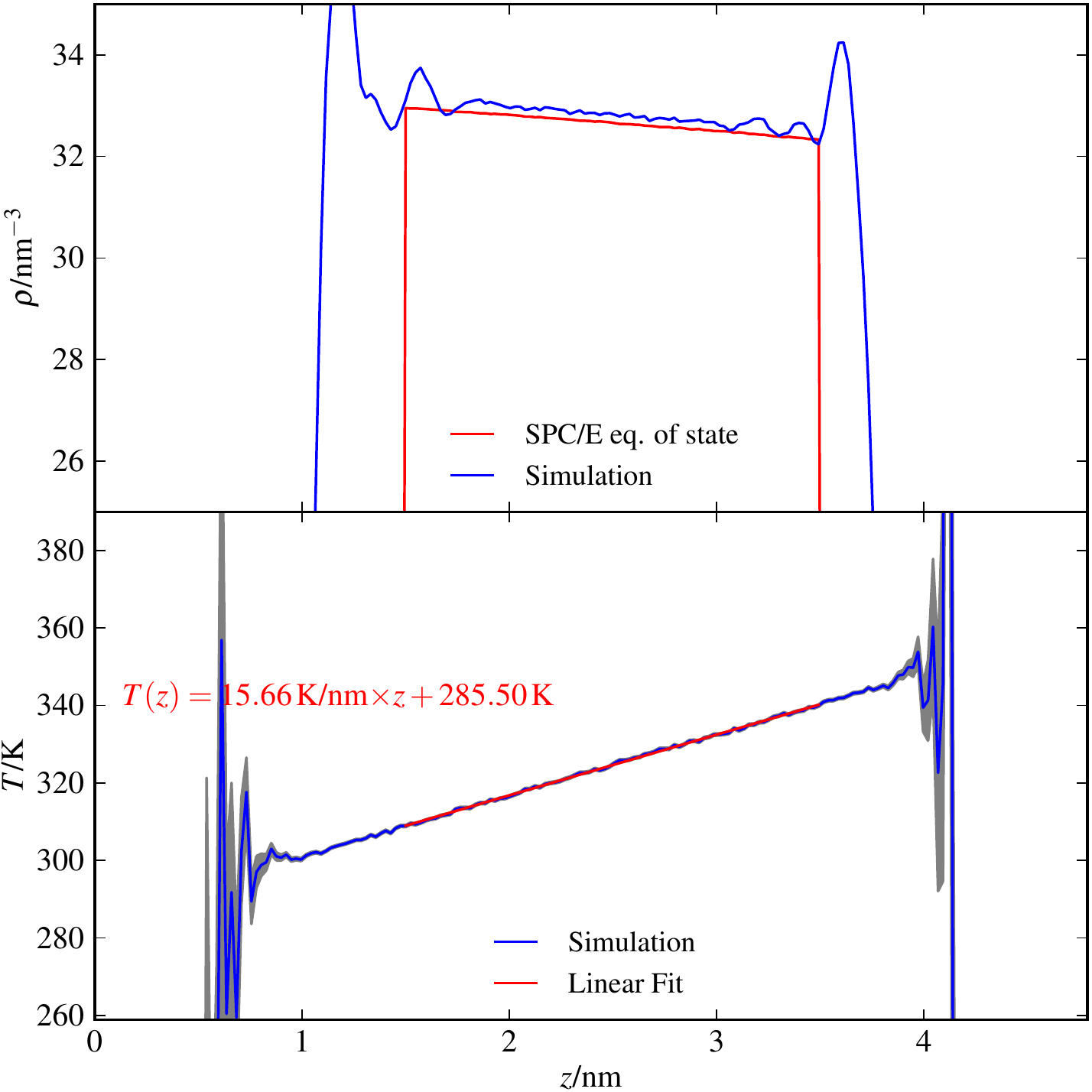}
    \includegraphics[width=10cm]{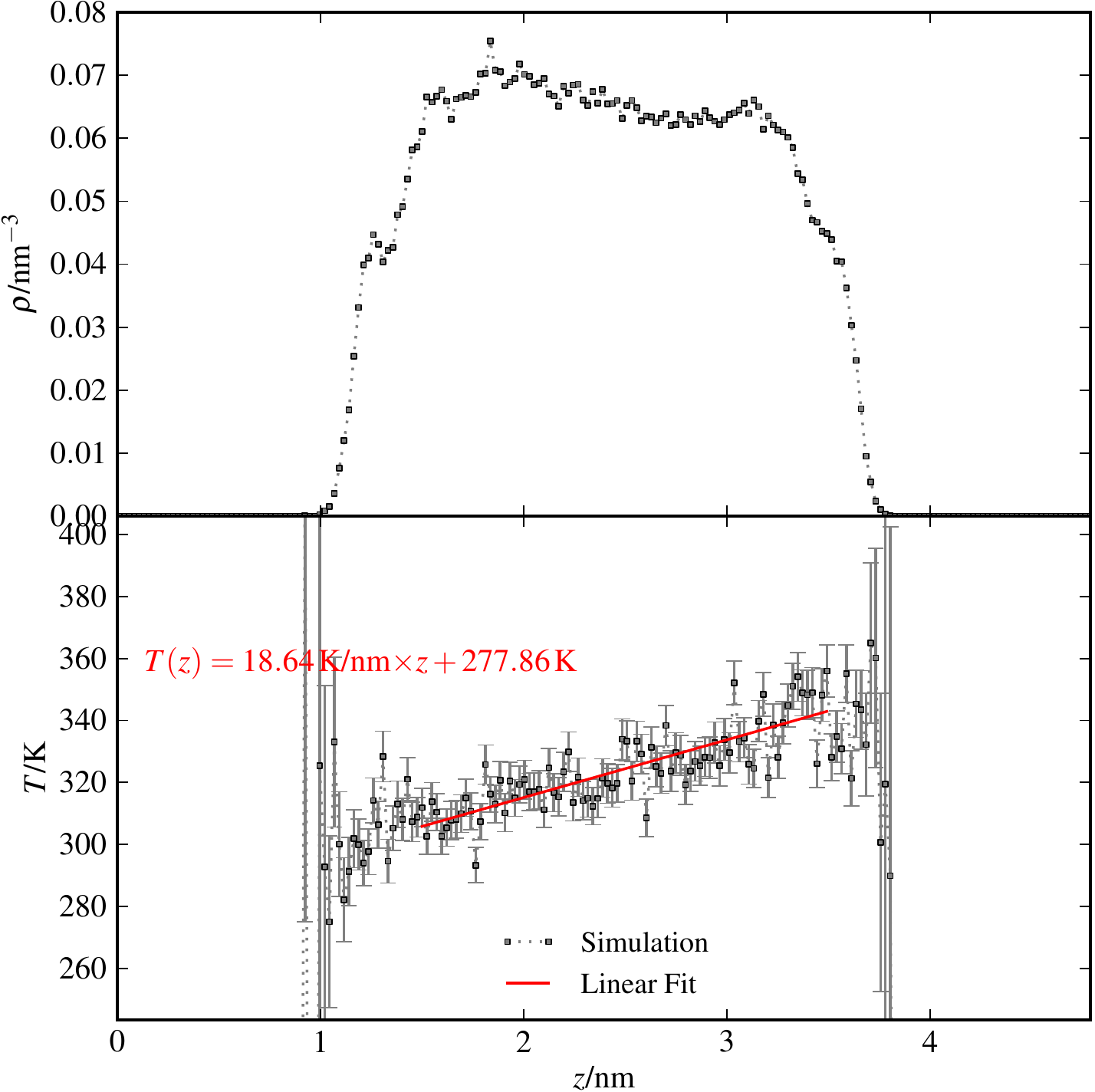}
 \caption{Solvent (\textbf{top}) and solute (\textbf{bottom}) densities and temperatures from a simulation of \textit{\textbf{argon}} in a thermostatted system of $T_c=300\unit K$, $T_h=350\unit K$ with a run time of 200ns}
 \label{fig:temp_ar_300_350}
\end{figure}

\begin{figure}[t!]
 \centering
    \includegraphics[width=\textwidth]{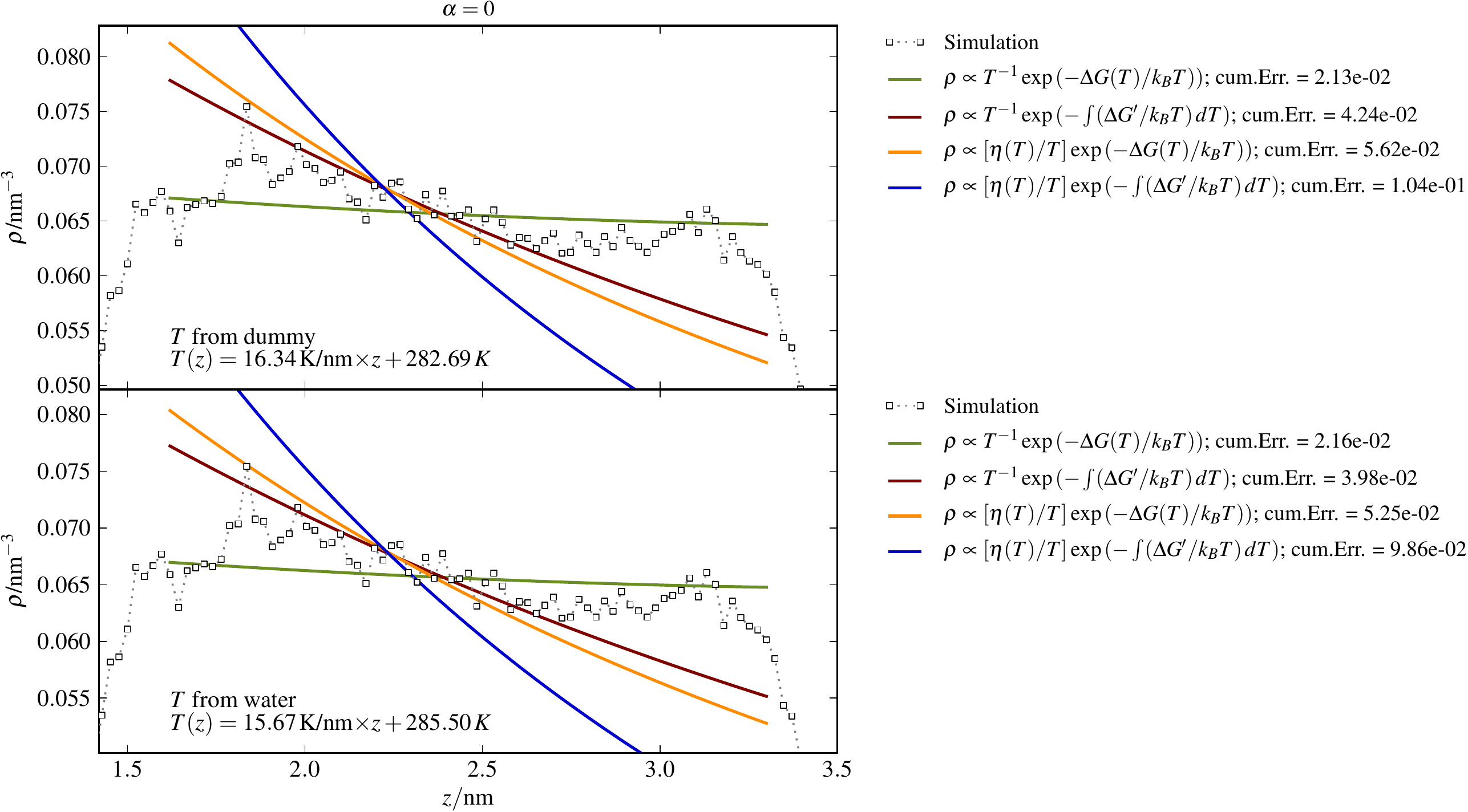}
    \includegraphics[width=\textwidth]{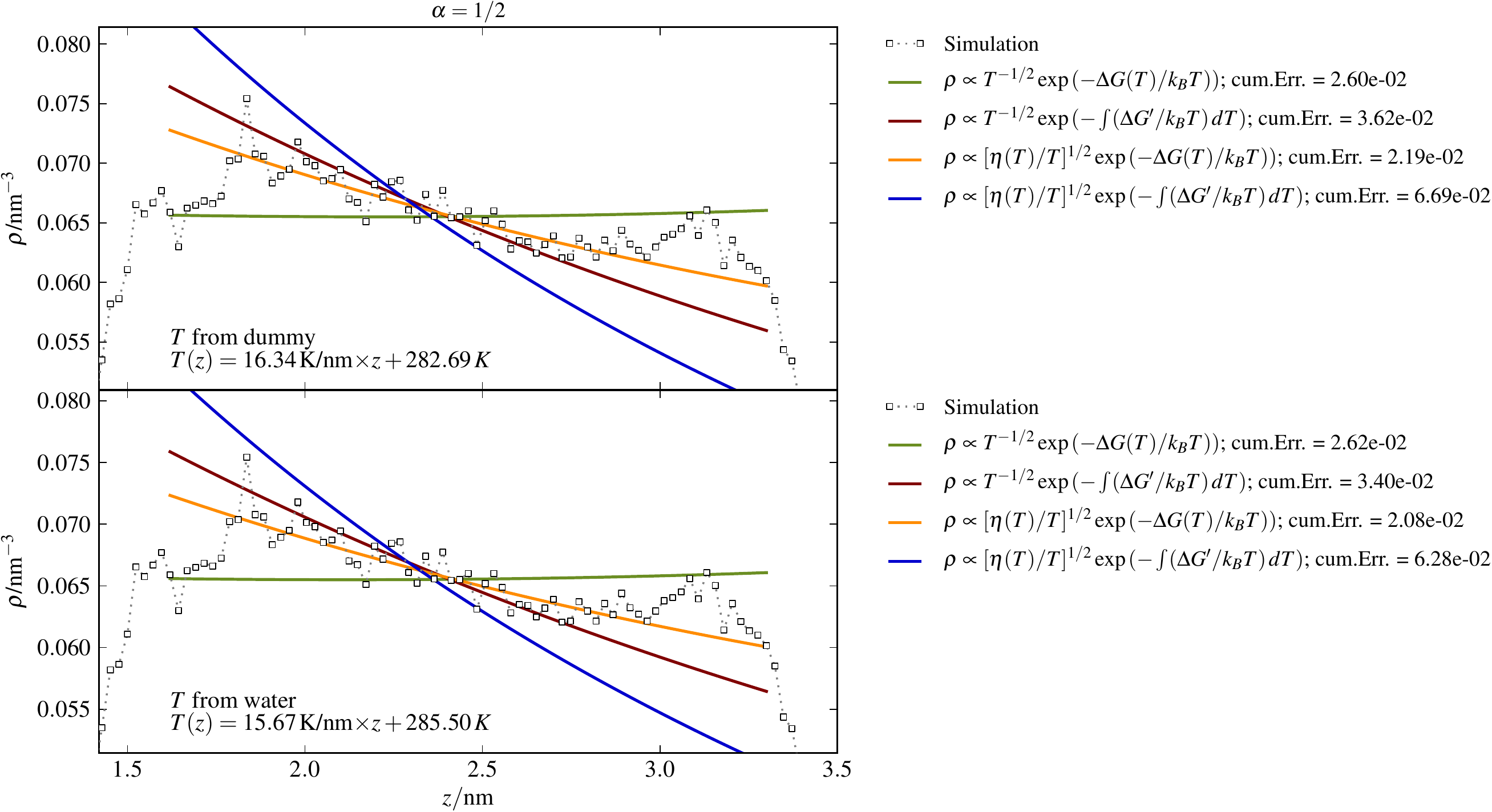}
 \caption{Density from a thermophoretic simulation of \textit{\textbf{argon}} running for $200$ns and temperature values of $T_c=300\unit K$, $T_h=350\unit K$, compared to all eight hypotheses.}
 \label{fig:dens_ar_300_350}
\end{figure}

\begin{figure}[t!]
 \centering
    \includegraphics[width=10cm]{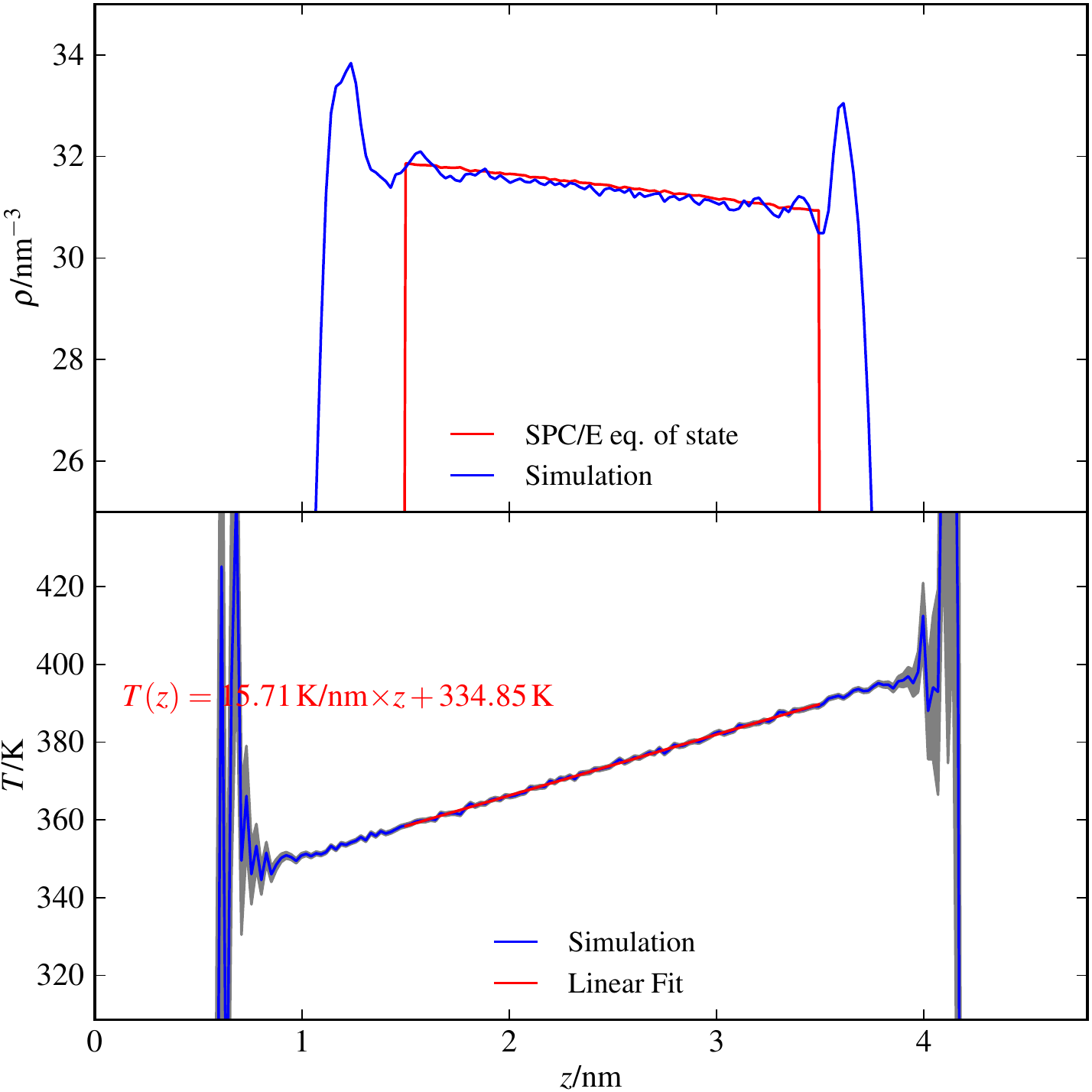}
    \includegraphics[width=10cm]{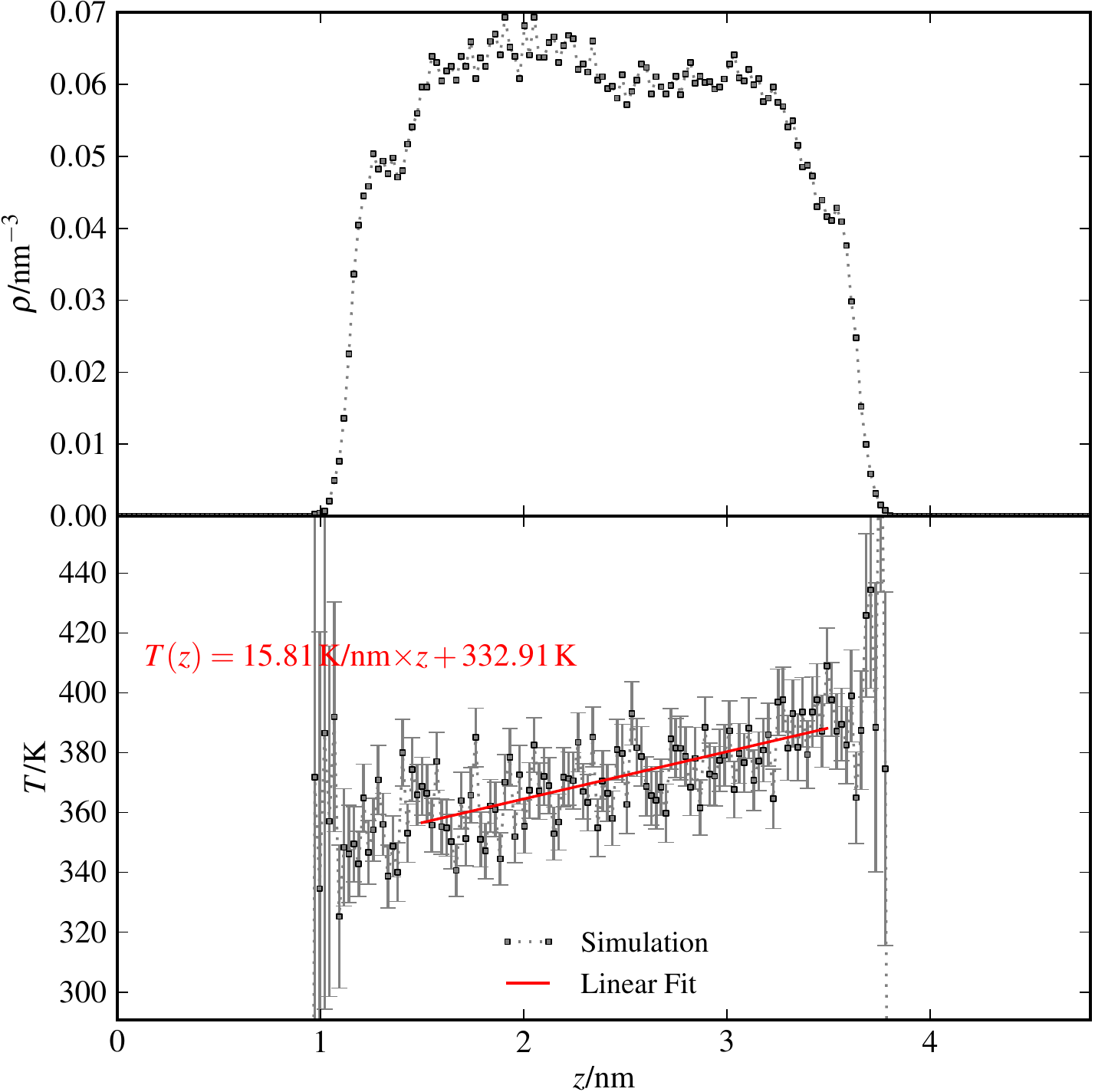}
 \caption{Solvent (\textbf{top}) and solute (\textbf{bottom}) densities and temperatures from a simulation of \textit{\textbf{argon}} in a thermostatted system of $T_c=350\unit K$, $T_h=400\unit K$ with a run time of 200ns.}
 \label{fig:temp_ar_350_400}
\end{figure}

\begin{figure}[t!]
 \centering
    \includegraphics[width=\textwidth]{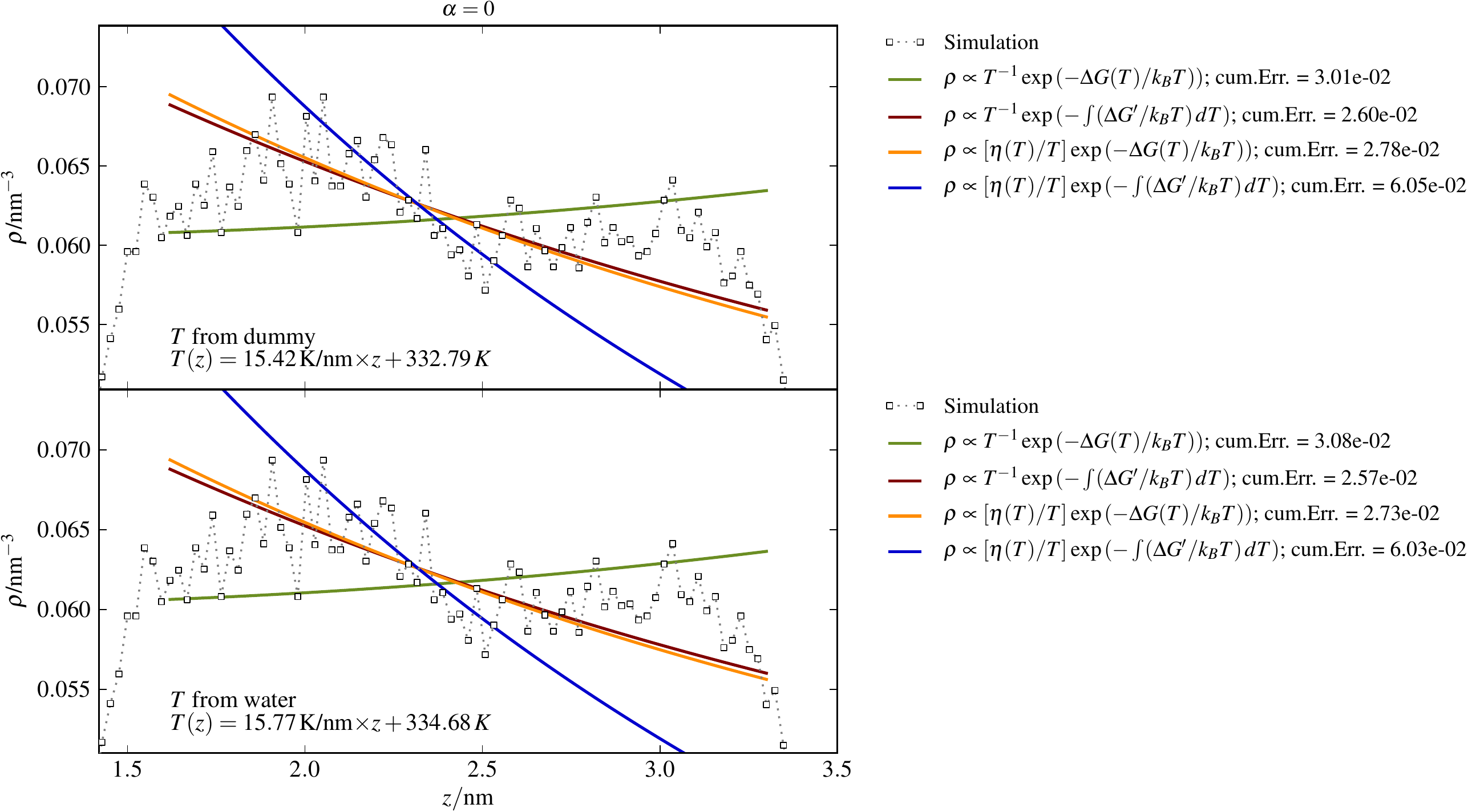}
    \includegraphics[width=\textwidth]{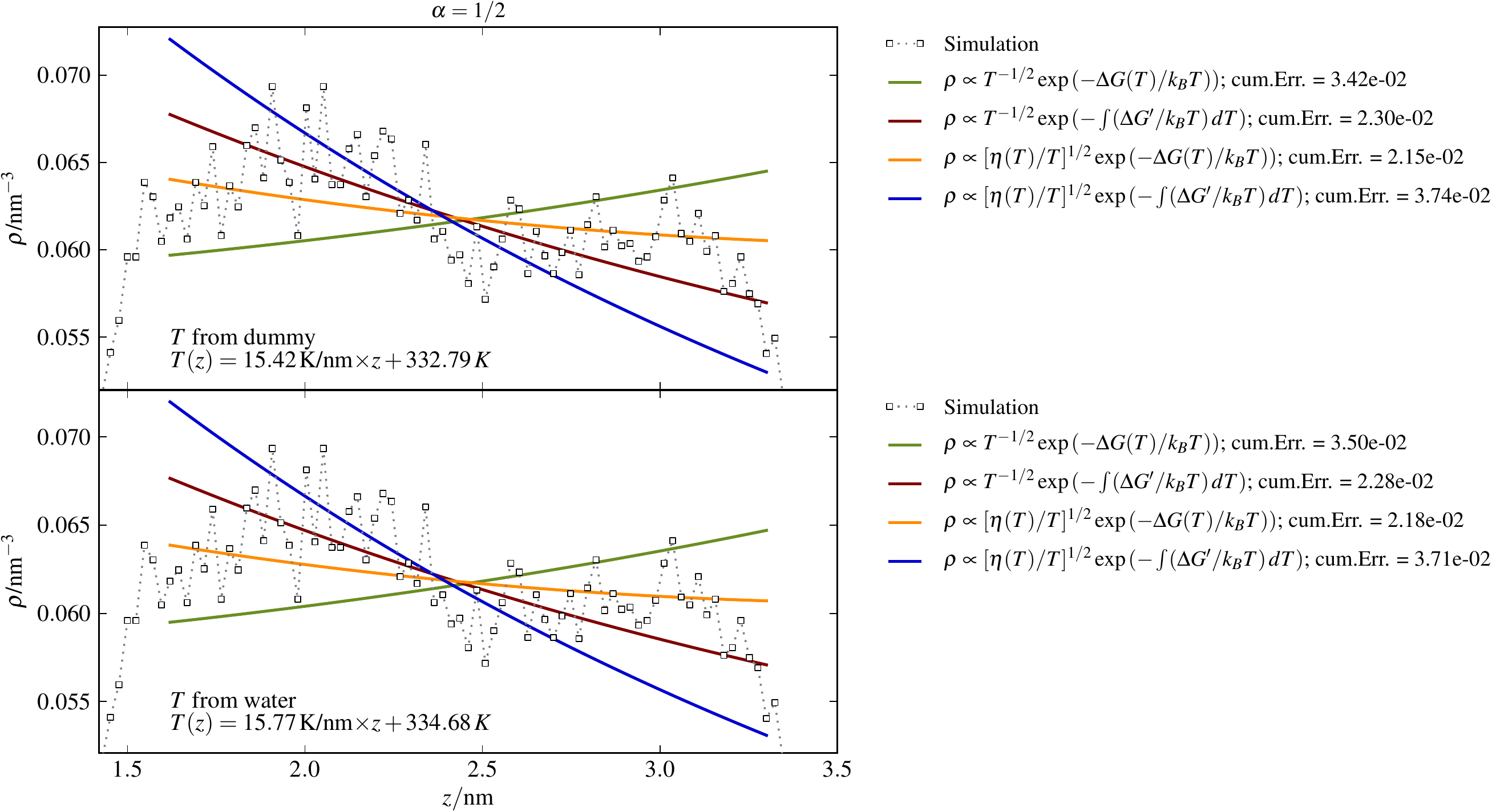}
 \caption{Density from a thermophoretic simulation of \textit{\textbf{argon}} running for $200$ns and temperature values of $T_c=350\unit K$, $T_h=400\unit K$, compared to all eight hypotheses.}
 \label{fig:dens_ar_350_400}
\end{figure}

\begin{figure}[t!]
 \centering
    \includegraphics[width=10cm]{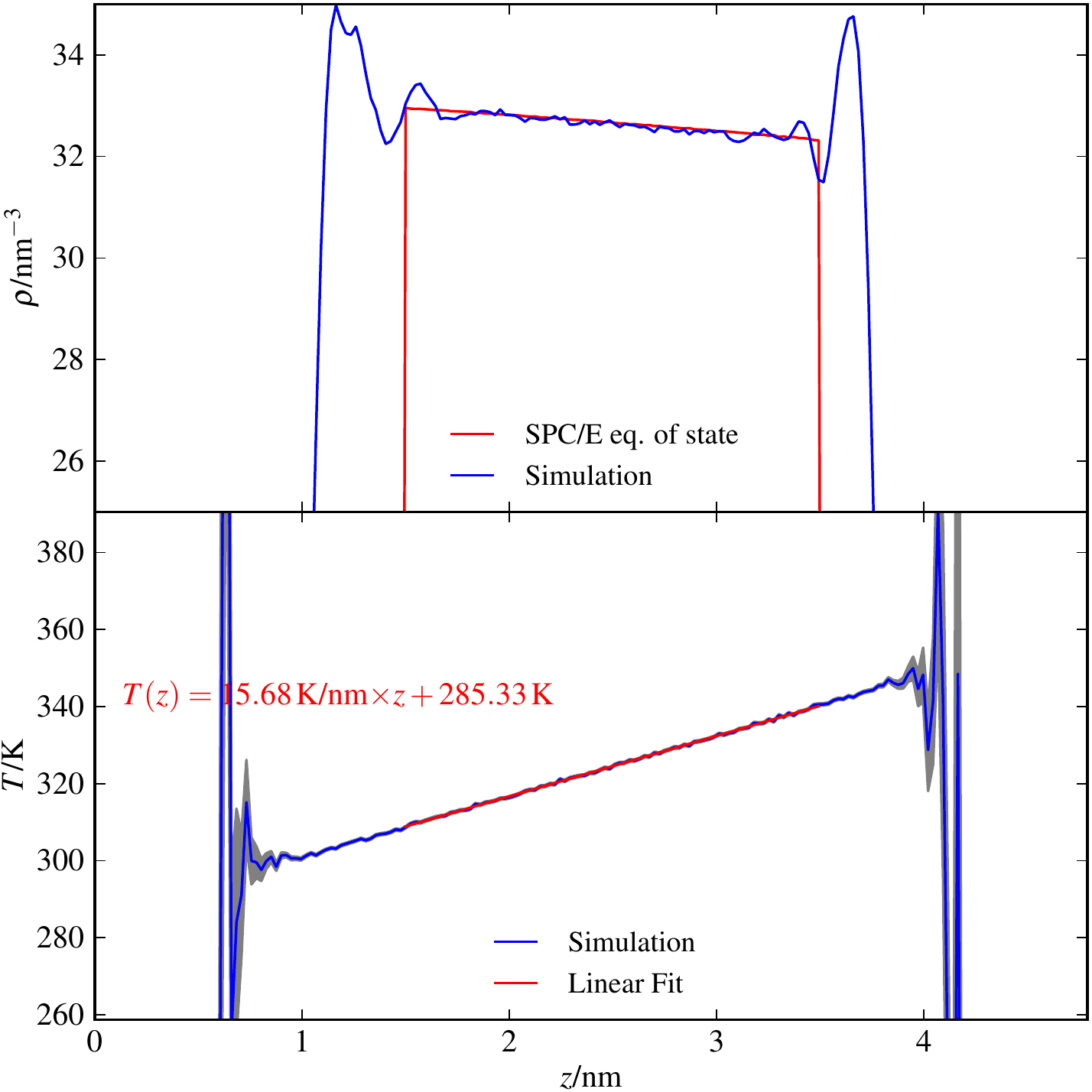}
    \includegraphics[width=10cm]{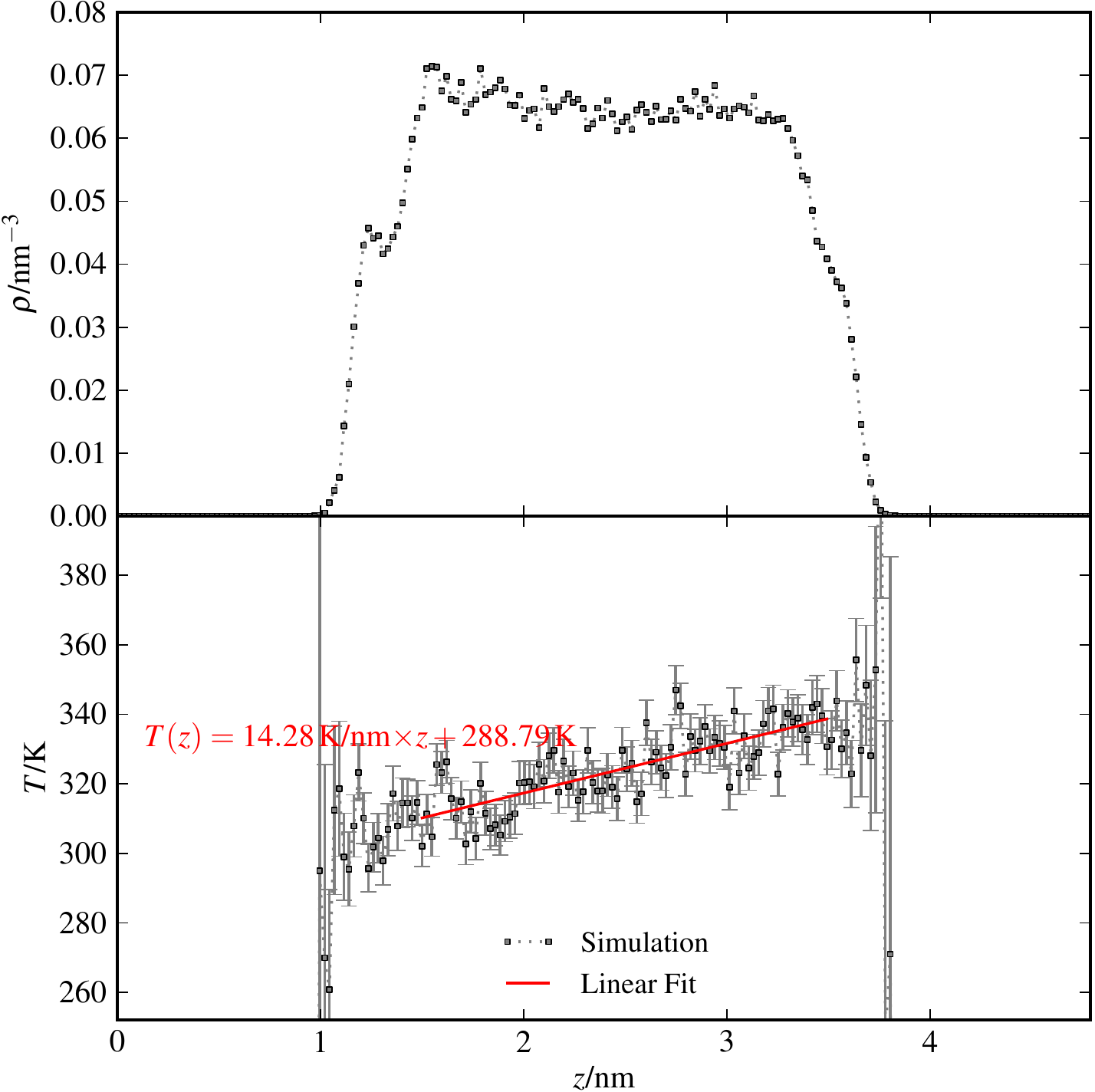}
 \caption{Solvent (\textbf{top}) and solute (\textbf{bottom}) densities and temperatures from a simulation of \textit{\textbf{krypton}} in a thermostatted system of $T_c=300\unit K$, $T_h=350\unit K$ with a run time of $\simeq350$ns.}
 \label{fig:temp_kr_300_350}
\end{figure}

\begin{figure}[t!]
 \centering
    \includegraphics[width=\textwidth]{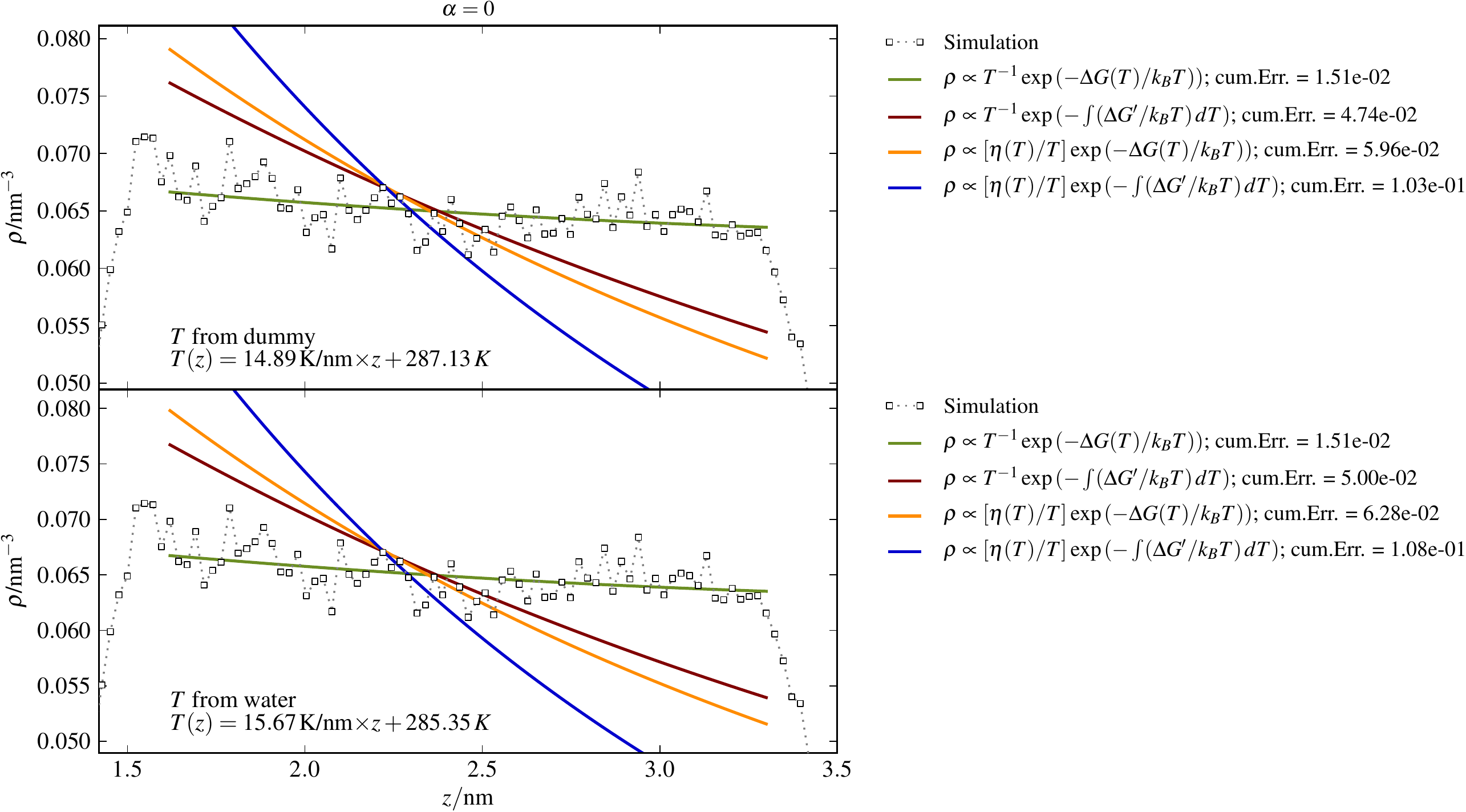}
    \includegraphics[width=\textwidth]{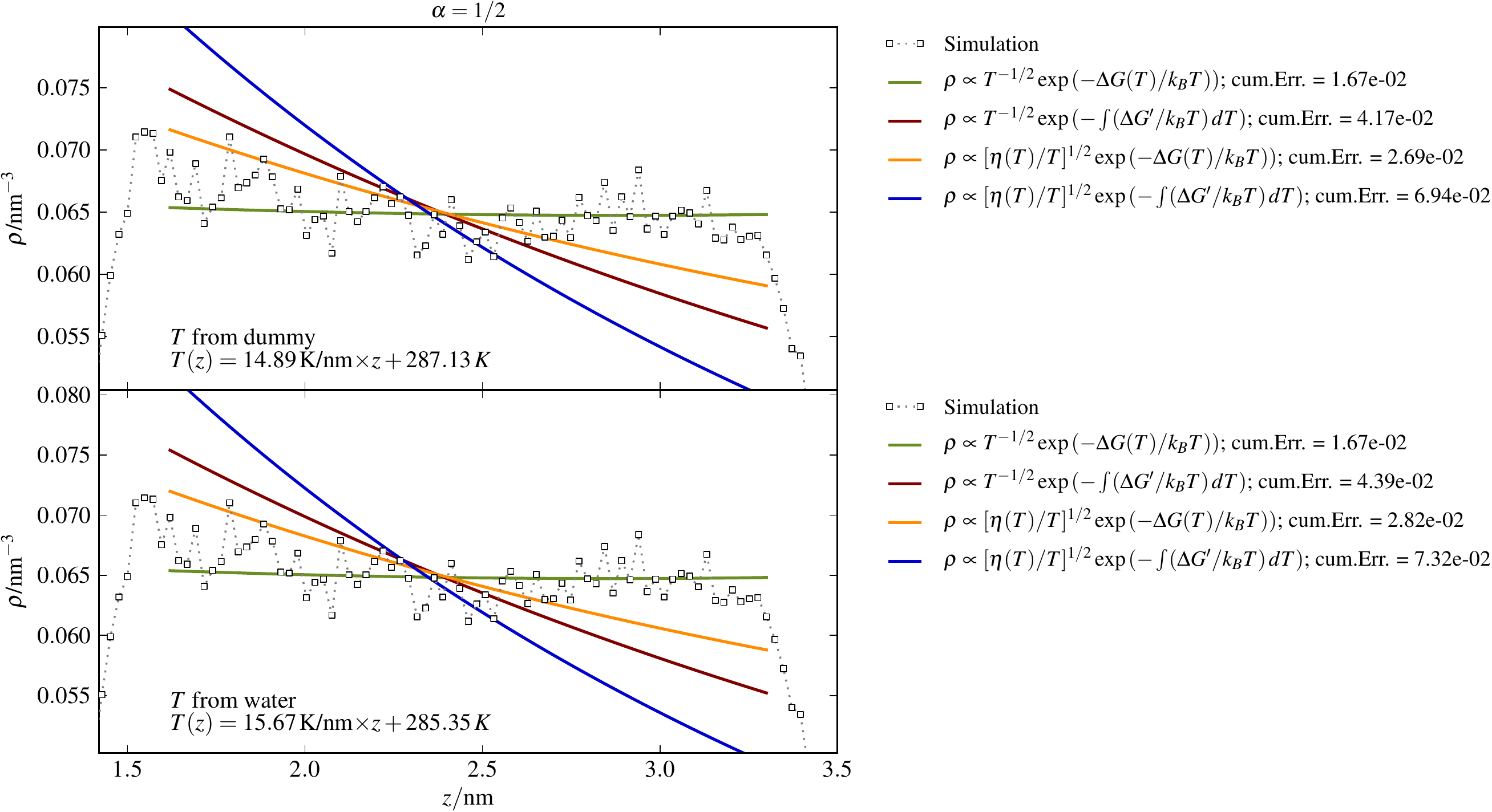}
 \caption{Density from a thermophoretic simulation of \textit{\textbf{krypton}} running for $\simeq350$ns and temperature values of $T_c=300\unit K$, $T_h=350\unit K$, compared to all eight hypotheses.}
 \label{fig:dens_kr_300_350}
\end{figure}

\begin{figure}[t!]
 \centering
    \includegraphics[width=10cm]{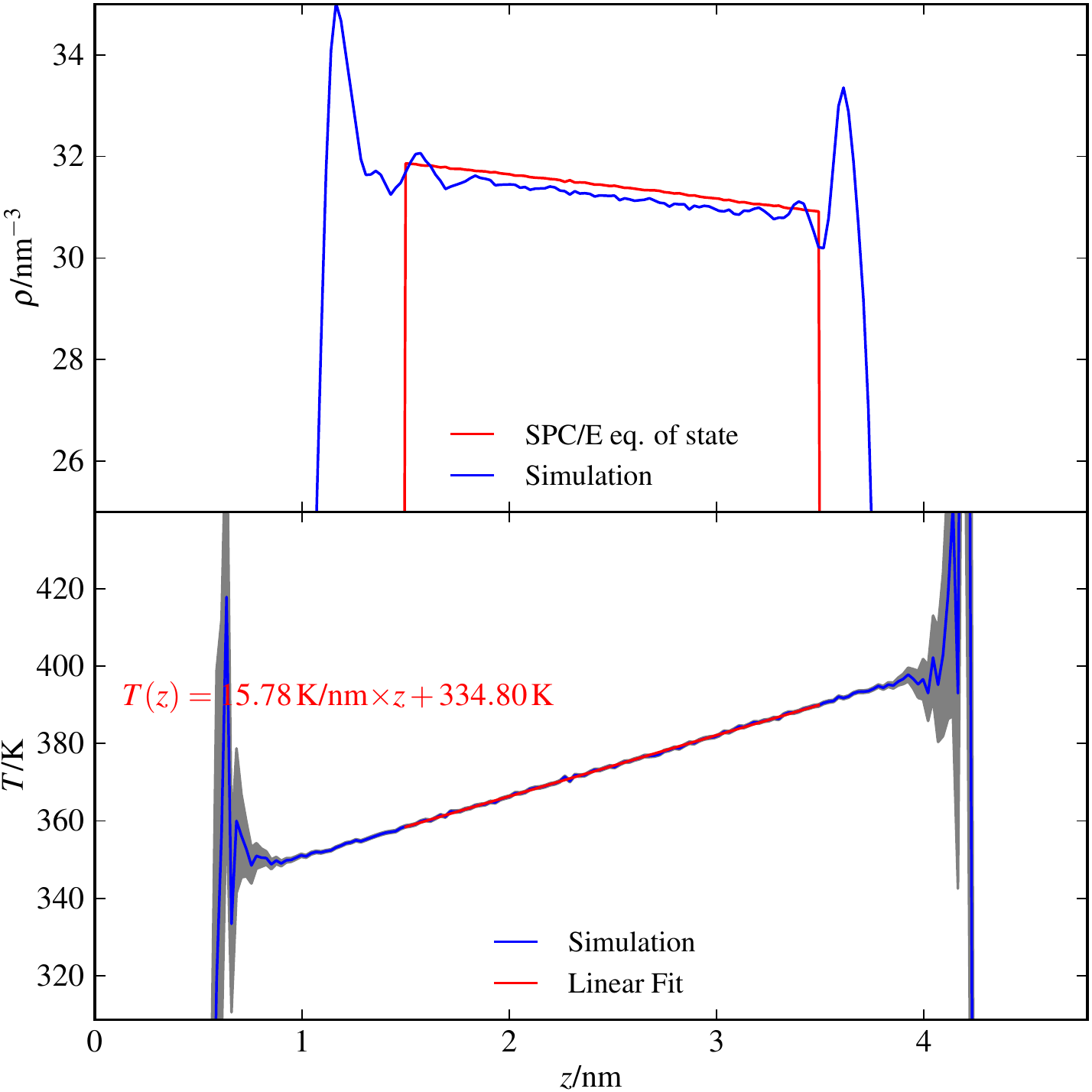}
    \includegraphics[width=10cm]{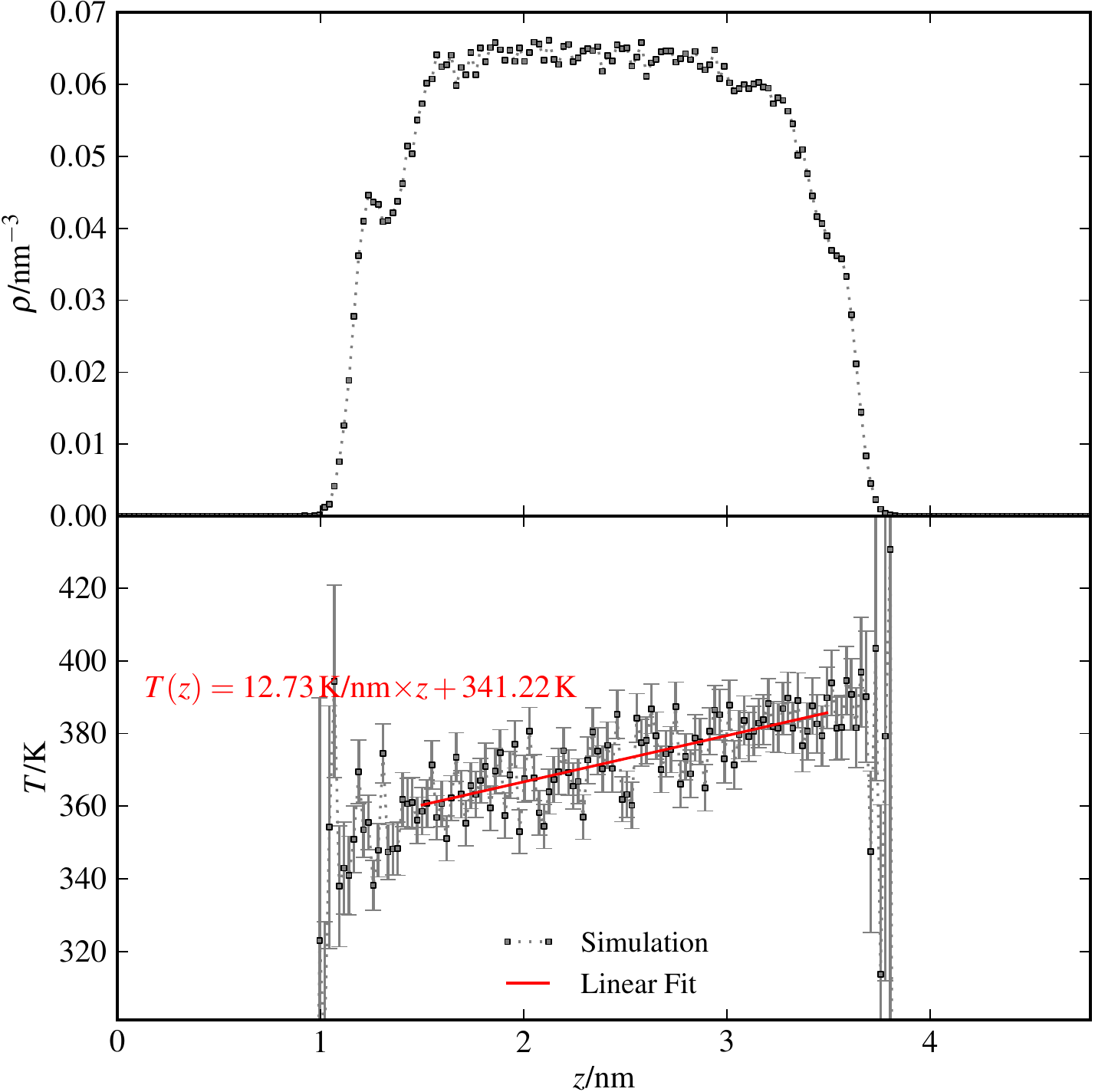}
 \caption{Solvent (\textbf{top}) and solute (\textbf{bottom}) densities and temperatures from a simulation of \textit{\textbf{krypton}} in a thermostatted system of $T_c=350\unit K$, $T_h=400\unit K$ with a run time of $\simeq434$ns.}
 \label{fig:temp_kr_350_400}
\end{figure}

\begin{figure}[t!]
 \centering
    \includegraphics[width=\textwidth]{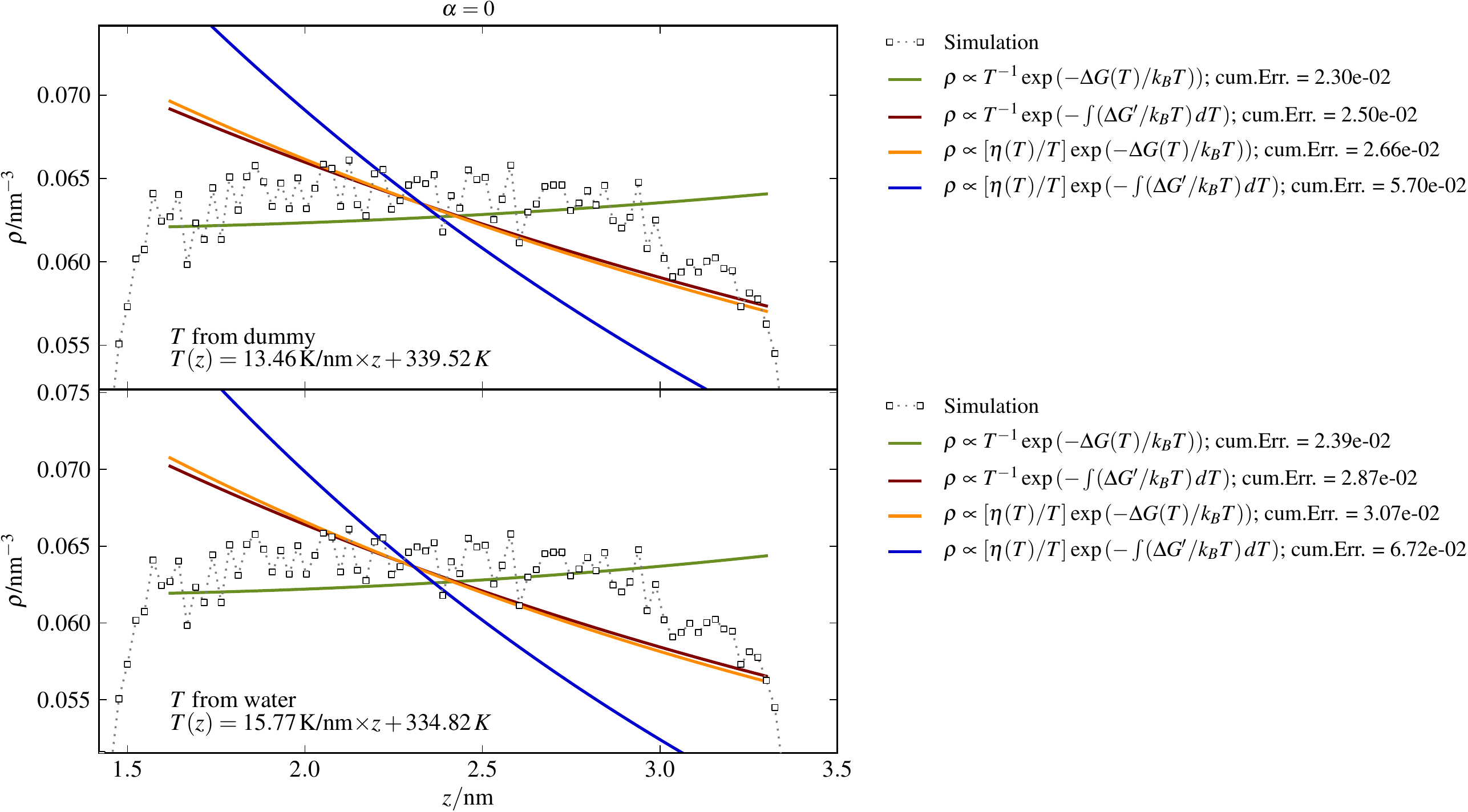}
    \includegraphics[width=\textwidth]{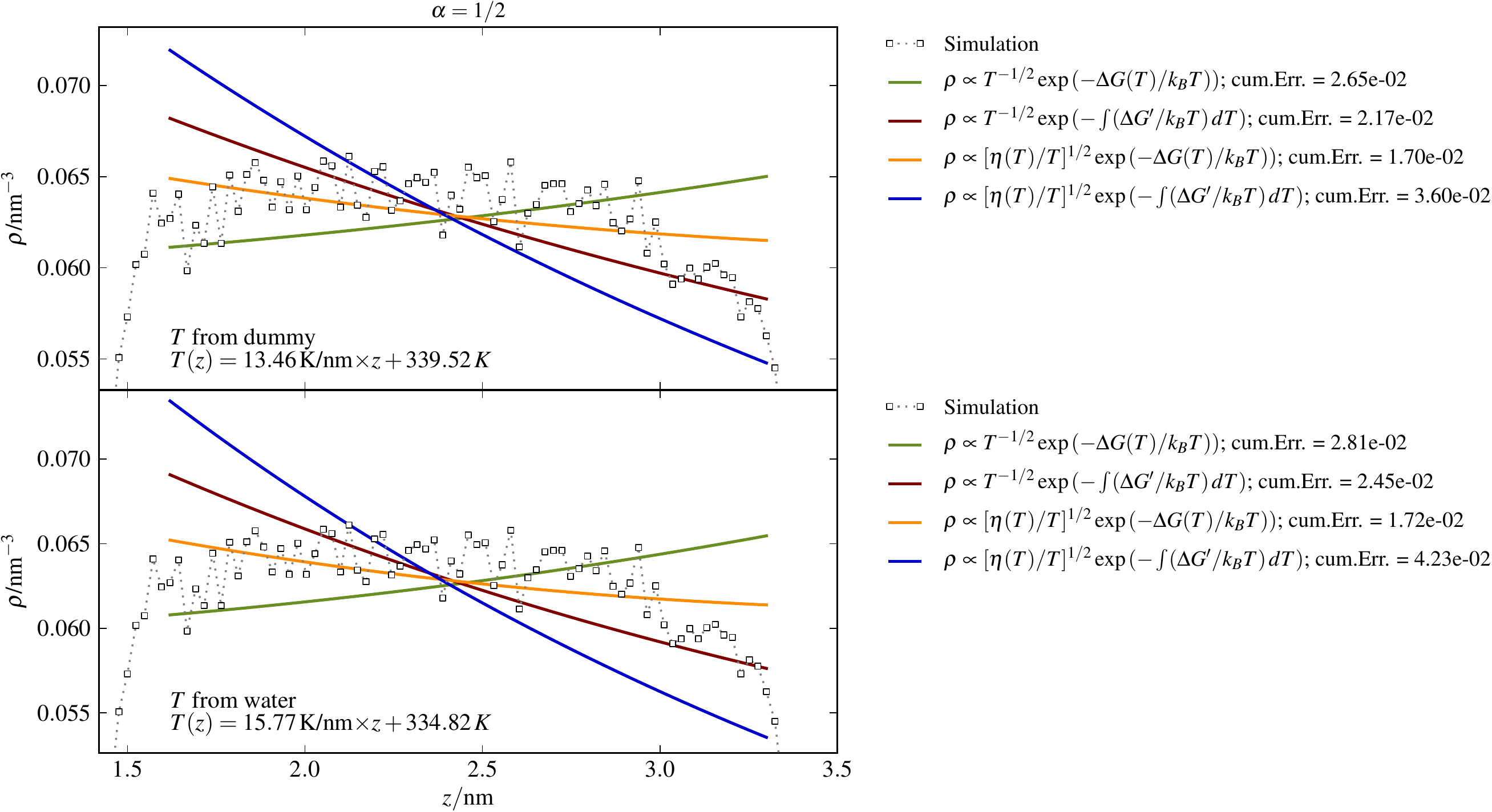}
 \caption{Density from a thermophoretic simulation of \textit{\textbf{krypton}} running for $\simeq434$ns and temperature values of $T_c=350\unit K$, $T_h=400\unit K$, compared to all eight hypotheses.}
 \label{fig:dens_kr_350_400}
\end{figure}

\begin{figure}[t!]
 \centering
    \includegraphics[width=10cm]{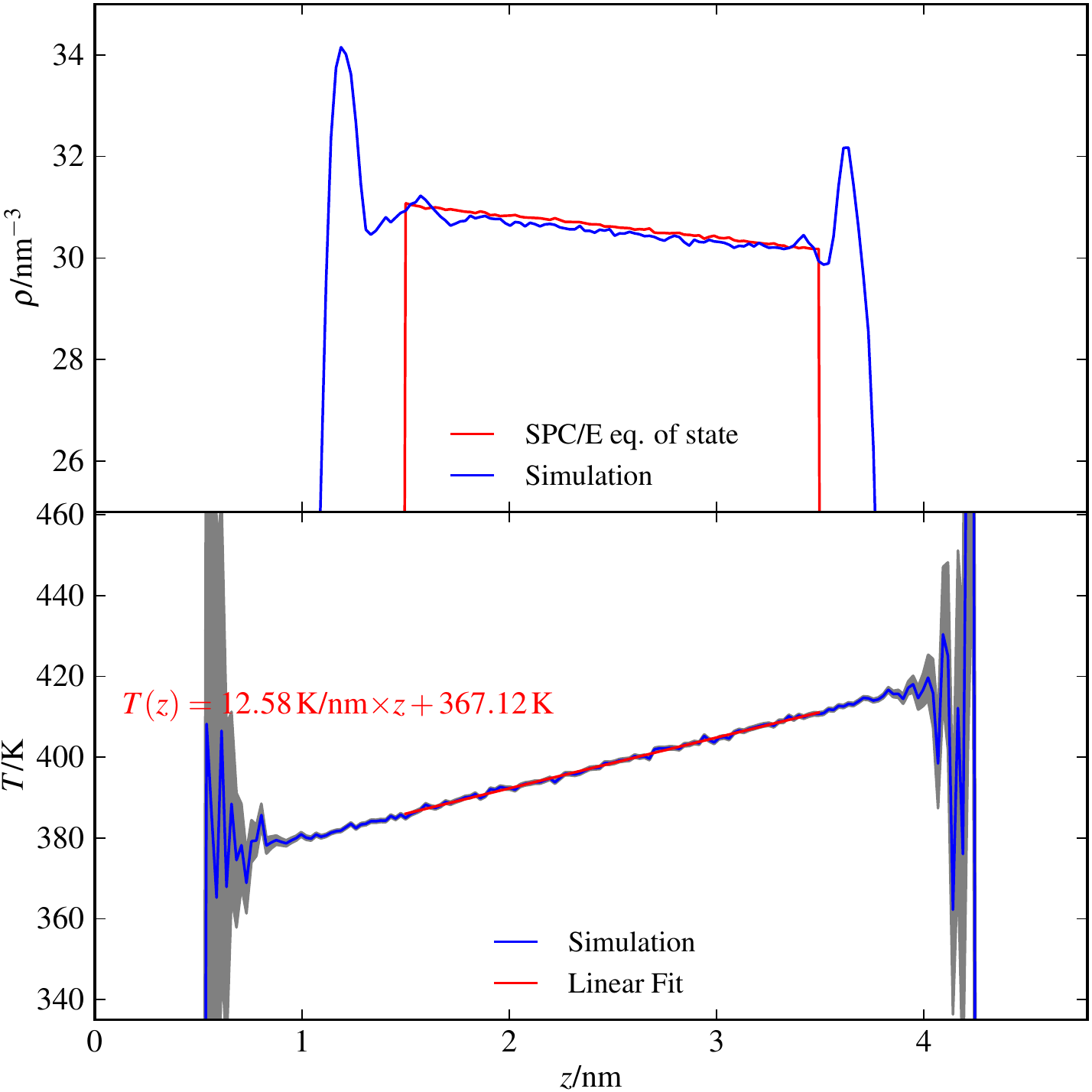}
    \includegraphics[width=10cm]{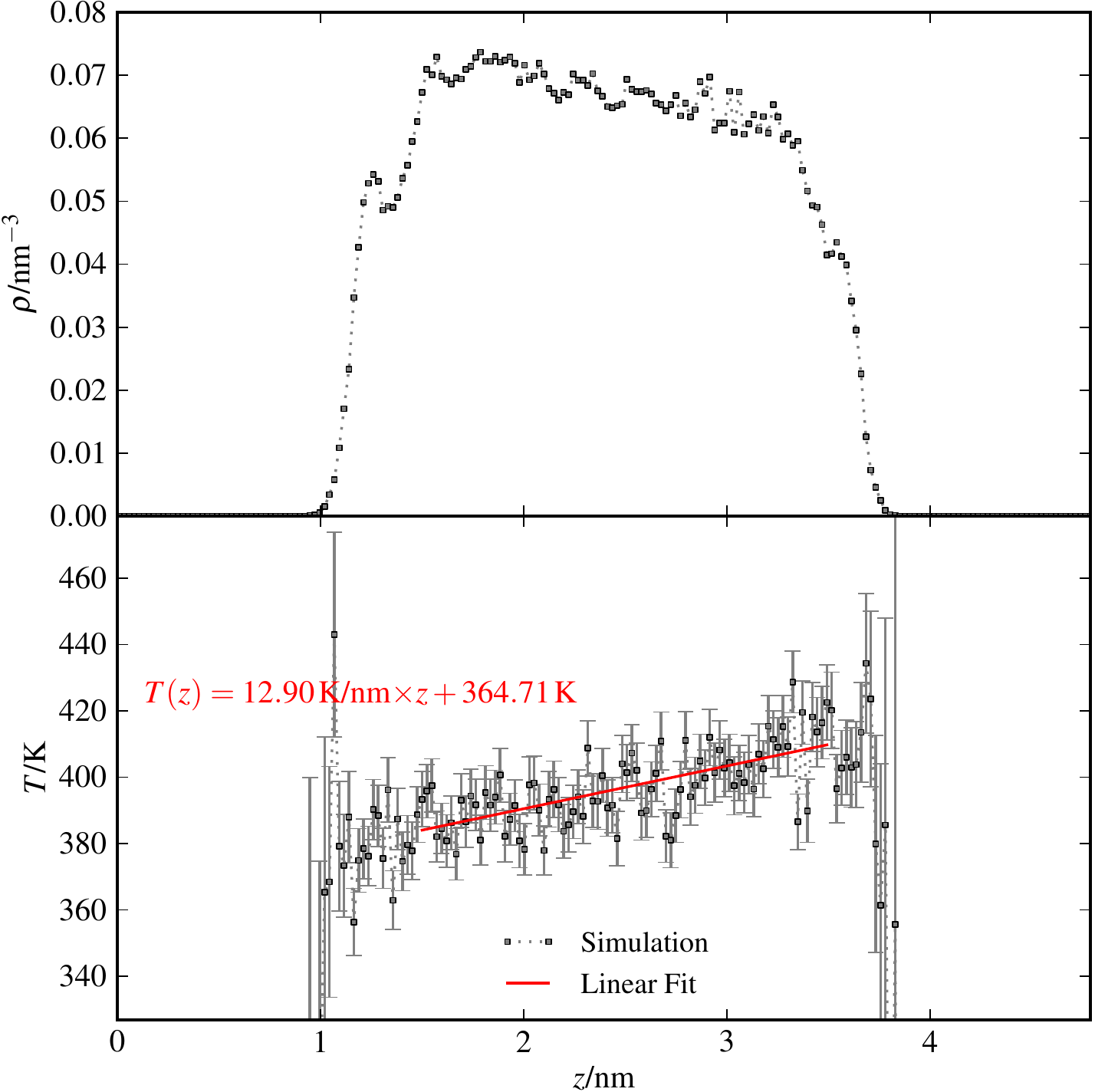}
 \caption{Solvent (\textbf{top}) and solute (\textbf{bottom}) densities and temperatures from a simulation of \textit{\textbf{krypton}} in a thermostatted system of $T_c=380\unit K$, $T_h=420\unit K$ with a run time of $\simeq307$ns.}
 \label{fig:temp_kr_380_420}
\end{figure}

\begin{figure}[t!]
 \centering
    \includegraphics[width=\textwidth]{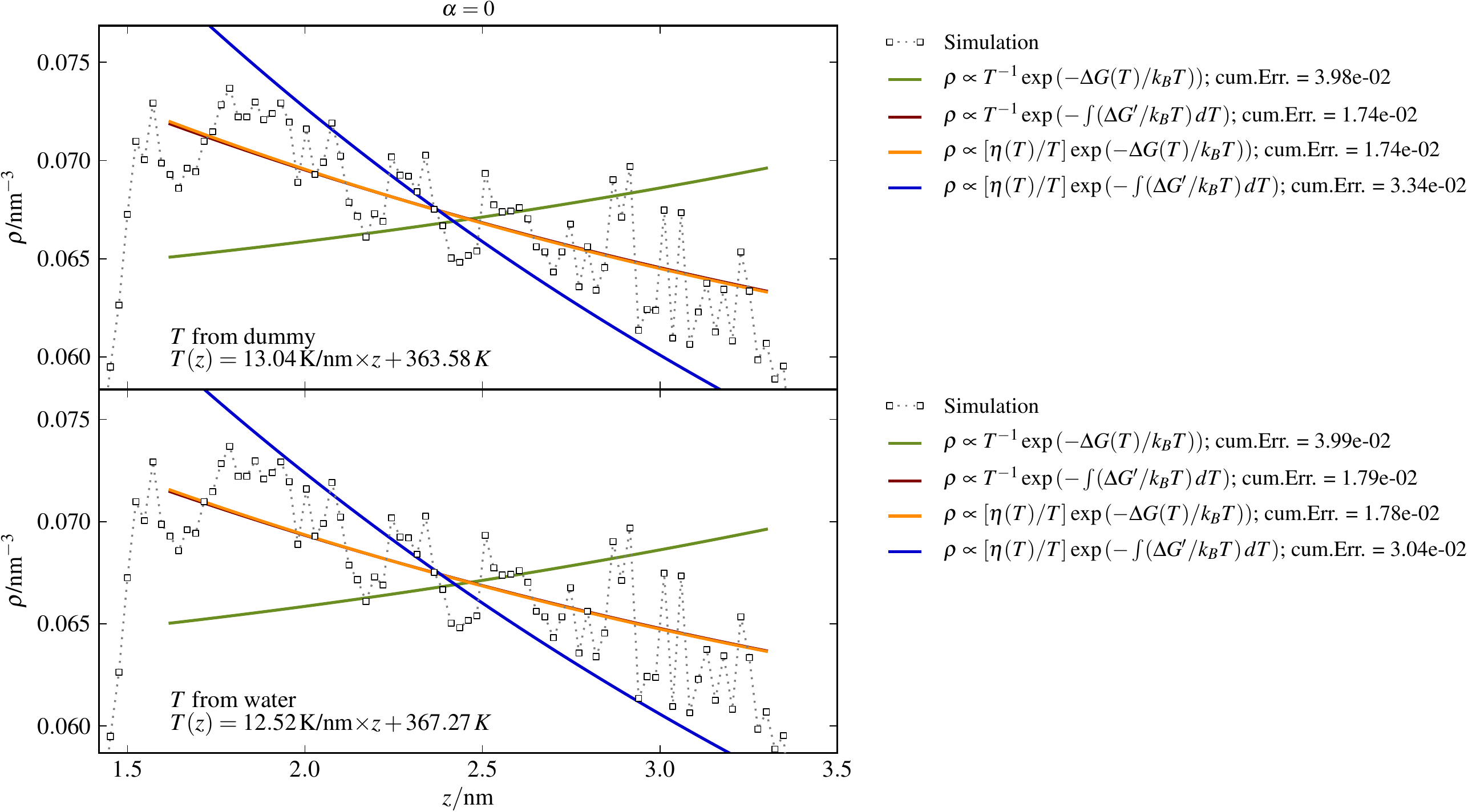}
    \includegraphics[width=\textwidth]{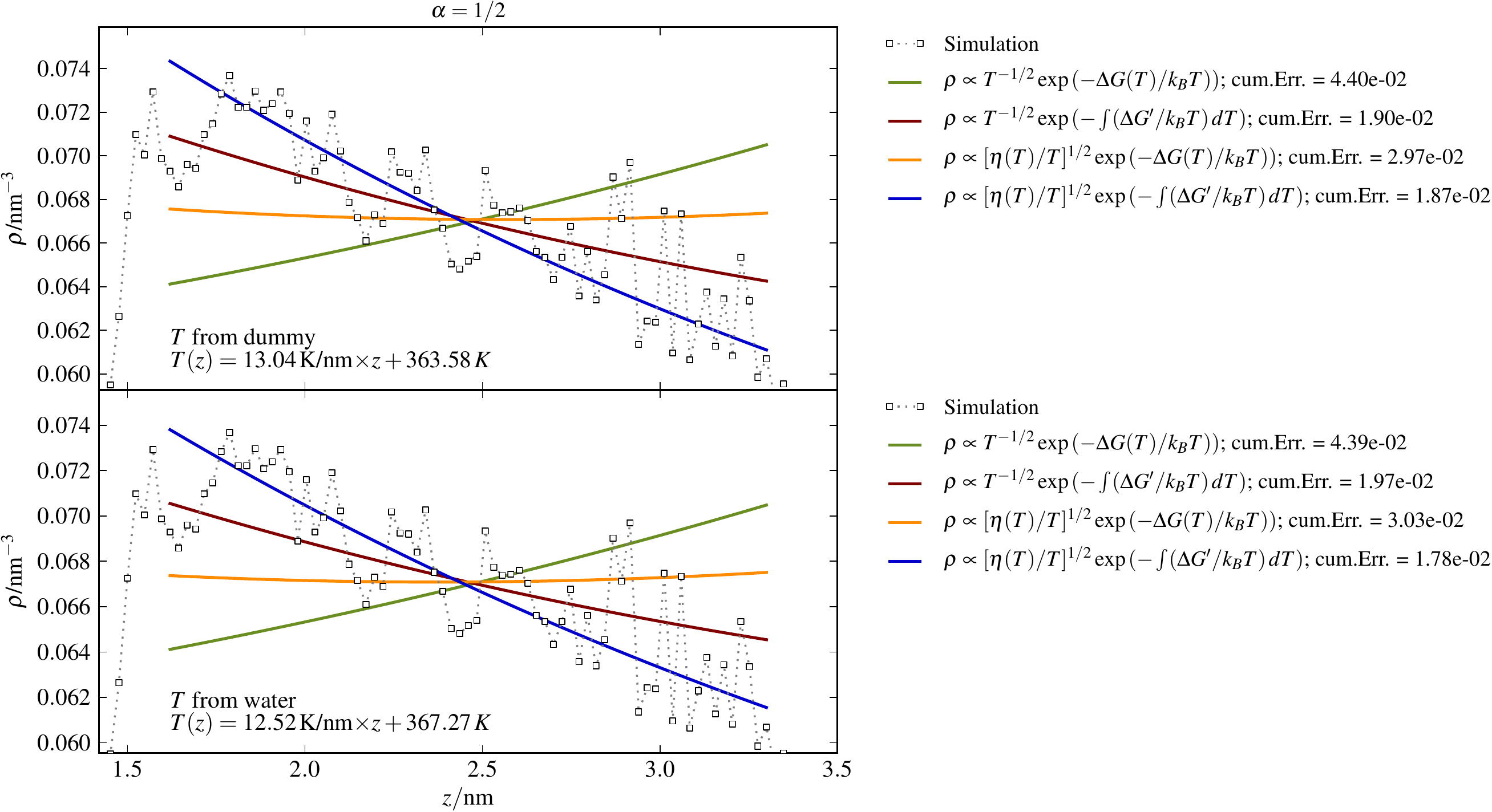}
 \caption{Density from a thermophoretic simulation of \textit{\textbf{krypton}} running for $\simeq307$ns and temperature values of $T_c=380\unit K$, $T_h=420\unit K$, compared to all eight hypotheses.}
 \label{fig:dens_kr_380_420}
\end{figure}

\begin{figure}[t!]
 \centering
    \includegraphics[width=10cm]{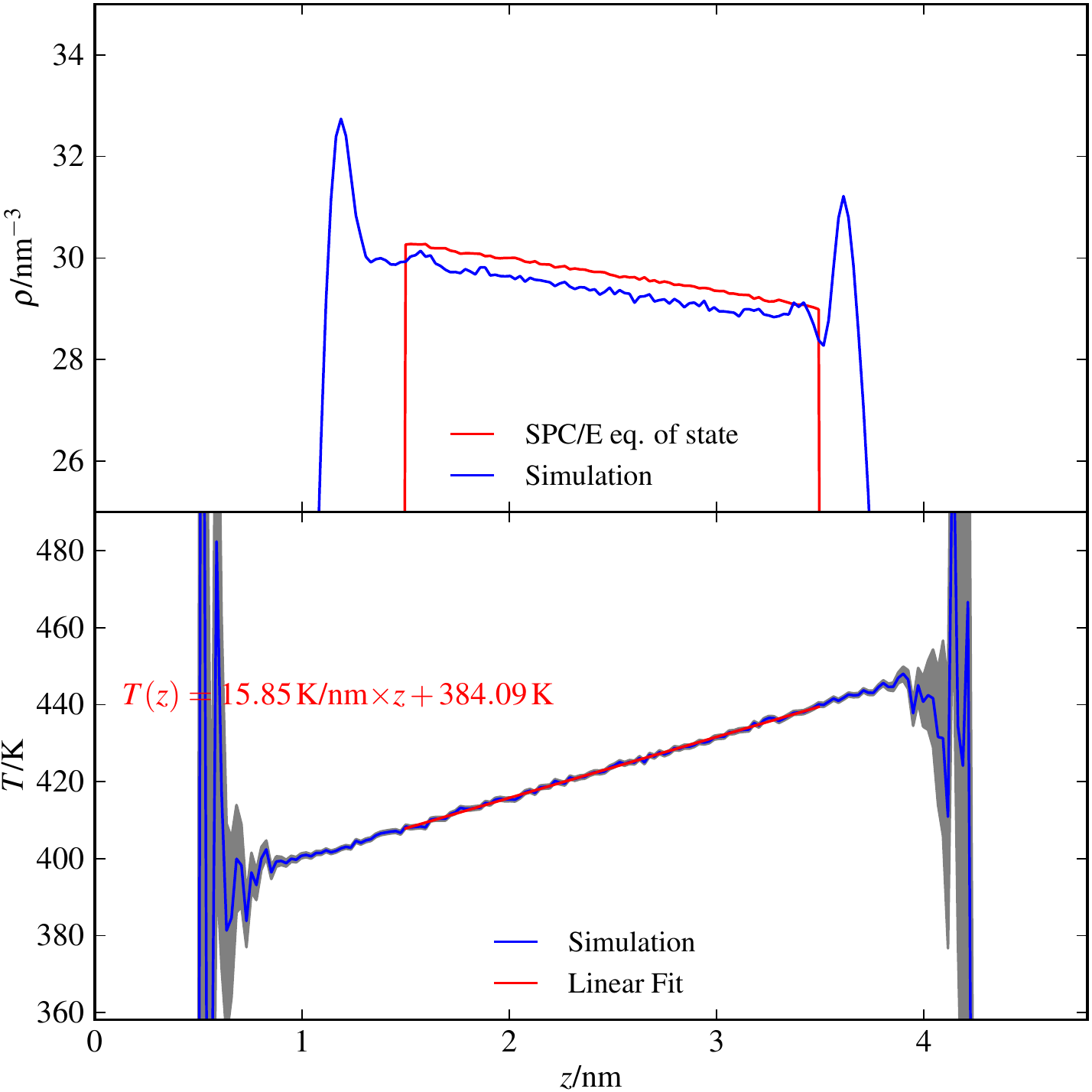}
    \includegraphics[width=10cm]{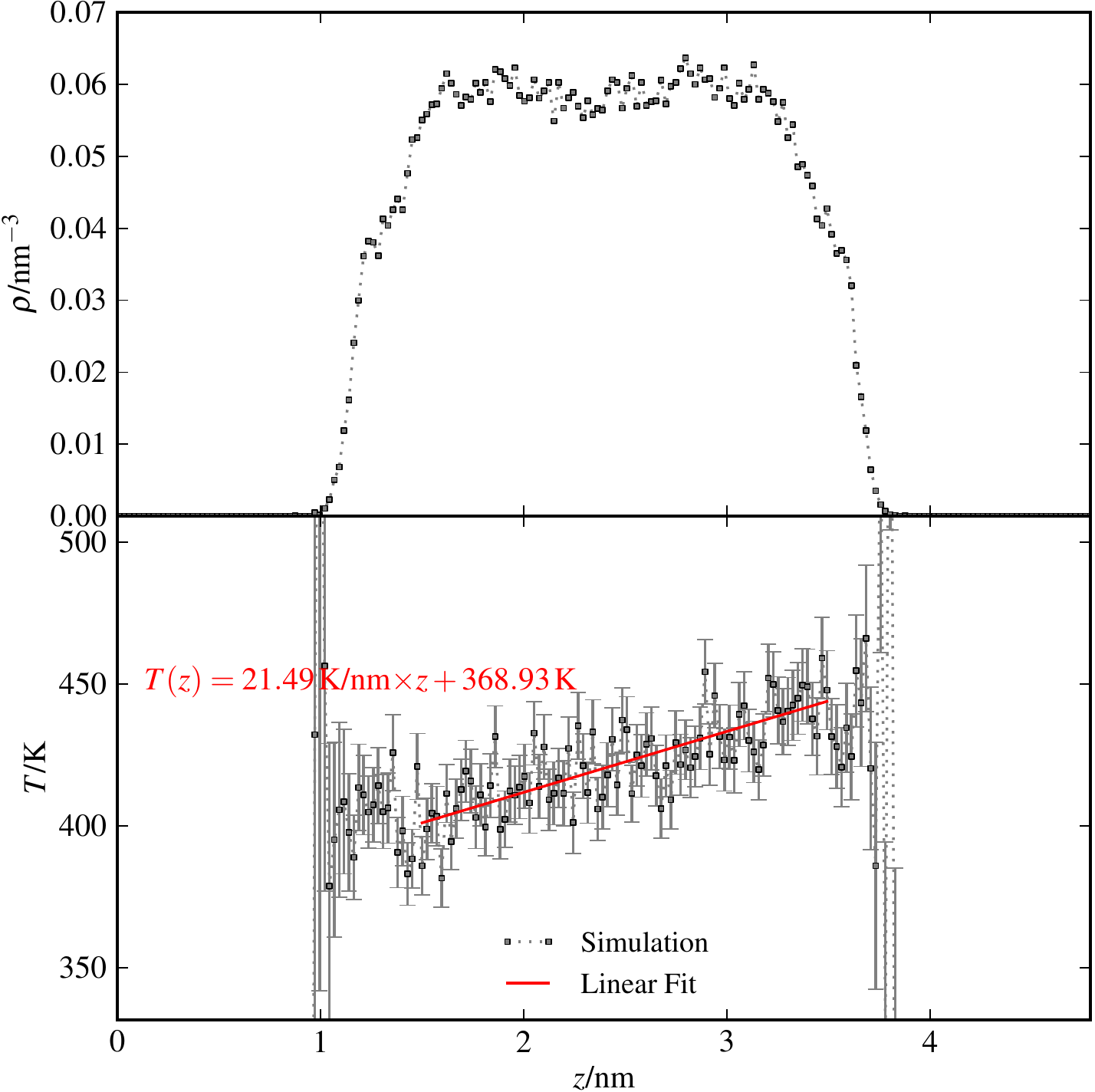}
 \caption{Solvent (\textbf{top}) and solute (\textbf{bottom}) densities and temperatures from a simulation of \textit{\textbf{krypton}} in a thermostatted system of $T_c=400\unit K$, $T_h=450\unit K$ with a run time of $200$ns.}
 \label{fig:temp_kr_400_450}
\end{figure}

\begin{figure}[t!]
 \centering
    \includegraphics[width=\textwidth]{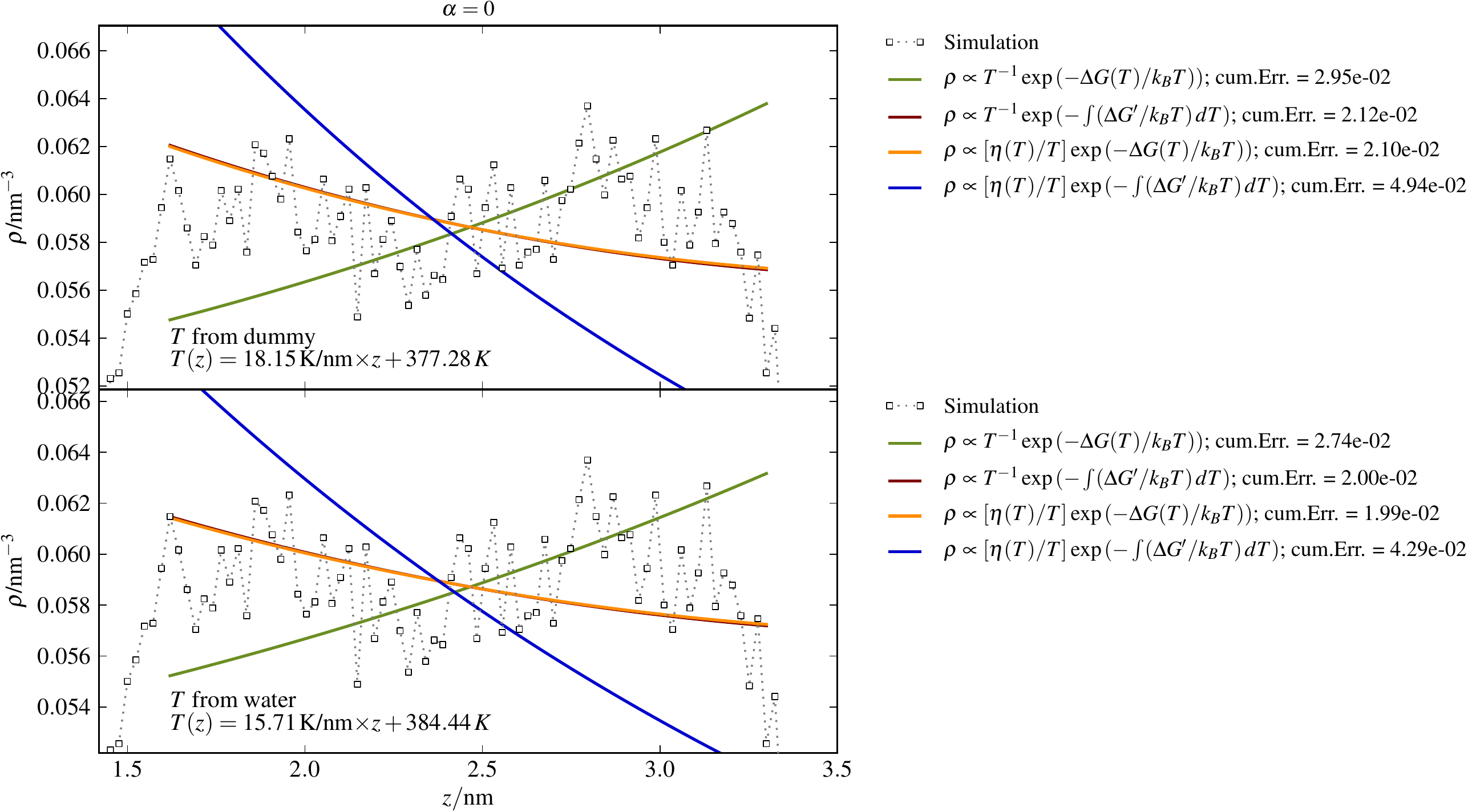}
    \includegraphics[width=\textwidth]{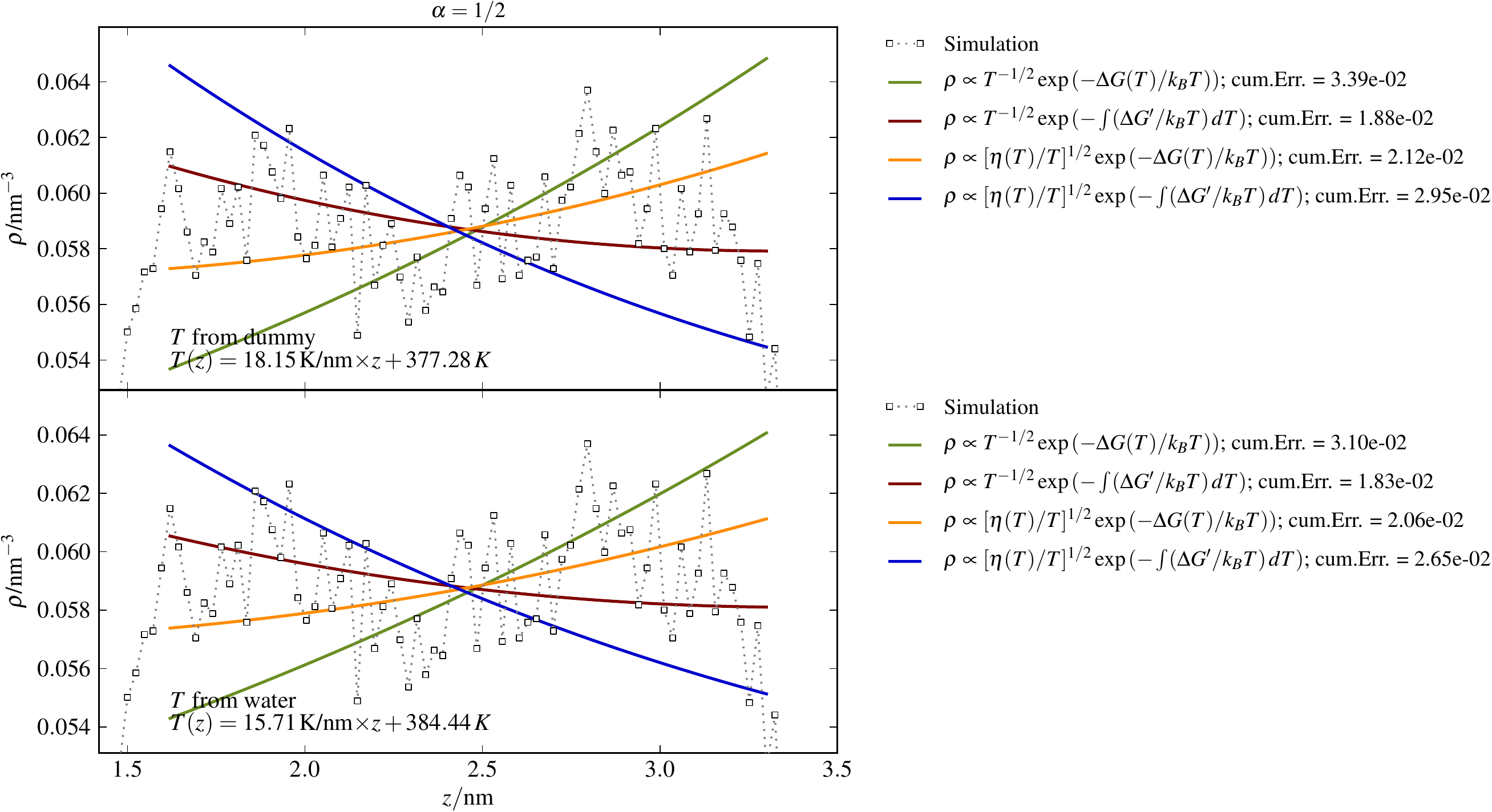}
 \caption{Density from a thermophoretic simulation of \textit{\textbf{krypton}} running for $200$ns and temperature values of $T_c=400\unit K$, $T_h=450\unit K$, compared to all eight hypotheses.}
 \label{fig:dens_kr_400_450}
\end{figure}

\newpage
\ 
\newpage
\ 
\newpage
\ 
\newpage
\ 
\newpage
\ 
\newpage
\

\section{Soret Coefficient}

\begin{figure}
 \centering
    \includegraphics[width=12cm]{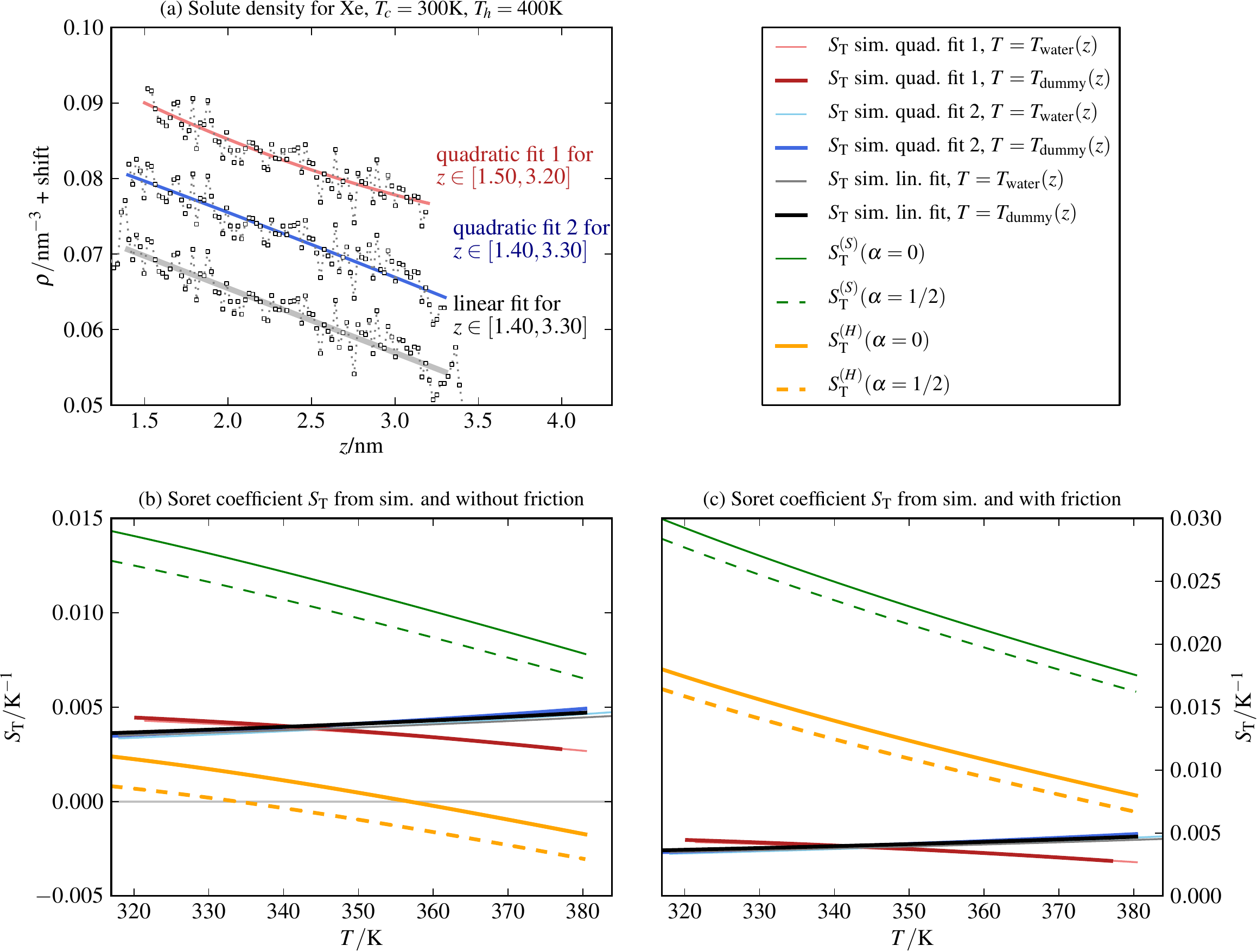}
 \caption{Supposed Soret coefficients from analytic considerations in \sec{\ref{sec:interacting_particles}} (i.e. \eq{\ref{eq:soret_definitions}}, bottom left, and \eq{\ref{eq:soret_definitions}} + \eq{\ref{eq:soret_fric}}) and from fits of the simulation density for \textit{\textbf{xenon}} in a thermostatted system of $T_c=300\unit K$, $T_h=400\unit K$.}
 \label{fig:soret_xe_300_400}
\end{figure}

\begin{figure}
 \centering
    \includegraphics[width=12cm]{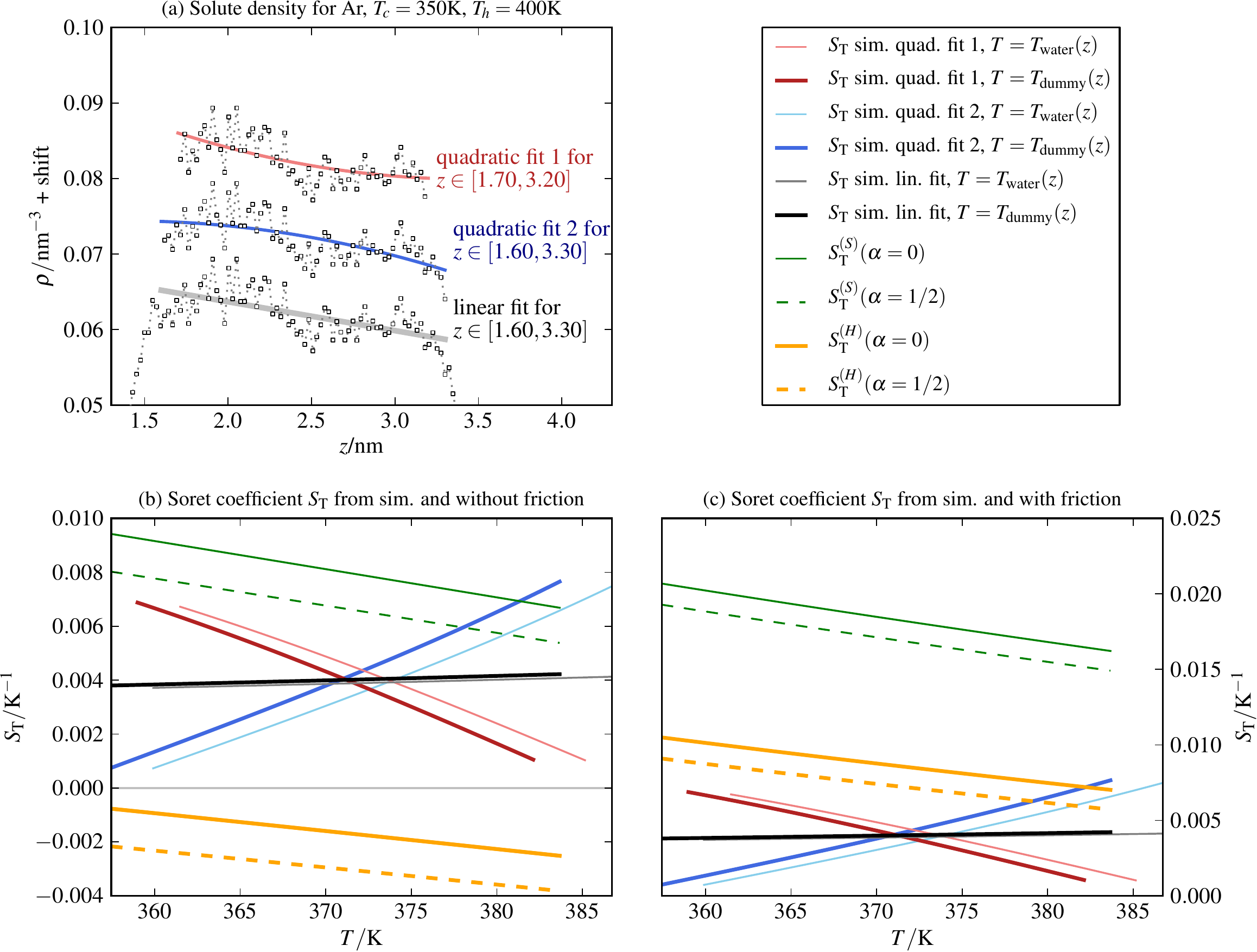}
 \caption{Supposed Soret coefficients from analytic considerations in \sec{\ref{sec:interacting_particles}} (i.e. \eq{\ref{eq:soret_definitions}}, bottom left, and \eq{\ref{eq:soret_definitions}} + \eq{\ref{eq:soret_fric}}) and from fits of the simulation density for \textit{\textbf{argon}} in a thermostatted system of $T_c=350\unit K$, $T_h=400\unit K$.}
 \label{fig:soret_ar_350_400}
\end{figure}

\begin{figure}
 \centering
    \includegraphics[width=12cm]{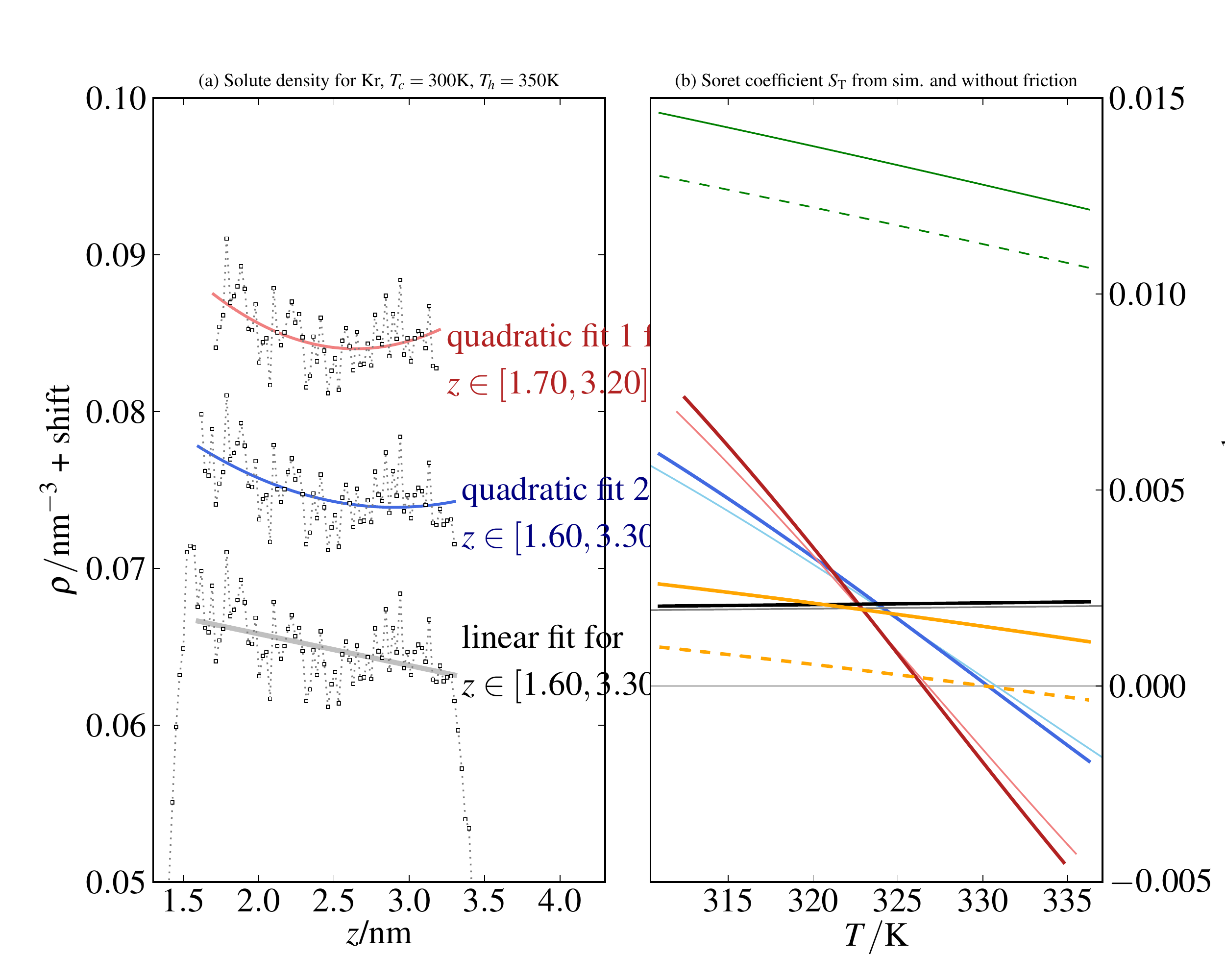}
 \caption{Supposed Soret coefficients from analytic considerations in \sec{\ref{sec:interacting_particles}} (i.e. \eq{\ref{eq:soret_definitions}}, bottom left, and \eq{\ref{eq:soret_definitions}} + \eq{\ref{eq:soret_fric}}) and from fits of the simulation density for \textit{\textbf{krypton}} in a thermostatted system of $T_c=300\unit K$, $T_h=350\unit K$.}
 \label{fig:soret_kr_300_350}
\end{figure}

\begin{figure}
 \centering
    \includegraphics[width=12cm]{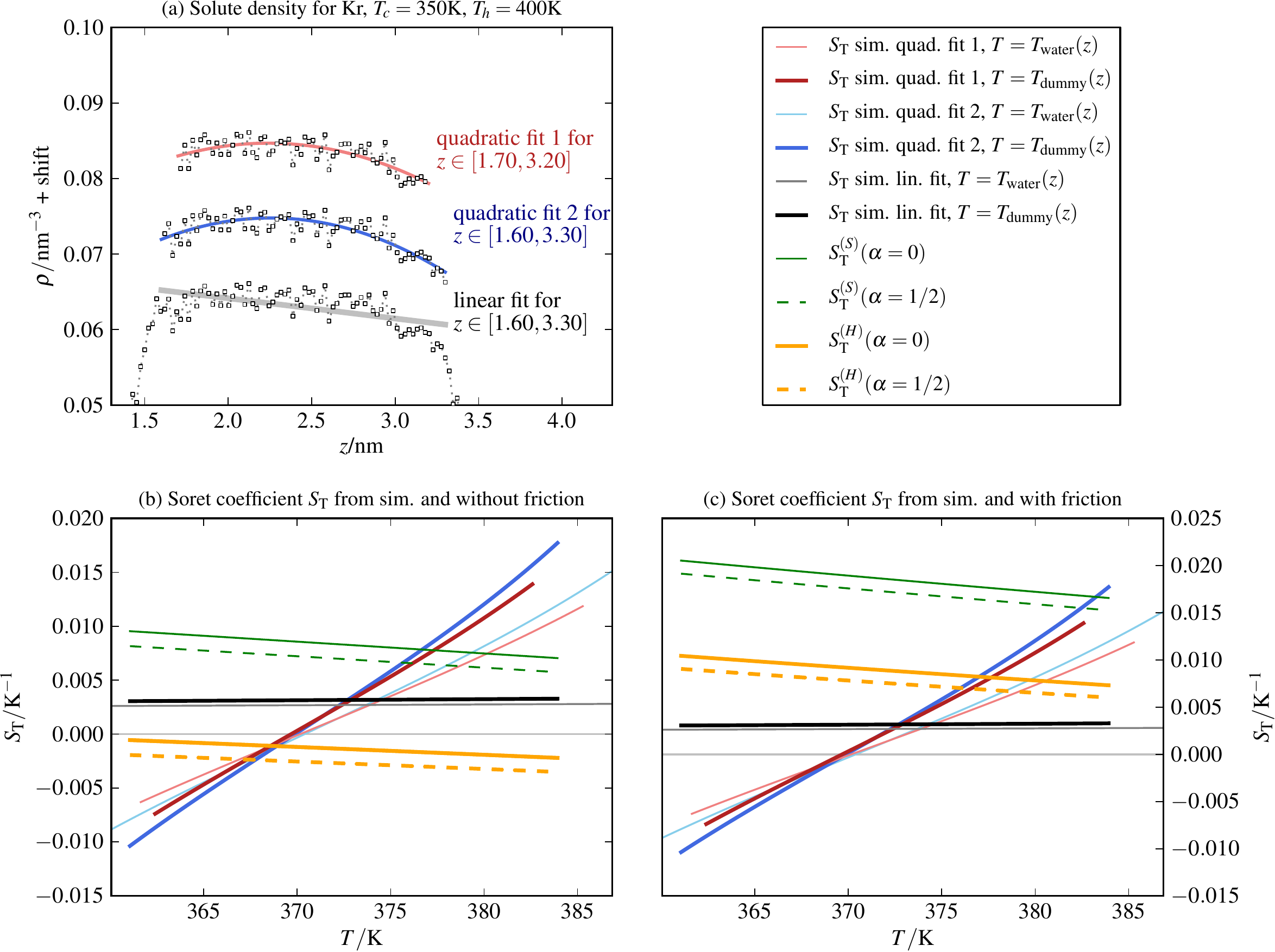}
 \caption{Supposed Soret coefficients from analytic considerations in \sec{\ref{sec:interacting_particles}} (i.e. \eq{\ref{eq:soret_definitions}}, bottom left, and \eq{\ref{eq:soret_definitions}} + \eq{\ref{eq:soret_fric}}) and from fits of the simulation density for \textit{\textbf{krypton}} in a thermostatted system of $T_c=350\unit K$, $T_h=400\unit K$.}
 \label{fig:soret_kr_350_400}
\end{figure}

\begin{figure}
 \centering
    \includegraphics[width=12cm]{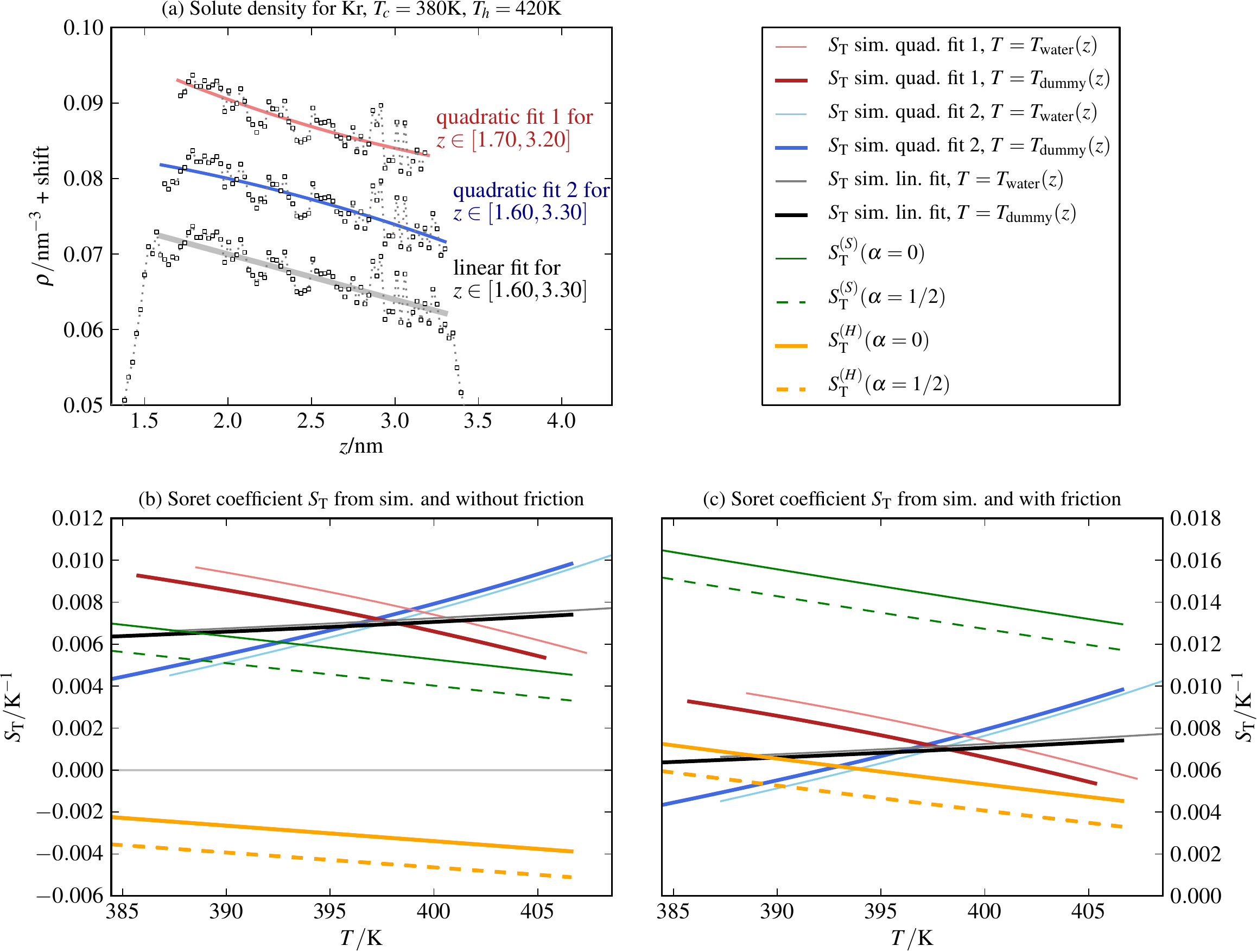}
 \caption{Supposed Soret coefficients from analytic considerations in \sec{\ref{sec:interacting_particles}} (i.e. \eq{\ref{eq:soret_definitions}}, bottom left, and \eq{\ref{eq:soret_definitions}} + \eq{\ref{eq:soret_fric}}) and from fits of the simulation density for \textit{\textbf{krypton}} in a thermostatted system of $T_c=380\unit K$, $T_h=420\unit K$.}
 \label{fig:soret_kr_380_420}
\end{figure}

\begin{figure}
 \centering
    \includegraphics[width=12cm]{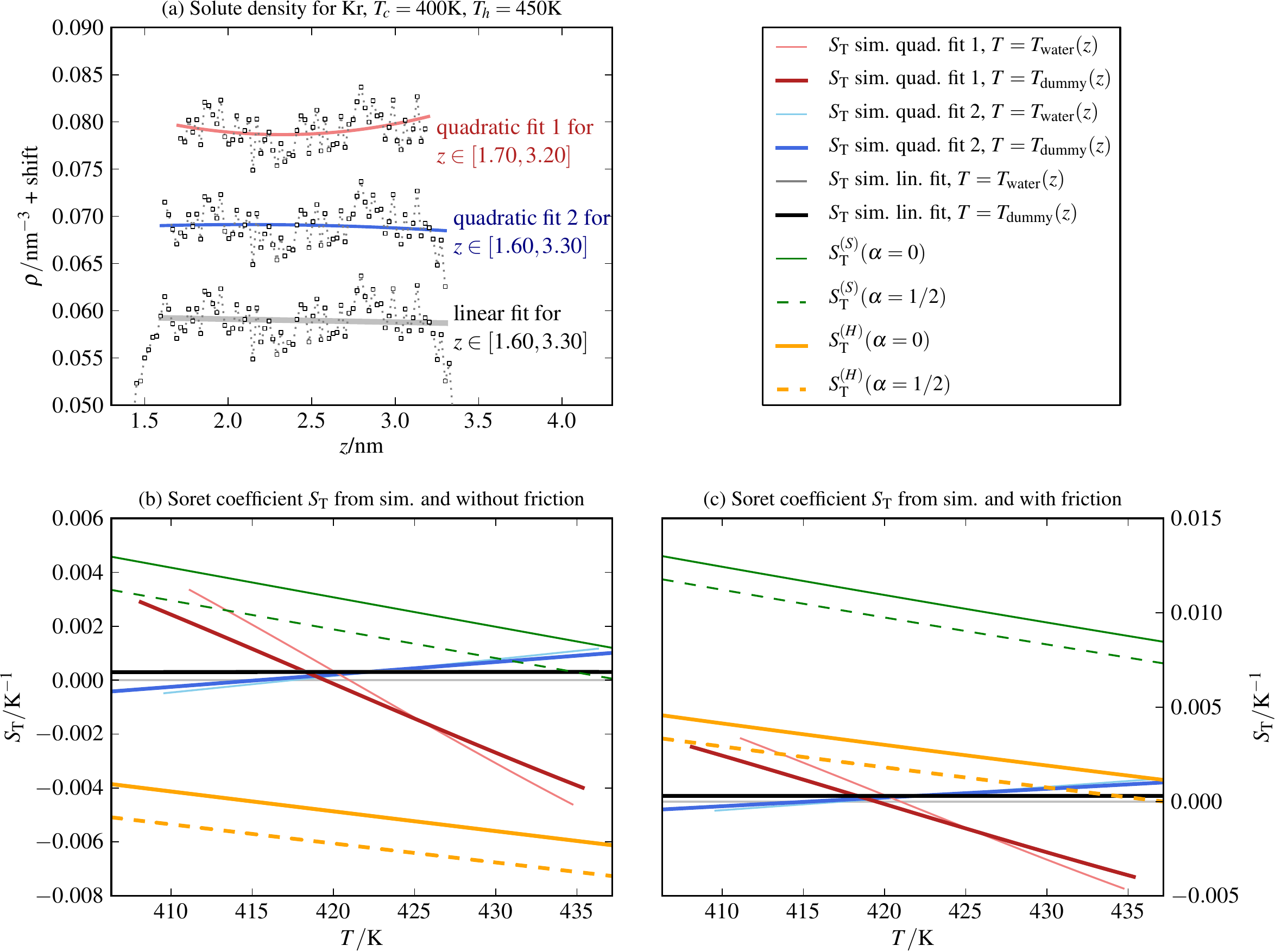}
 \caption{Supposed Soret coefficients from analytic considerations in \sec{\ref{sec:interacting_particles}} (i.e. \eq{\ref{eq:soret_definitions}}, bottom left, and \eq{\ref{eq:soret_definitions}} + \eq{\ref{eq:soret_fric}}) and from fits of the simulation density for \textit{\textbf{krypton}} in a thermostatted system of $T_c=400\unit K$, $T_h=450\unit K$.}
 \label{fig:soret_kr_400_450}
\end{figure}

%% file: chapters/appendixC.tex
\chapter{Tolman Length for the Interfacial Solvation Free Energy}
\label{app:tolman_length}

As was described in \sec{\ref{sec:solute_hydrophobicity}}, one can calculate solvation free energies by means of an interfacial interaction energy which in turn can be used for the calculation of the Soret coefficient as defined by \eq{\ref{eq:soret_definitions}}. To this end, one can employ \eq{\ref{eq:vism_free_energy}} with the temperature dependent properties of water $\gammalv(T)$, $\rho(T)$, $\epsilon_\ell(T)$ and $\delta(T)$. The surface tension $\gammalv(T)$, the density $\rho(T)$ and the static permittivity $\epsilon_\ell(T)$ are given in \sec{\ref{sec:thermodynamics}}. The only remaining unknown temperature dependent quantity is the tolman length $\delta(T)$, which will consequently be approximated in the following. It will be modeled using the VISM free energy for an uncharged sphere of radius $R$ and experimental values for the solvation of noble gases.

\section{Hydration Entropy of Noble Gases}

\begin{figure}[tb]
 \centering
    \includegraphics[width=9cm]{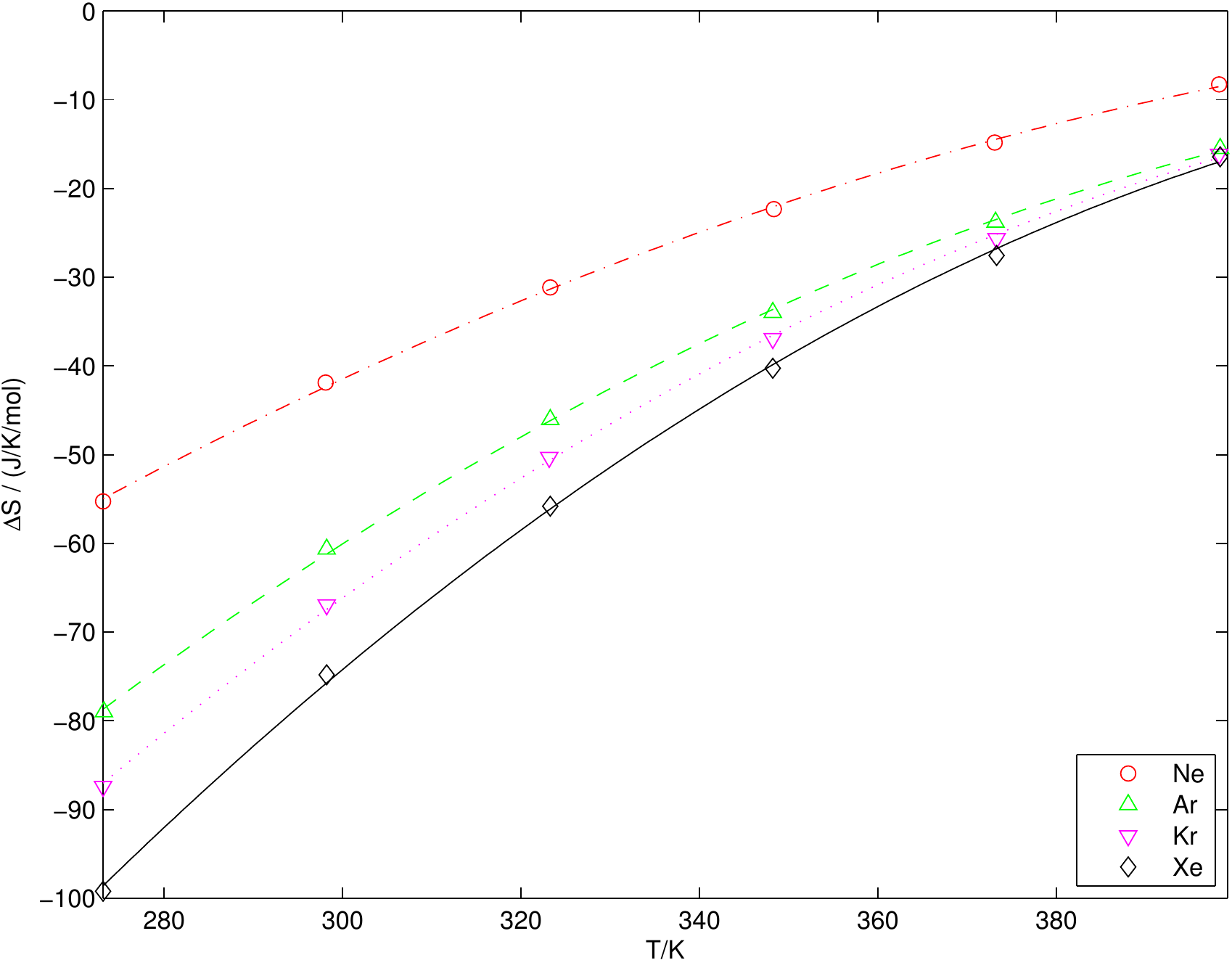}
 \caption{Quadratic fits for the hydration entropy of various noble gases, experimental data from \cite{Graziano:2003}. Fit parameters can be found in \tab{\ref{tab:hyd_entropy_fits}}}
 \label{fig:hyd_entropy_fits}
\end{figure}

As will be seen in the next paragraph, the hydration entropy for noble gases is needed as a function of temperature to obtain the Tolman length $\delta$. To this end, it was graphically extracted from \cite{Graziano:2003}, \fig{1} %and \fig{2}
and modeled with a polynomial fit
\begin{equation}
 \label{eq:hyd_entropy}
 \Delta S = b_2T^2+b_1T+b_0,
\end{equation}
where the values of the fit parameters are given in \tab{\ref{tab:hyd_entropy_fits}} and the fits can be seen in \fig{\ref{fig:hyd_entropy_fits}}.
\begin{table}[htb]
\begin{center}
\begin{tabular}{lccc}
   \hline\hline 
    & $b_2/\text J \text K^{-3} \text{mol}^{-1} $ & $b_1/\text J \text K^{-2} \text{mol}^{-1}$ & $b_0/\text J \text K^{-1} \text{mol}^{-1}$ \\
   \hline  
   \input{tables/hyd_entropy_fit_table_molar.tex}
    \hline
    & $b_2/\text J \text K^{-3}$ & $b_1/\text J \text K^{-2}$ & $b_0/\text J \text K^{-1}$ \\
    \hline
   \input{tables/hyd_entropy_fit_table_nonmolar.tex}
    \hline
\end{tabular}
\caption{Fit parameters for the hydration entropy of various noble gases}
\label{tab:hyd_entropy_fits}
\end{center}
\end{table}

% ********************
% ********************
% ********************
% ********************
% ********************

\section{Tolman Length}
\begin{figure}[htb]
 \centering
    \includegraphics[width=9cm]{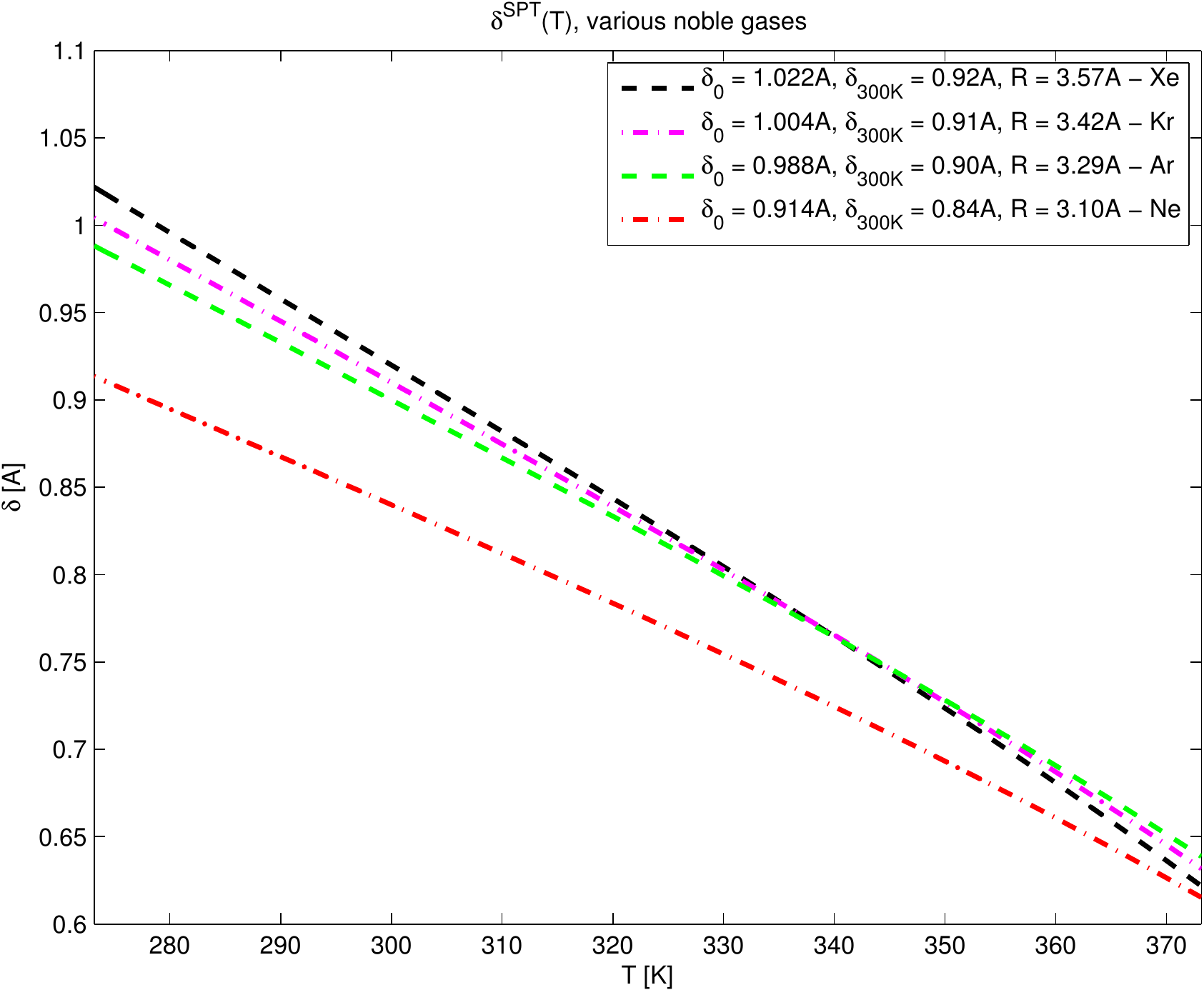}
 \caption{Results for $\delta(T)$ from the experimental hydration entropy. $\delta(T_0)$ has been fixed using data fitted from SPT given in \tab{1} of \cite{Dzubiella:2006}. 
%For the radius of helium, $R = 3\mathring A$, we used a seemingly sufficient approximation obtained through SPT from \tab{2} in \cite{Graziano:2003} and $\delta(300K)=0.81\mathring A$ as an assumption.
}
 \label{fig:delta_JCP2006}
\end{figure}

Using the VISM free energy approximation \eq{\ref{eq:vism_free_energy}} and the thermodynamic relation \eq{\ref{eq:entropy}} for an uncharged spherical particle of radius $R$ yields
\begin{equation}
 \label{eq:soret}
 \Delta S = -4\pi R^2 \l[\gamma'-2(\delta\gamma)'/R\r]
\end{equation}
where the prime denotes the derivative with respect to $T$.
Solving for $(\gamma\delta)'$ yields
\begin{align}
  (\gamma\delta)'&=\frac1{8\pi R} \l[\Delta S+4\pi R^2\gamma' \r],
\end{align}
which can now be easily integrated from $T_0 = 273.15 \text K$ to obtain
\begin{align}
 \delta(T) =& \frac{\gamma(T_0)\delta(T_0)}{\gamma(T)} + \frac R2 \l[1-\frac{\gamma(T_0)}{\gamma(T)}\r] +
              \frac1{8\pi R \gamma(T)}\,
              \Bigg[ \frac{b_2}3 \l(T^3-T_0^3\r) + 
                     \frac{b_1}2 \l(T^2-T_0^2\r) +
                           b_0   \l(T  -T_0  \r) \Bigg] \nonumber\\
           =& \frac R2 +  \frac{\gamma(T_0)}{\gamma(T)} \l[\delta(T_0)-\frac R2\r] +
              \frac1{8\pi R \gamma(T)}\,
              \Bigg[ \frac{b_2}3 \l(T^3-T_0^3\r) + 
                     \frac{b_1}2 \l(T^2-T_0^2\r) +
                           b_0   \l(T  -T_0  \r) \Bigg].
\label{eq:delta_entropy}
\end{align}
To fix the integration constant $\delta(T_0)$, one needs a $\delta$ measured in $[0,100]^\circ\text C$ from experiments or model simulations. For scaled particle theory (SPT) one finds the results given in \fig{\ref{fig:delta_JCP2006}}.

\begin{figure}[h!]
 \centering
    \includegraphics[width=12cm]{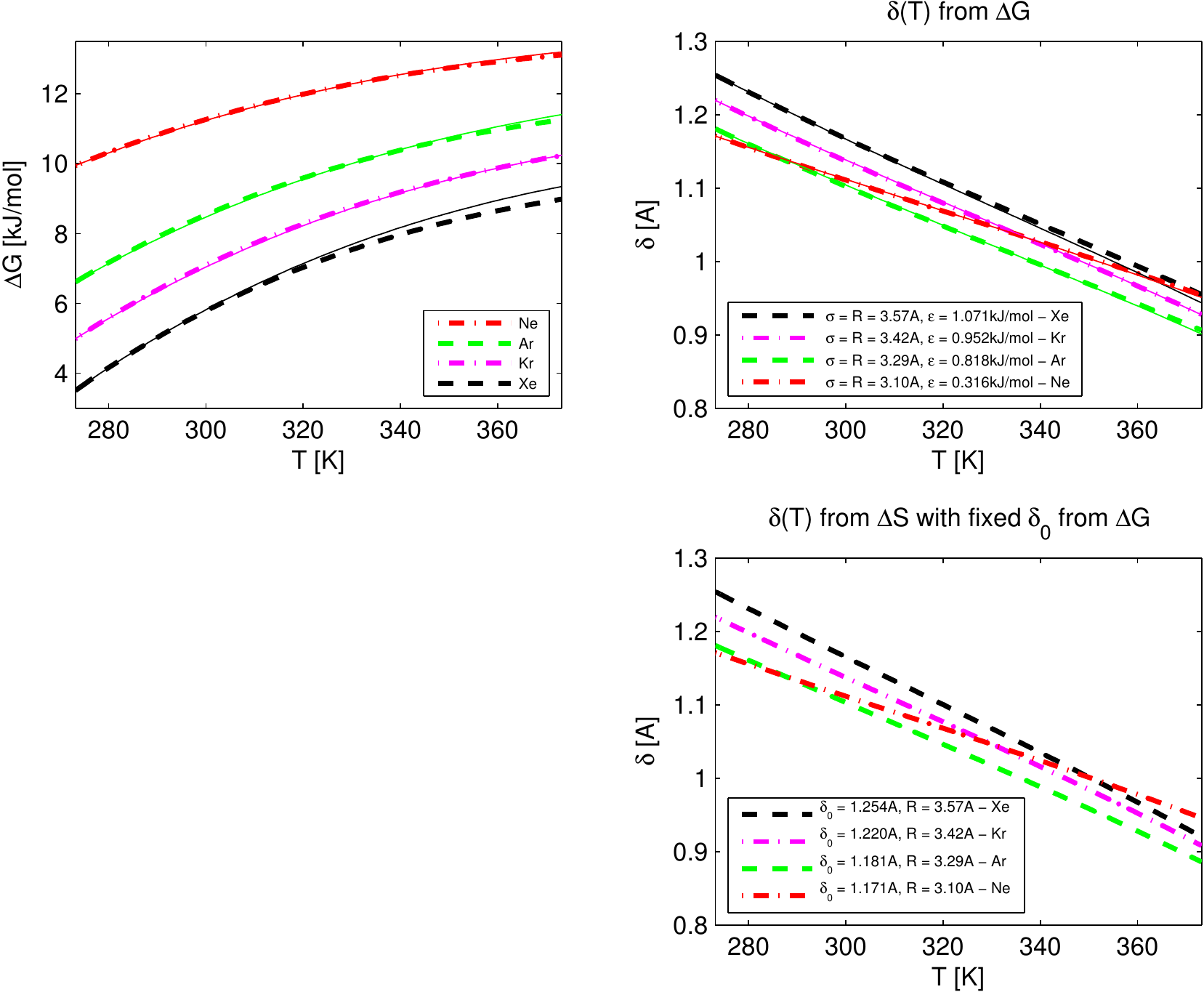}
 \caption{\textbf{Top left:} Hydration free energy for noble gases in water, thick dashed lines from \eq{\ref{eq:mu_final}}, solid lines represent the integrated entropy with fixed constant. \textbf{Top right:} Results for $\delta(T)$ derived from free energy, \eq{\ref{eq:delta_free_energy}}. \textbf{Bottom right:} $\delta(T)$ derived from hydration entropy, \eq{\ref{eq:delta_entropy}}, with fixed $\delta(T_0)$ from \eq{\ref{eq:delta_free_energy}}.}
 \label{fig:free_energy_exp}
\end{figure}

Another approach to obtain $\delta(T)$ is to solve \eq{18} from \cite{Dzubiella:2006} for $\delta$ and enter a free energy from experiments. For an uncharged particle of radius $R$, \eq{18} yields
\begin{align}
 \Delta G(T) &= 4\pi R^2\gamma(T)\l[1-\frac{2\delta}{R}\r] + 16\pi\varepsilon\rho(T)\l[\frac{\sigma^{12}}{9R^9}-\frac{\sigma^6}{3R^3}\r], \\
 \delta(T) &= -\frac{\Delta G(T)}{8\pi R\gamma(T)} + \frac{2\varepsilon\rho(T)}{R\gamma(T)}\l[\frac{\sigma^{12}}{9R^9}-\frac{\sigma^6}{3R^3}\r] + \frac R2
\label{eq:delta_free_energy}
\end{align}
where $\sigma$ and $\varepsilon$ are LJ-parameters.

Experimental data has been taken from Henry's constant as displayed by \eq{5} in \cite{Fernandez:1989}, whereas a correction has to be applied \cite{Paschek:2004}. The chemical potential $\Delta\mu$ (which is quantitatively equivalent to the hydration free energy here) is given by
\begin{align}
  \label{eq:mu}
  \Delta\mu = R T \l[\ln k_H^\infty-\ln P\r],
\end{align}
where $R$ is the universal gas constant, $P$ is the gas pressure and $k_H^\infty$ is Henry's constant. As shown in \cite{Fernandez:1989}, the temperature dependence of Henry's constant can be modeled by
\begin{align}
  \label{eq:k_approx}
  \ln k_H^\infty \approx f(T) &= K_0 + K_1\,\frac{1000}{T} + K_2\,\l(\frac{1000}{T}\r)^2,
\end{align}
with constants given in \tab{\ref{tab:henrys_constants}} and temperature in K.
This function results in dimension $\ln\text{GPa}$, thus has to be corrected with the gas pressure as indicated in \eq{\ref{eq:mu}}.
Following \cite{Paschek:2004}, in equilibrium and assuming an ideal gas, the gas pressure can be estimated to be 
\begin{align}
  P = \rho(T) k_B T,
\end{align}
where $\rho$ is the number density of water given in $\unit{m^-3}$. The pressure has dimension Pa -- in order to scale Henry's constant, one needs to change the pressure's dimension to GPa, yielding the formula
\begin{align}
  \label{eq:mu_final}
  \Delta\mu \approx R T \Big[f(T)-\ln\l(\rho(T) k_B T\times10^{-9}\r)\Big]
\end{align}
as a representation for the experimental values.

These results are compared with the integrated fitted entropy, whereas the constant $\Delta G(T_0)$ for the latter is fixed by \eq{\ref{eq:mu_final}}. The results for $\Delta G$ and $\delta(T)$ can be seen in \fig{\ref{fig:free_energy_exp}}. 
\begin{table}[t!]
\begin{center}
\begin{tabular}{lccc}
   \hline\hline 
     &$K_0$ & $K_1$ & $K_2$ \\
    \hline
     Ne & -5.9825 &5.5176& -0.8886\\
     Ar & -7.8972 &7.0178& -1.2649\\
     Kr & -7.5642 &6.8773& -1.3047\\
     Xe & -9.3604 &7.9654& -1.5167\\
    \hline
\end{tabular}
\caption{Parameters for the approximation of Henry's constant, \eq{\ref{eq:k_approx}}}
\label{tab:henrys_constants}
\end{center}
\end{table}
\begin{figure}[t!]
 \centering
    \includegraphics[width=12cm]{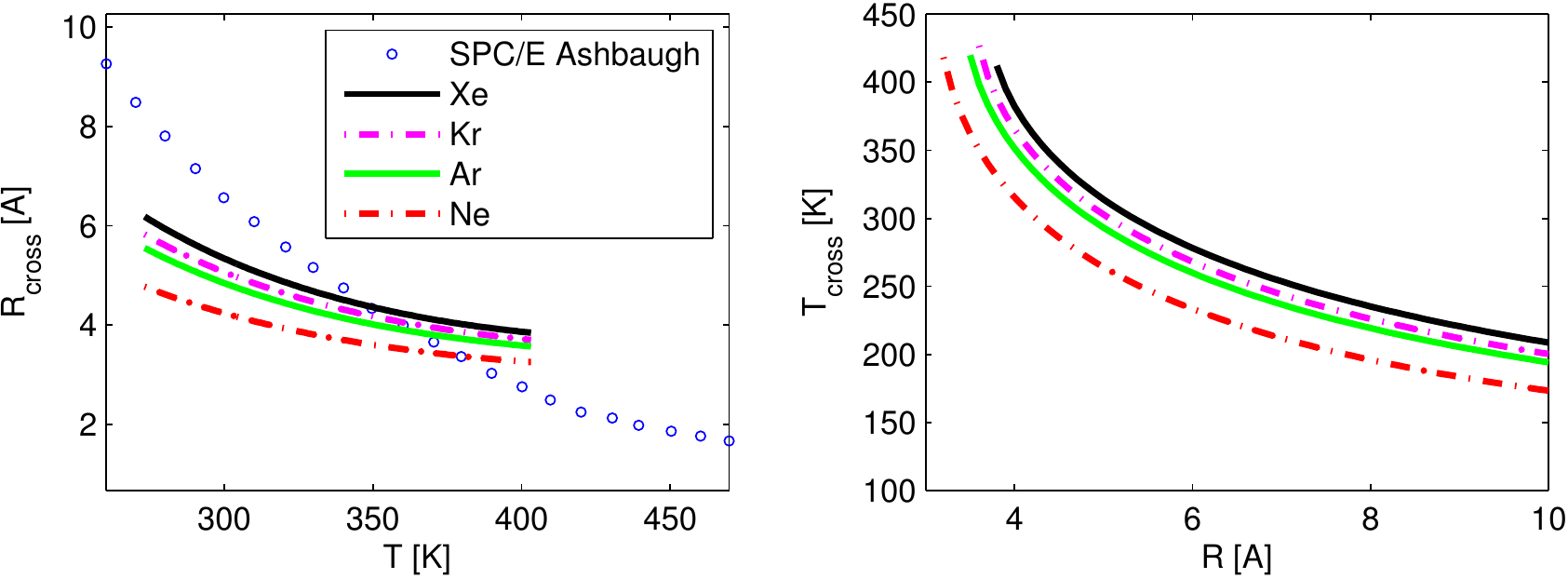}
 \caption{Comparison of quantities determining the switch of sign of the hydration entropy per surface area. \textbf{Left:} Crossover radius dependent on the temperature. \textbf{Right:} Crossover temperature dependent on the cavity radius}
 \label{fig:R_T_cross}
\end{figure}

\section{Entropy Sign Change}
As was shown for SPC/E-water in \cite{Ashbaugh:2009}, the hydration entropy per surface area of cavities is showing a sign change with increasing cavity radius. In first order curvature correction and ignoring electrostatic interactions, the hydration entropy per surface area reads
\begin{align}
 \frac{\partial \Delta S}{\partial A} = \gamma'-2(\delta\gamma)'/R.
\end{align}
To find the radius of the sign change, one has to equate this with 0 and solve for $R$, yielding
\begin{align}
  R_{\text{cross}} = 2(\delta\gamma)'/\gamma'.
\end{align}
Both numerator and denominator are functions established above. The results are shown and compared to those of \cite{Ashbaugh:2009} in \fig{\ref{fig:R_T_cross}}. Inverting the function yields a crossover temperature $T_{\text{cross}}$ dependending on the cavity radius $R$, which can be seen in \fig{\ref{fig:R_T_cross}}, as well.

%% file: tables/hyd_entropy_fit_table_molar.tex
Ne & $-1.334396\times10^{-3}$ & $1.267162$ & $-3.015192\times10^{2}$\\
Ar & $-1.945660\times10^{-3}$ & $1.809257$ & $-4.277216\times10^{2}$\\
Kr & $-2.177453\times10^{-3}$ & $2.025416$ & $-4.778052\times10^{2}$\\
Xe & $-2.605129\times10^{-3}$ & $2.401321$ & $-5.601495\times10^{2}$\\

%% file: tables/hyd_entropy_fit_table_nonmolar.tex
Ne & $-2.215817\times10^{-27}$ & $2.104172\times10^{-24}$ & $-5.006844\times10^{-22}$\\
Ar & $-3.230845\times10^{-27}$ & $3.004342\times10^{-24}$ & $-7.102484\times10^{-22}$\\
Kr & $-3.615746\times10^{-27}$ & $3.363283\times10^{-24}$ & $-7.934142\times10^{-22}$\\
Xe & $-4.325918\times10^{-27}$ & $3.987488\times10^{-24}$ & $-9.301501\times10^{-22}$\\

%% file: acknowledgements.tex
\newpage
\thispagestyle{empty}
\quad
\selectlanguage{german}
% \begin{Large}
\section*{Danksagung}
% \end{Large}

\vspace{1.0cm}

Bei der Entstehung dieser Arbeit wurde ich von einer Vielzahl von Menschen unterst\"utzt. Zuerst genannt seien meine Betreuer Prof. Dr. Dzubiella und Dr. Angioletti-Uberti, die mich \"uber ein Jahr lang intensiv betreuten und sich nicht f\"ur die Beantwortung jeder noch so kleinen Frage zu schade waren. Prof. Dr. Dzubiella danke ich insbesondere f\"ur die freundliche Aufnahme in seine Forschungsgruppe und die Bereitstellung der Materialien die n\"otig f\"ur meine Forschung waren. Auch die restlichen Mitglieder der Forschungsgruppe, insbesondere Dr. Heyda, waren stets f\"ur die Beantwortung von Fragen und f\"ur fachliche Diskussionen zu haben, wof\"ur ich ebenfalls \"uberaus dankbar bin.

Neben der fachlichen Unterst\"utzung ist die emotionale Unterst\"utzung w\"ahrend des Verfassens solch einer Arbeit nicht zu vernachl\"assigen. Ohne die r\"uckhaltlose Unterst\"utzung meiner Eltern Berndt und Carola sowie meines Stiefvaters Ralf w\"are es mir nicht m\"oglich gewesen, mein Studium so erfolgreich abzuschlie\ss{}en. Auch meinem Bruder Daniel sei gedankt f\"ur Momente der Zerstreuung. Doch wurde mir nicht nur von meiner leiblichen, sondern auch von meiner gew\"ahlten Familie der R\"ucken gedeckt. Stellvertretend f\"ur alle anderen seien an dieser Stelle Marc, Flo, Chris und Jakob gedankt, die sich immer ohne Gemurre meine Gemurre zu Gem\"ute f\"uhrten und mir in den letzten Monaten stets mit guten Worten und Tatkraft zur Seite standen. Nicht zuletzt danke ich meiner Freundin Teddy, die besonders in den letzten Woche vor der Abgabe meine Hochs und Tiefs geduldig mitmachte und mich mit ihrer Ruhe zur Ruhe brachte, wenn ich dachte, Ruhe w\"are unangebracht.

Ich danke ebenfalls der Studienstiftung des Deutschen Volkes, die mich sowohl finanziell als auch ideell unterst\"utzt und gef\"ordert hat.
\selectlanguage{english}